\documentclass{aa}
\usepackage{txfonts}
\usepackage{natbib}
\bibpunct{(}{)}{;}{a}{}{,}
\usepackage{graphicx}
\usepackage{placeins}
\usepackage[]{rotating}
\usepackage{ulem}

\newcommand{\teff}{$T_{\rm{eff}}$}
\newcommand{\logg}{$\log g$}
\newcommand{\lL}{\ifmmode \log \frac{L}{L_{\sun}} \else $\log \frac{L}{L_{\sun}}$\fi}

\newcommand{\vsini}{$V$~sin$i$}

\newcommand{\vturb}{$\varv_{\rm turb}$}
\newcommand{\vmac}{$\varv_{\rm mac}$}

\newcommand{\kms}{km~s$^{-1}$}
\newcommand{\msun}{M$_{\sun}$}

\begin{document}

\title{X-Shooting ULLYSES: Massive stars at low metallicity }
\subtitle{V. Effect of metallicity on surface abundances of O stars}

\author{F. Martins\inst{1}
\and J.-C. Bouret\inst{2}
\and D.J. Hillier\inst{3}
\and S.A. Brands\inst4
\and P.A. Crowther\inst{5}
\and A. Herrero\inst{6,7}
\and F. Najarro\inst{8}
\and D. Pauli\inst{9}
\and J. Puls\inst{10}
\and V. Ramachandran \inst{11}
\and A.A.C. Sander\inst{11}
\and J.S. Vink\inst{12}
\and the XshootU collaboration
}

\institute{
  LUPM, Universit\'e de Montpellier, CNRS, Place Eug\`ene Bataillon, F-34095 Montpellier, France 
  \and
  Aix-Marseille Univ, CNRS, CNES, LAM, Marseille, France 
  \and
  Department of Physics and Astronomy \& Pittsburgh Particle Physics, Astrophysics and Cosmology Center (PITT PACC), University of Pittsburgh, 3941 O’Hara Street, Pittsburgh, PA 15260, USA 
  \and
  Astronomical Institute Anton Pannekoek, University of Amsterdam, Science Park 904, 1098 XH, Amsterdam, The Netherlands
  \and
  Dept of Physics \& Astronomy, University of Sheffield, Hounsfield Road, Sheffield S3 7RH, UK 
  \and
  Instituto de Astrofísica de Canarias, C. Vía Láctea, s/n, 38205 La Laguna, Tenerife, Spain 
  \and
  Departamento de Astrofísica, Universidad de La Laguna, Avenida Astrofísico Francisco Sánchez, s/n, 38205 La Laguna, Tenerife, Spain
  \and
  Departamento de Astrofísica, Centro de Astrobiología (CSIC-INTA), Ctra. Torrejón a Ajalvir km 4, 28850, Torrejón de Ardoz, Spain
  \and
  Institut f{\"u}r Physik und Astronomie, Universit{\"a}t Potsdam, Karl-Liebknecht-Str. 24/25, 14476 Potsdam, Germany
  \and
  LMU M\"{u}nchen, Universit\"{a}ts-Sternwarte, Scheinerstr. 1, 81679 M\"{u}nchen, Germany 
  \and
  Zentrum f{\"u}r Astronomie der Universit{\"a}t Heidelberg, Astronomisches Rechen-Institut, M{\"o}nchhofstr. 12-14, 69120 Heidelberg, Germany
  \and
Armagh Observatory and Planetarium, College Hill, BT61 9DG Armagh, UK
}

\offprints{Fabrice Martins\\ \email{fabrice.martins@umontpellier.fr}}

\date{Received / Accepted }

\abstract
{Massive stars rotate faster, on average, than lower mass stars. Stellar rotation triggers hydrodynamical instabilities which transport angular momentum and chemical species from the core to the surface. Models of high-mass stars that include these processes predict that chemical mixing is stronger at lower metallicity.}
{We aim to test this prediction by comparing the surface abundances of massive stars at different metallicities.}
{We performed a spectroscopic analysis of single O stars in the Magellanic Clouds (MCs) based on the ULLYSES and XshootU surveys. We determined the fundamental parameters and helium, carbon, nitrogen, and oxygen surface abundances of 17 LMC and 17 SMC non-supergiant O6-9.5 stars. We complemented these determinations by literature results for additional MCs and also Galactic stars to increase the sample size and metallicity coverage. We investigated the differences in the surface chemical enrichment at different metallicities and compared them with predictions of three sets of evolutionary models.}
{Surface abundances are consistent with CNO-cycle nucleosynthesis. The maximum surface nitrogen enrichment is stronger in MC stars than in Galactic stars. Nitrogen enrichment is also observed in stars with higher surface gravities in the SMC than in the Galaxy. This trend is predicted by models that incorporate chemical transport caused by stellar rotation. The distributions of projected rotational velocities in our samples are likely biased towards slow rotators. }
{A metallicity dependence of surface abundances is demonstrated. The analysis of larger samples with an unbiased distribution of projected rotational velocities is required to better constrain  the treatment of chemical mixing and angular momentum transport in massive single and binary stars.}


\authorrunning{Martins et al.}
\titlerunning{X-Shooting ULLYSES -- VII. Surface abundances of O stars}

\maketitle

\section{Introduction}
\label{s_intro}

Stars more massive than 8~\msun\ evolve beyond the core carbon-burning phase and produce most metals heavier than oxygen and up to the iron group \citep{cm86,langer12}. They are hot and consequently produce photons that can ionise hydrogen and heavier elements. They drive strong stellar winds \citep{puls08,vink22}. Massive stars are thus important sources of feedback in terms of chemistry, energetics, and ionising flux. Their explosions as core-collapse supernovae leave compact objects that may merge to produce gravitational wave events \citep{abbott21}. Understanding their properties and evolution is relevant for many astrophysical fields \citep{massey13,eldridge22}. 

Spectroscopic analysis of massive stars provides their present-day fundamental parameters. It requires sophisticated atmosphere models that include a non-LTE treatment of radiative transfer, spherical extension caused by stellar winds, and line-blanketing to account for the effects of metals on the atmospheric structure and emergent spectrum \citep{hm98,puls05,sander15}. 

Variations in surface abundances while stars evolve probe the efficiency of mixing processes in stellar interior. The products of nucleosynthesis in the core are transported to the surface by these mechanisms. During the main-sequence it is mainly helium, carbon, nitrogen, and oxygen that are affected since nucleosynthesis takes place through the CNO-cycle. Among other processes stellar rotation triggers instabilities that transport of both angular momentum and chemical species. The implementation of these mixing processes in stellar evolution codes has led to the prediction of three main trends with physical parameters \citep[e.g.][]{mm00,brott11}: first, stars rotating faster transport more efficiently chemical species from their core to their surface; second, mixing is more efficient in more massive stars; third, stars with a lower metal content are more chemically processed at their surface. 

\citet{hunter08} and \citet{hunter09} investigate the relation between surface nitrogen enrichment and rotational velocity in B-type stars of the SMC, LMC, and Galaxy. They report a general trend of higher nitrogen content for higher projected rotational velocity, but also stress that a number of objects are chemically enriched while rotating slowly. Simulations of stellar populations based on the evolutionary tracks of \citet{brott11} indicate a difficulty for these tracks to account for these slowly rotating N-rich B stars. Qualitatively similar results are found for additional objects: the LMC O stars studied by \citet{rivero12} and \citet{grin17}, the LMC B dwarfs and giants analysed by \citet{dufton18}, and the SMC B stars studied by \citet{dufton20} all show a fraction of chemically enriched slow rotators that the models of \citet{brott11} can not explain. Alternatively \citet{martins15} and \citet{martins17} find that the tracks computed by \citet{ek12} can reproduce most of the Galactic stars they studied, although some outliers exist. These analyses are not based on population synthesis though, so a quantitative assessment of the ability of these models to explain the chemical and rotational properties of Galactic O stars remains to be performed. Using a similar approach, \citet{bouret13} and \citet{bouret21} reach the same qualitative conclusion for SMC O stars: models from \citet{georgy13} can account for the abundances and rotational velocities of most of the O stars. There are two main things that would advance these analyses. First, a systematic investigation of the differences in the predictions of various evolutionary models needs to be performed. Second, proper population synthesis simulations for all types of evolutionary models (i.e. models that include rotational mixing with different prescriptions, as well as binary models) are required before the relation between surface chemical enrichment and rotational velocity can be quantitatively probed. 

Regarding the effect of stellar mass on chemical enrichment, \citet{martins17} report an average increase by $\sim$0.7 dex in surface nitrogen enrichment between Galactic B and O stars, based on their spectroscopic analysis of late O giants and the results of \citet{hunter09} and \citet{nieva12}. In the SMC \citet{bouret21} confirm that more massive O stars have a higher N/C ratio than less massive ones. 

\citet{bouret21} also performed a principal component analysis based on the abundances of C, N, and O in stars of the SMC and the Milky Way. They report a trend of more efficient mixing at lower metallicity, as predicted. However the significance of the results is limited mostly by the limited size of the SMC sample which is not commensurate to the reference Galactic sample.

The physical understanding of this metallicity effect is the following. Early simulations of the evolution of low metallicity stars showed that the zero-age main-sequence (ZAMS) is displaced to the blue part of the Hertzsprung-Russell (HR) diagram compared to solar metallicity stars \citep{eleid83,bertelli84,schaller92}. The reduction of the metal content, especially the CNO abundances, leads to a reduced energy release by the CNO cycle and thus to a reduced pressure support. As a consequence, stars contract until the associated temperature increase boosts the efficiency of the CNO cycle providing the extra energy release necessary to recover hydrostatic equilibrium. Therefore stars are more compact at low metallicity \citep[e.g. Fig.~1 of][]{ek08}. 

Chemical transport is affected by the change in the structure of the star at low metallicity. It is dominated by horizontal shear turbulence and treated as a diffusive process, parameterised by a coefficient D$_{shear}$, in evolutionary calculations. D$_{shear}$ is proportional to the gradient of the angular velocity \citep[see Eq.~16 of][]{mm00}. At lower metallicity this gradient is stronger, since the star is more compact, and consequently chemical transport is more efficient \citep{mm01,mm02,brott11,georgy13}. In particular chemical enrichment or depletion, depending on the species, is stronger at the surface of low metallicity stars. 

In this study we further investigate the metallicity dependence of chemical transport caused by stellar rotation. We leverage on the availability of homogeneous data -- in terms of spectral coverage and instruments used -- from the ULLYSES\footnote{\url{https://ullyses.stsci.edu/}} (Roman-Duval et al., in prep.) and XshootU\footnote{X-shooting ULLYSES  - \url{https://massivestars.org/xshootu/}} \citep{xshooI} surveys that provide UV and optical spectroscopy of about 200 massive stars in the Magellanic Clouds (MCs). We determine the surface abundance of He, C, N, and O for a sample of O stars in the MCs. We complement these abundances by those of published studies in the MCs and the Galaxy. We very much emphasise
that our goal is not to interpret the surface properties of large samples of stars that could probe all types of mixing and transport processes taking place in all possible sorts of massive stars (single and binary). We merely want to test if stars at different metallicities have different surface chemical composition, as predicted by models that include mixing caused by stellar rotation.

\section{Samples and method}
\label{s_2}

\subsection{Samples}
\label{s_sample}

The primary goal of this study is to investigate the metallicity dependence of chemical mixing. For that we relied on samples of stars in the SMC, the LMC, and the Galaxy. The SMC and LMC samples were partly built from the XshootU and ULLYSES surveys. Complementary stars were selected to increase the sample sizes (see below). To avoid the effect of mass on surface abundances to be present in our analysis, we selected stars in a relatively narrow mass range. 

\subsubsection{XshootU sample}
\label{s_sampleX}

We selected O6 to O9.5 objects from the XshootU sample, i.e. stars with an X-shooter spectrum. We excluded stars for which the presence of a companion was clear from the spectra (presence of double sets of lines). The sample is made of 17 stars in both the LMC and the SMC. We avoided luminous supergiants because no large enough Galactic sample of such objects was available for comparisons of surface abundances at the time of the present study. Only a few class I stars were included because their stellar parameters are similar to giants and bright giants (luminosity classes III and II) of the sample (see Sect.~\ref{s_ev}). In the following we will refer to these MC samples as the XshootU samples. The members of the samples are listed in Tables~\ref{tab_lmc} and \ref{tab_smc}. The observational data for this sample is described at length in \citet{xshooI}. The optical spectra were retrieved from the XshootU database presented by \citet{xshooII}. They have a spectral resolution of about 7000 and a signal-to-noise ratio (SNR) of about 200. The UV spectra were downloaded from the  MAST archive\footnote{\url{https://archive.stsci.edu/hlsp/ullyses}}. Their spectral resolution is $\sim$15000 and the SNR of the order 100. Spectra were manually normalised by choosing spectral regions free of lines and applying a spline cubic function to define the continuum. 

We stress that the spectral types and luminosity classes listed in Tables~\ref{tab_lmc} and \ref{tab_smc} have been collected from literature. Some are based on poor quality data and are likely uncertain. A complete revision of spectral classification of the entire ULLYSES samples will be provided in a subsequent publication of the series (Ma\'{i}z Apell\'{a}niz et al., in prep.). Spectral types can be used to infer coarse stellar parameters, but stars with similar parameters can have different spectral types and luminosity classes \citep{sergio14b,mp17}. In addition spectral classification depends on metallicity since some of the classification criteria depend on wind sensitive lines (and thus on the metal content) and on metallic lines \citep{sota11,martins18,mp21}. For the present study the accuracy of spectral classification is not an issue since we aim at obtaining homogeneous samples in terms of evolutionary state. We demonstrate in the following that this goal is achieved. 

\subsubsection{Literature samples}
\label{s_samplecomp}

To increase the number of stars for which surface abundances are determined, we complemented our results with those of \citet{bouret13}, \citet{bouret21}, and \citet{grin17}. The former two studies include SMC stars while the latter study focuses on LMC targets. All stars are presumably single. We included only stars that cover the same region of the \logg\ - \teff\ diagram as the XshootU samples. By doing so we ensured that we kept a relatively narrow mass range for the entire sample, thus minimising the effect of mass on surface abundances as explained above (see also Sect.~\ref{s_ev}). 

In addition to literature results in the MCs we also considered stars in the Galaxy to extend the metallicity range over which abundances can be compared. We selected stars from the studies of \citet{martins15}, \citet{martins17}, \citet{cazorla17a}, and \citet{markova18}. As for the complementary stars in the MCs, the Galactic stars were chosen to cover the same mass range as the XshootU samples. The position of the sample stars in evolutionary diagrams will be discussed in Sect.~\ref{s_ev}.
These various samples defined from the literature will be referred as the complementary samples in the remainder of the manuscript.
The members of these additional samples are given in Table~\ref{tab_compsamp}. We stress that we did not re-determine the fundamental stellar parameters and surface abundances for these stars, but took the values published in the studies listed in Table~\ref{tab_compsamp}. 
The methods adopted by \citet{bouret13}, \citet{martins15}, \citet{martins17}, and \citet{bouret21} are very similar to the one used in the present study. In particular they rely on the code CMFGEN. \citet{grin17} and \citet{markova18} use atmosphere models and synthetic spectra computed with the code FASTWIND \citep{puls05,rivero11} and rely on optical spectra only.

\subsection{Stellar parameters}
\label{s_stelpar}

For the present study we determined the surface parameters of the stars in the XshootU samples. 
We performed a spectroscopic analysis with the atmosphere code CMFGEN \citep{hm98}. Non-LTE radiative transfer is performed in a spherical geometry. Stellar winds are included by connecting a pseudo-hydrostatic velocity structure to a $\beta$ velocity law\footnote{$v = v_{\infty} \times\ (1-R/r)^{\beta}$, $v$ being the velocity, $R$ the stellar radius, and $r$ the radial coordinate}. Line-blanketing is taken into account. Once the atmosphere model has converged, a formal solution of the radiative transfer is performed and leads to the emergent spectrum that can be compared to observational data to infer stellar parameters. We proceeded as follows in the present analysis.

We relied on two grids of models designed to cover the main sequence and early-post main-sequence in the LMC and SMC, respectively. The grids will be presented in Marcolino et al. (in prep.) and we refer the reader to this publication for details. Models in the grids all assume a microturbulent velocity (\vturb) of 10~\kms\ in the photosphere and a fixed chemical composition\footnote{This implies that He/H is kept constant in the determination of the effective temperature.}, that is scaled solar abundances from \citet{a09} assuming a metallicity of 1/2 and 1/5~Z$_{\odot}$ for the LMC and SMC respectively. The steps in \teff\ and \logg\ are 1000~K and 0.1~dex. Mass loss rates are from \citet{vink01} and no clumping is taken into account.

The first step of the analysis was the determination of the projected rotational velocity (\vsini) and of the macroturbulent velocity (\vmac). For this we used the Fourier transform method \citep{gray,sergio07} to isolate \vsini. We relied on \ion{He}{i}~4713\footnote{This line is broadened by the Stark effect, but in most cases it is the only clean line available for the determination of \vsini. Alternative lines usually considered for the determination of \vsini\ are \ion{Si}{iii} lines that are at best weak in the spectrum of the stars we consider here}. and when present on \ion{Si}{iii}~4553. Once obtained, we estimated the effective temperature and surface gravity of the star from its spectral type using the calibration of \citet{martins05}. We then selected the model from the grid with the closest parameters to these values. 
We convolved it by a rotational profile parameterised by \vsini. Convolution by a Gaussian profile was also performed to account for instrumental resolution. 
Finally we also included a convolution by a radial-tangential profile to take macroturbulence into account \citep{sergio14}. We then compared the resulting line profile of \ion{He}{i}~4713 or \ion{Si}{iii}~4553 to the observed profile and select \vmac\ that best reproduced the line shape, especially the far wings. The typical uncertainties on the measurements of \vsini\ and \vmac\ are 10 and 20~\kms\ respectively.

Once obtained the values of \vsini\ and \vmac\ were adopted to convolve the entire grids of synthetic spectra. This yielded theoretical spectra that can be quantitatively compared to the observed ones to estimate the stellar parameters. For that purpose we selected various sets of lines that are sensitive to \teff\ and \logg. For the chosen set of lines a $\chi^2$ analysis was performed allowing us to determine the best fit value and the uncertainties. The lines we used are the following:

\begin{itemize}
    \item Effective temperature: we relied entirely on helium lines. For the spectral types we considered, both \ion{He}{i} and \ion{He}{ii} features are present in the spectra so the ionisation balance method can be robustly used. The full set of lines that can be potentially used is: \ion{He}{i}-\ion{He}{ii}~4026, \ion{He}{i}~4143, \ion{He}{ii}~4200, \ion{He}{i}~4388, \ion{He}{i}~4471, \ion{He}{ii}~4542, \ion{He}{i}~4713, \ion{He}{i}~4920, \ion{He}{ii}~5412, \ion{He}{i}~5876, \ion{He}{i}~6678, \ion{He}{i}~7065. Given the predominance of \ion{He}{i} lines in this list, we gave more weight to the \ion{He}{ii} in order to avoid  that the results are systematically biased towards low \teff. The weight can be adjusted and we found that the results depend little on its value. A value of 2 was adopted for the present analysis.

    \item Surface gravity: the wings of Balmer lines were the main indicators as is customary for OB-type stars. We relied on H8, H$\epsilon$, H$\delta$, H$\gamma$, H$\beta$.
    
\end{itemize}

\begin{figure}[]
\centering
\includegraphics[width=0.49\textwidth]{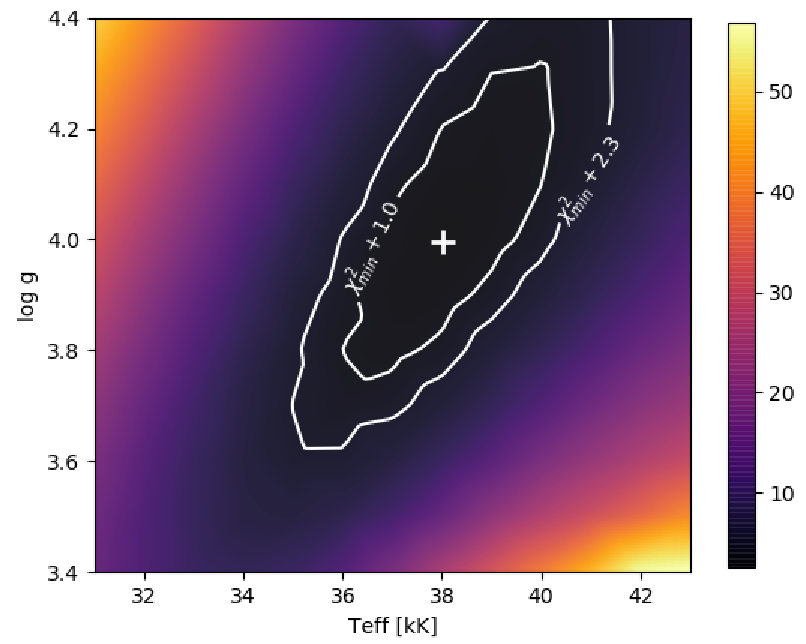}
\caption{Example of determination of \teff\ and \logg\ for star AV207 in the SMC. The colour scale indicates the value of the $\chi^2$ function. The 1 and 2.3~$\sigma$ contours are shown by the white lines. The white cross marks the minimum of the $\chi^2$ function.} 
\label{Tg_av207}
\end{figure}

The set of helium and hydrogen lines we used depended on the target. Data quality was sometimes insufficient to clearly isolate the line profile of a particular line. Nebular contamination may affect some lines. In addition depending on the position in the parameter space, some lines were too weak to be used. Fig.~\ref{Tg_av207} shows an example of the determination of \teff\ and \logg. We see that the two parameters are highly correlated and so are their error bars. This is expected since our method relies on the measurement of the ionisation balance at the atmospheric depth where lines are formed, and ionisation depends on both temperature and density, hence gravity. For these reasons we do not quote formal errors on \teff\ and \logg\ but refer to Fig.~\ref{Tg_av207} for representative values. In the following we adopt the $\chi^2_{\rm min} + 1.0$ contour as representative of our error measurements.

The luminosity was subsequently adjusted for the fixed set of \teff\ and \logg. We computed the bolometric corrections associated to the effective temperature \citep{mp06}, the extinction from the colour excess E(B-V) and determined the luminosity from the assumed distance and the photometry of each star, taken from Table~B.1 of \citet{xshooI}. 
We did not determine wind properties (mass loss rates, wind terminal velocity, and clumping) that will be investigated in other publications of the XshootU series. We simply assumed the classical mass loss rates of \citet{vink01}. At the metallicity of the LMC and SMC the winds remain weak enough not to affect most photospheric lines used in the present study.

The results of the present analysis are summarised in the first columns of Tables~\ref{tab_lmc}
 and \ref{tab_smc}. In those Tables we also provide the surface gravity corrected for the effect of centrifugal acceleration, $g_c = g + \frac{V~sini^2}{R}$. When \vsini\ is large the difference between \logg\ and $\log g_c$ can reach almost 0.2 dex as in the case of AzV~251. The best fits are given in appendix \ref{ap_1}.

\subsection{Surface abundances}
\label{s_ab}

Once the fundamental stellar parameters were determined we moved on to the investigation of the surface abundances. We considered helium, carbon, nitrogen, and oxygen. Abundances were determined from the fit of selected lines. Line strength depends on many parameters which control the level populations of the upper and lower levels of the transitions, as well as the intensity of the radiation field at the transition's wavelength. Ionisation and excitation processes, whether collisional or radiative, are partly fixed when the temperature and gravity are determined. Radiation field is affected by opacities that depend on elemental abundances. All of this sets the level populations. The line intensity further depends on broadening mechanisms, in particular the turbulent ones. Macroturbulence was constrained in the early phase of the analysis (see above) but microturbulence was not. We constrained it in parallel to surface abundances, in the same spirit as that of the curve of growth method \citep[e.g.][]{sergio20}.  

\begin{figure}[]
\centering
\includegraphics[width=0.49\textwidth]{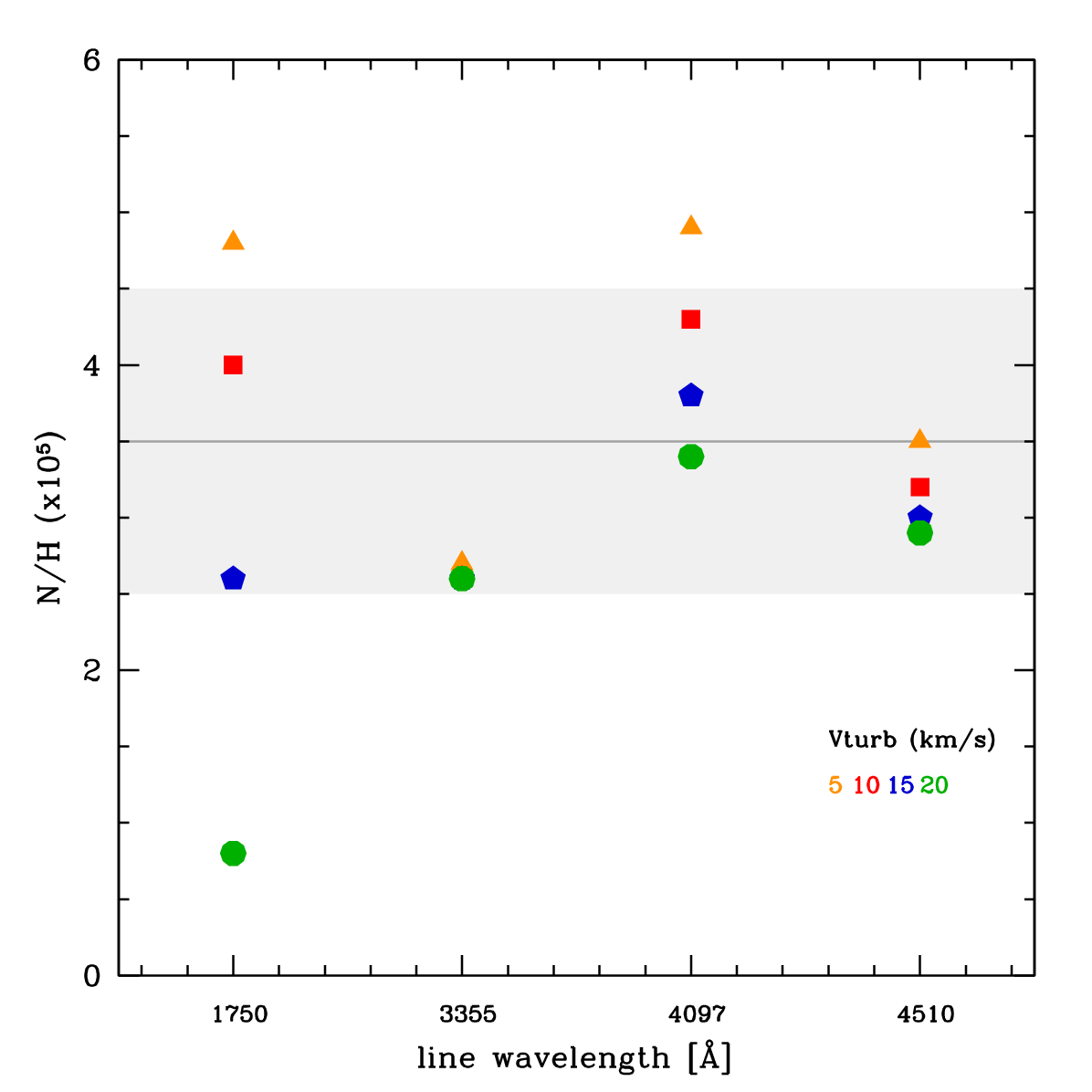}
\caption{Example of determination of the nitrogen abundance and \vturb\ for star AV47. The horizontal axis shows the wavelengths of the different nitrogen lines used, labelled by their wavelength. The vertical axis gives the best fitting abundance (N/H by number) for each line. Different symbols and colours correspond to different microturbulent velocities. All points for line \ion{N}{iii}~3355 overlap. The grey line and area highlight the final abundance value and its uncertainty. See text for discussion.} 
\label{cog_N_av47}
\end{figure}

We proceeded as follows. Adopting the stellar parameters previously determined, we ran new CMFGEN models varying the abundances and \vturb. For the latter, we used values of 5, 10, 15, and 20 \kms. We computed the synthetic spectrum for each of these values, and for various abundances. Typically five sets of abundances with a ratio of 50 to 100 between the smallest and largest values were selected. Both new atmosphere models and synthetic spectra were computed with additional sets of abundances, while microturbulence was changed only in the formal solution of the radiative transfer leading to the synthetic spectrum. For each of these new models we quantified the goodness of fit of individual lines of a given element by means of a $\chi^2$ analysis. For each line and for a given microturbulent velocity we determined the abundance leading to the best fit. The resulting abundance measurements had a dispersion (among various lines) that depended on the adopted microturbulence. We then adopted the final \vturb\ that lead to the smallest dispersion and for that value, we fixed the final abundance by taking the average of the individual line determinations. The uncertainty was set to the dispersion of the measurements. When two values of \vturb\ gave similar results the final choice was made by visually inspecting the fits and selecting by eye the most relevant value. An example of this process is given in Fig.~\ref{cog_N_av47}. In that case both \vturb\ = 10 and 15 \kms\ (the sets of red and blue points) lead to similar dispersion. The first value was adopted from the final fit (see bottom panel of Fig.~\ref{fit_av47}). We stress that uncertainties in the fundamental parameters are not propagated into error on the abundance measurement, because of our method. This should not significantly affect our estimate of uncertainties on surface abundances, as was discussed by \citet{martins12b} and \citet{martins15}.

The set of lines used for the abundance determination depended on the star, the data quality and the robustness of the normalisation. Here is the list of lines that were considered:

\begin{itemize}
    \item helium:  \ion{He}{i}~4026, \ion{He}{ii}~4200, \ion{He}{i}~4388, \ion{He}{i}~4471, \ion{He}{ii}~4542, \ion{He}{i}~4713, \ion{He}{i}~4920, \ion{He}{ii}~ 5412, \ion{He}{i}~5876, \ion{He}{i}~6678, \ion{He}{i}~7065

    \item carbon: \ion{C}{iv}~1169, \ion{C}{iii}~1176, \ion{C}{iii}~4068-70

    \item nitrogen: \ion{N}{iii}~1183-85, \ion{N}{iii}~1748-52, \ion{N}{iv}~3478-83, \ion{N}{iii}~3355, \ion{N}{iii}~4097, \ion{N}{iii}~4510-20

    \item oxygen: \ion{O}{iv}~1339, \ion{O}{iv}~1343, \ion{O}{iii}~3261, \ion{O}{iii}~3265, \ion{O}{iii}~3760, \ion{O}{iii}~3963, \ion{O}{iii}~5593
    
\end{itemize}

\noindent We excluded \ion{N}{iii}~4634-40-42, \ion{C}{iii}~4647-50-51, and \ion{C}{iii}~5696 from the analysis because these lines have been shown to depend critically on atomic data, line-blanketing, and stellar winds \citep{rivero11,martins12}. We also excluded
strong resonance lines (\ion{N}{v}~1240 and \ion{C}{iv}~1550) that depend on X-ray emission and mainly on wind parameters \citep{pauldrach94}. In addition they are often saturated (at least for LMC stars) and thus not sensitive to abundance variations.

 The results of the abundance analysis are gathered in Table~\ref{tab_lmc} and \ref{tab_smc}. The bottom panels of the figures shown in appendix~\ref{ap_1} highlight the fits to lines from carbon, nitrogen, and oxygen.

\begin{sidewaystable*}[ht]
\begin{center}
\caption{Stellar parameters for LMC stars} \label{tab_lmc}
\begin{tabular}{lrrrrrrrrrccc}
\hline
Star	               &   ST          & \teff      & \logg\ & $\log g_c$ & \lL\ & \vsini\ & \vmac\ & \vturb\ & He/H & C/H & N/H & O/H \\
                       &               & [kK]       &        &      &  [\kms] &  [\kms] &  [\kms] &     & [$10^{-5}$] & [$10^{-5}$] & [$10^{-5}$] \\
\hline
Sk-66$^{\circ}$ 18 & O6\,V & 38.0 & 3.70 & 3.71 & 5.42 & 100 & 0 & 15 & 0.10$\pm$0.010 & 7.0$\pm$1.0 & 25.0$\pm$6.0 & 12.0$\pm$3.0  \\
Sk-71$^{\circ}$ 19 & O6\,III & 39.0 & 4.10 & 4.16 & 5.12 & 311 & 0 & 15 & 0.10$\pm$0.010 & 5.0$\pm$3.0 & 20.0$\pm$4.0 & --  \\
N11 018  & O6\,II(f+) & 37.0 & 3.60 & 3.60 & 5.67 & 50 & 70 & 15 & 0.085$\pm$0.010 & 5.5$\pm$1.0 & 20.0$\pm$5.0 & 11.2$\pm$3.0 \\
Sk-71$^{\circ}$ 50 & O6.5\,III & 35.0 & 3.50 & 3.57 & 5.21 & 210 & 0 & 20 & 0.10$\pm$0.010 & 7.0$\pm$1.0 & 25.0$\pm$6.0 & 13.0$\pm$6.0  \\
Sk-66$^{\circ}$ 152 & O7\,Ib(f) & 32.9 & 3.20 & 3.26 & 5.60 & 180 & 45 & 15 & 0.12$\pm$0.02 & 7.48$\pm$1.49 & 20.7$\pm$1.34 & 19.1$\pm$3.25 \\ 
N11 032 & O7\,II(f) & 34.9 &	3.50 & 3.51 & 5.34 & 60 & 70 & 10 & 0.085$\pm$0.035 & 10.6$\pm$1.1 & 69.7$\pm$0.7 & $<$24.3 \\
Sk-69$^{\circ}$ 50 & O7(n)(f)p & 35.0 & 3.40 & 3.46 & 5.45 & 197 & 80 & 15 & 0.14$\pm$0.02 & 5.9$\pm$2.6 & 44.5$\pm$10.0 & 18.5$\pm$9.84   \\
Sk-68$^{\circ}$ 16 & O7\,III & 36.9 & 3.70 & 3.71 & 5.74 & 83 & 40 &  10 & 0.12$\pm$0.03 & 9.59$\pm$1.48 & 12.5$\pm$2.1 & 15.5$\pm$4.0 \\
N11 049 & O7.5\,V & 37.0 & 4.00 & 4.02 & 5.11 & 187 & 21 & 10 & 0.085$\pm$0.015 & 10.8$\pm$0.8 & 3.69$\pm$1.0 & 24.3$\pm$0.6 \\
Sk-67$^{\circ}$ 101 & O8\,II((f)) & 34.0 & 3.60 & 3.61 & 5.56 & 85 & 78 & 10 & 0.10$\pm$0.010 & 8.35$\pm$1.25 & 4.37$\pm$2.1 & 7.35$\pm$4.81 \\
BI 173 & O8\,III & 33.9 & 3.50 & 3.52 & 5.53 & 146 & 92 & 15 & 0.085$\pm$0.015 & 12.2$\pm$2.67 & 6.38$\pm$0.67 & 9.00$\pm$0.17 \\
Sk-67$^{\circ}$ 261 & O8.5\,III & 27.0 & 3.00 & 3.01 & 5.48 & 60 & 50 & 10 & 0.10$\pm$0.02 & 5.57$\pm$1.4 & 7.01$\pm$ 0.7 & 22.9$\pm$3.8 \\  
Sk-67$^{\circ}$ 191 & O8\,V & 34.1 & 3.60 & 3.62 & 5.27 & 123 & 74 & 15 & 0.10$\pm$0.02 & 4.88$\pm$0.84 & 12.6$\pm$4.55 & 9.95$\pm$4.00 \\  
Sk-66$^{\circ}$ 171 & O9\,Ia & 30.0 & 3.00 & 3.02 & 5.71 & 98 & 20 & 15 & 0.15$\pm$0.010 & 13.4$\pm$1.32 & 5.18$\pm$2.18 & 29.0$\pm$7.5 \\
Sk-71$^{\circ}$ 8 & O9\,II & 28.0 & 3.30 & 3.31 & 5.20 & 69 & 31 & 15 & 0.13$\pm$0.020 & $<$3.87 & 6.98$\pm$1.90 & 13.7$\pm$4.40 \\
BI 128 & O9\,V & 32.3 & 3.50 & 3.51 & 5.08 & 72 & 0 & 15 & 0.11$\pm$0.010 & 12.8$\pm$3.02 & 2.92$\pm$0.61 & 26.1$\pm$6.54 \\
Sk-70$^{\circ}$ 13 & O9\,V & 29.0 & 3.20 & 3.21 & 5.64 & 75 & 70 & 10 & 0.085$\pm$0.015 & 4.80$\pm$1.74 & 1.96$\pm$1.67 & 13.5$\pm$4.12  \\
\hline 
\end{tabular}
\tablefoot{Columns are star's ID, spectral type from \citet{xshooI}, effective temperature, surface gravity, surface gravity corrected for centrifugal acceleration, luminosity, projected rotational velocity, macroturbulent velocity, microturbulent velocity, abundance of He, C, N, and O by number. Typical uncertainties on \vsini\ and \vmac\ are of the order 10-20~\kms. For \vturb\ 5~\kms\ is a reasonable error as explained in Sect.~\ref{s_ab}. Uncertainties on \teff\ and \logg\ are correlated and typical values are shown in Fig.~\ref{Tg_av207}. An error of about 0.15~dex on luminosity is a typical value.}
\end{center}
\end{sidewaystable*}

\begin{sidewaystable*}[ht]
\begin{center}
\caption{Stellar parameters for SMC stars} \label{tab_smc}
\begin{tabular}{lrrrrrrrrrrrr}
\hline
Star	               &   ST          & \teff      & \logg\ & $\log g_c$ & \lL\ & \vsini\ & \vmac\ & \vturb\ & He/H & C/H & N/H & O/H \\
                       &               & [kK]       &    &    &      &  [\kms] &  [\kms] &  [\kms] &  & [$10^{-5}$] & [$10^{-5}$] & [$10^{-5}$]     \\
\hline
AzV6	& O9\,III	        & 36.0  & 4.20  & 4.20 &	 5.84 & 60  & 50 & 10   & 0.085$\pm$0.010	& 6.7$\pm$2.0  & 1.0 $\pm$0.5  &  10.0$\pm$5.0 \\
AzV15	& O6.5\,II(f)     & 38.0  & 3.70   & 3.71 &   5.72 & 90  & 90 & 10   & 0.085$\pm$0.010 & 3.0$\pm$1.0  & 12.0 $\pm$3.0 &  13.0$\pm$1.0 \\
AzV47	& O8\,III((f))    & 38.0  & 4.30   & 4.30    &   5.60 & 65  & 50 & 10   & 0.085$\pm$0.010 & 5.0$\pm$2.5  & 3.5$\pm$1.0  &   2.0$\pm$1.0 \\
AzV69	& OC7.5\,III((f)) & 36.0  & 3.60   & 3.61     &   5.60 & 85  & 70 & 15   & 0.085$\pm$0.010 & 6.0$\pm$1.5 & 0.5$\pm$0.2  &   10.0$\pm$3.0 \\
AzV80	& O7\,III	        & 38.0  & 3.70   &  3.78    &   5.71 & 350 & 0  & 10   & 0.085$\pm$0.010 & $<$5.0           & $>$20.0 	        &   --                  \\
AzV95	& O7\,III((f))    & 37.0  & 3.70   &  3.71    &   5.43 & 55  & 50 & 15   & 0.092$\pm$0.010 & 1.4$\pm$0.5  & 6.5$\pm$2.0  &   8.0$\pm$3.0 \\
AzV148	& O8.5\,V	        & 31.0  & 3.60   &  3.60    &   5.16 & 35  & 0  & 10   & 0.090$\pm$0.010 & 3.0$\pm$0.5 & 2.6$\pm$0.7  &   10.0$\pm$5.0 \\
AzV186	& O8.5\,III((f))  & 36.0  & 3.80   &  3.80    &   5.63 & 55  & 10 & 10   & 0.130$\pm$0.02  & 2.0$\pm$1.0 & 10.0$\pm$5.0  &  10.0$\pm$5.0 \\
AzV207	& O7\,III((f))    & 38.0  & 4.00   &  4.00    &   5.21 & 75  & 70 & 15   & 0.085$\pm$0.010 & 5.0$\pm$3.0 & 10.0$\pm$4.0 &   6.0$\pm$2.0 \\
AzV243	& O6\,V	        & 41.0  & 4.00   &  4.00    &   5.50 & 60  & 40 & 10   & 0.085$\pm$0.010 & 10.0$\pm$3.0 & 7.0$\pm$3.0   &   7.0$\pm$2.0 \\
AzV251	& O7\,V	        & 36.0  & 3.90   &  4.09     &   5.01 & 500 & 0  & 15   & 0.095$\pm$0.010 & $<$10.0           & $>$10.0	  &   5.0$\pm$4.0 \\
AzV267	& O8\,V	        & 37.0  & 4.00   &  4.08    &   4.93 & 325 & 0  & 15   & 0.13$\pm$0.02	& $<$5.0           & 8.0 $\pm$2.0   &   $<$10.0 \\
AzV307	& O9\,III	        & 29.0  & 3.40   &  3.40    &   5.12 & 45  & 30 & 15   & 0.10$\pm$0.02   & 1.2$\pm$0.4& 6.0$\pm$2.0   &   10.0$\pm$3.0 \\
AzV327	& O9.5\,II-Ibw    & 31.0  & 3.40   &  3.40    &   5.55 & 60  & 40 & 10   & 0.12$\pm$0.02   & 1.4$\pm$0.2& 6.0$\pm$1.0   &   6.0$\pm$2.0 \\
AzV440	& O7.5\,III	    & 37.0  & 4.20   &  4.20    &   5.03 & 35  & 30 & 5    & 0.085$\pm$0.010 & 5.0$\pm$2.0 & 3.0$\pm$1.0   &   7.0$\pm$2.0 \\
AzV446	& O6.5\,V	        & 42.0  & 4.40   &  4.40    &   5.22 & 35  & 0  & 5    & 0.085$\pm$0.010 & 1.0 $\pm$0.9& 5.0$\pm$2.0   &   12.0$\pm$7.0 \\
AzV469	& O8.5\,II((f))   & 33.0  & 3.50   &  3.50    &   5.58 & 80  & 30 & 20   & 0.16 $\pm$0.02 	& 5.0 $\pm$3.0 & 15.0 $\pm$5.0  &   20.0$\pm$3.0 \\
\hline
\end{tabular}
\tablefoot{Columns are star's ID, spectral type from \citet{xshooI}, effective temperature, surface gravity, surface gravity corrected for centrifugal acceleration, luminosity, projected rotational velocity, macroturbulent velocity, microturbulent velocity, abundance of He, C, N, and O by number. Uncertainties are described in Table.~\ref{tab_lmc}.}
\end{center}
\end{sidewaystable*}

\subsection{Comparison to previous studies}
\label{s_prev}

Table ~\ref{tab_prev} gathers the results from previous studies for some stars of our XshootU sample. Effective temperatures agree very well for all stars. The dispersion is usually about 1000~K or even less. Surface gravities can differ significantly but in that case it is mainly due to the adoption of \logg\ in some studies. For instance, \citet{bouret13} fixed \logg\ when no optical spectra was available. This is the case for AzV446: Bouret et al. adopted \logg\ = 4.0 as a classical value for dwarfs, while the use of Balmer lines to measure \logg\ leads to higher values for the present study and for \citet{massey05}. AzV47 is an exception since optical spectra are available in all studies that included that star. We found that the UVES spectrum used by \citet{bouret21} is quite different from the X-shooter spectrum we used in the present analysis. In particular the Balmer lines are narrower in the former study, explaining the lower surface gravity (3.75 vs 4.3). The reason for this difference is not clear. One can only speculate that normalisation issues and low SNR cause most of it.

Luminosities are consistent between various studies, with differences of the order 0.1 dex, at most 0.2 dex. Some differences are seen in the determination of projected rotational velocities. Large discrepancies are explained by macroturbulence: studies that do take it into account provide lower \vsini\ than studies that assume rotation incorporates all the macro-broadening mechanisms. 

Helium surface abundance determinations are usually consistent in the sense that when a star is He-rich, all studies agree on enrichment, and when it is not, all studies also agree that it is not. Two exceptions are AzV95 and AzV243 for which \citet{mokiem06} find an enrichment in He while both \citet{bouret13} and the present study do not find evidence for enrichment. 

Carbon, nitrogen, and oxygen abundances have been mostly determined by \citet{bouret13,bouret21}, in addition to the present study. Only a few other determinations exist for some of the XshootU sample stars. In general abundances are consistent within the error bars between the results of Bouret et al. and the present ones. However, some differences may exist. For instance, the carbon content of AzV15 differs by almost 0.5 dex between both studies. Similarly the surface nitrogen abundance of AzV47 is $\sim$0.4 dex higher in the present analysis than in Bouret et al. This stresses that using the same atmosphere models, the same fitting strategy ($\chi^2$ analysis) but slightly different observational sets and spectral lines can affect the final values. This highlights that when discussing surface abundances, general trends are more robust than individual measurements.

\section{Surface abundances and metallicity}
\label{s_resab}

In this section we present our results and discuss how they  compare to predictions of stellar evolution for single stars with rotation. We focus on the variation of surface chemistry with metallicity. We discuss the results using three sets of evolutionary models: the Geneva models \citep{ek12,georgy13,eggen21}, the Bonn models (\citealt{brott11} - see also \citealt{szecsi22} for interpolated tracks), and the Stromlo models \citep{grasha21}. Other models calculated with alternative stellar evolution codes exist in the literature. Those of \citet{lc18} do not cover the SMC and LMC metallicity, and are not considered here. The three sets of models include stellar rotation using different formalisms, and we refer the reader to the original publications listed above for the details. In short the main differences are a purely diffusive approach for the treatment of angular momentum transport in the Bonn and Stromlo models, while an advecto-diffusive scheme is implemented in the Geneva models. In addition, the Bonn models include the effect of magnetic field on angular momentum transport (but not on chemical mixing). Finally different models assume different degree of core-overshooting that affect the shape of evolutionary tracks, as well as different initial values of carbon, nitrogen, and oxygen  abundances.

\begin{table}[!ht]
\begin{center}
\caption{Baseline abundances adopted in this study.} \label{tab_baseline}
\begin{tabular}{lrrrrrrrrccc}
\hline
Galaxy	 &   C  & N   & O    \\ 
\hline
MW      & 8.43 & 7.83 & 8.69 \\            
LMC     & 8.01 & 7.03 & 8.40 \\
SMC     & 7.42 & 6.66 & 8.05 \\
\hline 
\end{tabular}
\tablefoot{Units are 12+$log(X/H)$. References are \citet{a09} for the MW and \citet{xshooI} for the MCs. For the MCs the values are an average from various literature studies.}
\end{center}
\end{table}

\subsection{Evolutionary diagrams}
\label{s_ev}

\begin{figure*}[t]
\centering
\includegraphics[width=0.33\textwidth]{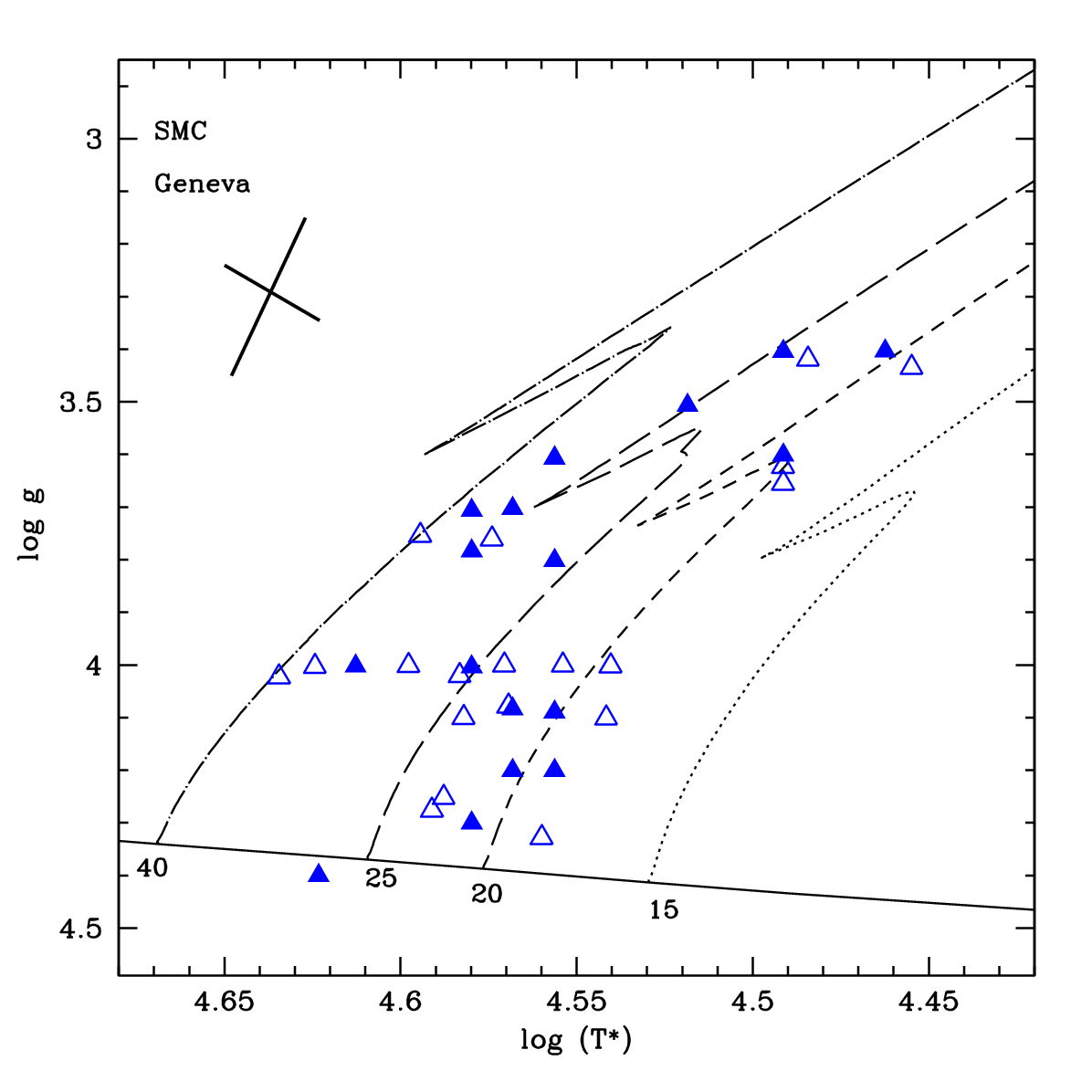}
\includegraphics[width=0.33\textwidth]{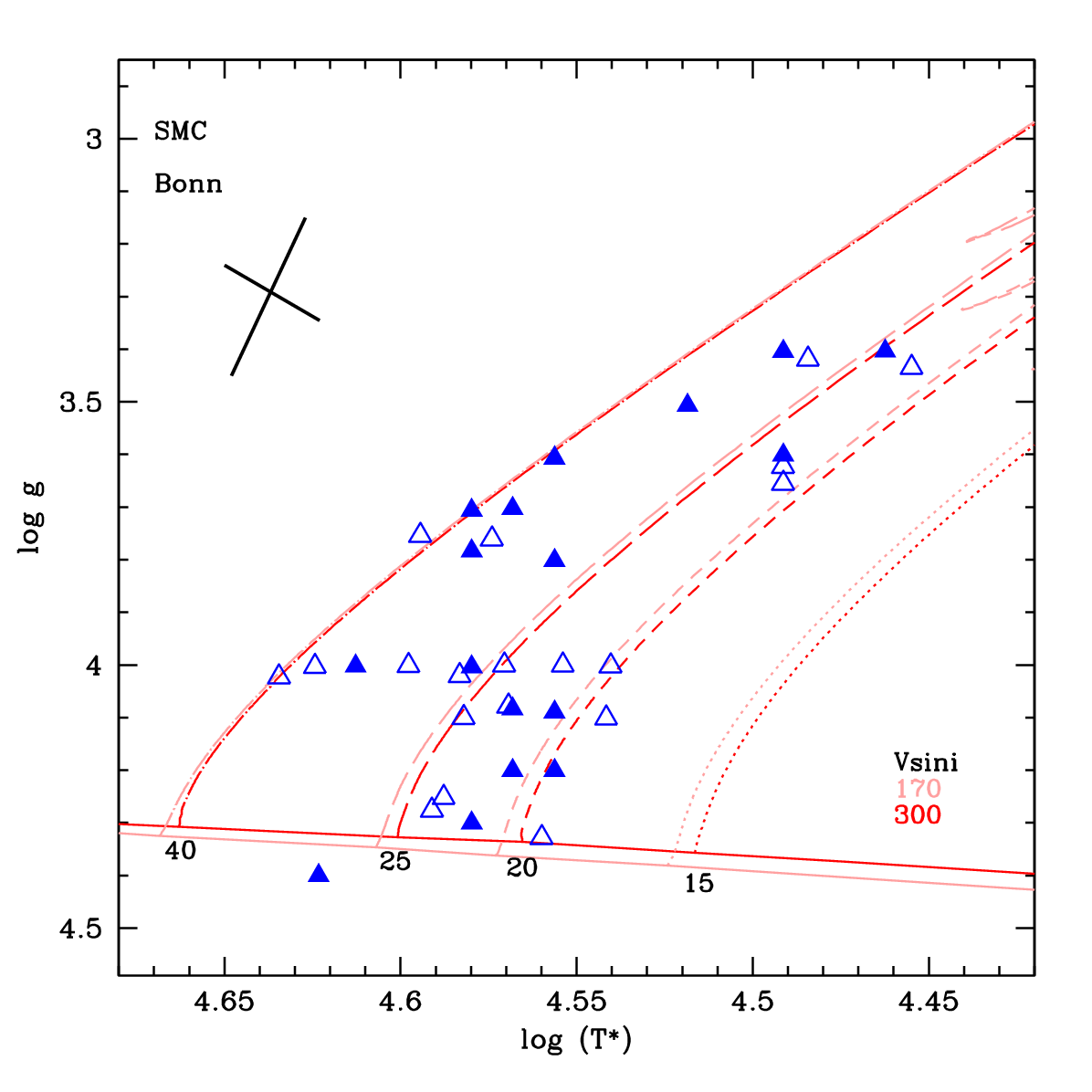}
\includegraphics[width=0.33\textwidth]{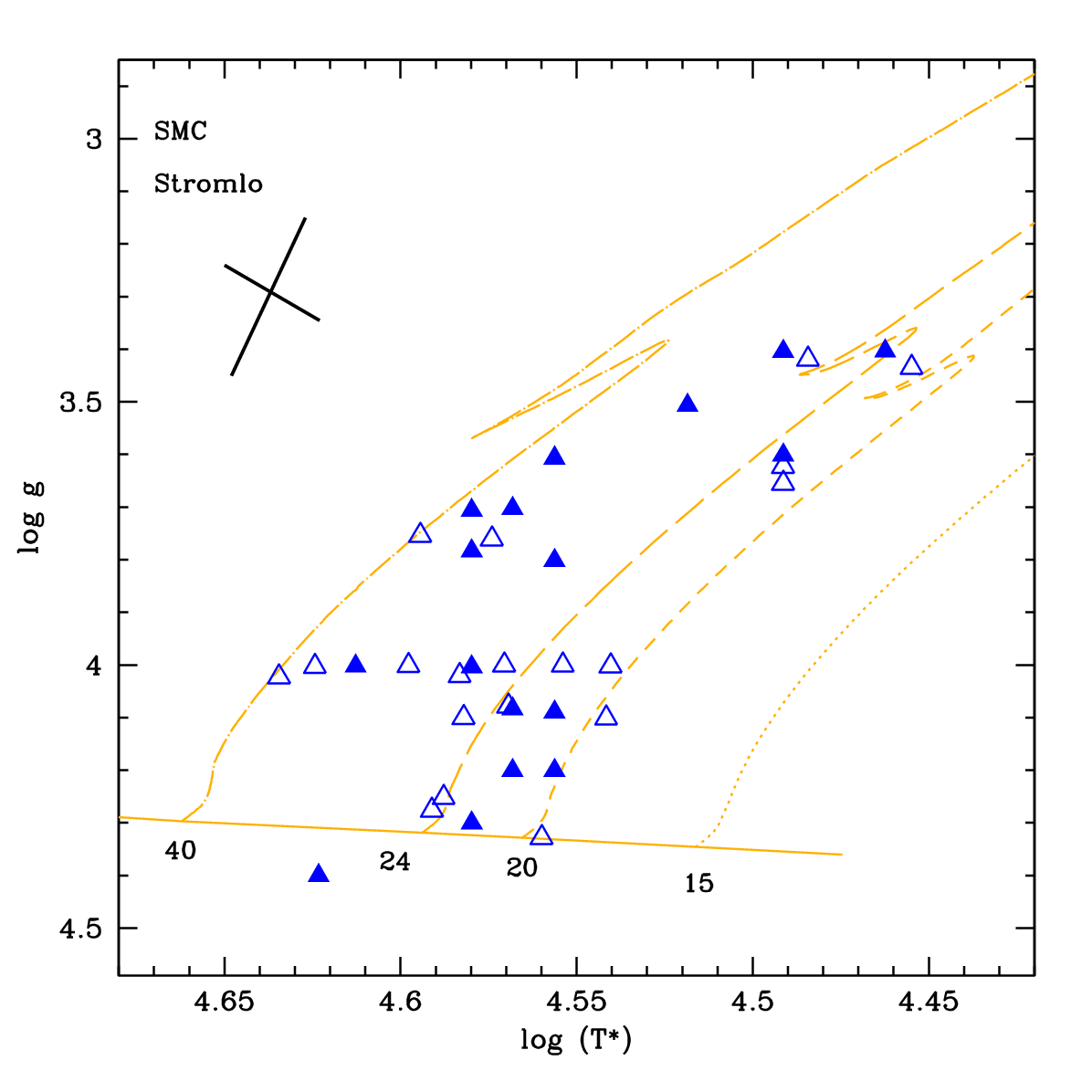}\\
\includegraphics[width=0.33\textwidth]{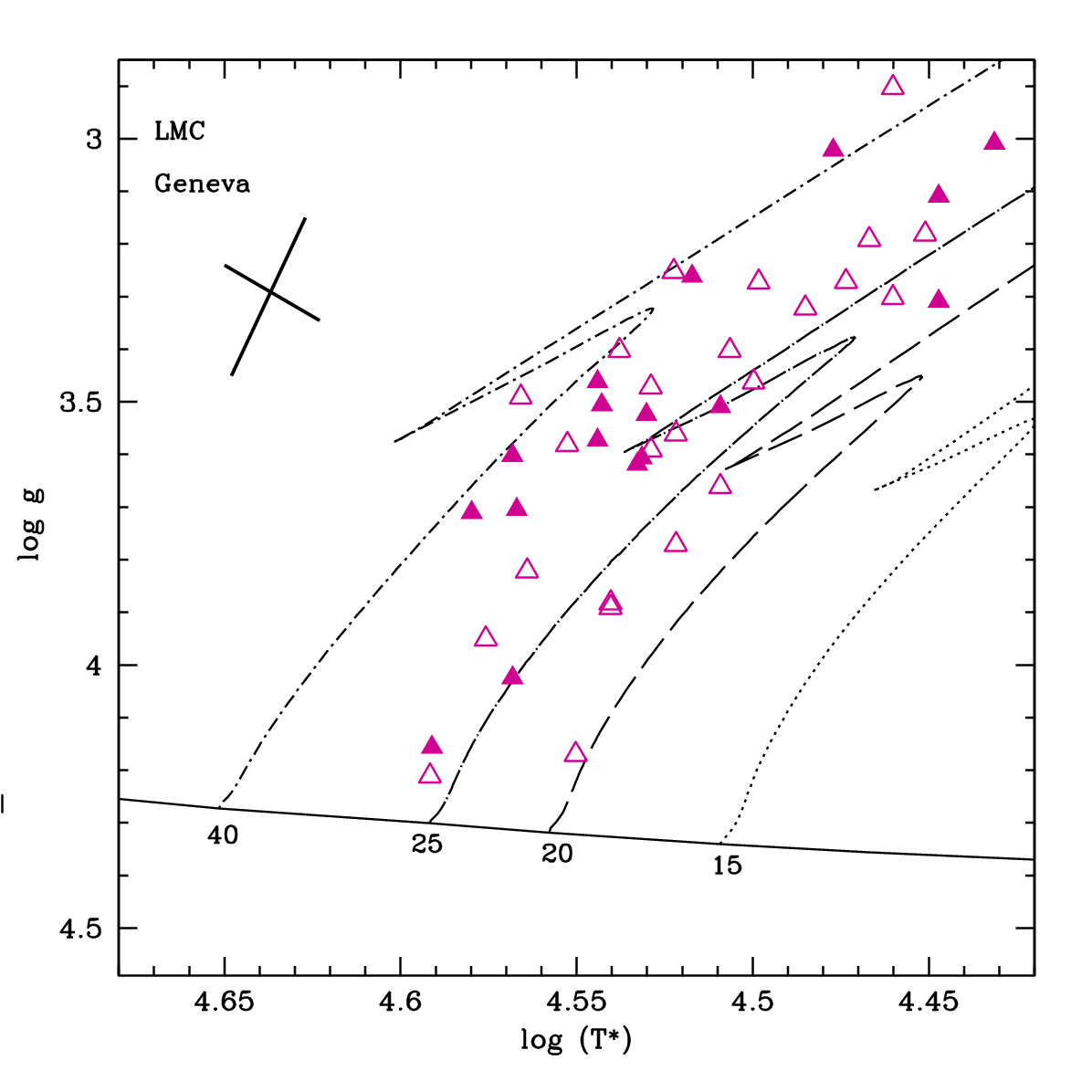}
\includegraphics[width=0.33\textwidth]{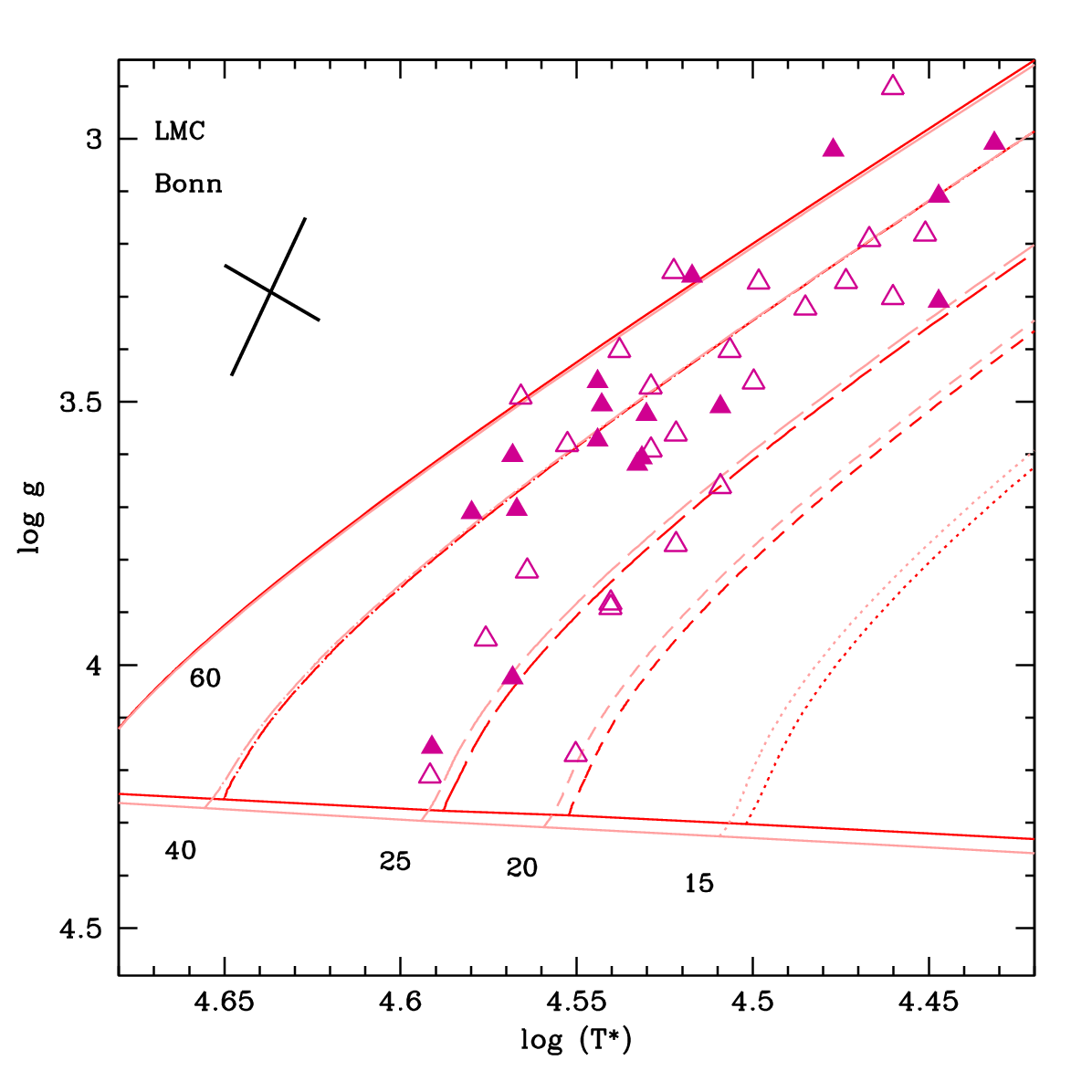}
\includegraphics[width=0.33\textwidth]{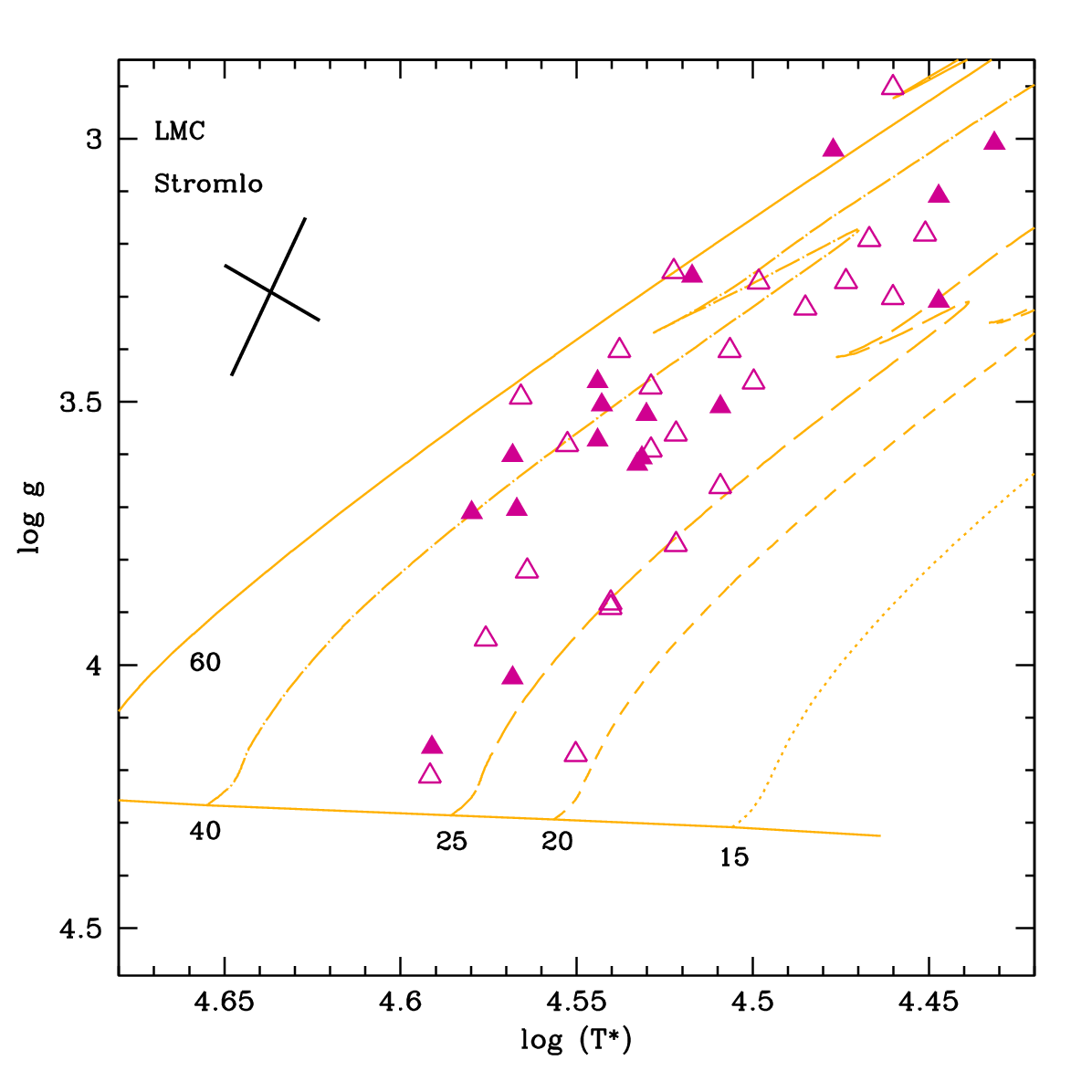}\\
\includegraphics[width=0.33\textwidth]{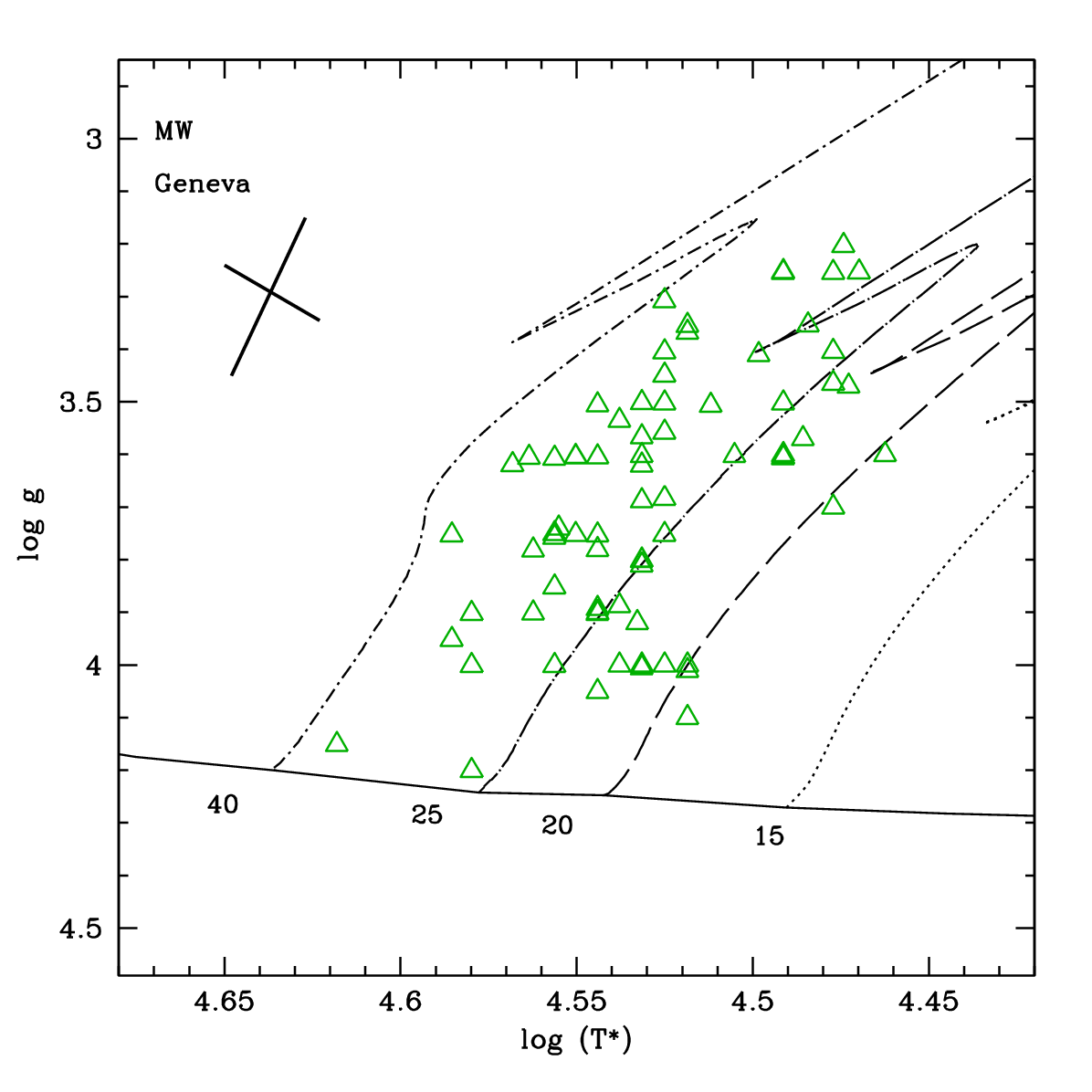}
\includegraphics[width=0.33\textwidth]{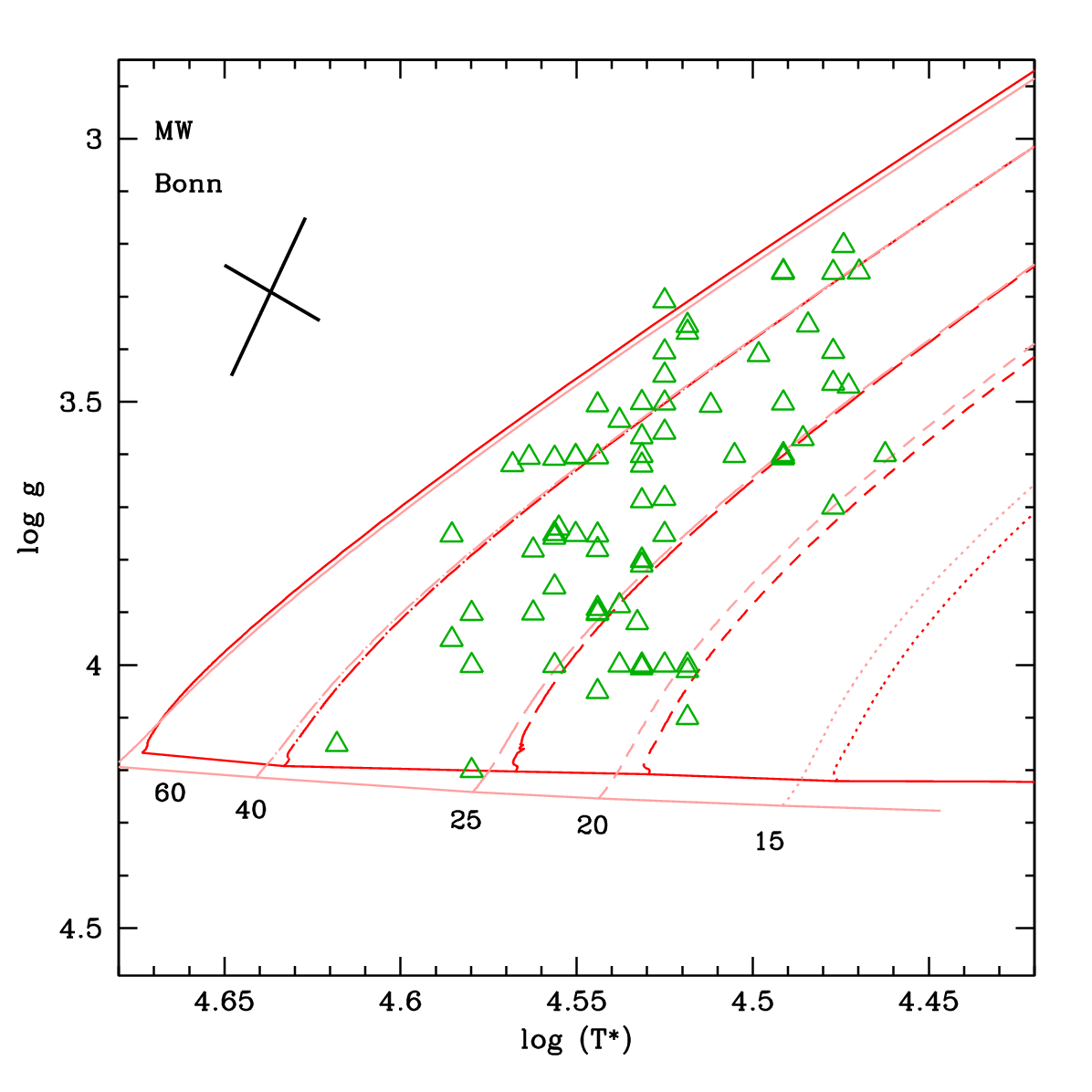}
\includegraphics[width=0.33\textwidth]{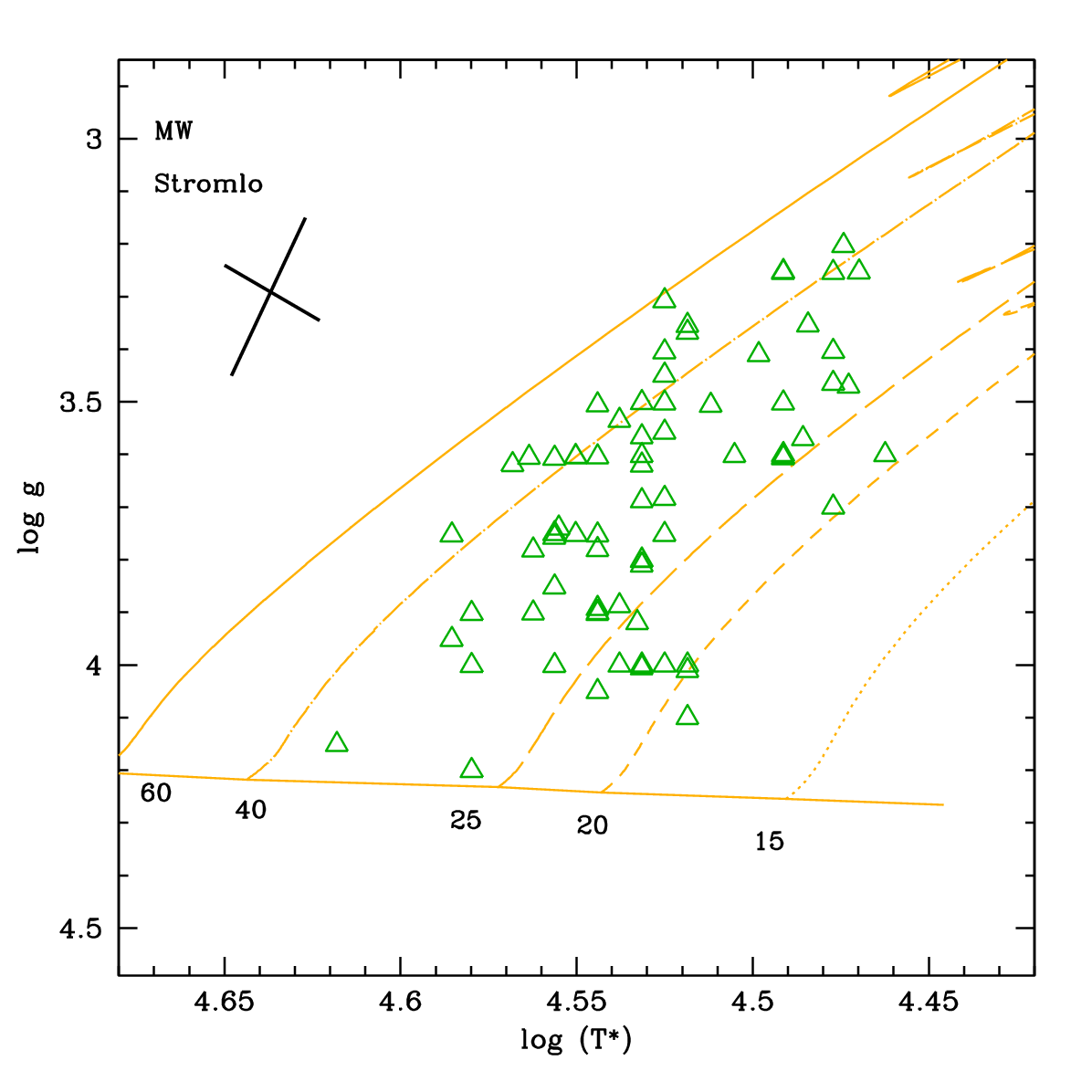}\\
\caption{$\log$\teff\ - \logg\ diagram for SMC (LMC, Galactic) sample stars in the top (middle, bottom) panels. Filled (open) symbols are for the XshootU (complementary) sample. Left panels are for Geneva tracks (\citet{eggen21} for the LMC, \citealt{georgy13} for the SMC, and \citet{ek12} for the Galaxy), middle panels for tracks from \citet{brott11}, and right panels for tracks from \citet{grasha21}. We show tracks for initial rotational velocities of about 300~\kms\ and, in the case of those of Brott et al., also 170~\kms\ (in pink). Surface gravity has been corrected for the effect of rotation in the sample stars. The cross in each panel indicates typical error bars that are taken from Fig.~\ref{Tg_av207}.} 
\label{loggteff}
\end{figure*}

Figure~\ref{loggteff} shows the $\log$\teff - \logg\ diagram for the XshootU and complementary sample stars. We overplot evolutionary tracks to provide a qualitative estimate of the mass and evolutionary state of the stars. In the three galaxies the sources are located on a relatively narrow band from the ZAMS to the end of the main-sequence. The SMC stars have initial masses between 20 and 40~\msun\ independently of the choice of the tracks used for comparison. In the LMC and the Galaxy, the Geneva tracks indicate a similar initial mass range while the Bonn and Stromlo tracks favour a wider mass range, from 20 to 60~\msun.

The inferred evolutionary status of the stars depends much more on the choice of the tracks. If we simply separate stars into two categories whether they are on the main sequence (MS) or beyond it, we find that in the Milky Way all stars fall in the first category. In the LMC, the morphology of the Geneva and Stromlo tracks imply that a fraction of the stars are post-MS objects, while the Bonn tracks favour a MS status. In the SMC, most stars are MS objects, with the exception of a couple of objects that may be post-MS objects depending on the tracks.

\subsection{Nucleosynthesis}
\label{s_cno}

\begin{figure}[]
\centering
\includegraphics[width=0.49\textwidth]{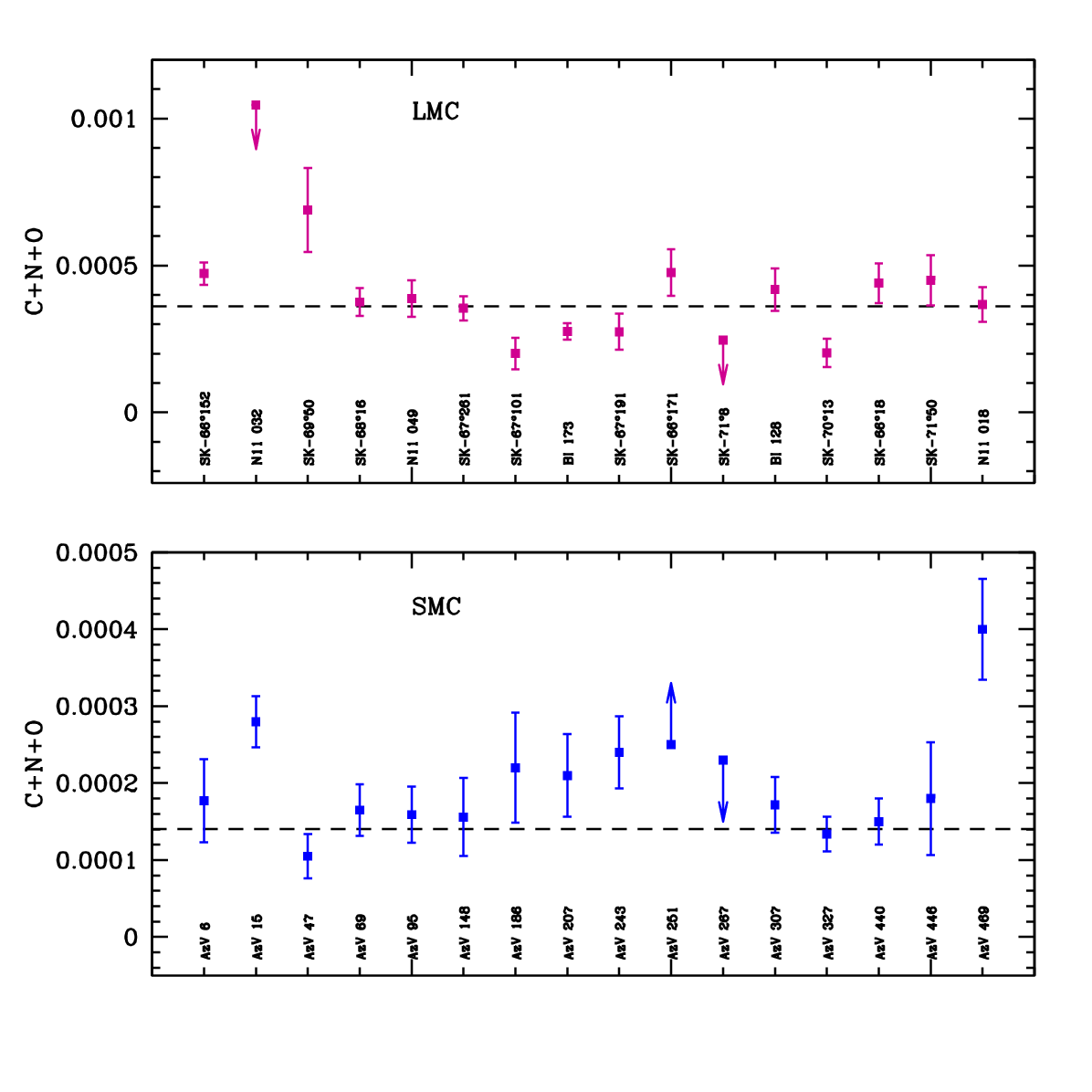}
\caption{Sum of the number fraction of carbon, nitrogen, and oxygen for the LMC (top panel) and SMC (bottom panel) stars. The horizontal dashed lines mark the C+N+O value in the LMC and SMC according to \citet{xshooI}. Stars for which the abundance determination of one of the three elements could not be performed are excluded from the figure.} 
\label{sumCNO}
\end{figure}

Figure \ref{sumCNO} shows the sum of the carbon, nitrogen, and oxygen abundances for stars for which a determination of the abundances of the three elements could be performed. Nucleosynthesis through the CNO cycle predicts that the total amount of these three elements should remain constant since they act as catalysts. In the LMC and SMC stars this is what is observed, most values of the C+N+O content being close to the baseline abundance of each galaxy (see Table~\ref{tab_baseline}). Small (i.e. less than 2$\sigma$ deviations occur in a few cases but are not unexpected because of 1) potential remaining bias in the abundance determination and 2) a spread in the baseline abundance depending on the star's locations in the galaxy. Two stars appear as outliers in the SMC. AzV469 in the SMC has a C+N+O content about three times larger than the baseline abundance. Fig.~\ref{fit_av469} shows a good fit to most lines. In particular the nitrogen and oxygen lines are strong and imply a large C+N+O content. AzV469 may thus have experienced a peculiar chemical evolution. The other outlier (more than 2$\sigma$ deviation from baseline C+N+O) is star AzV15. In that case we attribute the potential discrepancy to the abundance determination since C, N, and O lines are weak and relatively noisy (see Fig.~\ref{fit_av15}). We also refer to Sect.~\ref{s_Zenrich} for a discussion on a potential spread in chemical abundances in the SMC.

In the LMC SK-70$^{\circ}$13 and SK-67$^{\circ}$101 have marginally small values of the C+N+O abundance. Figs.~\ref{fit_skm67d101} and \ref{fit_skm70d13} indicate that the C, N, and O lines are not perfectly fitted which mainly explains the relatively small deviation compared to the baseline abundances. SK-69$^{\circ}$50 lies above the baseline value but the error bars are the largest of the LMC sample, making this object only a weak outlier. 

For consistency we show the C+N+O sum for Galactic stars of the complementary sample in Fig.~\ref{sumCNOmw}. We see that 21 out of the 36 objects in this figure have values consistent with the solar case. Thirteen stars lie in between the solar and LMC reference values. 


\smallbreak

\begin{figure}[]
\centering
\includegraphics[width=0.49\textwidth]{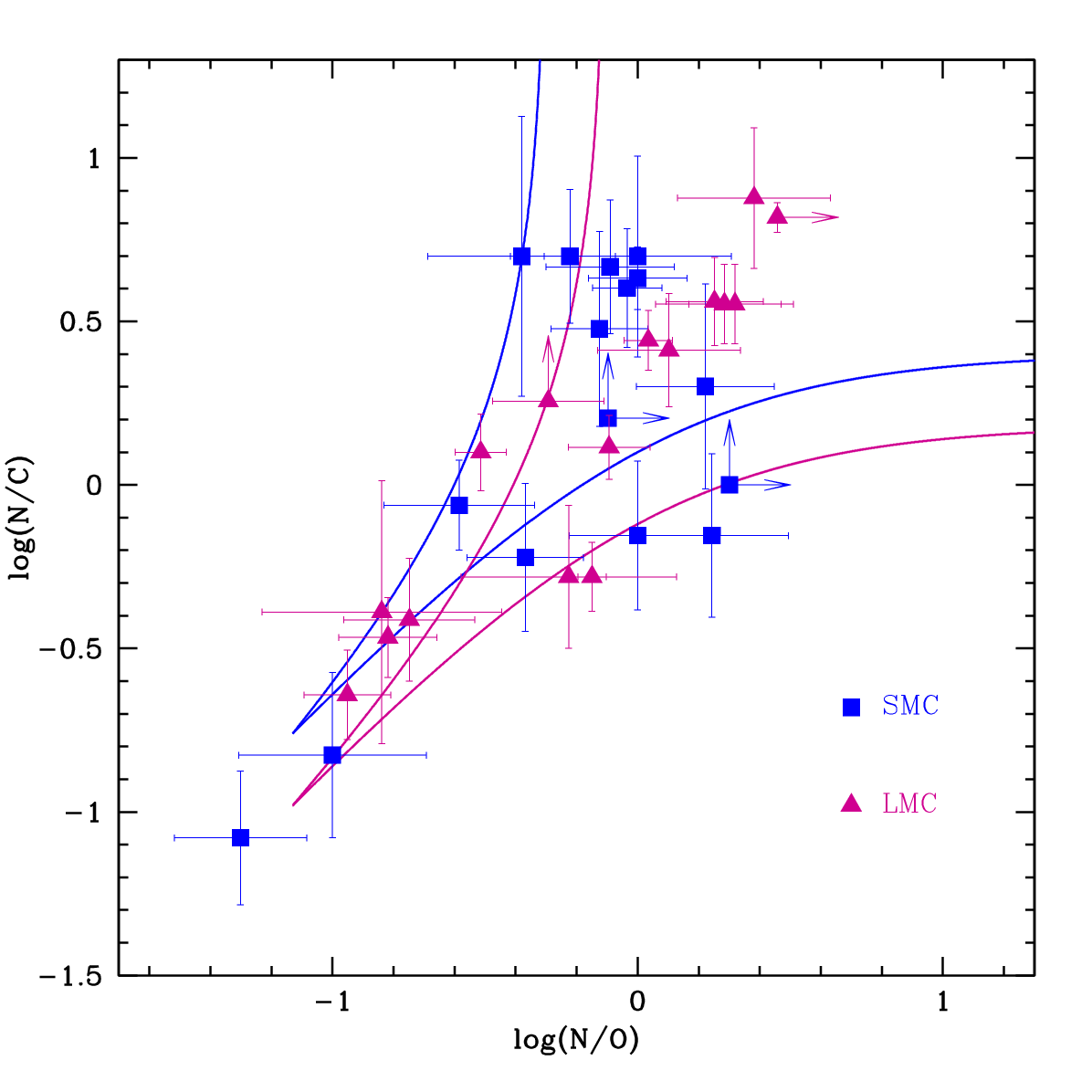}
\caption{Logarithm of N/C versus logarithm of N/O (all number ratios) for the XshootU SMC (LMC) stars in blue (purple). Solid lines are the theoretical values for CNO (top) and CN (bottom) equilibrium.} 
\label{nonc_nono}
\end{figure}

Figure~\ref{nonc_nono} shows the ratios of N/H over C/H versus N/H over O/H for the XshootU LMC and SMC sample stars. The results for the complementary samples have been presented in the publications from which the results were retrieved. The baseline values of the ratios, from which the solid lines expand, are taken from \citet{xshooI}. All stars nicely follow the trend of increasing N/C for higher N/O, in full agreement with the expectation that these elements are processed by the CNO-cycle in the interior of the stars. Within the error bars the vast majority of stars are also located in between the two theoretical curves, strengthening this conclusion. One exception in the SMC is star AzV69, which has a very low N/H ratio (blue square at the bottom left of the figure). This object has been discussed by \citet{hil03}. It is an OC star, a class of O stars with unprocessed abundances \citep{walborn71,martins16OC}. The other SMC outlier is star AzV47 which lies slightly below the expected limit. In the LMC SK-67$^{\circ}$ 261 has N/C marginally above the expected value for its N/O ratio.
We stress that the exact position and shape of the curves shown in Fig.~\ref{nonc_nono} depend on the initial N/C and N/O ratios. \citet{bouret21} show that variations in these quantities lead to shifts in the positions of these curves. Consequently stars that are outliers for a given set of initial abundances may be accommodated by a slightly different initial composition.

\begin{figure}[]
\centering
\includegraphics[width=0.49\textwidth]{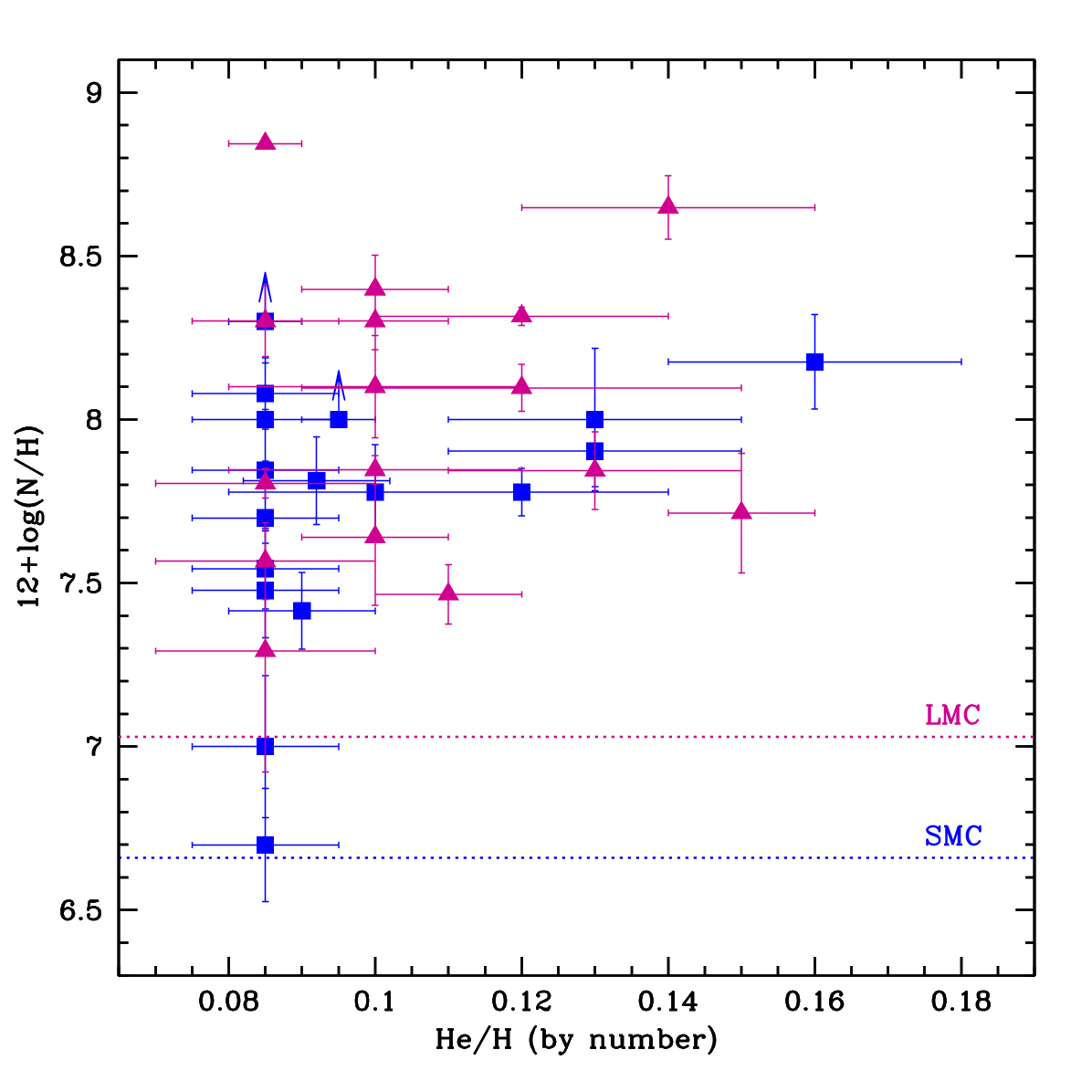}
\caption{Nitrogen abundance (in units of 12+$log$(N/H)) versus He/H number ratio for XshootU SMC (LMC) stars in blue (purple). Horizontal dotted lines mark the baseline nitrogen abundances in the SMC and LMC.} 
\label{NHeobs}
\end{figure}

The nitrogen surface abundance is shown as a function of the surface He/H number ratio in Fig.~\ref{NHeobs}. Stars with no significant helium enrichment can show a large range of surface nitrogen abundance. Contrarily, stars that show helium enrichment are also nitrogen rich. This is consistent with CNO-burning and chemical mixing \citep[see also][]{rivero12}. Nitrogen can be quickly brought to the surface even if the increase of the surface helium content has not yet happened. Indeed the fractional change relative to the baseline values is much larger for nitrogen than for helium. When a sufficient fraction of helium produced in the core has been brought to the surface so the He/H ratio is affected, the other products of CNO-burning are also seen, in particular a large nitrogen enrichment.

\subsection{Metallicity effect on chemical enrichment}
\label{s_Zenrich}

In this section we investigate the effect of metallicity on the surface abundances. We consider not only the stars analysed in the present study, and part of the XshootU project, but also all complementary samples described in Sect.~\ref{s_sample}.

\begin{figure*}[]
\centering
\includegraphics[width=0.33\textwidth]{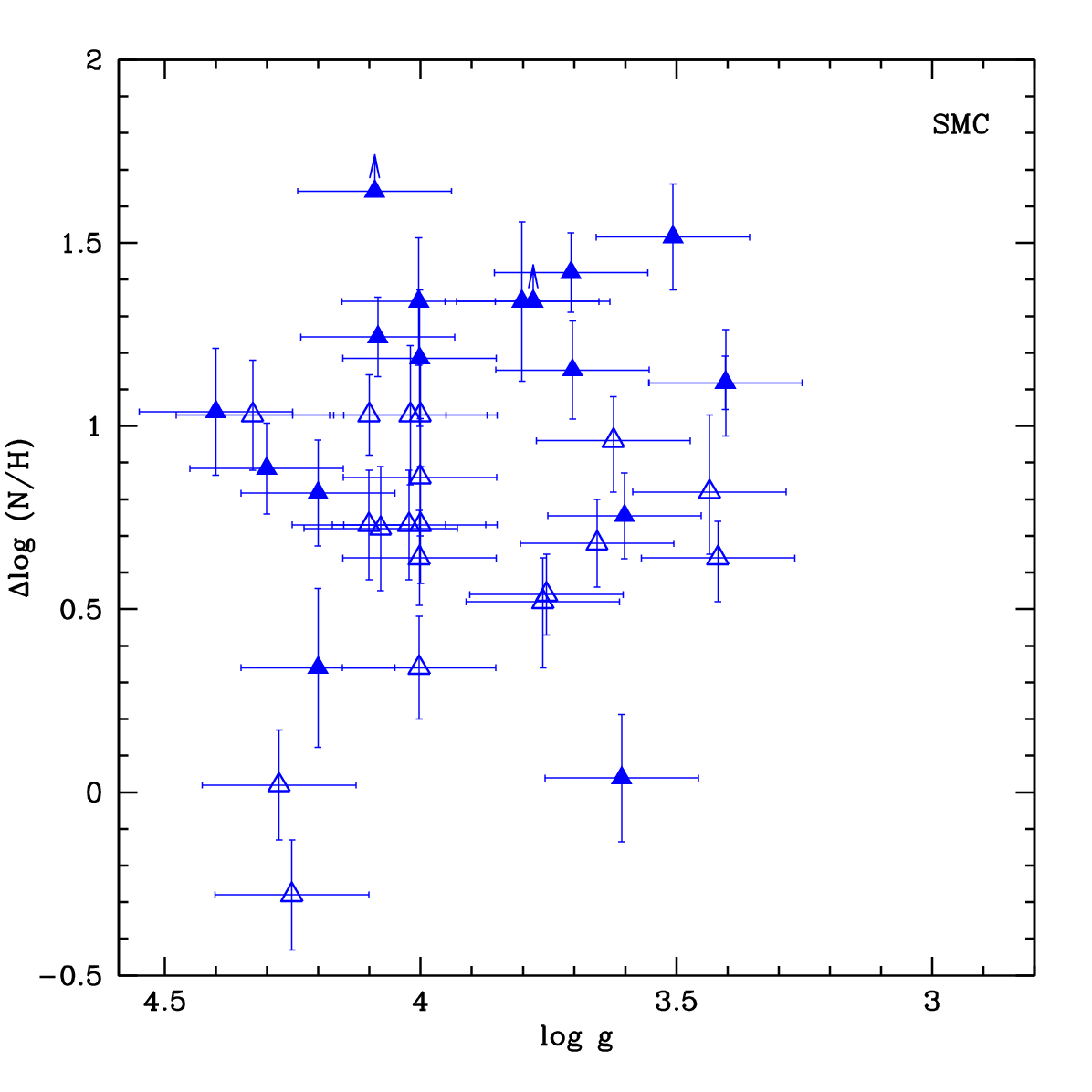}
\includegraphics[width=0.33\textwidth]{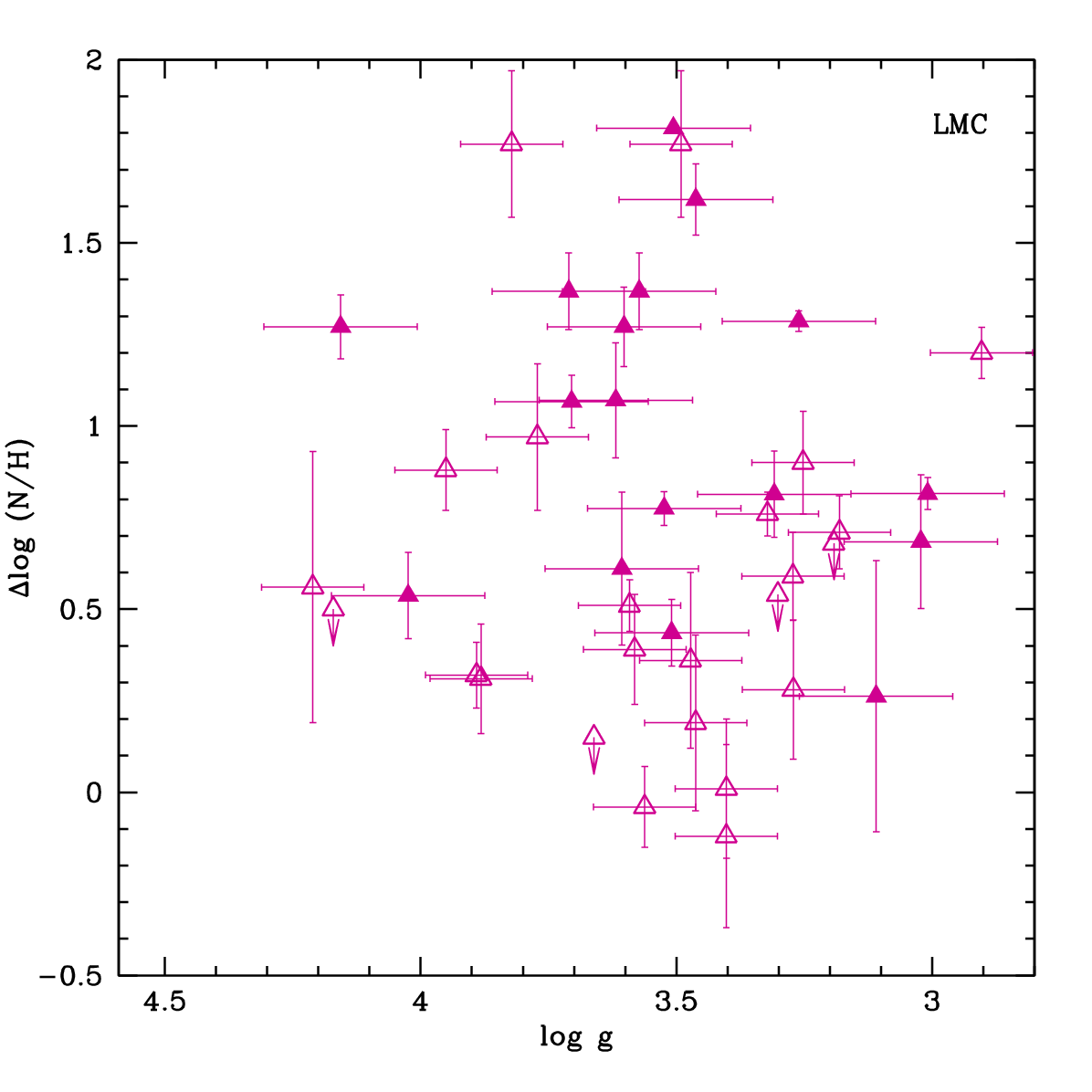}
\includegraphics[width=0.33\textwidth]{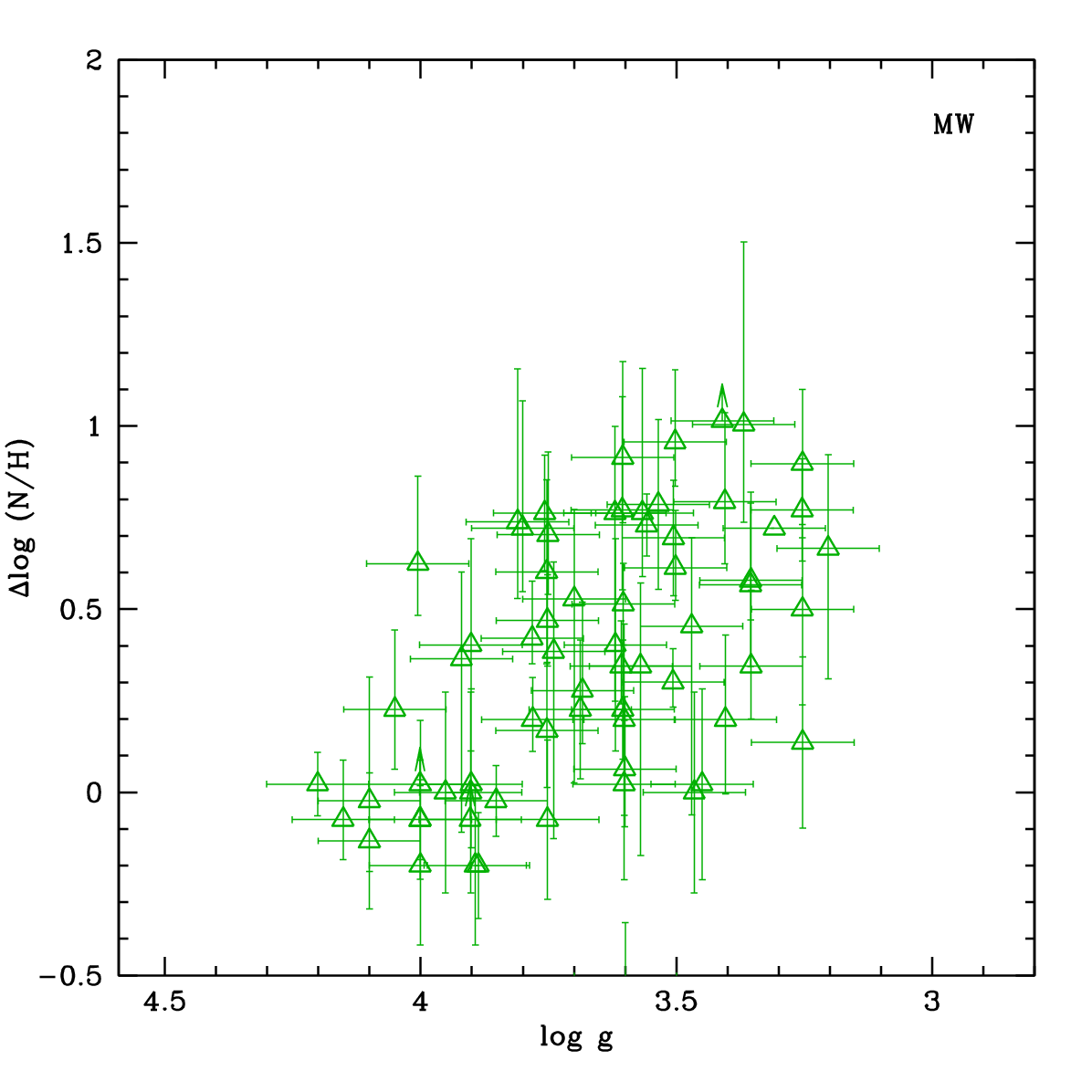}
\caption{logarithm of N/H minus the baseline value as a function of surface gravity for the SMC (left panel), the LMC (middle panel), the MW (right panel). Surface gravity has been corrected for the effect of rotation. XshootU sample stars are shown by filled symbols, complementary sample stars are shown by open symbols.}  
\label{abobscompN}
\end{figure*}

\begin{figure*}[]
\centering
\includegraphics[width=0.33\textwidth]{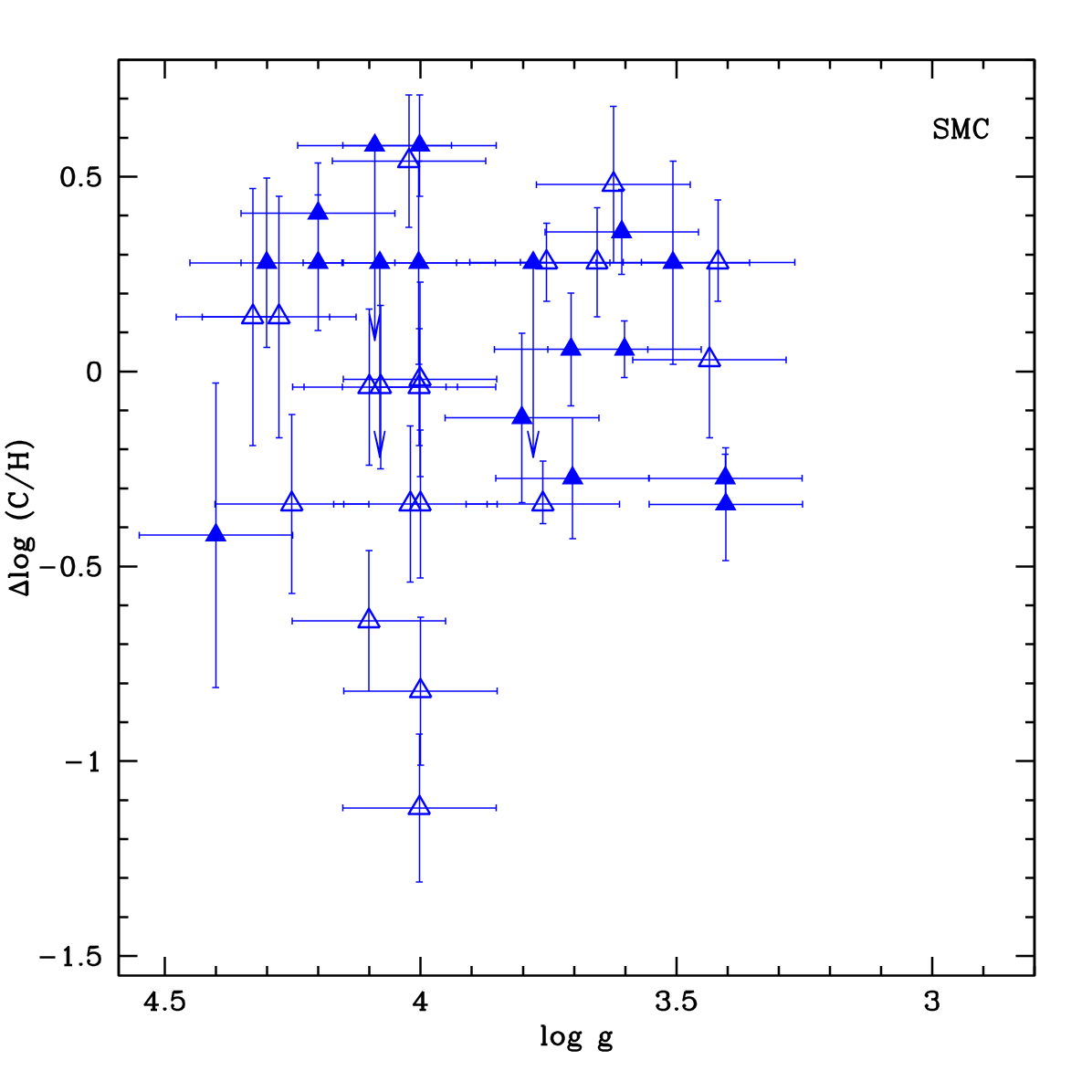}
\includegraphics[width=0.33\textwidth]{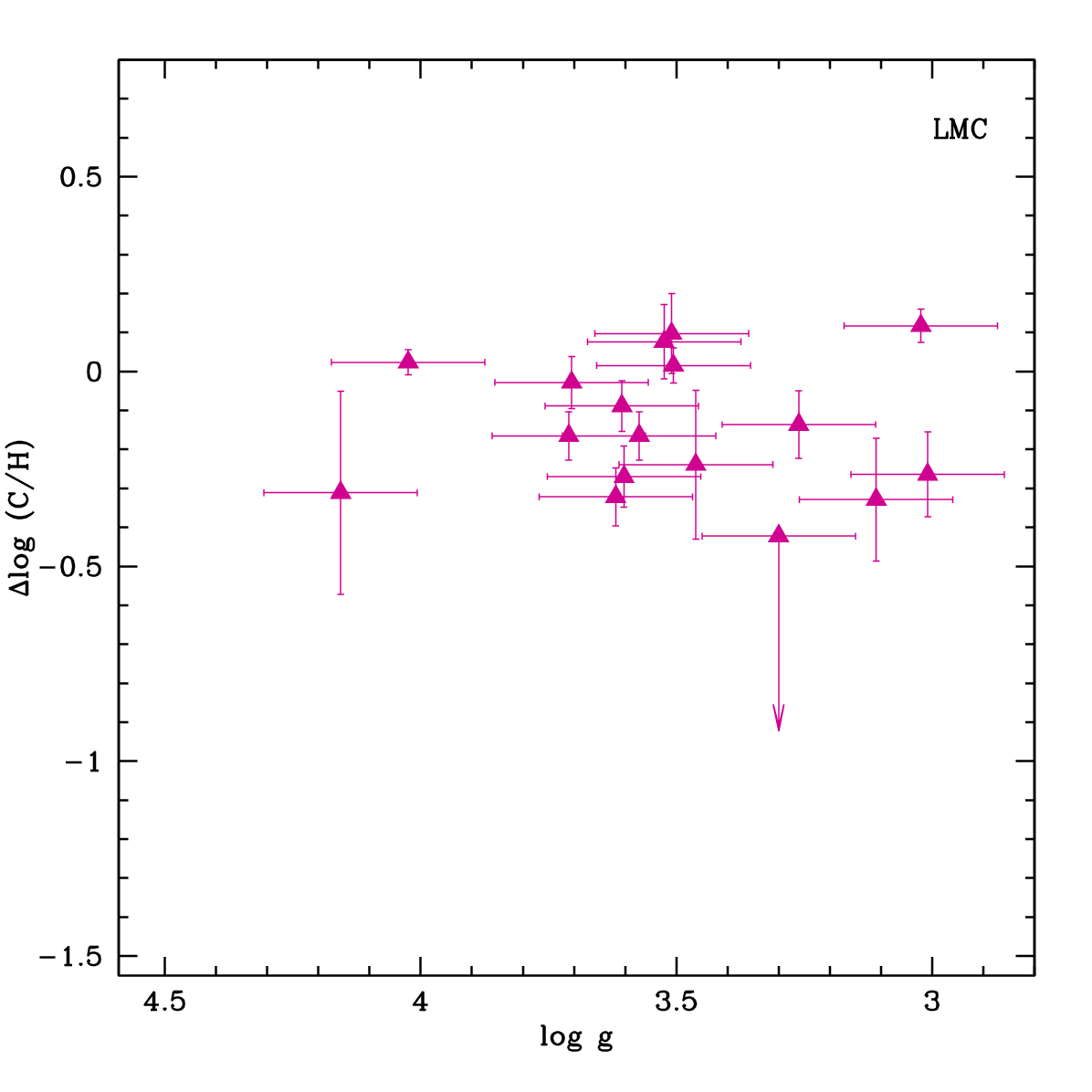}
\includegraphics[width=0.33\textwidth]{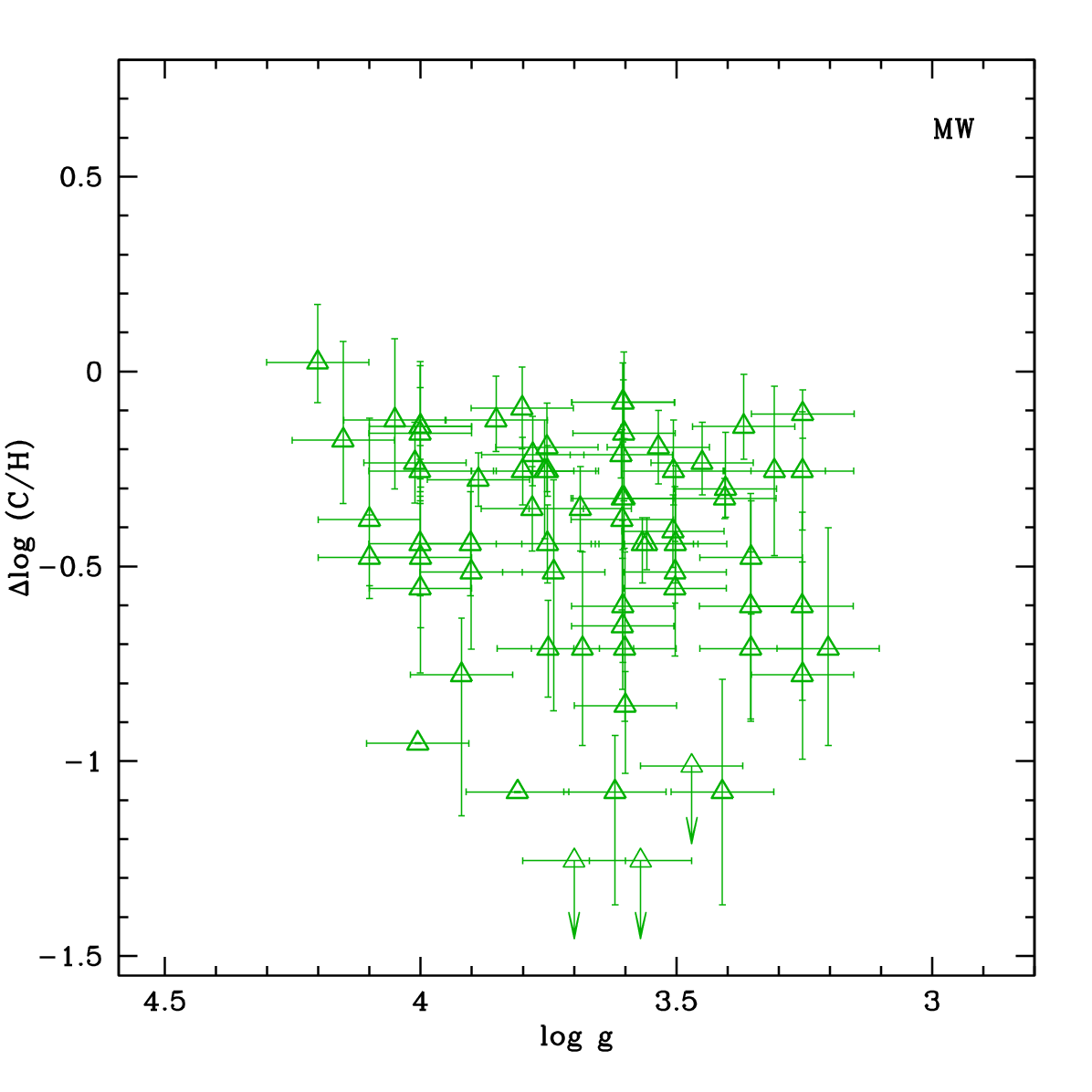}
\caption{Same as Fig.~\ref{abobscompN} but for the difference between C/H and the baseline carbon abundance. }  
\label{abobscompC}
\end{figure*}

\begin{figure*}[]
\centering
\includegraphics[width=0.33\textwidth]{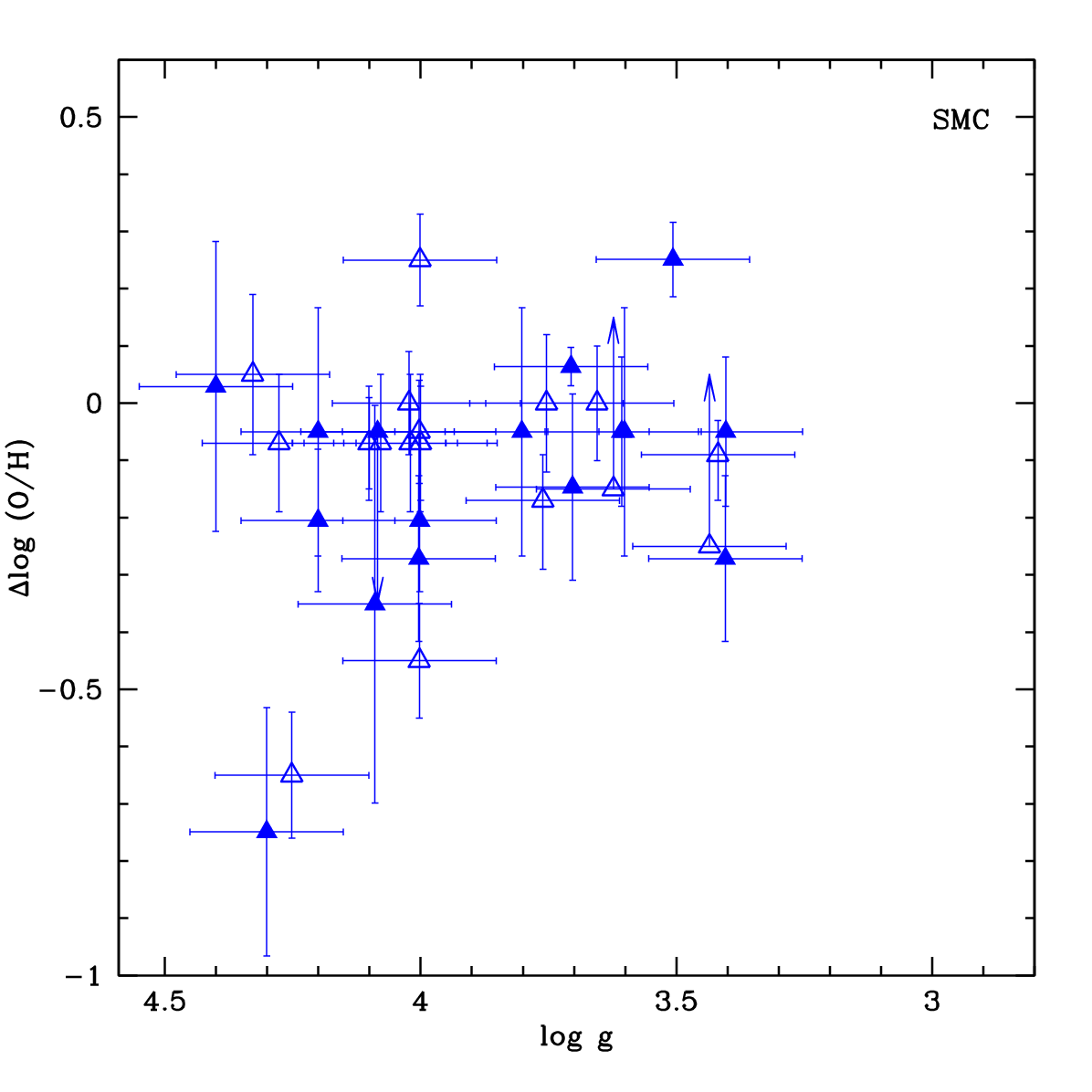}
\includegraphics[width=0.33\textwidth]{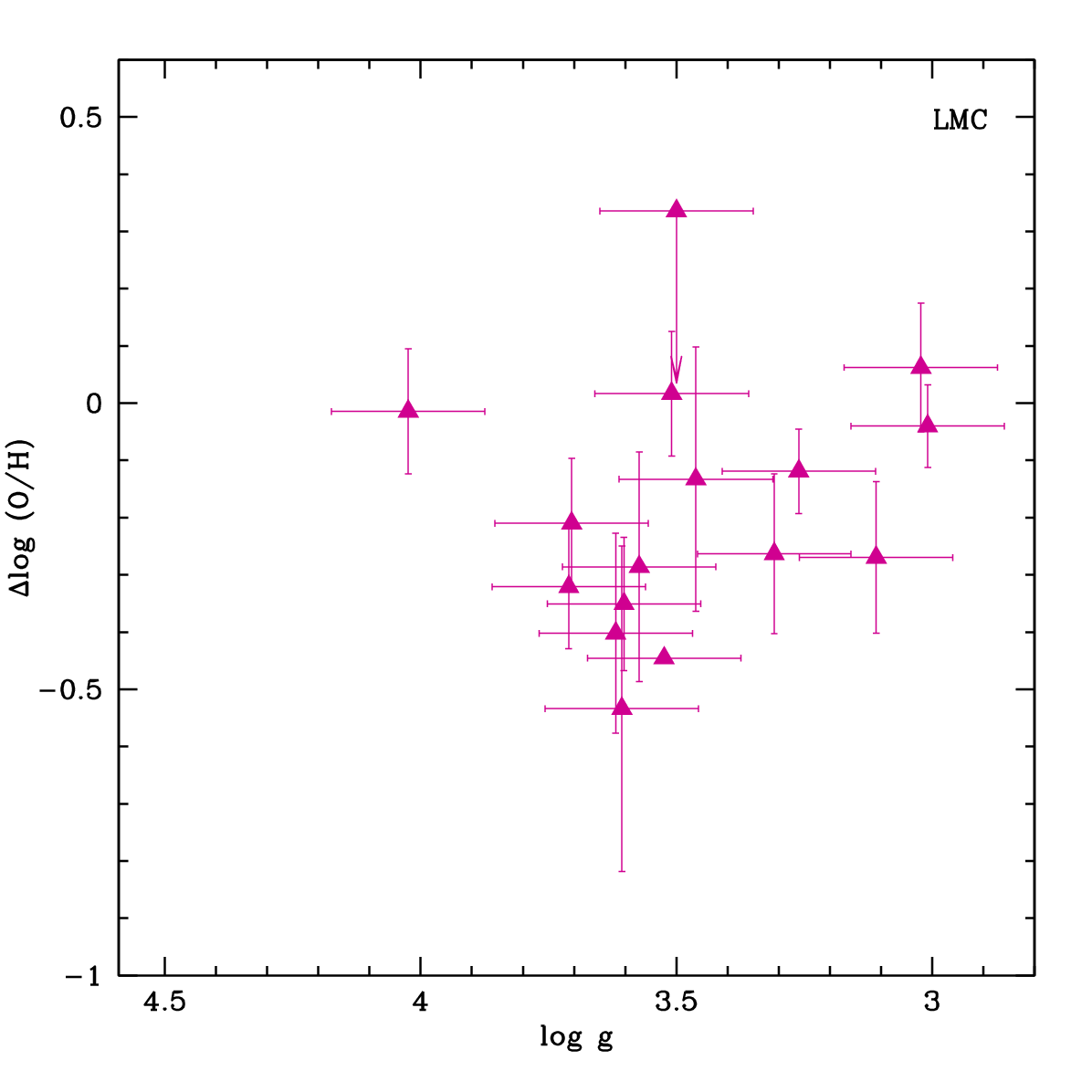}
\includegraphics[width=0.33\textwidth]{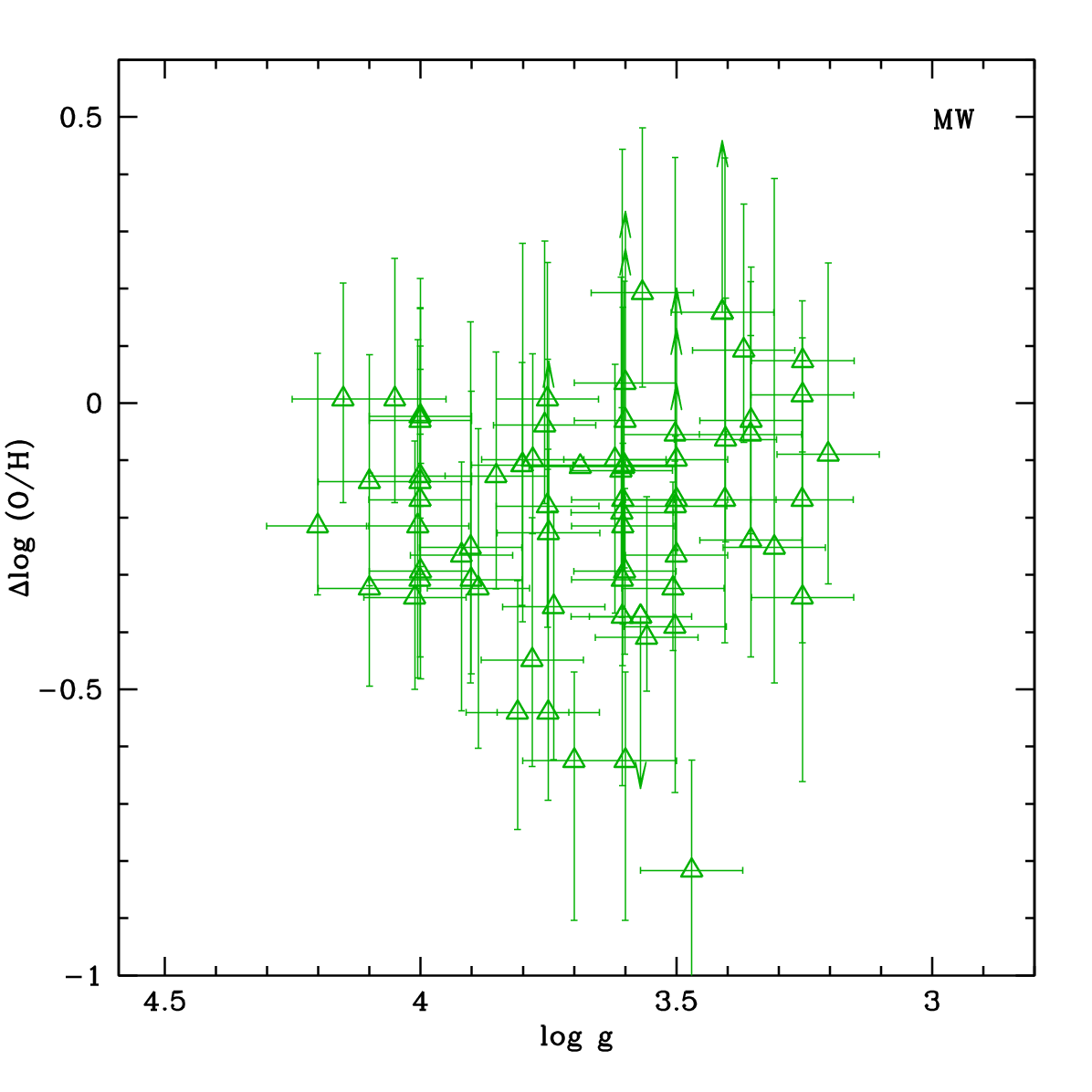}
\caption{Same as Fig.~\ref{abobscompN} but for the difference between O/H and the baseline oxygen abundance. }  
\label{abobscompO}
\end{figure*}

Figure~\ref{abobscompN} shows the difference between the present-day and baseline N/H ratio, as a function of surface gravity. We used the baseline abundances in the LMC, SMC, and Milky Way reported in Table~\ref{tab_baseline}. 
In Figure~\ref{abobscompN} we detect a trend of surface enrichment with lower \logg. This is expected of mixing processes that transport the products of CNO-burning to the surface while stars evolve and reach lower surface gravity. The trend is clearly observed in the Galaxy. In the MCs significant enrichment is already seen at relatively early stages, i.e. high surface gravities. 
The second major conclusion is that surface nitrogen enrichment reaches a higher maximum value in the SMC than in the Galaxy, and this happens at higher surface gravity. In the Milky Way $\Delta log(N/H)$ ranges from about 0 to $\sim$1.0. In the SMC the enrichment is about twice as large, reaching $\Delta log(N/H)$=1.5. In the LMC even larger values of $\Delta log(N/H)$ are observed, up to 1.8. But the scatter is large compared to that of the SMC and MW.
These results are qualitatively in line with the predictions of all models that are presented in Fig.~\ref{abmod}: a stronger nitrogen enrichment is predicted at lower metallicity, and enrichment is observed at higher gravity for lower metallicity models. However the level and path to the highest enrichment is different from model to model. 

Turning to the C/H ratio shown in Fig.~\ref{abobscompC}, Galactic stars are significantly carbon depleted at lower surface gravity. This is expected from CNO-cycle nucleosynthesis and chemical transport to the surface, as seen in all models of Fig.~\ref{abmod}. However we note that depletion reaches levels lower than any model of Fig.~\ref{abmod} predicts. This may be partly explained by the possible sub-solar metallicity of the Galactic stars, as indicated by the C+N+O values (Fig.~\ref{sumCNOmw}). But even in that case depletion is stronger than model predictions at low metallicity. Both incorrect model predictions and abundance determinations could explain this trend, but at present no clear answer can be given. In the LMC, carbon is on average depleted. The maximum depletion seems not as strong as in the Galaxy. The identification of a metallicity trend is difficult. Compared to N/H the variation of C/H is much lower (about a factor of 3) but since the error bars remain of the same order of magnitude, trends are somewhat hidden.
The SMC stars show a peculiar behaviour: the dispersion is very large; some stars are carbon-depleted, but at least half of the objects are actually carbon-enhanced compared to the baseline abundance. This is opposite to all theoretical predictions which all show a slow decrease of the carbon surface abundance as a star evolves (see middle panels of Fig.~\ref{abmod}). This surprising behaviour in the SMC may be due to 1) systematic offsets in the carbon abundance determination or 2) improper baseline carbon abundance. If explanation 1 was correct one would expect the same trend at all metallicities, which is not observed. We visually checked that lower C/H values  for the problematic objects translated into too weak \ion{C}{iii}~1176 and \ion{C}{III}~4070 lines compared to the observed spectra. One may wonder whether the lower metallicity of the SMC produces weaker lines that are more difficult to analyse. This trend exists but as seen in Figs~\ref{fit_av6} to \ref{fit_av469} the carbon lines remain relatively well detected and are useful indicators of the carbon content. For explanation 2 listed above, an offset by $\sim$0.3 dex would be needed (i.e. the baseline carbon abundance in the SMC would need to be two times larger than currently estimated). In that case the most C-rich objects should also be the least nitrogen-enhanced, which is not observed: stars with high C/H show a wide range of nitrogen enrichment. Hence a systematic offset in the initial carbon content is not likely. \citet{rama21} report a significant spread in the surface chemical composition of three binary stars in the bridge region of the SMC. One has LMC-like abundances. If such a degree of inhomogeneity was widespread in the SMC this could explain part of the issue regarding C/H in the present sample. But as we will see below this is not corroborated by surface oxygen abundances.
From this we tentatively conclude that surface carbon abundances are presently not reliable to test the effect of metallicity on chemical mixing. We also see in the middle panels of Fig.~\ref{abmod} that stronger depletion at lower Z is predicted by the Geneva models, and to a lesser extent by the Stromlo models. However the Bonn models do not predict any clear trend.

Finally the variation of the oxygen abundance as a function of metallicity is shown in Fig.~\ref{abobscompO}. On average oxygen is depleted in most stars, in the three galaxies. This is again expected of nucleosynthesis. However given the uncertainties and the magnitude of the variations ($\Delta log(O/H) \sim$ 0.2-0.3) no conclusion can be drawn regarding any metallicity trend. The Stromlo models show such a trend but with a magnitude of the order 0.3 dex. The Geneva and Bonn models do not show a clear trend. Thus oxygen cannot be used to investigate the metallicity dependence of chemical mixing. We also note the same issue as C/H for Galactic stars: oxygen depletion reaches levels lower than model predictions. However the problem is less severe for oxygen since only a handful of stars show oxygen depletion clearly stronger than model predictions.

From this analysis we conclude that surface nitrogen enrichment is  the best indicator of any metallicity trend. The nitrogen enrichment is stronger in the SMC than in the Galaxy, in qualitative agreement with model predictions. For the intermediate case of the LMC, no conclusion can be drawn at present because of the large dispersion of the measurements. The metallicity difference between the LMC and the Galaxy on one side, and the LMC and the SMC on the other side, is probably too small to lead to effects that can be distinguished with the accuracy of the measurements we can achieve with the method used for abundance determinations.

\subsection{Projected rotational velocities}
\label{s_vsini}

\begin{figure*}[]
  \centering
\includegraphics[width=0.33\textwidth]{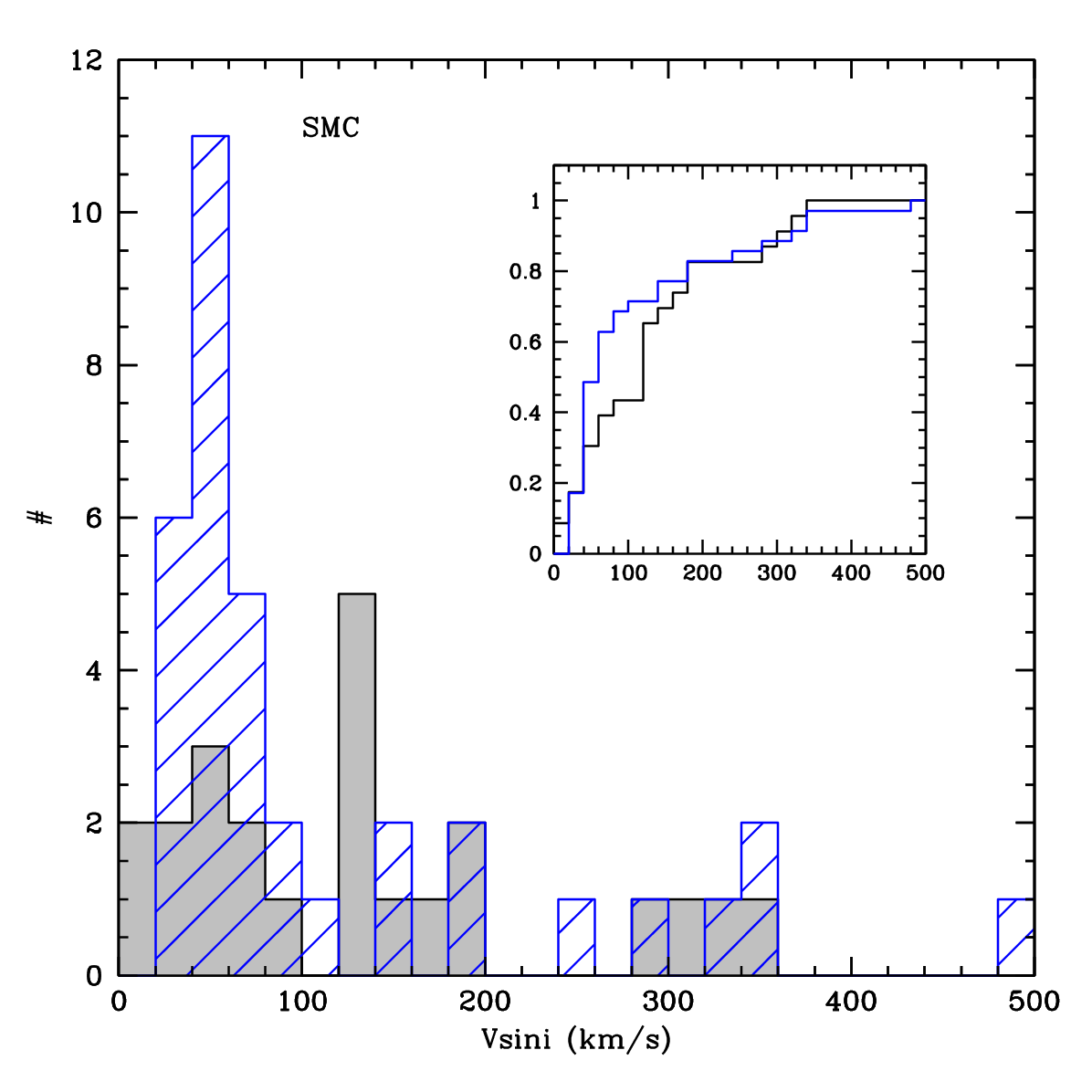}
\includegraphics[width=0.33\textwidth]{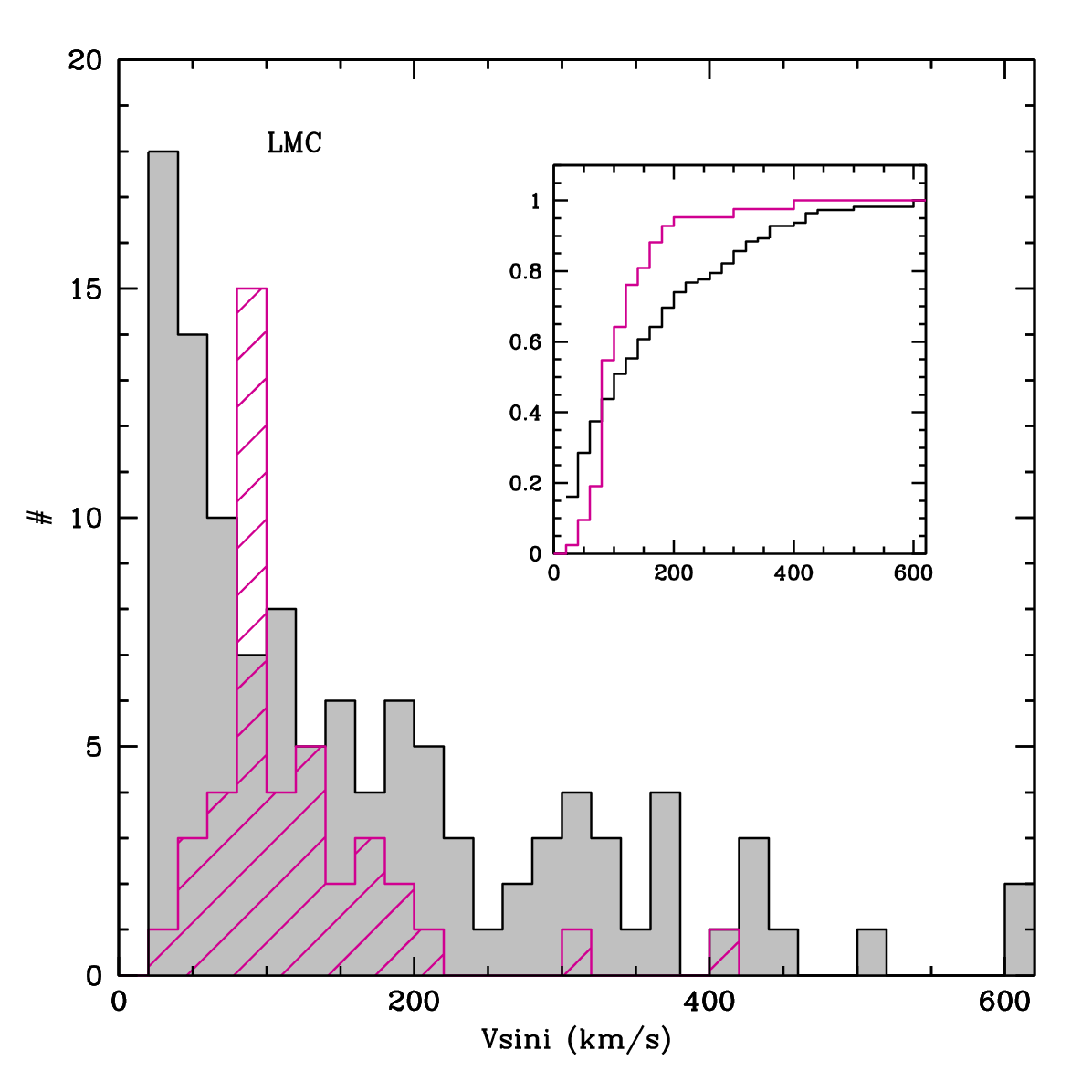}  
\includegraphics[width=0.33\textwidth]{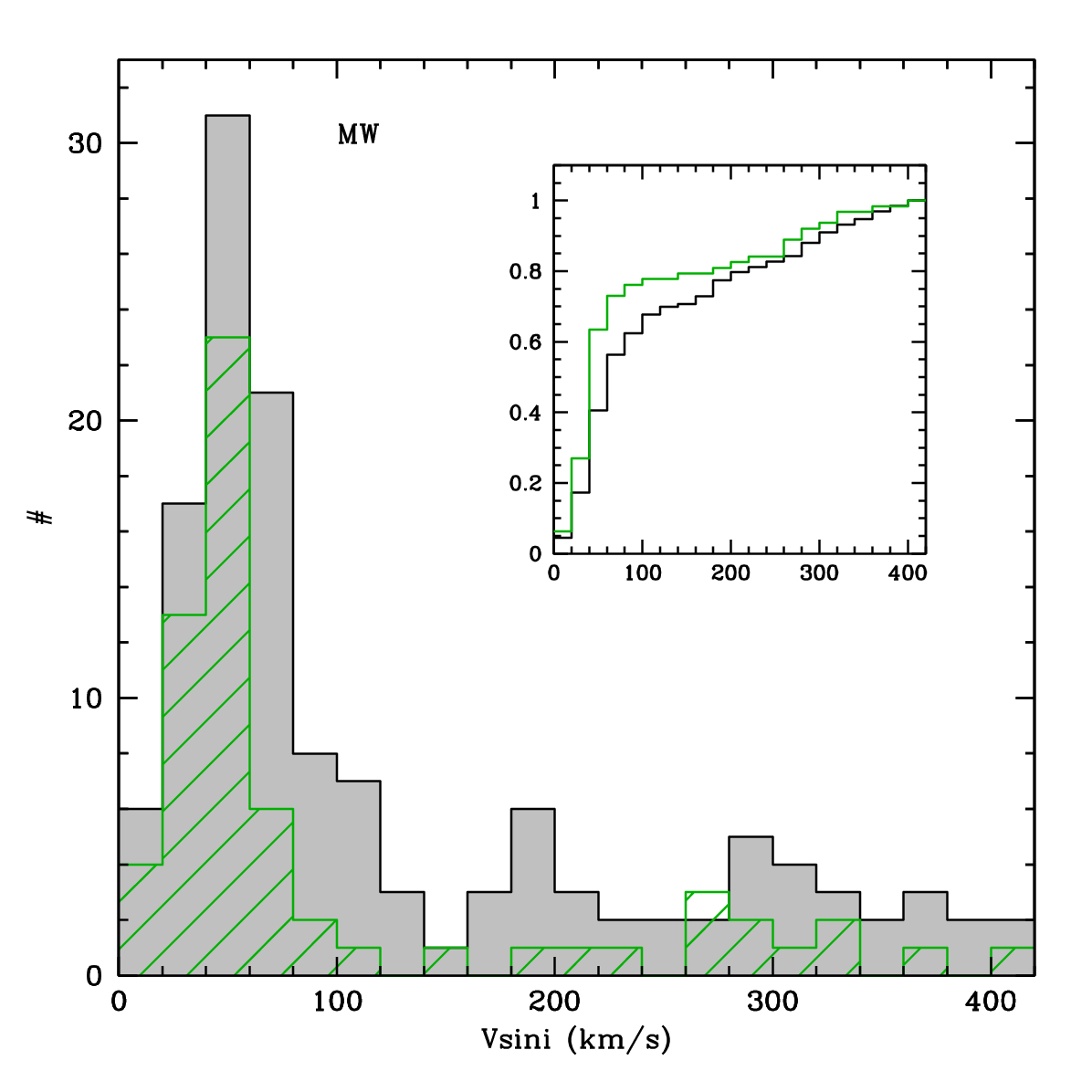}
\caption{Projected rotational velocity distribution in the SMC (left), LMC (middle), and the Galaxy (right). The blue, purple, and green histograms corresponds to the XshootU plus complementary samples. The grey histograms show the O6-9.5 V-II stars of \citet{mokiem06} in the SMC, \citet{ramirez13} in the LMC, and of \citet{holgado22} in the Galaxy. The inserts in each panel show the cumulative distribution functions.}  
\label{nb_vsini}
\end{figure*}

A classical diagram to test the predictions of evolutionary models that includes chemical and angular momentum transport caused by rotation is the N/H versus \vsini\ diagram. Several studies have used this diagram over the last fifteen years \citep{hunter08,hunter09,brott11b,rivero12,bouret13,grin17,dufton18,dufton20,markova18,bouret21}. As recalled in Section~\ref{s_intro} and stressed by \citet{maeder14} surface chemical enrichment does not depend only on \vsini, and thus comparing evolutionary models and observational properties requires population synthesis models \citep{brott11b}. 
Various sets of models treating differently transport and mixing processes have to be considered. It is also equally important to ensure that the samples to which these models are compared to are not biased and are representative of the global properties of massive stars.

In this Section we discuss the distribution of projected rotational velocities of our sample to investigate any potential bias. We show in Fig.~\ref{nb_vsini} the number of stars as a function of projected rotational velocity. In the LMC and MW we use a reference sample that contains a larger number of stars and no bias in the selection except the availability of high resolution optical spectra (i.e. the samples are magnitude limited). The reference samples are further refined to include only O6-9.5 stars with luminosity classes V to II. As such, we have the same distributions of spectral types and luminosity classes as in the samples we studied in the previous sections. We have also excluded all known binaries from the reference samples. 

In the Galaxy we use the sample of \citet{holgado22} as a reference. It is the largest sample of Galactic O stars for which projected rotational velocities were determined in a homogeneous way, using the Fourier transform method. We see in the right panel of Fig.~\ref{nb_vsini} that the distribution of \vsini\ in this reference sample is different from that that of the sample used in the present study (i.e. the XshootU + complementary sample, that for which surface abundance determinations exist). The latter contains 80\% of stars with projected rotational velocities below 80~\kms, while in the reference sample only 60\% have such low \vsini. In other words the  sample with available abundance determinations contains relatively more slow rotators than the reference sample.

The same conclusion applies even more strongly to the LMC stars for which surface nitrogen abundances have been determined. In the middle panel of Fig.~\ref{nb_vsini} we use the sample of \citet{ramirez13} as a reference. It includes most O stars of 30~Dor. As for the MW we see that the distributions of \vsini\ in the sample  for which N/H is available (purple lines) is biased towards low \vsini\ compared to the reference sample. 

In the SMC no large reference sample exist. In the left panel of Fig.~\ref{nb_vsini} we show how our XshootU plus complementary sample compares to that of \citet{mokiem06}. The latter sample contains less stars than the former but shows a flatter distribution. However the statistical significance of the difference is low given the small number of objects considered. \citet{penny09} also  determine projected rotational velocities for O stars in the SMC. But when considering only the relevant spectral types and luminosity classes, the remaining sample is even smaller than that of \citet{mokiem06}. Consequently we cannot safely answer the question of whether our XshootU plus complementary sample is biased towards slow rotators or not in the SMC.

\vspace{0.5cm}

From these investigations we conclude that the samples of stars for which surface abundances and projected rotational velocities are both available in the literature (including the present study)  appear to be biased towards slow rotators. The velocity distributions of the reference samples of \citet{ramirez13} and \citet{holgado22} show a high velocity component, above $\sim$150-200~\kms. \citet{demink13} predicted that such high velocities can be caused by interactions in binary systems. 
\citet{britavskiy23} report that stars with \vsini\ $>$ 200~\kms\ in the sample of \citet{holgado22} host less spectroscopic binaries than the slow velocity sub-sample. At the same time the high velocity sub-sample contains a fraction of runaway stars about twice as large as that of the slow velocity sub-sample. They argue that this feature can be explained by the scenario of \citet{demink13} in which the high runaway velocity is acquired because of the supernova explosion of the companion in binary systems, with subsequent disruption of the system \citep{blauw61}. However this is in contradiction with the theoretical work of \citet{renzo19} who predict that less than 2.6\% of binaries should produce single runaway stars. In that case the majority of runaway stars should be the result of dynamical interaction \citep{poveda67}. The origin of the high velocity tail of the \vsini\ distribution is thus not firmly established.

It thus appears that comparing the predictions of stellar evolution models with observed trends to constrain their ability to reproduce surface abundances and projected rotational velocities is a complex task.
It requires large samples including stars with high \vsini, which are lacking in the current samples for which both surface abundances and \vsini\ have been determined. Additionally it requires populations synthesis models that include both single and binary evolutionary channels with fair accounts of their (somehow uncertain) relative contributions. Different models that account differently for the transport of angular momentum and chemical species should be tested too, since they behave rather differently. This general task is beyond the scope of the present analysis that merely focuses on investigating the effect of metallicity on surface chemical enrichment.

\section{Conclusion}
\label{s_conc}

We have performed a spectroscopic analysis of 17 LMC and 17 SMC O-type stars using data from the ULLYSES and XshootU surveys. All stars are presumably single. We have determined the fundamental parameters (\teff, \logg, \lL) and the He, C, N, and O surface abundances. We have complemented these results by published values for additional stars in the MCs and the Galaxy. Our goal was to investigate the effect of metallicity on chemical transport triggered by stellar rotation. We show that the surface abundances of the stars analysed in this study are consistent with the products of nucleosynthesis through the CNO cycle. A clear increase of the surface nitrogen content, sometimes accompanied by an increase in the helium content, is detected as stars evolve to lower surface gravities, that is away from the ZAMS. The evolution of surface carbon and oxygen abundances is more difficult to constrain because of the small variations relative to the error bars on abundance measurements. The maximum increase of surface N/H relative to the baseline value is larger at lower metallicity. Nitrogen enrichment is observed at higher \logg\ in the SMC than in the Galaxy, in agreement with the predictions of evolutionary models including rotational mixing. No conclusion can be drawn for LMC objects due to the large dispersion in N/H measurements.
Our sample of stars for which both surface abundances and projected rotational velocities have been determined appear to be biased towards slow rotators, at least in the LMC and the Galaxy.
Future studies aiming at testing model predictions should rely on: 1) large samples with unbiased rotational velocity distributions; 2) large grids of evolutionary and population synthesis models that take into account various implementations of the physics of chemical mixing and angular momentum transport, as well as a mix of single and binary stars.

\section*{Acknowledgments}

We thank an anonymous referee for a positive report. We thank Rolf Kuiper, Laurent Mahy, Andrea Mehner, Grace Telford, Matheus Bernini for text corrections and suggestions of clarifications. Based on observations obtained with the NASA/ESA Hubble Space Telescope, retrieved from the Mikulski Archive for Space Telescopes (MAST) at the Space Telescope Science Institute (STScI). STScI is operated by the Association of Universities for Research in Astronomy, Inc. under NASA contract NAS 5-26555. DP acknowledges financial support by the Deutsches Zentrum f\"ur Luft und Raumfahrt (DLR) grant FKZ 50 OR 2005. AACS and VR acknowledge support by the Deutsche Forschungsgemeinschaft (DFG, German Research Foundation) in the form of an Emmy Noether Research Group -- Project-ID 445674056 (SA4064/1-1, PI Sander). AACS and VR further acknowledge support from the Federal Ministry of Education and Research (BMBF) and the Baden-Württemberg Ministry of Science as part of the Excellence Strategy of the German Federal and State Governments. F.N. acknowledges grants PID2019-105552RB-C41 and  PID2022-137779OB-C41 funded by the Spanish MCIN/AEI/ 10.13039/501100011033.

\bibliographystyle{aa}
\bibliography{xshoo_ab}

\begin{thebibliography}{78}
\expandafter\ifx\csname natexlab\endcsname\relax\def\natexlab#1{#1}\fi

\bibitem[{{Abbott} {et~al.}(2021){Abbott}, {Abbott}, {Abraham}, {Acernese},
  {Ackley}, {Adams}, {Adams}, {Adhikari}, {Adya}, {Affeldt}, {Agarwal},
  {Agathos}, {Agatsuma}, {Aggarwal}, {Aguiar}, {Aiello}, {Ain}, {Ajith},
  {Akutsu}, {Aleman}, {Allen}, {Allocca}, {Altin}, {Amato}, {Anand},
  {Ananyeva}, {Anderson}, {Anderson}, {Ando}, {Angelova}, {Ansoldi}, {Antelis},
  {Antier}, {Appert}, {Arai}, {Arai}, {Arai}, {Araki}, {Araya}, {Araya},
  {Areeda}, {Ar{\`e}ne}, {Aritomi}, {Arnaud}, {Aronson}, {Arun}, {Asada},
  {Asali}, {Ashton}, {Aso}, {Aston}, {Astone}, {Aubin}, {Aufmuth}, {Aultoneal},
  {Austin}, {Babak}, {Badaracco}, {Bader}, {Bae}, {Bae}, {Baer}, {Bagnasco},
  {Bai}, {Baiotti}, {Baird}, {Bajpai}, {Ball}, {Ballardin}, {Ballmer}, {Bals},
  {Balsamo}, {Baltus}, {Banagiri}, {Bankar}, {Bankar}, {Barayoga}, {Barbieri},
  {Barish}, {Barker}, {Barneo}, {Barone}, {Barr}, {Barsotti}, {Barsuglia},
  {Barta}, {Bartlett}, {Barton}, {Bartos}, {Bassiri}, {Basti}, {Bawaj},
  {Bayley}, {Baylor}, {Bazzan}, {B{\'e}csy}, {Bedakihale}, {Bejger},
  {Belahcene}, {Benedetto}, {Beniwal}, {Benjamin}, {Benkel}, {Bennett},
  {Bentley}, {Benyaala}, {Bergamin}, {Berger}, {Bernuzzi}, {Berry},
  {Bersanetti}, {Bertolini}, {Betzwieser}, {Bhandare}, {Bhandari},
  {Bhattacharjee}, {Bhaumik}, {Bidler}, {Bilenko}, {Billingsley}, {Birney},
  {Birnholtz}, {Biscans}, {Bischi}, {Biscoveanu}, {Bisht}, {Biswas}, {Bitossi},
  {Bizouard}, {Blackburn}, {Blackman}, {Blair}, {Blair}, {Blair}, {Bobba},
  {Bode}, {Boer}, {Bogaert}, {Boldrini}, {Bondu}, {Bonilla}, {Bonnand},
  {Booker}, {Boom}, {Bork}, {Boschi}, {Bose}, {Bose}, {Bossilkov}, {Boudart},
  {Bouffanais}, {Bozzi}, {Bradaschia}, {Brady}, {Bramley}, {Branch},
  {Branchesi}, {Brau}, {Breschi}, {Briant}, {Briggs}, {Brillet}, {Brinkmann},
  {Brockill}, {Brooks}, {Brooks}, {Brown}, {Brunett}, {Bruno}, {Bruntz},
  {Bryant}, {Buikema}, {Bulik}, {Bulten}, {Buonanno}, {Buscicchio}, {Buskulic},
  {Byer}, {Cadonati}, {Caesar}, {Cagnoli}, {Cahillane}, {Cain}, {Calder{\'o}n
  Bustillo}, {Callaghan}, {Callister}, {Calloni}, {Camp}, {Canepa},
  {Cannavacciuolo}, {Cannon}, {Cao}, {Cao}, {Cao}, {Capocasa}, {Capote},
  {Carapella}, {Carbognani}, {Carlin}, {Carney}, {Carpinelli}, {Carullo},
  {Carver}, {Casanueva Diaz}, {Casentini}, {Castaldi}, {Caudill},
  {Cavagli{\`a}}, {Cavalier}, {Cavalieri}, {Cella}, {Cerd{\'a}-Dur{\'a}n},
  {Cesarini}, {Chaibi}, {Chakravarti}, {Champion}, {Chan}, {Chan}, {Chan},
  {Chan}, {Chandra}, {Chanial}, {Chao}, {Charlton}, {Chase},
  {Chassande-Mottin}, {Chatterjee}, {Chaturvedi}, {Chatziioannou}, {Chen},
  {Chen}, {Chen}, {Chen}, {Chen}, {Chen}, {Chen}, {Chen}, {Chen}, {Cheng},
  {Cheong}, {Cheung}, {Chia}, {Chiadini}, {Chiang}, {Chierici}, {Chincarini},
  {Chiofalo}, {Chiummo}, {Cho}, {Cho}, {Choate}, {Choudhary}, {Choudhary},
  {Christensen}, {Chu}, {Chu}, {Chu}, {Chua}, {Chung}, {Ciani}, {Ciecielag},
  {Cie{\'s}lar}, {Cifaldi}, {Ciobanu}, {Ciolfi}, {Cipriano}, {Cirone}, {Clara},
  {Clark}, {Clark}, {Clarke}, {Clearwater}, {Clesse}, {Cleva}, {Coccia},
  {Cohadon}, {Cohen}, {Cohen}, {Colleoni}, {Collette}, {Colpi}, {Compton},
  {Constancio}, {Conti}, {Cooper}, {Corban}, {Corbitt}, {Cordero-Carri{\'o}n},
  {Corezzi}, {Corley}, {Cornish}, {Corre}, {Corsi}, {Cortese}, {Costa},
  {Cotesta}, {Coughlin}, {Coughlin}, {Coulon}, {Countryman}, {Cousins},
  {Couvares}, {Covas}, {Coward}, {Cowart}, {Coyne}, {Coyne}, {Creighton},
  {Creighton}, {Criswell}, {Croquette}, {Crowder}, {Cudell}, {Cullen},
  {Cumming}, {Cummings}, {Cuoco}, {Cury{\l}o}, {Dal Canton}, {D{\'a}lya},
  {Dana}, {Daneshgaranbajastani}, {D'Angelo}, {Danilishin}, {D'Antonio},
  {Danzmann}, {Darsow-Fromm}, {Dasgupta}, {Datrier}, {Dattilo}, {Dave},
  {Davier}, {Davies}, {Davis}, {Daw}, {Dean}, {Debra}, {Deenadayalan},
  {Degallaix}, {de Laurentis}, {Del{\'e}glise}, {Del Favero}, {de Lillo}, {de
  Lillo}, {Del Pozzo}, {Demarchi}, {de Matteis}, {D'Emilio}, {Demos}, {Dent},
  {Depasse}, {de Pietri}, {De Rosa}, {de Rossi}, {Desalvo}, {de Simone},
  {Dhurandhar}, {D{\'\i}az}, {Diaz-Ortiz}, {Didio}, {Dietrich}, {di Fiore}, {di
  Fronzo}, {di Giorgio}, {di Giovanni}, {di Girolamo}, {di Lieto}, {Ding}, {di
  Pace}, {di Palma}, {di Renzo}, {Divakarla}, {Dmitriev}, {Doctor},
  {D'Onofrio}, {Donovan}, {Dooley}, {Doravari}, {Dorrington}, {Drago},
  {Driggers}, {Drori}, {Du}, {Ducoin}, {Dupej}, {Durante}, {D'Urso}, {Duverne},
  {Dwyer}, {Easter}, {Ebersold}, {Eddolls}, {Edelman}, {Edo}, {Edy}, {Effler},
  {Eguchi}, {Eichholz}, {Eikenberry}, {Eisenmann}, {Eisenstein}, {Ejlli},
  {Enomoto}, {Errico}, {Essick}, {Estell{\'e}s}, {Estevez}, {Etienne}, {Etzel},
  {Evans}, {Evans}, {Ewing}, {Fafone}, {Fair}, {Fairhurst}, {Fan}, {Farah},
  {Farinon}, {Farr}, {Farr}, {Farrow}, {Fauchon-Jones}, {Favata}, {Fays},
  {Fazio}, {Feicht}, {Fejer}, {Feng}, {Fenyvesi}, {Ferguson},
  {Fernandez-Galiana}, {Ferrante}, {Ferreira}, {Fidecaro}, {Figura}, {Fiori},
  {Fishbach}, {Fisher}, {Fittipaldi}, {Fiumara}, {Flaminio}, {Floden}, {Flynn},
  {Fong}, {Font}, {Fornal}, {Forsyth}, {Franke}, {Frasca}, {Frasconi},
  {Frederick}, {Frei}, {Freise}, {Frey}, {Fritschel}, {Frolov}, {Fronz{\'e}},
  {Fujii}, {Fujikawa}, {Fukunaga}, {Fukushima}, {Fulda}, {Fyffe}, {Gabbard},
  {Gadre}, {Gaebel}, {Gair}, {Gais}, {Galaudage}, {Gamba}, {Ganapathy},
  {Ganguly}, {Gao}, {Gaonkar}, {Garaventa}, {Garc{\'\i}a-N{\'u}{\~n}ez},
  {Garc{\'\i}a-Quir{\'o}s}, {Garufi}, {Gateley}, {Gaudio}, {Gayathri}, {Ge},
  {Gemme}, {Gennai}, {George}, {Gergely}, {Gewecke}, {Ghonge}, {Ghosh},
  {Ghosh}, {Ghosh}, {Ghosh}, {Ghosh}, {Giacomazzo}, {Giacoppo}, {Giaime},
  {Giardina}, {Gibson}, {Gier}, {Giesler}, {Giri}, {Gissi}, {Glanzer},
  {Gleckl}, {Godwin}, {Goetz}, {Goetz}, {Gohlke}, {Goncharov}, {Gonz{\'a}lez},
  {Gopakumar}, {Gosselin}, {Gouaty}, {Grace}, {Grado}, {Granata}, {Granata},
  {Grant}, {Gras}, {Grassia}, {Gray}, {Gray}, {Greco}, {Green}, {Green},
  {Gretarsson}, {Gretarsson}, {Griffith}, {Griffiths}, {Griggs}, {Grignani},
  {Grimaldi}, {Grimes}, {Grimm}, {Grote}, {Grunewald}, {Gruning}, {Guerrero},
  {Guidi}, {Guimaraes}, {Guix{\'e}}, {Gulati}, {Guo}, {Guo}, {Gupta}, {Gupta},
  {Gupta}, {Gustafson}, {Gustafson}, {Guzman}, {Ha}, {Haegel}, {Hagiwara},
  {Haino}, {Halim}, {Hall}, {Hamilton}, {Hammond}, {Han}, {Haney}, {Hanks},
  {Hanna}, {Hannam}, {Hannuksela}, {Hansen}, {Hansen}, {Hanson}, {Harder},
  {Hardwick}, {Haris}, {Harms}, {Harry}, {Harry}, {Hartwig}, {Hasegawa},
  {Haskell}, {Hasskew}, {Haster}, {Hattori}, {Haughian}, {Hayakawa}, {Hayama},
  {Hayes}, {Healy}, {Heidmann}, {Heintze}, {Heinze}, {Heinzel}, {Heitmann},
  {Hellman}, {Hello}, {Helmling-Cornell}, {Hemming}, {Hendry}, {Heng},
  {Hennes}, {Hennig}, {Hennig}, {Hernandez Vivanco}, {Heurs}, {Hild}, {Hill},
  {Himemoto}, {Hinderer}, {Hines}, {Hiranuma}, {Hirata}, {Hirose}, {Ho},
  {Hochheim}, {Hofman}, {Hohmann}, {Holgado}, {Holland}, {Hollows}, {Holmes},
  {Holt}, {Holz}, {Hong}, {Hopkins}, {Hough}, {Howell}, {Hoy}, {Hoyland},
  {Hreibi}, {Hsieh}, {Hsu}, {Huang}, {Huang}, {Huang}, {Huang}, {Huang},
  {Huang}, {H{\"u}bner}, {Huddart}, {Huerta}, {Hughey}, {Hui}, {Hui}, {Husa},
  {Huttner}, {Huxford}, {Huynh-Dinh}, {Ide}, {Idzkowski}, {Iess}, {Ikenoue},
  {Imam}, {Inayoshi}, {Inchauspe}, {Ingram}, {Inoue}, {Intini}, {Ioka}, {Isi},
  {Isleif}, {Ito}, {Itoh}, {Iyer}, {Izumi}, {Jaberianhamedan}, {Jacqmin},
  {Jadhav}, {Jadhav}, {James}, {Jan}, {Jani}, {Janssens}, {Janthalur},
  {Jaranowski}, {Jariwala}, {Jaume}, {Jenkins}, {Jeon}, {Jeunon}, {Jia},
  {Jiang}, {Jin}, {Johns}, {Jones}, {Jones}, {Jones}, {Jones}, {Jones},
  {Jonker}, {Ju}, {Jung}, {Jung}, {Junker}, {Kaihotsu}, {Kajita}, {Kakizaki},
  {Kalaghatgi}, {Kalogera}, {Kamai}, {Kamiizumi}, {Kanda}, {Kandhasamy},
  {Kang}, {Kanner}, {Kao}, {Kapadia}, {Kapasi}, {Karat}, {Karathanasis},
  {Karki}, {Kashyap}, {Kasprzack}, {Kastaun}, {Katsanevas}, {Katsavounidis},
  {Katzman}, {Kaur}, {Kawabe}, {Kawaguchi}, {Kawai}, {Kawasaki},
  {K{\'e}f{\'e}lian}, {Keitel}, {Key}, {Khadka}, {Khalili}, {Khan}, {Khan},
  {Khazanov}, {Khetan}, {Khursheed}, {Kijbunchoo}, {Kim}, {Kim}, {Kim}, {Kim},
  {Kim}, {Kim}, {Kimball}, {Kimura}, {King}, {Kinley-Hanlon}, {Kirchhoff},
  {Kissel}, {Kita}, {Kitazawa}, {Kleybolte}, {Klimenko}, {Knee}, {Knowles},
  {Knyazev}, {Koch}, {Koekoek}, {Kojima}, {Kokeyama}, {Koley}, {Kolitsidou},
  {Kolstein}, {Komori}, {Kondrashov}, {Kong}, {Kontos}, {Koper}, {Korobko},
  {Kotake}, {Kovalam}, {Kozak}, {Kozakai}, {Kozu}, {Kringel}, {Krishnendu},
  {Kr{\'o}lak}, {Kuehn}, {Kuei}, {Kumar}, {Kumar}, {Kumar}, {Kumar}, {Kume},
  {Kuns}, {Kuo}, {Kuo}, {Kuromiya}, {Kuroyanagi}, {Kusayanagi}, {Kwak},
  {Kwang}, {Laghi}, {Lalande}, {Lam}, {Lamberts}, {Landry}, {Landry}, {Lane},
  {Lang}, {Lange}, {Lantz}, {La Rosa}, {Lartaux-Vollard}, {Lasky}, {Laxen},
  {Lazzarini}, {Lazzaro}, {Leaci}, {Leavey}, {Lecoeuche}, {Lee}, {Lee}, {Lee},
  {Lee}, {Lee}, {Lee}, {Lehmann}, {Lema{\^\i}tre}, {Leon}, {Leonardi}, {Leroy},
  {Letendre}, {Levin}, {Leviton}, {Li}, {Li}, {Li}, {Li}, {Li}, {Li}, {Lin},
  {Lin}, {Lin}, {Lin}, {Lin}, {Linde}, {Linker}, {Linley}, {Littenberg}, {Liu},
  {Liu}, {Liu}, {Liu}, {Llorens-Monteagudo}, {Lo}, {Lockwood}, {Lollie},
  {London}, {Longo}, {Lopez}, {Lorenzini}, {Loriette}, {Lormand}, {Losurdo},
  {Lough}, {Lousto}, {Lovelace}, {L{\"u}ck}, {Lumaca}, {Lundgren}, {Luo},
  {Macas}, {Macinnis}, {MacLeod}, {MacMillan}, {Macquet}, {Maga{\~n}a
  Hernandez}, {Maga{\~n}a-Sandoval}, {Magazz{\`u}}, {Magee}, {Maggiore},
  {Majorana}, {Makarem}, {Maksimovic}, {Maliakal}, {Malik}, {Man}, {Mandic},
  {Mangano}, {Mango}, {Mansell}, {Manske}, {Mantovani}, {Mapelli},
  {Marchesoni}, {Marchio}, {Marion}, {Mark}, {M{\'a}rka}, {M{\'a}rka},
  {Markakis}, {Markosyan}, {Markowitz}, {Maros}, {Marquina}, {Marsat},
  {Martelli}, {Martin}, {Martin}, {Martinez}, {Martinez}, {Martinovic},
  {Martynov}, {Marx}, {Masalehdan}, {Mason}, {Massera}, {Masserot},
  {Massinger}, {Masso-Reid}, {Mastrogiovanni}, {Matas}, {Mateu-Lucena},
  {Matichard}, {Matiushechkina}, {Mavalvala}, {McCann}, {McCarthy},
  {McClelland}, {McClincy}, {McCormick}, {McCuller}, {McGhee}, {McGuire},
  {McIsaac}, {McIver}, {McManus}, {McRae}, {McWilliams}, {Meacher}, {Mehmet},
  {Mehta}, {Melatos}, {Melchor}, {Mendell}, {Menendez-Vazquez}, {Menoni},
  {Mercer}, {Mereni}, {Merfeld}, {Merilh}, {Merritt}, {Merzougui}, {Meshkov},
  {Messenger}, {Messick}, {Meyers}, {Meylahn}, {Mhaske}, {Miani}, {Miao},
  {Michaloliakos}, {Michel}, {Michimura}, {Middleton}, {Milano}, {Miller},
  {Millhouse}, {Mills}, {Milotti}, {Milovich-Goff}, {Minazzoli}, {Minenkov},
  {Mio}, {Mir}, {Mishkin}, {Mishra}, {Mishra}, {Mistry}, {Mitra}, {Mitrofanov},
  {Mitselmakher}, {Mittleman}, {Miyakawa}, {Miyamoto}, {Miyazaki}, {Miyo},
  {Miyoki}, {Mo}, {Mogushi}, {Mohapatra}, {Mohite}, {Molina}, {Molina-Ruiz},
  {Mondin}, {Montani}, {Moore}, {Moraru}, {Morawski}, {More}, {Moreno},
  {Moreno}, {Mori}, {Morisaki}, {Moriwaki}, {Mours}, {Mow-Lowry}, {Mozzon},
  {Muciaccia}, {Mukherjee}, {Mukherjee}, {Mukherjee}, {Mukherjee}, {Mukund},
  {Mullavey}, {Munch}, {Mu{\~n}iz}, {Murray}, {Musenich}, {Nadji}, {Nagano},
  {Nagano}, {Nagar}, {Nakamura}, {Nakano}, {Nakano}, {Nakashima}, {Nakayama},
  {Nardecchia}, {Narikawa}, {Naticchioni}, {Nayak}, {Nayak}, {Negishi}, {Neil},
  {Neilson}, {Nelemans}, {Nelson}, {Nery}, {Neunzert}, {Ng}, {Ng}, {Nguyen},
  {Nguyen}, {Nguyen}, {Nguyen Quynh}, {Ni}, {Nichols}, {Nishizawa}, {Nissanke},
  {Nocera}, {Noh}, {Norman}, {North}, {Nozaki}, {Nuttall}, {Oberling},
  {O'Brien}, {Obuchi}, {O'Dell}, {Ogaki}, {Oganesyan}, {Oh}, {Oh}, {Oh},
  {Ohashi}, {Ohishi}, {Ohkawa}, {Ohme}, {Ohta}, {Okada}, {Okutani}, {Okutomi},
  {Olivetto}, {Oohara}, {Ooi}, {Oram}, {O'Reilly}, {Ormiston}, {Ormsby},
  {Ortega}, {O'Shaughnessy}, {O'Shea}, {Oshino}, {Ossokine}, {Osthelder},
  {Otabe}, {Ottaway}, {Overmier}, {Pace}, {Pagano}, {Page}, {Pagliaroli},
  {Pai}, {Pai}, {Palamos}, {Palashov}, {Palomba}, {Pan}, {Panda}, {Pang},
  {Pang}, {Pankow}, {Pannarale}, {Pant}, {Paoletti}, {Paoli}, {Paolone},
  {Parisi}, {Park}, {Parker}, {Pascucci}, {Pasqualetti}, {Passaquieti},
  {Passuello}, {Patel}, {Patricelli}, {Payne}, {Pechsiri}, {Pedraza},
  {Pegoraro}, {Pele}, {Pe{\~n}a Arellano}, {Penn}, {Perego}, {Pereira},
  {Pereira}, {Perez}, {P{\'e}rigois}, {Perreca}, {Perri{\`e}s}, {Petermann},
  {Petterson}, {Pfeiffer}, {Pham}, {Phukon}, {Piccinni}, {Pichot},
  {Piendibene}, {Piergiovanni}, {Pierini}, {Pierro}, {Pillant}, {Pilo},
  {Pinard}, {Pinto}, {Piotrzkowski}, {Piotrzkowski}, {Pirello}, {Pitkin},
  {Placidi}, {Plastino}, {Pluchar}, {Poggiani}, {Polini}, {Pong}, {Ponrathnam},
  {Popolizio}, {Porter}, {Powell}, {Pracchia}, {Pradier}, {Prajapati},
  {Prasai}, {Prasanna}, {Pratten}, {Prestegard}, {Principe}, {Prodi},
  {Prokhorov}, {Prosposito}, {Prudenzi}, {Puecher}, {Punturo}, {Puosi},
  {Puppo}, {P{\"u}rrer}, {Qi}, {Quetschke}, {Quinonez}, {Quitzow-James},
  {Raab}, {Raaijmakers}, {Radkins}, {Radulesco}, {Raffai}, {Rail}, {Raja},
  {Rajan}, {Ramirez}, {Ramirez}, {Ramos-Buades}, {Rana}, {Rapagnani}, {Rapol},
  {Ratto}, {Ray}, {Raymond}, {Raza}, {Razzano}, {Read}, {Rees}, {Regimbau},
  {Rei}, {Reid}, {Reitze}, {Relton}, {Rettegno}, {Ricci}, {Richardson},
  {Richardson}, {Richardson}, {Ricker}, {Riemenschneider}, {Riles}, {Rizzo},
  {Robertson}, {Robie}, {Robinet}, {Rocchi}, {Rocha}, {Rodriguez},
  {Rodriguez-Soto}, {Rolland}, {Rollins}, {Roma}, {Romanelli}, {Romano},
  {Romel}, {Romero}, {Romero-Shaw}, {Romie}, {Rose}, {Rosi{\'n}ska},
  {Rosofsky}, {Ross}, {Rowan}, {Rowlinson}, {Roy}, {Roy}, {Rozza}, {Ruggi},
  {Ryan}, {Sachdev}, {Sadecki}, {Sadiq}, {Sago}, {Saito}, {Saito}, {Sakai},
  {Sakai}, {Sakellariadou}, {Sakuno}, {Salafia}, {Salconi}, {Saleem}, {Salemi},
  {Samajdar}, {Sanchez}, {Sanchez}, {Sanchez}, {Sanchis-Gual}, {Sanders},
  {Sanuy}, {Saravanan}, {Sarin}, {Sassolas}, {Satari}, {Sathyaprakash}, {Sato},
  {Sato}, {Sauter}, {Savage}, {Savant}, {Sawada}, {Sawant}, {Sawant}, {Sayah},
  {Schaetzl}, {Scheel}, {Scheuer}, {Schindler-Tyka}, {Schmidt}, {Schnabel},
  {Schneewind}, {Schofield}, {Sch{\"o}nbeck}, {Schulte}, {Schutz}, {Schwartz},
  {Scott}, {Scott}, {Seglar-Arroyo}, {Seidel}, {Sekiguchi}, {Sekiguchi},
  {Sellers}, {Sengupta}, {Sennett}, {Sentenac}, {Seo}, {Sequino}, {Sergeev},
  {Setyawati}, {Shaffer}, {Shahriar}, {Shams}, {Shao}, {Sharifi}, {Sharma},
  {Sharma}, {Shawhan}, {Shcheblanov}, {Shen}, {Shibagaki}, {Shikauchi},
  {Shimizu}, {Shimoda}, {Shimode}, {Shink}, {Shinkai}, {Shishido}, {Shoda},
  {Shoemaker}, {Shoemaker}, {Shukla}, {Shyamsundar}, {Sieniawska}, {Sigg},
  {Singer}, {Singh}, {Singh}, {Singha}, {Sintes}, {Sipala}, {Skliris},
  {Slagmolen}, {Slaven-Blair}, {Smetana}, {Smith}, {Smith}, {Somala}, {Somiya},
  {Son}, {Soni}, {Soni}, {Sorazu}, {Sordini}, {Sorrentino}, {Sorrentino},
  {Sotani}, {Soulard}, {Souradeep}, {Sowell}, {Spagnuolo}, {Spencer}, {Spera},
  {Srivastava}, {Srivastava}, {Staats}, {Stachie}, {Steer}, {Steinlechner},
  {Steinlechner}, {Stops}, {Stevenson}, {Stover}, {Strain}, {Strang},
  {Stratta}, {Strunk}, {Sturani}, {Stuver}, {S{\"u}dbeck}, {Sudhagar},
  {Sudhir}, {Sugimoto}, {Suh}, {Summerscales}, {Sun}, {Sun}, {Sunil}, {Sur},
  {Suresh}, {Sutton}, {Suzuki}, {Suzuki}, {Swinkels}, {Szczepa{\'n}czyk},
  {Szewczyk}, {Tacca}, {Tagoshi}, {Tait}, {Takahashi}, {Takahashi}, {Takamori},
  {Takano}, {Takeda}, {Takeda}, {Talbot}, {Tanaka}, {Tanaka}, {Tanaka},
  {Tanaka}, {Tanaka}, {Tanasijczuk}, {Tanioka}, {Tanner}, {Tao}, {Tapia},
  {Tapia San Martin}, {Tasson}, {Telada}, {Tenorio}, {Terkowski}, {Test},
  {Thirugnanasambandam}, {Thomas}, {Thomas}, {Thompson}, {Thondapu}, {Thorne},
  {Thrane}, {Tiwari}, {Tiwari}, {Tiwari}, {Toland}, {Tolley}, {Tomaru},
  {Tomigami}, {Tomura}, {Tonelli}, {Torres-Forn{\'e}}, {Torrie}, {Tosta E
  Melo}, {T{\"o}yr{\"a}}, {Trapananti}, {Travasso}, {Traylor}, {Tringali},
  {Tripathee}, {Troiano}, {Trovato}, {Trozzo}, {Trudeau}, {Tsai}, {Tsai},
  {Tsang}, {Tsang}, {Tsao}, {Tse}, {Tso}, {Tsubono}, {Tsuchida}, {Tsukada},
  {Tsuna}, {Tsutsui}, {Tsuzuki}, {Turconi}, {Tuyenbayev}, {Ubhi}, {Uchikata},
  {Uchiyama}, {Udall}, {Ueda}, {Uehara}, {Ueno}, {Ueshima}, {Ugolini},
  {Unnikrishnan}, {Uraguchi}, {Urban}, {Ushiba}, {Usman}, {Utina}, {Vahlbruch},
  {Vajente}, {Vajpeyi}, {Valdes}, {Valentini}, {Valsan}, {van Bakel}, {van
  Beuzekom}, {van den Brand}, {van den Broeck}, {Vander-Hyde}, {van der
  Schaaf}, {van Heijningen}, {Vanosky}, {van Putten}, {Vardaro}, {Vargas},
  {Varma}, {Vas{\'u}th}, {Vecchio}, {Vedovato}, {Veitch}, {Veitch},
  {Venkateswara}, {Venneberg}, {Venugopalan}, {Verkindt}, {Verma}, {Veske},
  {Vetrano}, {Vicer{\'e}}, {Viets}, {Villa-Ortega}, {Vinet}, {Vitale}, {Vo},
  {Vocca}, {von Reis}, {von Wrangel}, {Vorvick}, {Vyatchanin}, {Wade}, {Wade},
  {Wagner}, {Walet}, {Walker}, {Wallace}, {Wallace}, {Walsh}, {Wang}, {Wang},
  {Wang}, {Ward}, {Warner}, {Was}, {Washimi}, {Washington}, {Watchi}, {Weaver},
  {Wei}, {Weinert}, {Weinstein}, {Weiss}, {Weller}, {Wellmann}, {Wen},
  {We{\ss}els}, {Westhouse}, {Wette}, {Whelan}, {White}, {Whiting}, {Whittle},
  {Wilken}, {Williams}, {Williams}, {Williamson}, {Willis}, {Willke}, {Wilson},
  {Winkler}, {Wipf}, {Wlodarczyk}, {Woan}, {Woehler}, {Wofford}, {Wong}, {Wu},
  {Wu}, {Wu}, {Wu}, {Wysocki}, {Xiao}, {Xu}, {Yamada}, {Yamamoto}, {Yamamoto},
  {Yamamoto}, {Yamamoto}, {Yamashita}, {Yamazaki}, {Yang}, {Yang}, {Yang},
  {Yang}, {Yang}, {Yap}, {Yeeles}, {Yelikar}, {Ying}, {Yokogawa}, {Yokoyama},
  {Yokozawa}, {Yoon}, {Yoshioka}, {Yu}, {Yu}, {Yuzurihara}, {Zadro{\.z}ny},
  {Zanolin}, {Zappa}, {Zeidler}, {Zelenova}, {Zendri}, {Zevin}, {Zhan},
  {Zhang}, {Zhang}, {Zhang}, {Zhang}, {Zhang}, {Zhao}, {Zhao}, {Zhao}, {Zhao},
  {Zhou}, {Zhu}, {Zhu}, {Zimmerman}, {Zlochower}, {Zucker}, {Zweizig}, {Ligo
  Scientific Collaboration}, {VIRGO Collaboration}, \& {KAGRA
  Collaboration}}]{abbott21}
{Abbott}, R., {Abbott}, T.~D., {Abraham}, S., {et~al.} 2021, \apjl, 915, L5

\bibitem[{{Asplund} {et~al.}(2009){Asplund}, {Grevesse}, {Sauval}, \&
  {Scott}}]{a09}
{Asplund}, M., {Grevesse}, N., {Sauval}, A.~J., \& {Scott}, P. 2009, \araa, 47,
  481

\bibitem[{{Bertelli} {et~al.}(1984){Bertelli}, {Bressan}, \&
  {Chiosi}}]{bertelli84}
{Bertelli}, G., {Bressan}, A.~G., \& {Chiosi}, C. 1984, \aap, 130, 279

\bibitem[{{Blaauw}(1961)}]{blauw61}
{Blaauw}, A. 1961, \bain, 15, 265

\bibitem[{{Bouret} {et~al.}(2013){Bouret}, {Lanz}, {Martins}, {Marcolino},
  {Hillier}, {Depagne}, \& {Hubeny}}]{bouret13}
{Bouret}, J.~C., {Lanz}, T., {Martins}, F., {et~al.} 2013, \aap, 555, A1

\bibitem[{{Bouret} {et~al.}(2021){Bouret}, {Martins}, {Hillier}, {Marcolino},
  {Rocha-Pinto}, {Georgy}, {Lanz}, \& {Hubeny}}]{bouret21}
{Bouret}, J.~C., {Martins}, F., {Hillier}, D.~J., {et~al.} 2021, \aap, 647,
  A134

\bibitem[{{Britavskiy} {et~al.}(2023){Britavskiy}, {Sim{\'o}n-D{\'\i}az},
  {Holgado}, {Burssens}, {Ma{\'\i}z Apell{\'a}niz}, {Eldridge}, {Naz{\'e}},
  {Pantaleoni Gonz{\'a}lez}, \& {Herrero}}]{britavskiy23}
{Britavskiy}, N., {Sim{\'o}n-D{\'\i}az}, S., {Holgado}, G., {et~al.} 2023,
  \aap, 672, A22

\bibitem[{{Brott} {et~al.}(2011{\natexlab{a}}){Brott}, {de Mink}, {Cantiello},
  {Langer}, {de Koter}, {Evans}, {Hunter}, {Trundle}, \& {Vink}}]{brott11}
{Brott}, I., {de Mink}, S.~E., {Cantiello}, M., {et~al.} 2011{\natexlab{a}},
  \aap, 530, A115

\bibitem[{{Brott} {et~al.}(2011{\natexlab{b}}){Brott}, {Evans}, {Hunter}, {de
  Koter}, {Langer}, {Dufton}, {Cantiello}, {Trundle}, {Lennon}, {de Mink},
  {Yoon}, \& {Anders}}]{brott11b}
{Brott}, I., {Evans}, C.~J., {Hunter}, I., {et~al.} 2011{\natexlab{b}}, \aap,
  530, A116

\bibitem[{{Cazorla} {et~al.}(2017){Cazorla}, {Morel}, {Naz{\'e}}, {Rauw},
  {Semaan}, {Daflon}, \& {Oey}}]{cazorla17a}
{Cazorla}, C., {Morel}, T., {Naz{\'e}}, Y., {et~al.} 2017, \aap, 603, A56

\bibitem[{{Chiosi} \& {Maeder}(1986)}]{cm86}
{Chiosi}, C. \& {Maeder}, A. 1986, \araa, 24, 329

\bibitem[{{de Mink} {et~al.}(2013){de Mink}, {Langer}, {Izzard}, {Sana}, \& {de
  Koter}}]{demink13}
{de Mink}, S.~E., {Langer}, N., {Izzard}, R.~G., {Sana}, H., \& {de Koter}, A.
  2013, \apj, 764, 166

\bibitem[{{Dufton} {et~al.}(2020){Dufton}, {Evans}, {Lennon}, \&
  {Hunter}}]{dufton20}
{Dufton}, P.~L., {Evans}, C.~J., {Lennon}, D.~J., \& {Hunter}, I. 2020, \aap,
  634, A6

\bibitem[{{Dufton} {et~al.}(2018){Dufton}, {Thompson}, {Crowther}, {Evans},
  {Schneider}, {de Koter}, {de Mink}, {Garland}, {Langer}, {Lennon}, {McEvoy},
  {Ram{\'\i}rez-Agudelo}, {Sana}, {S{\'\i}mon D{\'\i}az}, {Taylor}, \&
  {Vink}}]{dufton18}
{Dufton}, P.~L., {Thompson}, A., {Crowther}, P.~A., {et~al.} 2018, \aap, 615,
  A101

\bibitem[{{Eggenberger} {et~al.}(2021){Eggenberger}, {Ekstr{\"o}m}, {Georgy},
  {Martinet}, {Pezzotti}, {Nandal}, {Meynet}, {Buldgen}, {Salmon},
  {Haemmerl{\'e}}, {Maeder}, {Hirschi}, {Yusof}, {Groh}, {Farrell}, {Murphy},
  \& {Choplin}}]{eggen21}
{Eggenberger}, P., {Ekstr{\"o}m}, S., {Georgy}, C., {et~al.} 2021, \aap, 652,
  A137

\bibitem[{{Ekstr{\"o}m} {et~al.}(2012){Ekstr{\"o}m}, {Georgy}, {Eggenberger},
  {Meynet}, {Mowlavi}, {Wyttenbach}, {Granada}, {Decressin}, {Hirschi},
  {Frischknecht}, {Charbonnel}, \& {Maeder}}]{ek12}
{Ekstr{\"o}m}, S., {Georgy}, C., {Eggenberger}, P., {et~al.} 2012, \aap, 537,
  A146

\bibitem[{{Ekstr{\"o}m} {et~al.}(2008){Ekstr{\"o}m}, {Meynet}, {Maeder}, \&
  {Barblan}}]{ek08}
{Ekstr{\"o}m}, S., {Meynet}, G., {Maeder}, A., \& {Barblan}, F. 2008, \aap,
  478, 467

\bibitem[{{El Eid} {et~al.}(1983){El Eid}, {Fricke}, \& {Ober}}]{eleid83}
{El Eid}, M.~F., {Fricke}, K.~J., \& {Ober}, W.~W. 1983, \aap, 119, 54

\bibitem[{{Eldridge} \& {Stanway}(2022)}]{eldridge22}
{Eldridge}, J.~J. \& {Stanway}, E.~R. 2022, \araa, 60, 455

\bibitem[{{Evans} {et~al.}(2004){Evans}, {Crowther}, {Fullerton}, \&
  {Hillier}}]{evans04}
{Evans}, C.~J., {Crowther}, P.~A., {Fullerton}, A.~W., \& {Hillier}, D.~J.
  2004, \apj, 610, 1021

\bibitem[{{Georgy} {et~al.}(2013){Georgy}, {Ekstr{\"o}m}, {Eggenberger},
  {Meynet}, {Haemmerl{\'e}}, {Maeder}, {Granada}, {Groh}, {Hirschi}, {Mowlavi},
  {Yusof}, {Charbonnel}, {Decressin}, \& {Barblan}}]{georgy13}
{Georgy}, C., {Ekstr{\"o}m}, S., {Eggenberger}, P., {et~al.} 2013, \aap, 558,
  A103

\bibitem[{{Grasha} {et~al.}(2021){Grasha}, {Roy}, {Sutherland}, \&
  {Kewley}}]{grasha21}
{Grasha}, K., {Roy}, A., {Sutherland}, R.~S., \& {Kewley}, L.~J. 2021, \apj,
  908, 241

\bibitem[{{Gray}(1976)}]{gray}
{Gray}, D.~F. 1976, {The observation and analysis of stellar photospheres}

\bibitem[{{Grin} {et~al.}(2017){Grin}, {Ram{\'\i}rez-Agudelo}, {de Koter},
  {Sana}, {Puls}, {Brott}, {Crowther}, {Dufton}, {Evans}, {Gr{\"a}fener},
  {Herrero}, {Langer}, {Lennon}, {van Loon}, {Markova}, {de Mink}, {Najarro},
  {Schneider}, {Taylor}, {Tramper}, {Vink}, \& {Walborn}}]{grin17}
{Grin}, N.~J., {Ram{\'\i}rez-Agudelo}, O.~H., {de Koter}, A., {et~al.} 2017,
  \aap, 600, A82

\bibitem[{{Heap} {et~al.}(2006){Heap}, {Lanz}, \& {Hubeny}}]{heap06}
{Heap}, S.~R., {Lanz}, T., \& {Hubeny}, I. 2006, \apj, 638, 409

\bibitem[{{Hillier} {et~al.}(2003){Hillier}, {Lanz}, {Heap}, {Hubeny}, {Smith},
  {Evans}, {Lennon}, \& {Bouret}}]{hil03}
{Hillier}, D.~J., {Lanz}, T., {Heap}, S.~R., {et~al.} 2003, \apj, 588, 1039

\bibitem[{{Hillier} \& {Miller}(1998)}]{hm98}
{Hillier}, D.~J. \& {Miller}, D.~L. 1998, \apj, 496, 407

\bibitem[{{Holgado} {et~al.}(2022){Holgado}, {Sim{\'o}n-D{\'\i}az}, {Herrero},
  \& {Barb{\'a}}}]{holgado22}
{Holgado}, G., {Sim{\'o}n-D{\'\i}az}, S., {Herrero}, A., \& {Barb{\'a}}, R.~H.
  2022, \aap, 665, A150

\bibitem[{{Hunter} {et~al.}(2009){Hunter}, {Brott}, {Langer}, {Lennon},
  {Dufton}, {Howarth}, {Ryans}, {Trundle}, {Evans}, {de Koter}, \&
  {Smartt}}]{hunter09}
{Hunter}, I., {Brott}, I., {Langer}, N., {et~al.} 2009, \aap, 496, 841

\bibitem[{{Hunter} {et~al.}(2008{\natexlab{a}}){Hunter}, {Brott}, {Lennon},
  {Langer}, {Dufton}, {Trundle}, {Smartt}, {de Koter}, {Evans}, \&
  {Ryans}}]{hunter08}
{Hunter}, I., {Brott}, I., {Lennon}, D.~J., {et~al.} 2008{\natexlab{a}}, \apjl,
  676, L29

\bibitem[{{Hunter} {et~al.}(2008{\natexlab{b}}){Hunter}, {Lennon}, {Dufton},
  {Trundle}, {Sim{\'o}n-D{\'\i}az}, {Smartt}, {Ryans}, \& {Evans}}]{hunter08a}
{Hunter}, I., {Lennon}, D.~J., {Dufton}, P.~L., {et~al.} 2008{\natexlab{b}},
  \aap, 479, 541

\bibitem[{{Langer}(2012)}]{langer12}
{Langer}, N. 2012, \araa, 50, 107

\bibitem[{{Limongi} \& {Chieffi}(2018)}]{lc18}
{Limongi}, M. \& {Chieffi}, A. 2018, \apjs, 237, 13

\bibitem[{{Maeder} \& {Meynet}(2001)}]{mm01}
{Maeder}, A. \& {Meynet}, G. 2001, \aap, 373, 555

\bibitem[{{Maeder} {et~al.}(2014){Maeder}, {Przybilla}, {Nieva}, {Georgy},
  {Meynet}, {Ekstr{\"o}m}, \& {Eggenberger}}]{maeder14}
{Maeder}, A., {Przybilla}, N., {Nieva}, M.-F., {et~al.} 2014, \aap, 565, A39

\bibitem[{{Markova} {et~al.}(2018){Markova}, {Puls}, \& {Langer}}]{markova18}
{Markova}, N., {Puls}, J., \& {Langer}, N. 2018, \aap, 613, A12

\bibitem[{{Martins}(2018)}]{martins18}
{Martins}, F. 2018, \aap, 616, A135

\bibitem[{{Martins} {et~al.}(2012){Martins}, {Escolano}, {Wade}, {Donati},
  {Bouret}, \& {MIMES Collaboration}}]{martins12b}
{Martins}, F., {Escolano}, C., {Wade}, G.~A., {et~al.} 2012, \aap, 538, A29

\bibitem[{{Martins} {et~al.}(2016){Martins}, {Foschino}, {Bouret}, {Barb{\'a}},
  \& {Howarth}}]{martins16OC}
{Martins}, F., {Foschino}, S., {Bouret}, J.~C., {Barb{\'a}}, R., \& {Howarth},
  I. 2016, \aap, 588, A64

\bibitem[{{Martins} {et~al.}(2015){Martins}, {Herv{\'e}}, {Bouret},
  {Marcolino}, {Wade}, {Neiner}, {Alecian}, {Grunhut}, \& {Petit}}]{martins15}
{Martins}, F., {Herv{\'e}}, A., {Bouret}, J.~C., {et~al.} 2015, \aap, 575, A34

\bibitem[{{Martins} \& {Hillier}(2012)}]{martins12}
{Martins}, F. \& {Hillier}, D.~J. 2012, \aap, 545, A95

\bibitem[{{Martins} \& {Palacios}(2017)}]{mp17}
{Martins}, F. \& {Palacios}, A. 2017, \aap, 598, A56

\bibitem[{{Martins} \& {Palacios}(2021)}]{mp21}
{Martins}, F. \& {Palacios}, A. 2021, \aap, 645, A67

\bibitem[{{Martins} \& {Plez}(2006)}]{mp06}
{Martins}, F. \& {Plez}, B. 2006, \aap, 457, 637

\bibitem[{{Martins} {et~al.}(2005){Martins}, {Schaerer}, \&
  {Hillier}}]{martins05}
{Martins}, F., {Schaerer}, D., \& {Hillier}, D.~J. 2005, \aap, 436, 1049

\bibitem[{{Martins} {et~al.}(2017){Martins}, {Sim{\'o}n-D{\'\i}az},
  {Barb{\'a}}, {Gamen}, \& {Ekstr{\"o}m}}]{martins17}
{Martins}, F., {Sim{\'o}n-D{\'\i}az}, S., {Barb{\'a}}, R.~H., {Gamen}, R.~C.,
  \& {Ekstr{\"o}m}, S. 2017, \aap, 599, A30

\bibitem[{{Massey}(2013)}]{massey13}
{Massey}, P. 2013, \nar, 57, 14

\bibitem[{{Massey} {et~al.}(2005){Massey}, {Puls}, {Pauldrach}, {Bresolin},
  {Kudritzki}, \& {Simon}}]{massey05}
{Massey}, P., {Puls}, J., {Pauldrach}, A.~W.~A., {et~al.} 2005, \apj, 627, 477

\bibitem[{{Massey} {et~al.}(2009){Massey}, {Zangari}, {Morrell}, {Puls},
  {DeGioia-Eastwood}, {Bresolin}, \& {Kudritzki}}]{massey09}
{Massey}, P., {Zangari}, A.~M., {Morrell}, N.~I., {et~al.} 2009, \apj, 692, 618

\bibitem[{{Meynet} \& {Maeder}(2000)}]{mm00}
{Meynet}, G. \& {Maeder}, A. 2000, \aap, 361, 101

\bibitem[{{Meynet} \& {Maeder}(2002)}]{mm02}
{Meynet}, G. \& {Maeder}, A. 2002, \aap, 390, 561

\bibitem[{{Mokiem} {et~al.}(2007){Mokiem}, {de Koter}, {Evans}, {Puls},
  {Smartt}, {Crowther}, {Herrero}, {Langer}, {Lennon}, {Najarro}, {Villamariz},
  \& {Vink}}]{mokiem07}
{Mokiem}, M.~R., {de Koter}, A., {Evans}, C.~J., {et~al.} 2007, \aap, 465, 1003

\bibitem[{{Mokiem} {et~al.}(2006){Mokiem}, {de Koter}, {Evans}, {Puls},
  {Smartt}, {Crowther}, {Herrero}, {Langer}, {Lennon}, {Najarro}, {Villamariz},
  \& {Yoon}}]{mokiem06}
{Mokiem}, M.~R., {de Koter}, A., {Evans}, C.~J., {et~al.} 2006, \aap, 456, 1131

\bibitem[{{Nieva} \& {Przybilla}(2012)}]{nieva12}
{Nieva}, M.~F. \& {Przybilla}, N. 2012, \aap, 539, A143

\bibitem[{{Pauldrach} {et~al.}(1994){Pauldrach}, {Kudritzki}, {Puls}, {Butler},
  \& {Hunsinger}}]{pauldrach94}
{Pauldrach}, A.~W.~A., {Kudritzki}, R.~P., {Puls}, J., {Butler}, K., \&
  {Hunsinger}, J. 1994, \aap, 283, 525

\bibitem[{{Penny} \& {Gies}(2009)}]{penny09}
{Penny}, L.~R. \& {Gies}, D.~R. 2009, \apj, 700, 844

\bibitem[{{Poveda} {et~al.}(1967){Poveda}, {Ruiz}, \& {Allen}}]{poveda67}
{Poveda}, A., {Ruiz}, J., \& {Allen}, C. 1967, Boletin de los Observatorios
  Tonantzintla y Tacubaya, 4, 86

\bibitem[{{Puls} {et~al.}(2005){Puls}, {Urbaneja}, {Venero}, {Repolust},
  {Springmann}, {Jokuthy}, \& {Mokiem}}]{puls05}
{Puls}, J., {Urbaneja}, M.~A., {Venero}, R., {et~al.} 2005, \aap, 435, 669

\bibitem[{{Puls} {et~al.}(2008){Puls}, {Vink}, \& {Najarro}}]{puls08}
{Puls}, J., {Vink}, J.~S., \& {Najarro}, F. 2008, \aapr, 16, 209

\bibitem[{{Ramachandran} {et~al.}(2021){Ramachandran}, {Oskinova}, \&
  {Hamann}}]{rama21}
{Ramachandran}, V., {Oskinova}, L.~M., \& {Hamann}, W.~R. 2021, \aap, 646, A16

\bibitem[{{Ram{\'\i}rez-Agudelo} {et~al.}(2013){Ram{\'\i}rez-Agudelo},
  {Sim{\'o}n-D{\'\i}az}, {Sana}, {de Koter}, {Sab{\'\i}n-Sanjul{\'\i}an}, {de
  Mink}, {Dufton}, {Gr{\"a}fener}, {Evans}, {Herrero}, {Langer}, {Lennon},
  {Ma{\'\i}z Apell{\'a}niz}, {Markova}, {Najarro}, {Puls}, {Taylor}, \&
  {Vink}}]{ramirez13}
{Ram{\'\i}rez-Agudelo}, O.~H., {Sim{\'o}n-D{\'\i}az}, S., {Sana}, H., {et~al.}
  2013, \aap, 560, A29

\bibitem[{{Renzo} {et~al.}(2019){Renzo}, {Zapartas}, {de Mink}, {G{\"o}tberg},
  {Justham}, {Farmer}, {Izzard}, {Toonen}, \& {Sana}}]{renzo19}
{Renzo}, M., {Zapartas}, E., {de Mink}, S.~E., {et~al.} 2019, \aap, 624, A66

\bibitem[{{Rivero Gonz{\'a}lez} {et~al.}(2011){Rivero Gonz{\'a}lez}, {Puls}, \&
  {Najarro}}]{rivero11}
{Rivero Gonz{\'a}lez}, J.~G., {Puls}, J., \& {Najarro}, F. 2011, \aap, 536, A58

\bibitem[{{Rivero Gonz{\'a}lez} {et~al.}(2012){Rivero Gonz{\'a}lez}, {Puls},
  {Najarro}, \& {Brott}}]{rivero12}
{Rivero Gonz{\'a}lez}, J.~G., {Puls}, J., {Najarro}, F., \& {Brott}, I. 2012,
  \aap, 537, A79

\bibitem[{{Sana} {et~al.}(2024){Sana}, {Tramper}, {Abdul-Masih}, {Blomme},
  {Dsilva}, {Maravelias}, {Martins}, {Mehner}, {Wofford}, {Banyard}, {Barbosa},
  {Bestenlehner}, {Hawcroft}, {Hillier}, {Todt}, {Larkin}, {Mahy}, {Najarro},
  {Ramachandran}, {Ramirez-Tannus}, {Rubio-Diez}, {Sander}, {Shenar}, {Vink},
  {Backs}, {Brands}, {Crowther}, {Decin}, {de Koter}, {Hamann}, {Kehrig},
  {Kuiper}, {Oskinova}, {Pauli}, {Sundqvist}, {Verhamme}, \& {the XSHOOT-U
  collaboration}}]{xshooII}
{Sana}, H., {Tramper}, F., {Abdul-Masih}, M., {et~al.} 2024, arXiv e-prints,
  arXiv:2402.16987

\bibitem[{{Sander} {et~al.}(2015){Sander}, {Shenar}, {Hainich},
  {G{\'\i}menez-Garc{\'\i}a}, {Todt}, \& {Hamann}}]{sander15}
{Sander}, A., {Shenar}, T., {Hainich}, R., {et~al.} 2015, \aap, 577, A13

\bibitem[{{Schaller} {et~al.}(1992){Schaller}, {Schaerer}, {Meynet}, \&
  {Maeder}}]{schaller92}
{Schaller}, G., {Schaerer}, D., {Meynet}, G., \& {Maeder}, A. 1992, \aaps, 96,
  269

\bibitem[{{Serebriakova} {et~al.}(2023){Serebriakova}, {Tkachenko}, {Gebruers},
  {Bowman}, {Van Reeth}, {Mahy}, {Burssens}, {IJspeert}, {Sana}, \&
  {Aerts}}]{sereb23}
{Serebriakova}, N., {Tkachenko}, A., {Gebruers}, S., {et~al.} 2023, \aap, 676,
  A85

\bibitem[{{Sim{\'o}n-D{\'\i}az}(2020)}]{sergio20}
{Sim{\'o}n-D{\'\i}az}, S. 2020, in Reviews in Frontiers of Modern Astrophysics;
  From Space Debris to Cosmology, 155--187

\bibitem[{{Sim{\'o}n-D{\'\i}az} \& {Herrero}(2007)}]{sergio07}
{Sim{\'o}n-D{\'\i}az}, S. \& {Herrero}, A. 2007, \aap, 468, 1063

\bibitem[{{Sim{\'o}n-D{\'\i}az} \& {Herrero}(2014)}]{sergio14}
{Sim{\'o}n-D{\'\i}az}, S. \& {Herrero}, A. 2014, \aap, 562, A135

\bibitem[{{Sim{\'o}n-D{\'\i}az} {et~al.}(2014){Sim{\'o}n-D{\'\i}az}, {Herrero},
  {Sab{\'\i}n-Sanjuli{\'a}n}, {Najarro}, {Garcia}, {Puls}, {Castro}, \&
  {Evans}}]{sergio14b}
{Sim{\'o}n-D{\'\i}az}, S., {Herrero}, A., {Sab{\'\i}n-Sanjuli{\'a}n}, C.,
  {et~al.} 2014, \aap, 570, L6

\bibitem[{{Sota} {et~al.}(2011){Sota}, {Ma{\'\i}z Apell{\'a}niz}, {Walborn},
  {Alfaro}, {Barb{\'a}}, {Morrell}, {Gamen}, \& {Arias}}]{sota11}
{Sota}, A., {Ma{\'\i}z Apell{\'a}niz}, J., {Walborn}, N.~R., {et~al.} 2011,
  \apjs, 193, 24

\bibitem[{{Sz{\'e}csi} {et~al.}(2022){Sz{\'e}csi}, {Agrawal}, {W{\"u}nsch}, \&
  {Langer}}]{szecsi22}
{Sz{\'e}csi}, D., {Agrawal}, P., {W{\"u}nsch}, R., \& {Langer}, N. 2022, \aap,
  658, A125

\bibitem[{{Vink}(2022)}]{vink22}
{Vink}, J.~S. 2022, \araa, 60, 203

\bibitem[{{Vink} {et~al.}(2001){Vink}, {de Koter}, \& {Lamers}}]{vink01}
{Vink}, J.~S., {de Koter}, A., \& {Lamers}, H.~J.~G.~L.~M. 2001, \aap, 369, 574

\bibitem[{{Vink} {et~al.}(2023){Vink}, {Mehner}, {Crowther}, {Fullerton},
  {Garcia}, {Martins}, {Morrell}, {Oskinova}, {St-Louis}, {ud-Doula}, {Sander},
  {Sana}, {Bouret}, {Kub{\'a}tov{\'a}}, {Marchant}, {Martins}, {Wofford}, {van
  Loon}, {Grace Telford}, {G{\"o}tberg}, {Bowman}, {Erba}, {Kalari},
  {Abdul-Masih}, {Alkousa}, {Backs}, {Barbosa}, {Berlanas}, {Bernini-Peron},
  {Bestenlehner}, {Blomme}, {Bodensteiner}, {Brands}, {Evans}, {David-Uraz},
  {Driessen}, {Dsilva}, {Geen}, {G{\'o}mez-Gonz{\'a}lez}, {Grassitelli},
  {Hamann}, {Hawcroft}, {Herrero}, {Higgins}, {John Hillier}, {Ignace},
  {Istrate}, {Kaper}, {Kee}, {Kehrig}, {Keszthelyi}, {Klencki}, {de Koter},
  {Kuiper}, {Laplace}, {Larkin}, {Lefever}, {Leitherer}, {Lennon}, {Mahy},
  {Ma{\'\i}z Apell{\'a}niz}, {Maravelias}, {Marcolino}, {McLeod}, {de Mink},
  {Najarro}, {Oey}, {Parsons}, {Pauli}, {Pedersen}, {Prinja}, {Ramachandran},
  {Ram{\'\i}rez-Tannus}, {Sabhahit}, {Schootemeijer}, {Reyero Serantes},
  {Shenar}, {Stringfellow}, {Sudnik}, {Tramper}, \& {Wang}}]{xshooI}
{Vink}, J.~S., {Mehner}, A., {Crowther}, P.~A., {et~al.} 2023, \aap, 675, A154

\bibitem[{{Walborn}(1971)}]{walborn71}
{Walborn}, N.~R. 1971, \apjl, 164, L67

\end{thebibliography}

\newpage

\begin{appendix}
\label{s_ap}

\FloatBarrier

\onecolumn

\section{Complementary samples}
\label{ap_compsamp}

\begin{figure}[th]
  \centering
\includegraphics[width=0.99\textwidth]{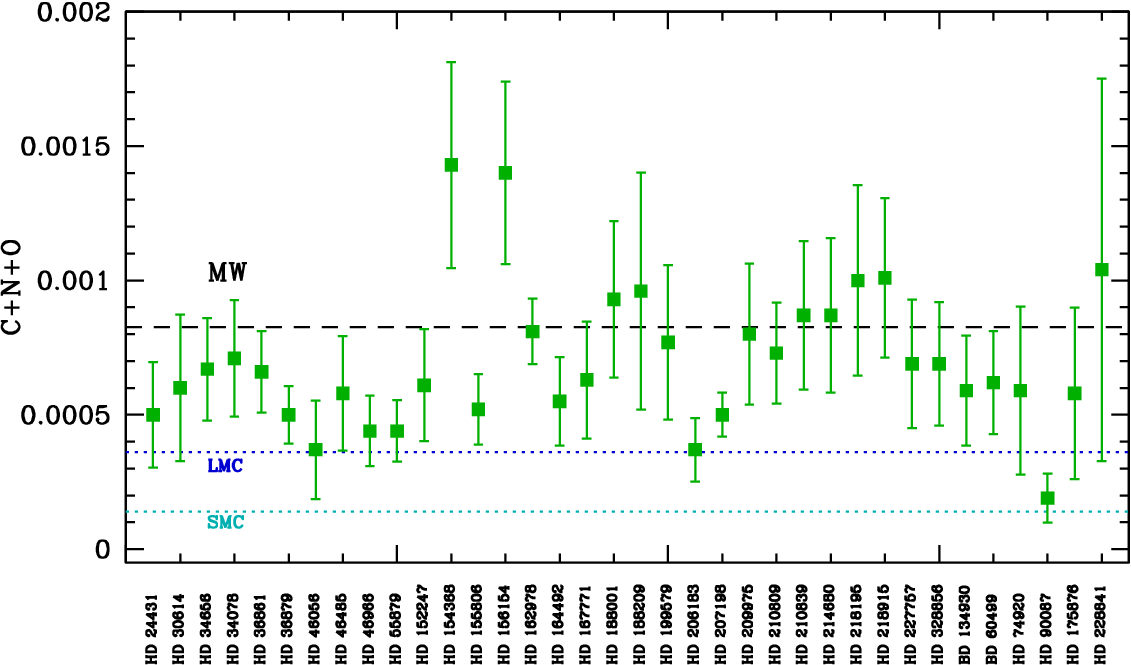}
\caption{Same as Fig.~\ref{sumCNO} but for the complementary Galactic sample stars. Stars with an upper or lower limit on at least one of the abundance measurements are excluded from this figure. The horizontal broken lines show the Galactic, LMC, and SMC values.}  
\label{sumCNOmw}
\end{figure}

Table~\ref{tab_compsamp} lists the members of the complementary samples described in Sect.~\ref{s_sample} and the references they were taken from. Figure~\ref{sumCNOmw} shows the sum of the carbon, nitrogen, and oxygen number fractions for the Galactic stars of the complementary samples. 

\begin{table}[p]
\begin{center}
\caption{Complementary samples.} \label{tab_compsamp}
\footnotesize
\begin{tabular}{lcr|lcr}
\hline
Star	&   ST          &  reference  & Star	&   ST          &  reference\\

\hline
        & SMC           &    &    & LMC & \\
\hline
AzV77	& O7\,III	        &  1  &  VFTS046   &    O9.7\,II((n))    & 3  \\
AzV439	& O9.5\,III       &  1  &	 VFTS076   &    O9.2\,III        & 3  \\
AzV170	& O9.7\,III       &  1  &	 VFTS087   &    O9.7\,Ib-II      & 3  \\
AzV43	& B0.5\,III       &  1  &	 VFTS103   &    O8.5\,III((f))   & 3  \\
AzV388	& O4\,V	        &  2  &	 VFTS151   &    O6.5\,II(f)p     & 3  \\
MPG324	& O4\,V           &  2  &	 VFTS153   &    O9\,III((n))     & 3  \\
MPG368  & O6\,V           &  2  &	 VFTS160   &    O9.5\,III((n))   & 3  \\
MPG113	& OC6\,Vz         &  2  &	 VFTS172   &    O9\,III((f))     & 3  \\
MPG356	& O6.5\,V         &  2  &	 VFTS178   &    O9.7\,Iab        & 3  \\
AzV429	& O7\,V           &  2  &	 VFTS185   &    O7.5\,III((f))   & 3  \\
MPG523	& O7\,Vz          &  2  &	 VFTS259   &    O6\,Iaf          & 3  \\
NGC346 046	& O7\,Vn      &  2  &	 VFTS306   &    O8.5\,II((f))    & 3  \\
NGC346 031	& O8\,Vz      &  2  &	 VFTS466   &    O9\,III          & 3  \\
AzV461	& O8\,V           &  2  &	 VFTS502   &    O9.7\,II         & 3  \\
MPG299	& O8\,Vn          &  2  &	 VFTS503   &    O9\,III          & 3  \\
MPG487	& O8\,V           &  2  &	 VFTS513   &    O6-7\,II(f)      & 3  \\
AzV468	& O8.5\,V         &  2  &	 VFTS546   &    O8-9\,III:((n))  & 3  \\
MPG682	& O9\,V           &  2  &	 VFTS664   &    O7\,II(f)        & 3  \\
MPG012	& B0\,IV          &  2  &	 VFTS669   &    O8\,Ib(f)        & 3  \\
	    &		        &     &	 VFTS764   &    O9.7\,Ia         & 3  \\       
       	& 		        &     &	 VFTS777   &    O9.2\,II         & 3  \\       
   	    & 		        &     &	 VFTS782   &    O8.5\,III        & 3  \\       
    	& 		        &     &	 VFTS807   &    O9.5\,III        & 3  \\       
       	&		        &     &	 VFTS819   &    O8\,III((f))     & 3  \\   
        
\hline
             &   MW    &     &  & MW     &  \\
\hline
BD+60 261    & O7.5\,III(n)(f)     &  4  &        HD167263     & O9.5\,III         &    5 \\
HD24912      & O7.5\,III(n)(f)     &  4  &        HD167771     & O7.5\,III(f)      &    5 \\
HD34656      & O7.5\,II(f)         &  4  &        HD188001     & O7.5\,Iabf        &    5 \\
HD35633      & O7.5\,II(n)(f)      &  4  &        HD188209     & O9.5\,Iab         &    5 \\
HD36861      & O8\,III(f)          &  4  &        HD189957     & O9.7\,III         &    5 \\
HD94963      & O7\,II(f)           &  4  &        HD192639     & O7.5\,Iabf        &    5 \\
HD97434      & O7.5\,III(n)(f)     &  4  &        HD193443     & O9\,III           &    5 \\
HD151515     & O7\,II(f)           &  4  &        HD199579     & O6.5\,V((f))z     &    5 \\
HD162978     & O8\,II(f)           &  4  &        HD201345     & ON9.2\,IV         &    5 \\
HD163800     & O7.5\,III(f)        &  4  &        HD203064     & O7.5\,IIIn((f))   &    5 \\
HD167659     & O7\,II-III(f)       &  4  &        HD206183     & O9.5\,IV-V        &    5 \\
HD171589     & O7\,II(f)           &  4  &        HD207198     & O8.5\,II          &    5 \\
HD175754     & O8\,II(n)((f))p     &  4  &        HD209975     & O9\,Ib            &    5 \\
HD186980     & O7.5\,III(f)        &  4  &        HD210809     & O9\,Iab           &    5 \\
HD203064     & O7.5\,IIIn((f))     &  4  &        HD210839     & O6\,I(n)fp        &    5 \\
HD14633      & ON8.5\,V            &  5  &        HD214680     & O9\,V             &    5 \\
HD24431      & O9\,III             &  5  &        HD218195     & O8.5\,IIINstr     &    5 \\
HD30614      & O9\,Ia              &  5  &        HD218915     & O9.2\,Iab         &    5 \\
HD34656      & O7.5\,II(f)         &  5  &        HD227757     & O9.5\,V           &    5 \\
HD34078      & O9.5\,V             &  5  &        HD258691     & O9\,V             &    5 \\
HD35619      & O7.5\,V((f))        &  5  &        HD328856     & O9.7\,II          &    5 \\
HD36861      & O8\,III((f))        &  5  &        BD-13 4930   & O9.5\,V           &    5 \\
HD36879      & O7V(n)((f))z        &  5  &        BD+60 499    & O9.5\,V           &    5 \\
HD38666      & O9.5\,V             &  5  &        HD63005      & O6.5\,IV(f)       &    6 \\
HD42088      & O6\,V               &  5  &        HD92504      & O8.5\,V(n)        &    6 \\ 
HD46056      & O8\,Vn              &  5  &        CPD-58 2620  & O7\,Vz            &    6 \\ 
HD46485      & O7\,V((f))nz        &  5  &        HD93222      & O7\,V((f))z       &    6 \\ 
HD46966      & O8.5\,IV            &  5  &        CD-44 4865   & O9.7\,III         &    6 \\ 
HD55879      & O9.7\,III           &  5  &        CD-43 4690   & O6.5\,III         &    6 \\ 
HD66788      & O8\,V               &  5  &        HD97848      & O8\,V             &    6 \\ 
HD152247     & O9.2\,III           &  5  &        HD302505     & O8.5\,III         &    6 \\ 
HD152249     & OC9\,Iab            &  5  &        HD15642      & O9.5\,III-IIIn    &    7 \\
HD153426     & O8.5\,III           &  5  &        HD74920      & O7.5\,IVn((f))    &    7 \\
HD154368     & O9\,Iab             &  5  &        HD90087      & O9.2\,III(n)      &    7 \\
HD154643     & O9.7\,III           &  5  &        HD117490     & ON9.5\,IIInn      &    7 \\
HD155806     & O7.5\,V((f))z       &  5  &        HD150574     & ON9\,III(n)       &    7 \\
HD156154     & O7.5\,Ib(f)         &  5  &        HD175876     & O6.5\,III(n)(f)   &    7 \\
HD162978     & O8\,II((f))         &  5  &        HD191423     & ON9\,II-IIInn     &    7 \\
HD164492     & O7.5\,Vz            &  5  &        HD228841     & O6.5\,Vn((f))     &    7 \\
\end{tabular}
\normalsize
\tablefoot{Columns are star's ID, spectral type and reference. References are 1- \citet{bouret21}, 2- \citet{bouret21}, 3- \citet{grin17}, 4- \citet{martins17}, 5- \citet{martins15}, 6- \citet{markova18}, 7- \citet{cazorla17a}.}
\end{center}
\end{table}

\FloatBarrier
\newpage

\section{Stellar parameters from present and literature studies}
\label{ap_compsamp}

Table~\ref{tab_prev} gathers the results of spectroscopic analysis for stars that have been previously studied in the literature. The present results are also included for clarity.

\begin{table*}[ht]
\begin{center}
\caption{Parameters from literature studies} \label{tab_prev}
\footnotesize
\begin{tabular}{lrrrrrrrrccc}
\hline
Star	               & \teff    & \logg\ & \lL\ & \vsini\ & He/H & $\epsilon$(C) & $\epsilon$(N) & $\epsilon$(O)  & reference\\
                       & [kK]     &        &      &  [\kms] &      &     &     &  & \\
\hline
AzV15                    & 38.0 & 3.7 & 5.72 & 90 & 0.085$\pm$0.005 & 7.48$\pm$0.15 & 8.08$\pm$0.11 & 8.11$\pm$0.33 & this study \\
                        & 39.0 & 3.61 & 5.83 & 120 & 0.10 & 7.00$\pm$0.11 & 7.78$\pm$0.13 & 7.90$\pm$0.13 & 1 \\
                        & 38.5 & 3.6 & 5.82 & 180 & 0.10 & -- & -- & -- & 2 \\
                        & 37.0 & 3.5 & -- & 100 & -- & 7.42 & 7.92 & -- & 3 \\
                        & 39.4 & 3.69 & 5.82 & 135 & 0.10$\pm$0.03 & -- & -- & -- & 4 \\
\hline
AzV47                    & 38.0 & 4.3 & 5.60 & 65 & 0.085$\pm$0.005 & 7.70$\pm$0.22 & 7.54$\pm$0.12 & 8.30$\pm$0.22 & this study \\
                        & 35.0 & 3.75 & 5.44 & 60 & 0.10 & 7.69$\pm$0.09 & 7.08$\pm$0.15 & 7.98$\pm$0.08 & 1 \\
                        & 35.0 & 3.75 & -- & 60 & -- & 7.42 & 7.22 & -- & 3 \\  
\hline
AzV69                    & 36.0 & 3.6 & 5.60 & 85 & 0.085$\pm$0.005 & 7.78$\pm$0.11 & 6.70$\pm$0.17 & 8.00$\pm$0.13 & this study \\
                        & 33.9 & 3.5 & 5.61 & 70 & 0.10 & 7.56$\pm$0.15 & 6.34$\pm$0.17 & 8.23$\pm$0.12 & 1 \\
                        & 35.0 & 3.4 & -- & 70 & -- & 7.52 & 6.70 & -- & 3 \\  
                        & 33.9 & 3.5 & 5.61 & -- & 0.10 & 7.56 & 6.34 & 8.23 & 5 \\
\hline
AzV95                   & 37.0 & 3.7 & 5.43 & 55 & 0.092$\pm$0.008 & 7.15$\pm$0.16 & 7.81$\pm$0.13 & 7.30$\pm$0.17 & this study \\
                        & 38.0 & 3.7 & 5.46 & 55 & 0.10 & 7.30$\pm$0.10 & 7.60$\pm$0.11 & 7.96$\pm$0.10 & 1 \\
                        & 35.0 & 3.4 & -- & 80 & -- & 7.42 & 7.62 & -- & 3 \\  
                        & 38.3 & 3.66 & 5.56 & 68 & 0.13$\pm$0.04 & -- & -- & -- & 4 \\
\hline
AzV148                   & 31.0 & 3.6 & 5.16 & 35 & 0.090$\pm$0.005 & 7.48$\pm$0.07 & 7.41$\pm$0.12 & 8.00$\pm$0.22 & this study \\
                        & 32.3 & 4.0$^a$ & 4.84 & 60 & 0.09 & 7.69$\pm$0.12 & 7.26$\pm$0.23 & 7.99$\pm$0.11 & 1 \\
\hline
AzV186                  & 36.0 & 3.8 & 5.63 & 55 & 0.13$\pm$0.02 & 7.30$\pm$0.22 & 8.00$\pm$0.22 & 8.00$\pm$0.22 & this study \\
                        & 34.5 & 3.4 & 5.40 & 78 & -- & -- & -- & -- & 6 \\  
\hline
AzV207                  & 38.0 & 4.0 & 5.21 & 75 & 0.085$\pm$0.005 & 7.70$\pm$0.26 & 8.00$\pm$0.17 & 7.78$\pm$0.14 & this study \\
                        & 37.0 & 3.72 & 5.32 & 120 & 0.1 & -- & -- & -- & 7 \\
\hline
AzV243                  & 41.0 & 4.0 & 5.50 & 60 & 0.085$\pm$0.005 & 8.00$\pm$0.14 & 7.85$\pm$0.19 & 7.85$\pm$0.12 & this study \\
                        & 39.6 & 3.9 & 5.59 & 60 & 0.09 & 7.20$\pm$0.15 & 7.50$\pm$0.24 & 7.90$\pm$0.10 & 1 \\
                        & 42.6 & 3.94 & 5.68 & 59 & 0.12$\pm$0.02 & -- & -- & -- & 4 \\
\hline
AzV267                   & 37.0 & 4.0 & 4.93 & 325 & 0.13$\pm$0.02 & $<$7.70 & 7.90$\pm$0.11 & $<$7.30 & this study \\
                        & 35.7 & 4.04 & 4.90 & 220 & 0.10 & 7.30$\pm$0.14 & 7.48$\pm$0.23 & 7.98$\pm$0.13 & 1 \\ 
\hline
AzV307                   & 29.0 & 3.4 & 5.12 & 45 & 0.10$\pm$0.02 & 7.08$\pm$0.14 & 7.78$\pm$0.14 & 8.00$\pm$0.14 & this study \\
                        & 30.0 & 3.5 & 5.15 & 60 & 0.12 & 7.38$\pm$0.17 & 7.56$\pm$0.18 & 7.95$\pm$0.10 & 1 \\   
\hline
AzV327                    & 31.0 & 3.4 & 5.55 & 60 & 0.12$\pm$0.02 & 7.15$\pm$0.06 & 7.78$\pm$0.07 & 7.78$\pm$0.14 & this study \\
                        & 30.0 & 3.12 & 5.54 & 95 & 0.15 & 7.30$\pm$0.11 & 8.08$\pm$0.25 & 7.66$\pm$0.13 & 1 \\
                        & 30.0 & 3.25 & -- & 60 & -- & 7.12 & 7.22 & -- & 3 \\
\hline
AzV440                    & 37.0 & 4.2 & 5.03 & 35 & 0.085$\pm$0.005 & 7.70$\pm$0.17 & 7.48$\pm$0.14 & 8.00$\pm$0.30 & this study \\
                        & 37.0 & 4.01 & 5.28 & 100 & 0.2 & -- & -- & -- & 7 \\
\hline
AzV446                    & 42.0 & 4.4 & 5.22 & 35 & 0.085$\pm$0.005 & 7.00$\pm$0.39 & 7.70$\pm$0.17 & 8.08$\pm$0.25 & this study \\
                        & 39.7 & 4.0$^a$ & 5.25 & 30 & 0.09 & 7.20$\pm$0.13 & 7.48$\pm$0.20 & 7.98$\pm$0.11 & 1 \\
                        & 41.0 & 4.15 & 5.29 & 95 & 0.15 & -- & -- & -- & 7 \\
\hline
AzV469                    & 33.0 & 3.5 & 5.58 & 80 & 0.16$\pm$0.02 & 7.69$\pm$0.26 & 8.17$\pm$0.14 & 8.30$\pm$0.29 & this study \\
                        & 34.0 & 3.41 & 5.70 & 81 & 0.17$\pm$0.03 & -- & -- & -- & 4 \\
                        & 32.0 & 3.13 & 5.64 & 120 & 0.20 & -- & -- & -- & 7 \\
                        & 33.0 & 3.4 & 5.5 & 80 & 0.20 & 7.1 & 8.2 & -- & 8 \\
\hline
N11 032                & 34.9 & 3.5 & 5.34 & 60 & 0.085$\pm$0.005 & 8.03$\pm$0.05 & 8.84$\pm$0.01 & $<$8.39 & this study \\
                        & 36.0 & 3.5 & 5.15 & 60 & 0.09 & -- & 7.87$\pm$0.15 & -- & 9 \\
                        & 35.2 & 3.45 & 5.43 & 65 & 0.09$\pm$0.02 & -- & -- & -- & 10 \\
\hline
BI 173                & 33.9 & 3.5 & 5.53 & 146 & 0.085$\pm$0.015 & 8.09$\pm$0.10 & 7.80$\pm$0.05 & 7.95$\pm$0.01 & this study \\
                        & 34.5 & 3.4 & 5.60 & 200 & 0.10 & -- & -- & -- & 2 \\
\hline
Sk-66$^{\circ}$ 18    & 38.0 & 3.7 & 5.42 & 100 & 0.10$\pm$0.01 & 7.85$\pm$0.06 & 8.40$\pm$0.10 & 8.08$\pm$0.11 & this study \\
                        & 39.7 & 3.76 & 5.52 & 70 & 0.14 & -- & 8.48$\pm$0.15 & -- & 9 \\
                        & 40.2 & 3.76 & 5.55 & 82 & 0.14 & -- & -- & -- & 10 \\
\hline
Sk-70$^{\circ}$ 13  & 29.0 & 3.2 & 5.64 & 75 & 0.085$\pm$0.015 & 7.68$\pm$0.16 & 7.29$\pm$0.37 & 8.13$\pm$0.13  \\                                         & 26.5 & 2.96 & -- & 60 & -- & -- & -- & -- & 11\\
\hline 
\end{tabular}
\tablefoot{$a$: adopted value. References are 1- \citet{bouret13,bouret21}, 2- \citet{massey09}, 3- \citet{heap06}, 4- \citet{mokiem06}, 5- \citet{hil03}, 6- \citet{hunter08a}, 7- \citet{massey05}, 8- \citet{evans04}, 9- \citet{rivero12}, 10- \citet{mokiem07} , 11- \citet{sereb23}.}
\normalsize
\end{center}
\end{table*}

\FloatBarrier
\newpage

\section{Surface abundances in various models}
\label{ap_abmod}

\begin{figure}[th]
\centering
\includegraphics[width=0.32\textwidth]{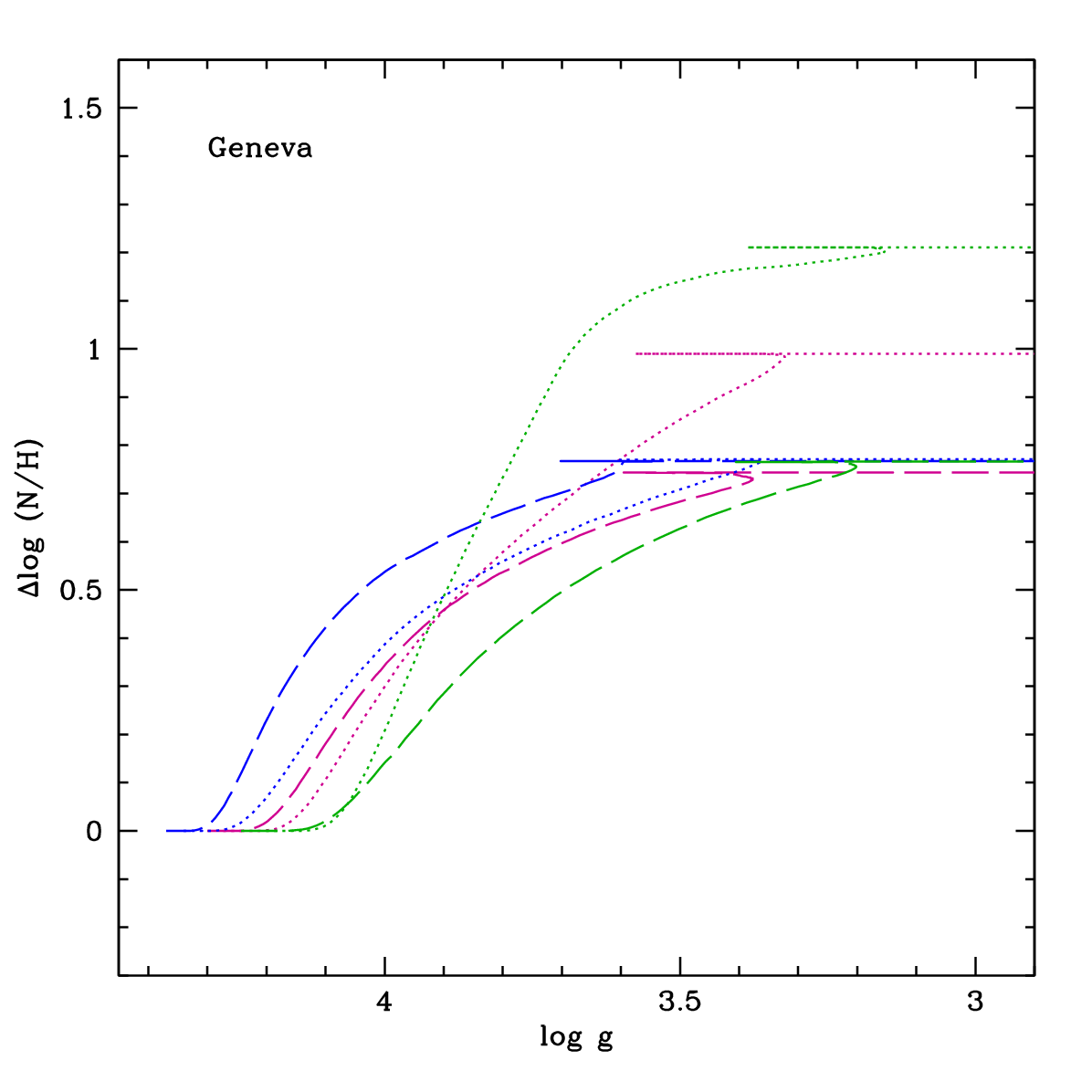}
\includegraphics[width=0.32\textwidth]{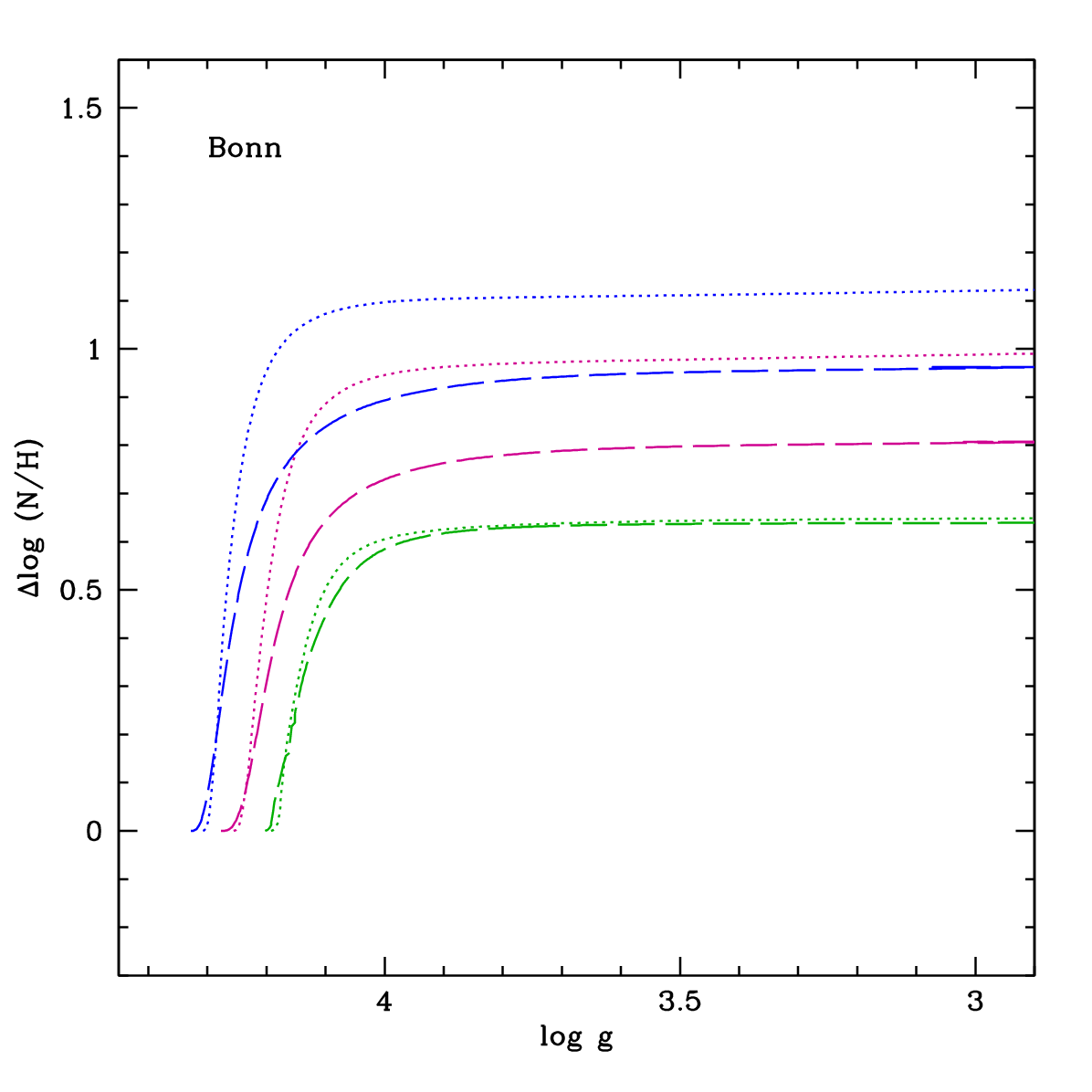}
\includegraphics[width=0.32\textwidth]{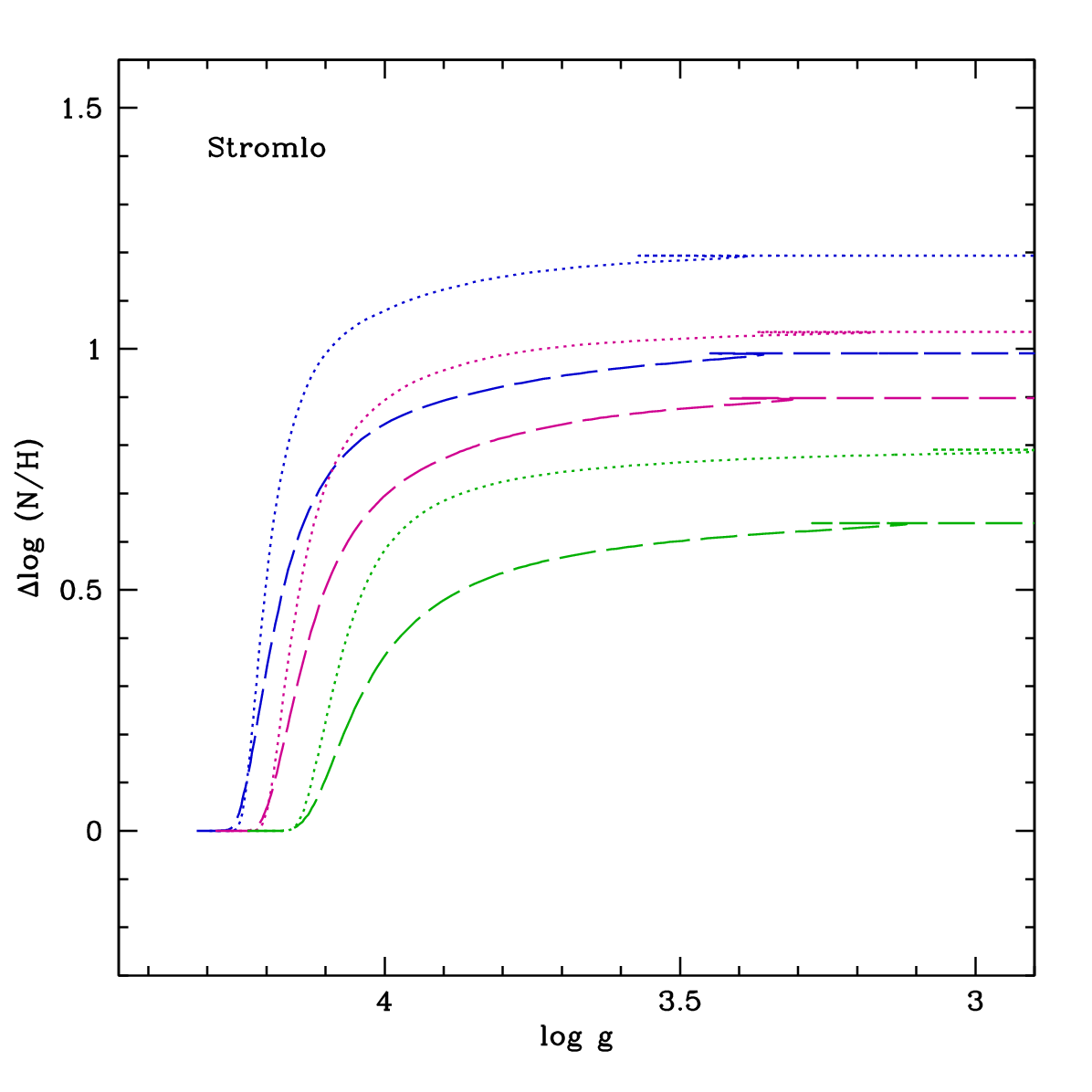}\\
\includegraphics[width=0.32\textwidth]{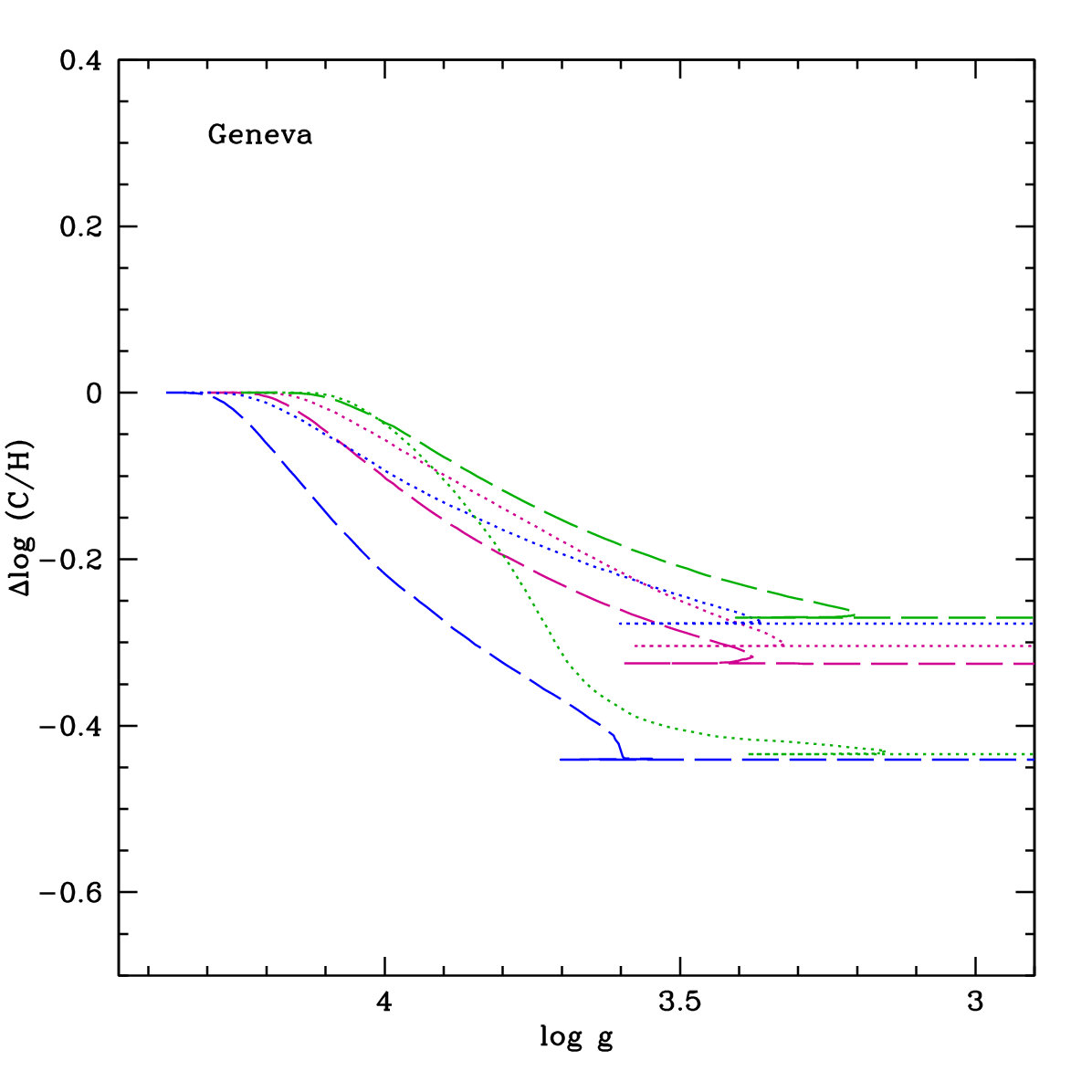}
\includegraphics[width=0.32\textwidth]{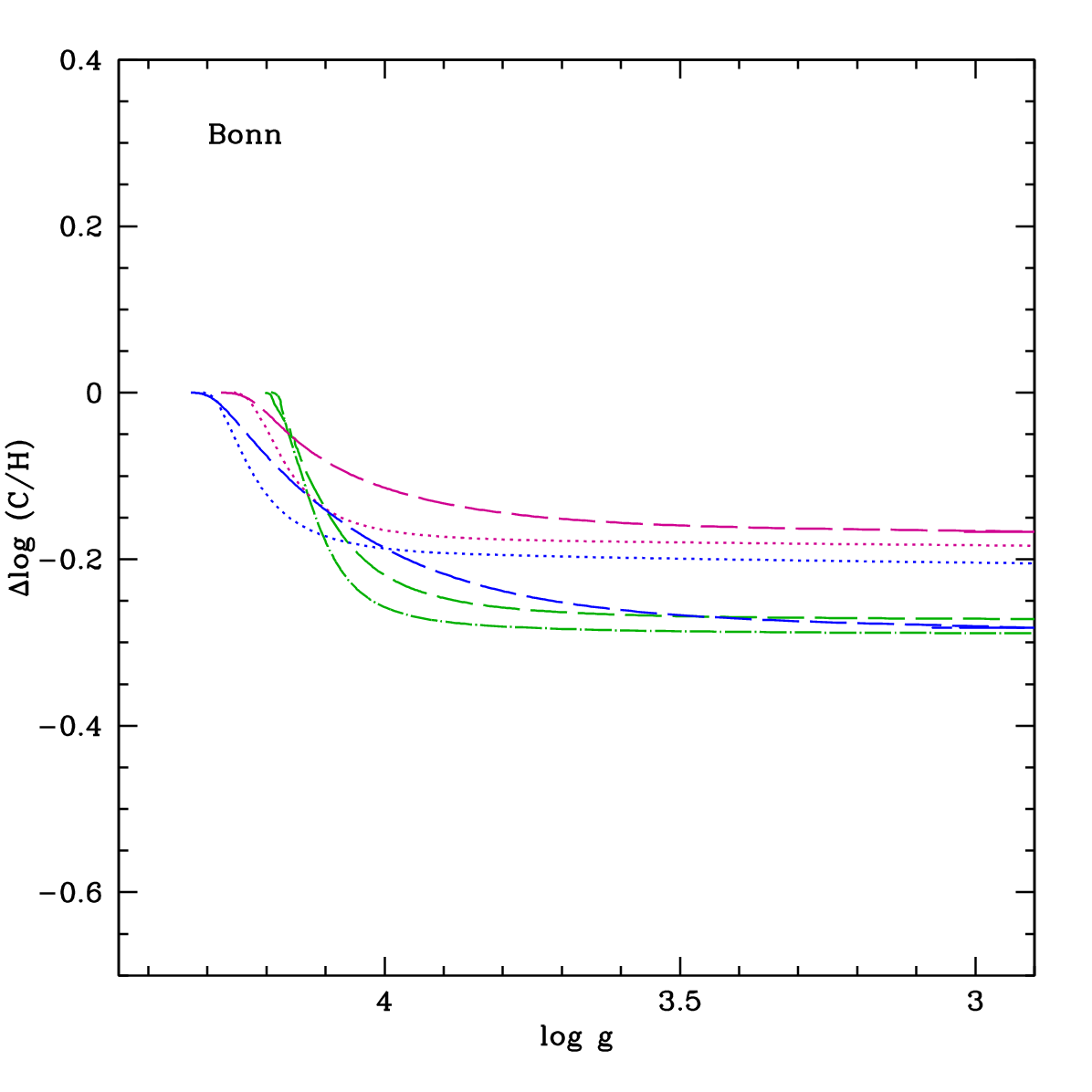}
\includegraphics[width=0.32\textwidth]{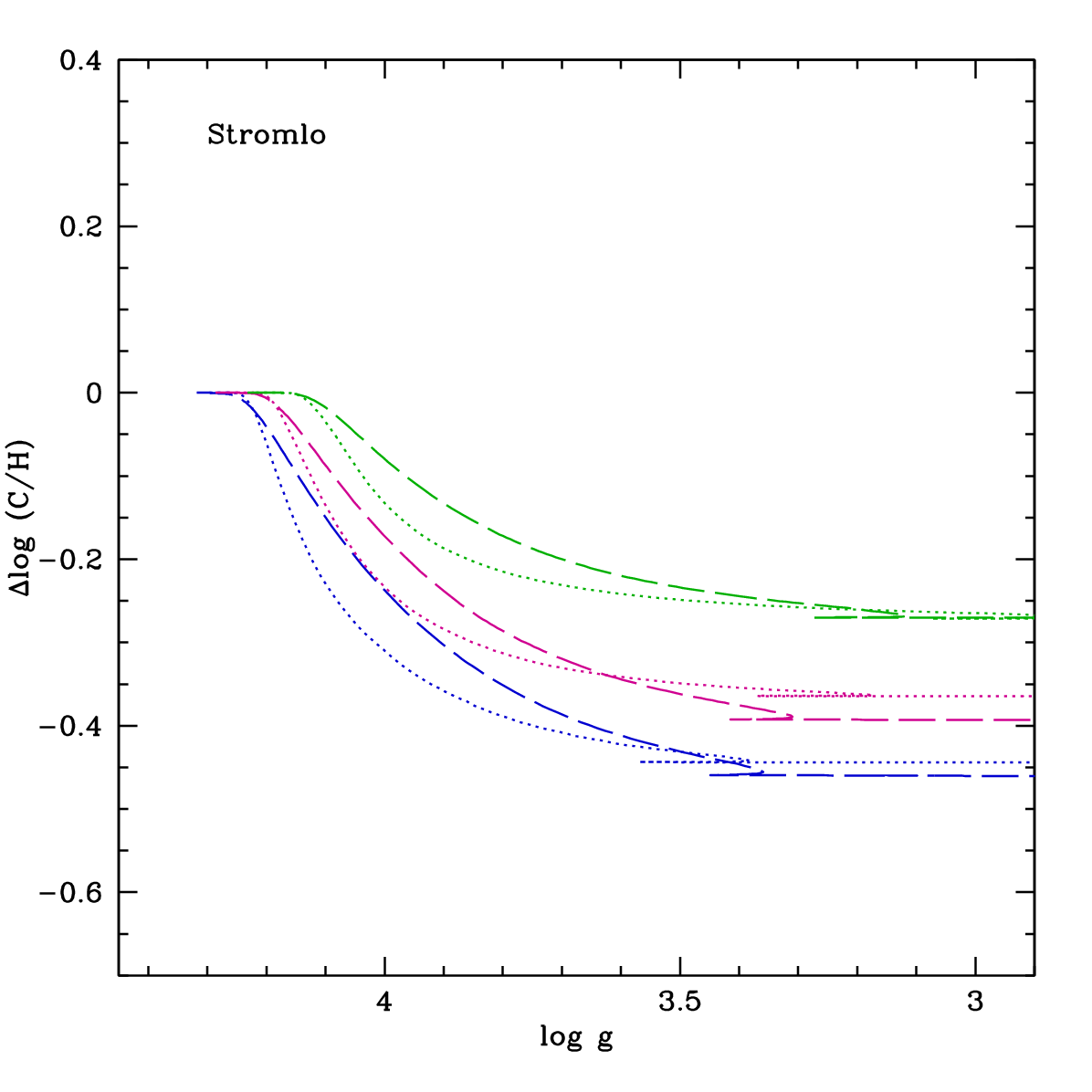}\\
\includegraphics[width=0.32\textwidth]{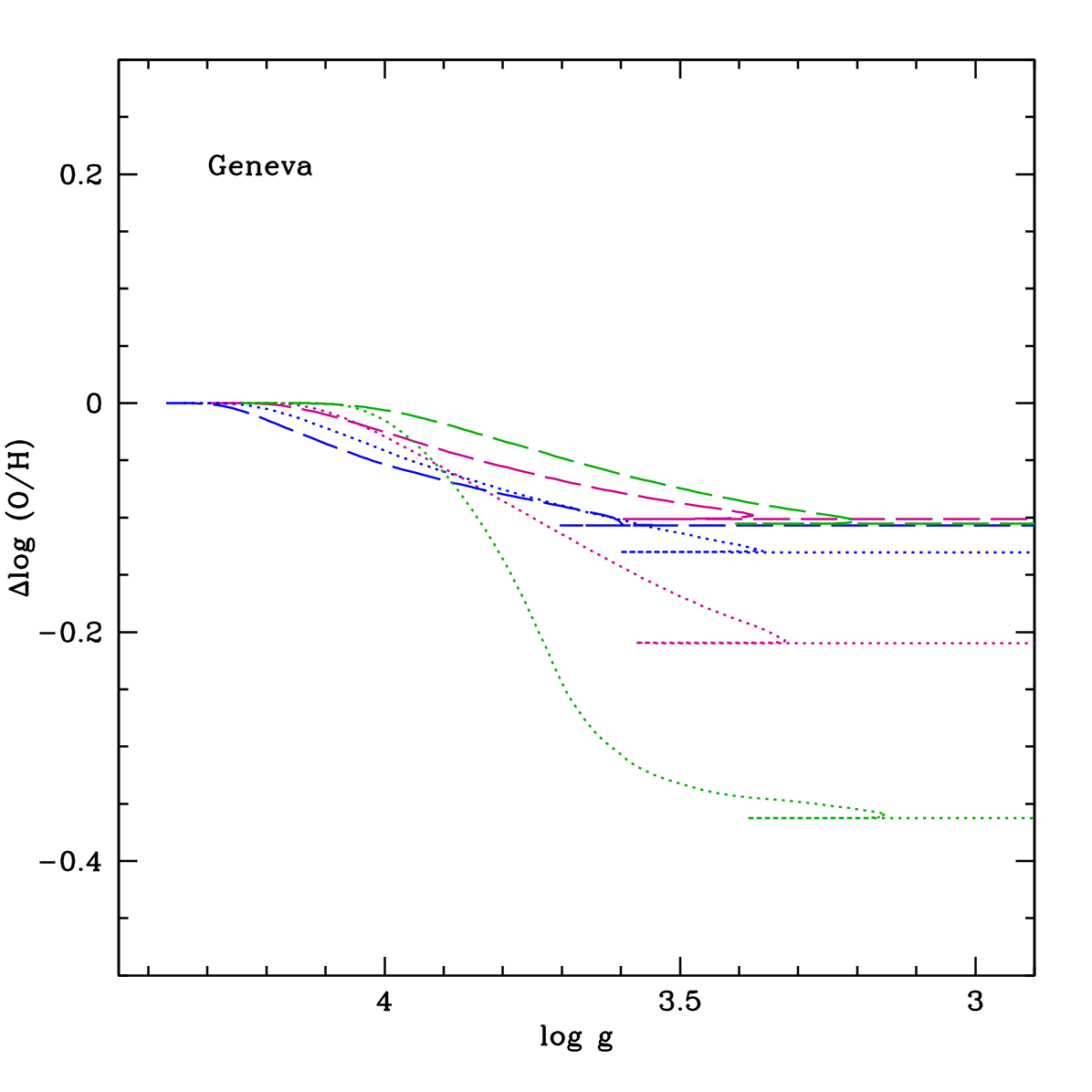}
\includegraphics[width=0.32\textwidth]{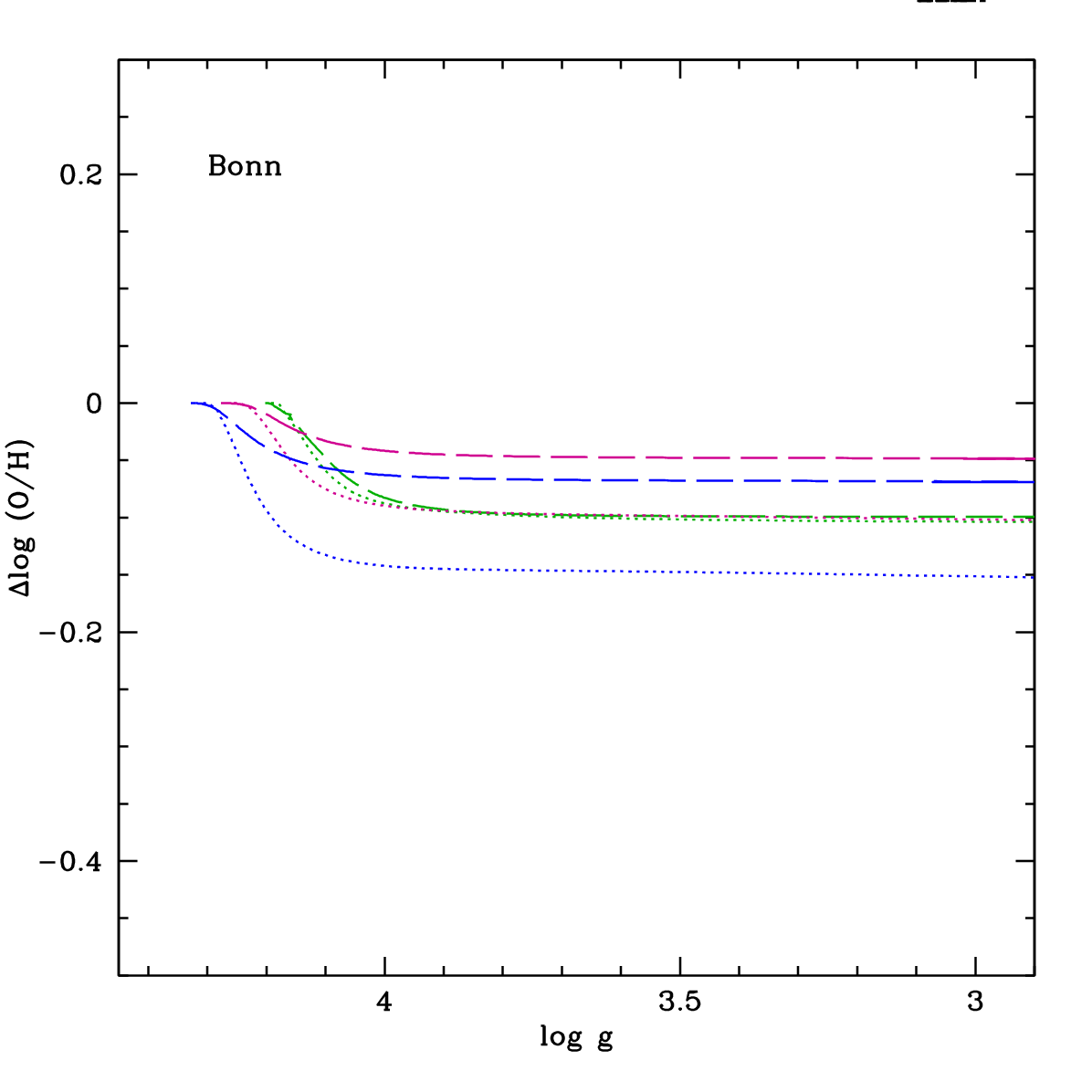}
\includegraphics[width=0.32\textwidth]{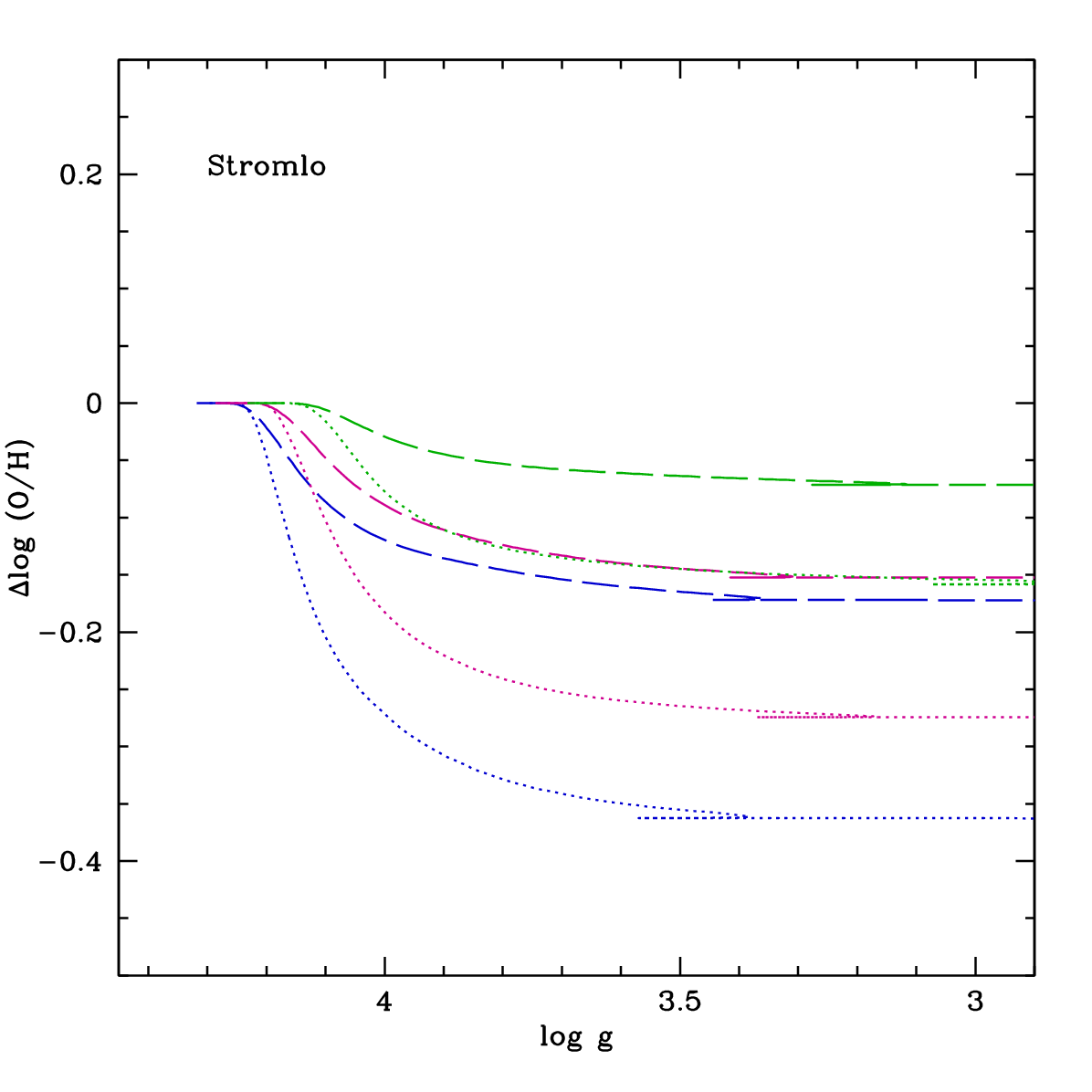}
\caption{Difference between present and initial surface abundance as a function of surface gravity. Top, middle, and bottom rows show differences in nitrogen, carbon, and oxygen abundances respectively. The Geneva, Bonn, and Stromlo models are shown in the left, middle, and right column respectively. For each set of models the 25 and 40~\msun\ tracks are shown by the dashed and dotted lines respectively. Green, magenta, and blue lines correspond to Galactic, LMC, and SMC chemical composition.} 
\label{abmod}
\end{figure}

Figs.~\ref{abmod} shows the predictions of the three sets of models regarding the evolution of surface abundances with surface gravity, taken as a proxy for time. 

\clearpage
\newpage

\section{Best fits}
\label{ap_1}

In this Section we gather the figures showing the best fit models (in red) compared to the observed UV and optical spectra of the XshootU sample stars. 

We summarise the main issues as follows. For some stars (BI~173, SK-67$^{^{\circ}}$101, SK-67$^{^{\circ}}$191, N11~018, AzZV15, AzV95, AzV207) \ion{N}{iv}~3478-83 is slightly under-predicted compared to other nitrogen lines. We found that increasing the effective temperature by 500 to 1000~K improves the fit at the cost of degrading the fit of the helium lines. Details of the line-blanketing such as microturbulence (not only in the spectrum computation, but also in the atmospheric calculations) affect the shape of the \ion{N}{iv} lines.

In some cases we encountered an inconsistency in the amount of broadening inferred from the UV and optical spectra. For stars AzV148 and AzV307 fitting the UV spectra would require a stronger broadening, while for SK-70$^{^{\circ}}$13 the model overestimates the broadening of the optical spectrum. We double-checked the determination of the projected rotational velocity and verified that the instrumental broadening was correctly taken into account in those cases. The reason for the discrepancy remains unclear.

\begin{figure*}[th]
\centering
\includegraphics[trim= 0.cm 0.cm 0.cm 0.cm ,clip,width=0.49\textwidth]{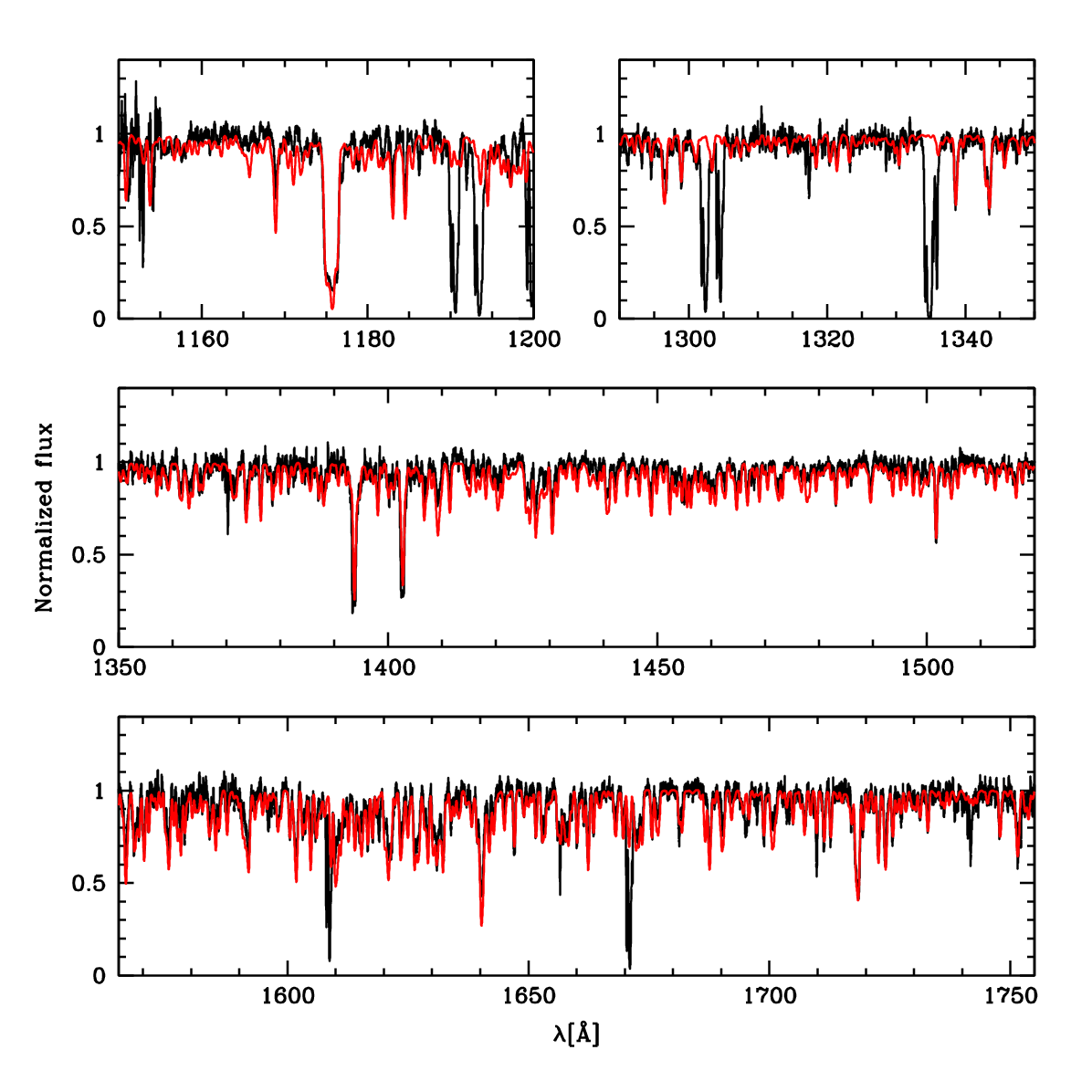}
\includegraphics[trim= 0.cm 0.cm 0.cm 0.cm ,clip,width=0.49\textwidth]{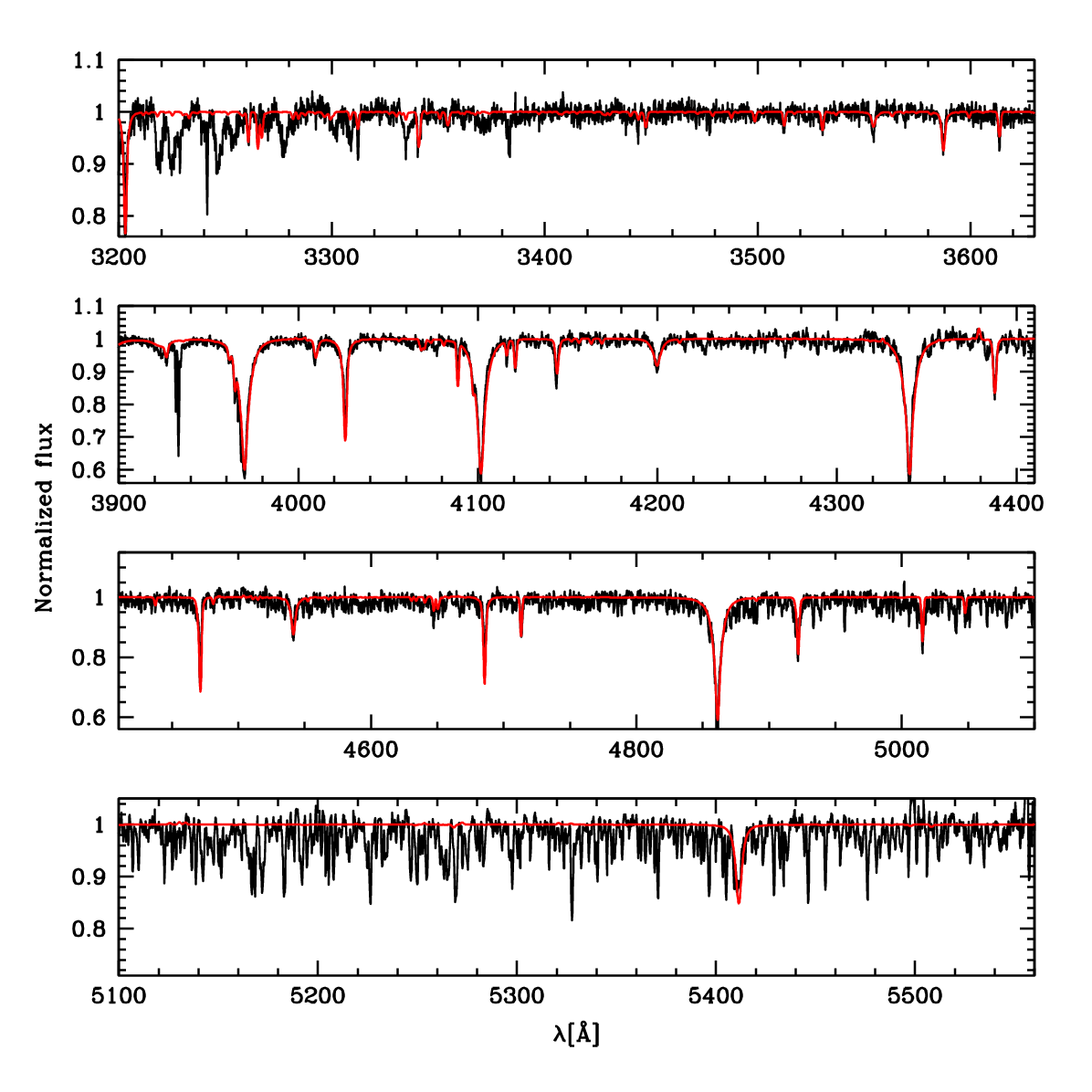}\\
\includegraphics[trim= 0.cm 0.cm 0.cm 0.cm ,clip,width=0.49\textwidth]{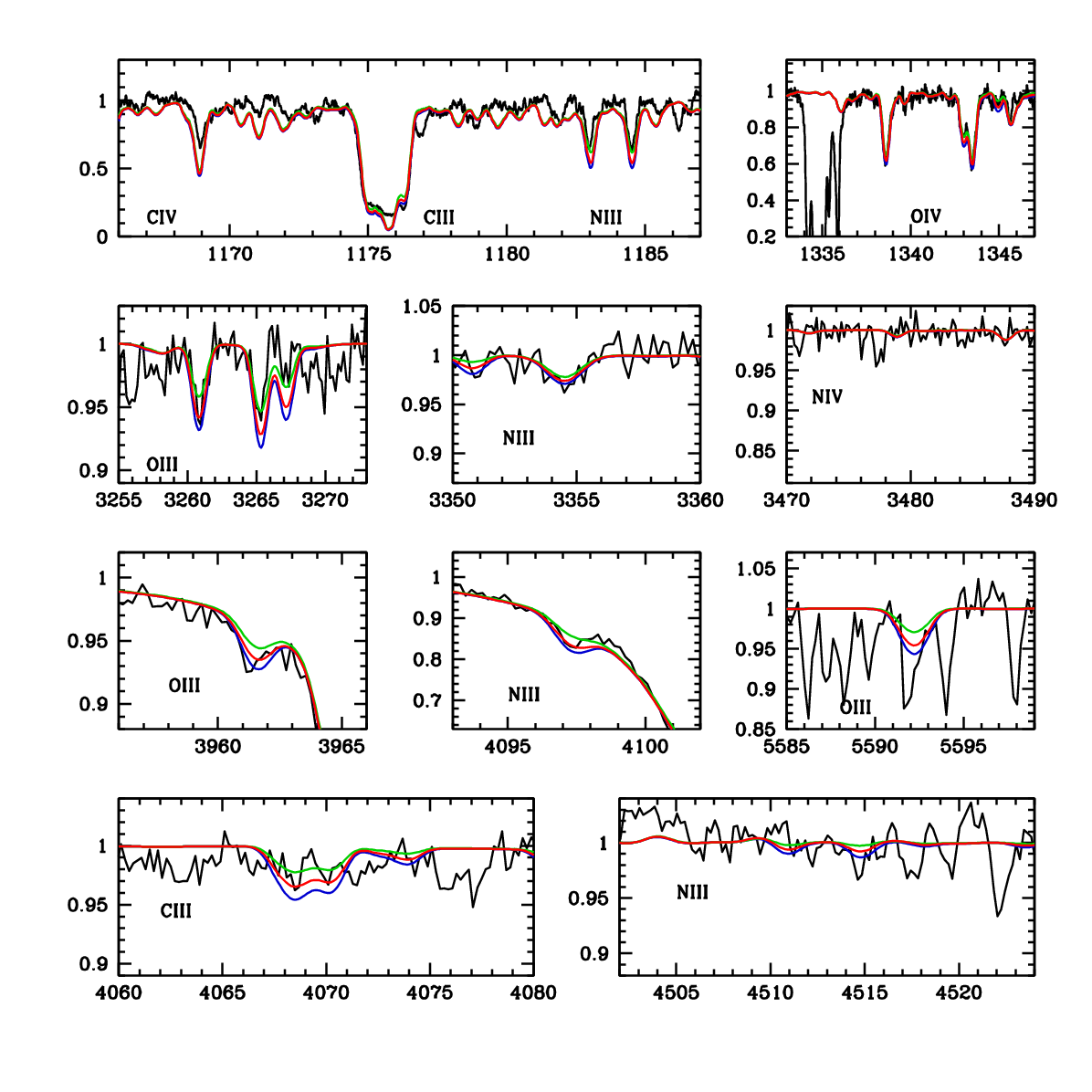}
\caption{Best fit model (red) of the spectrum (black) of AzV6. The top left (top right) panel shows the UV (optical) range. The bottom panel shows a zoom on CNO lines. In that panel the blue (green) line show the effect of increasing (decreasing) the abundances by the uncertainties quoted in Table.\ref{tab_smc}.} 
\label{fit_av6}
\end{figure*}

\begin{figure*}[ht]
\centering
\includegraphics[width=0.49\textwidth]{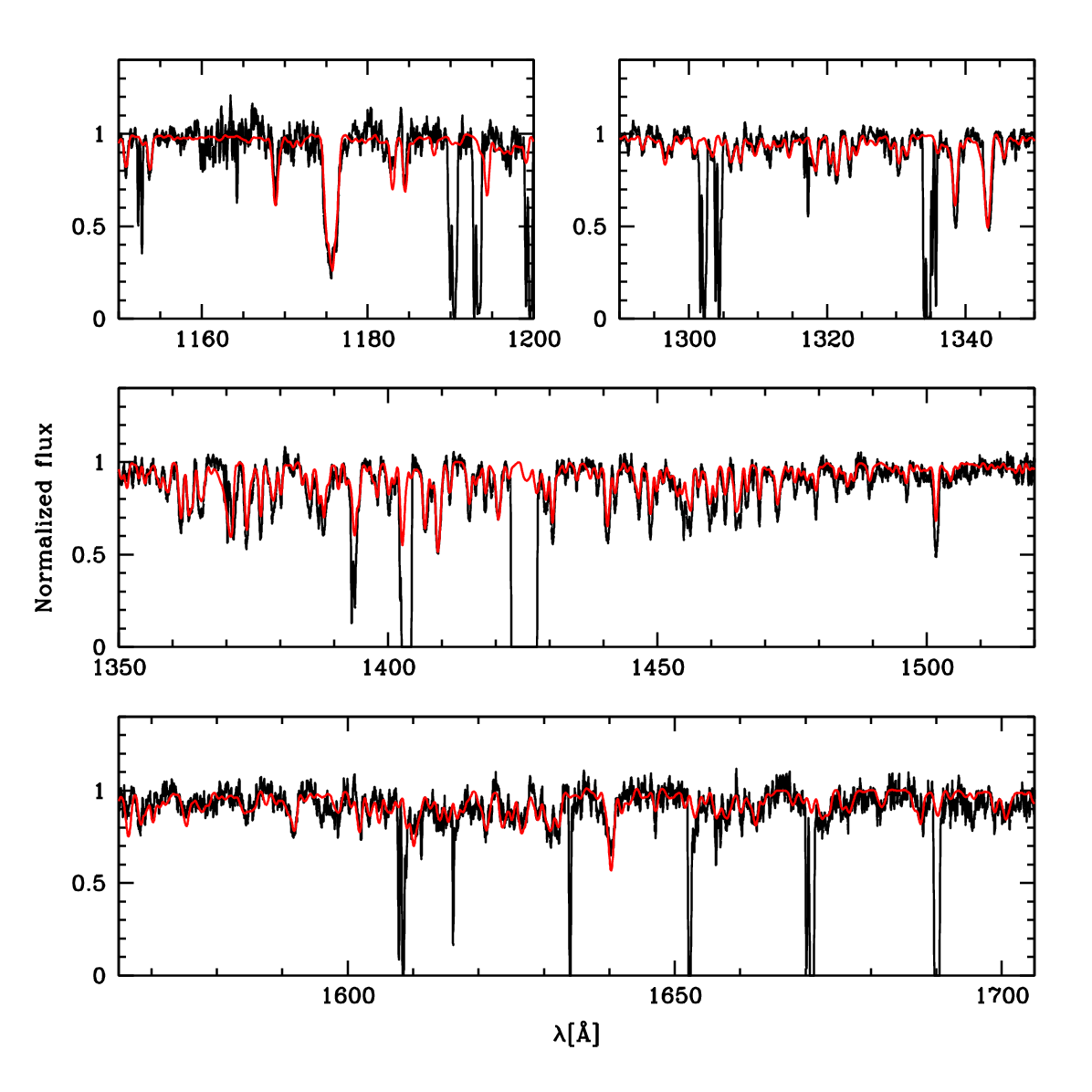}
\includegraphics[width=0.49\textwidth]{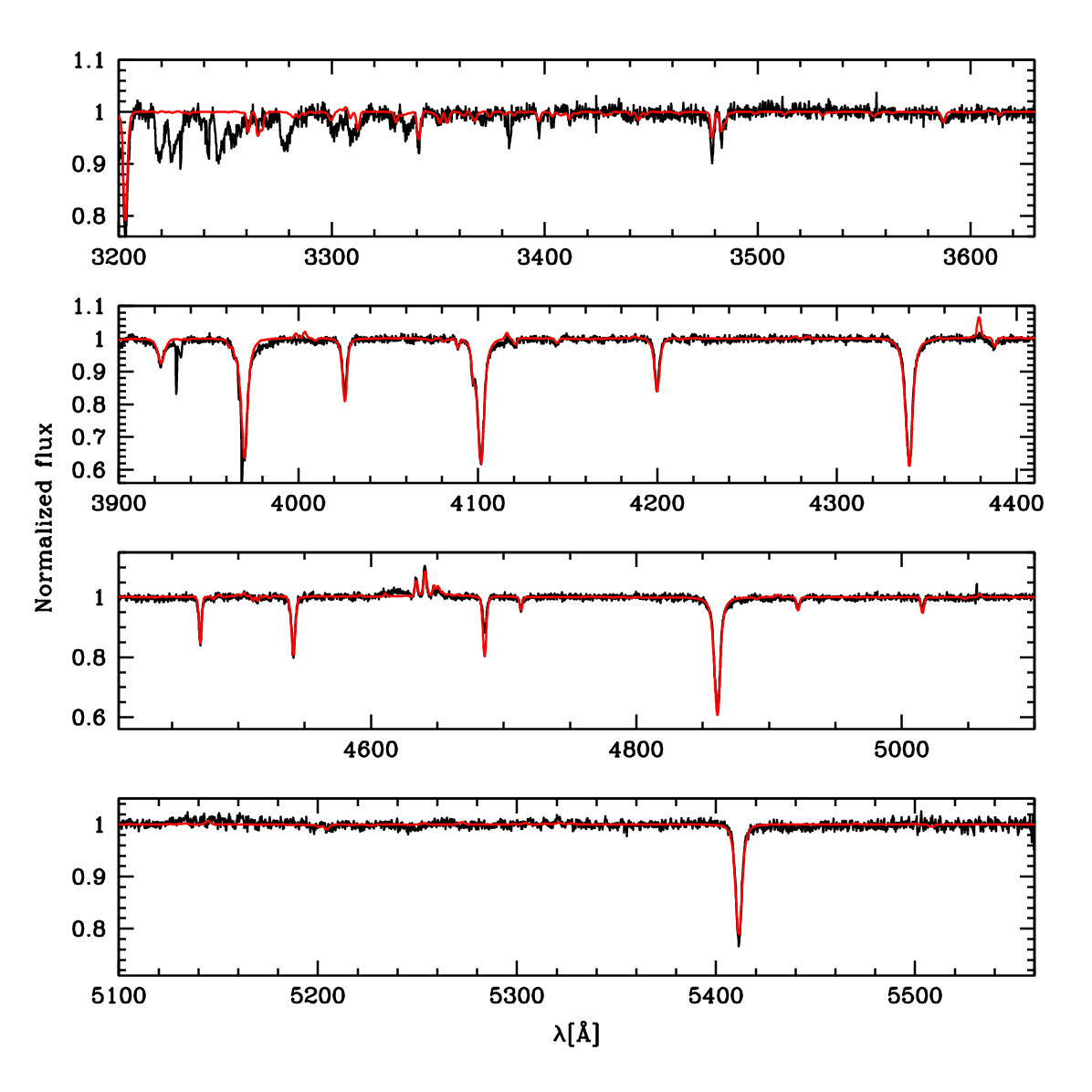}\\
\includegraphics[width=0.75\textwidth]{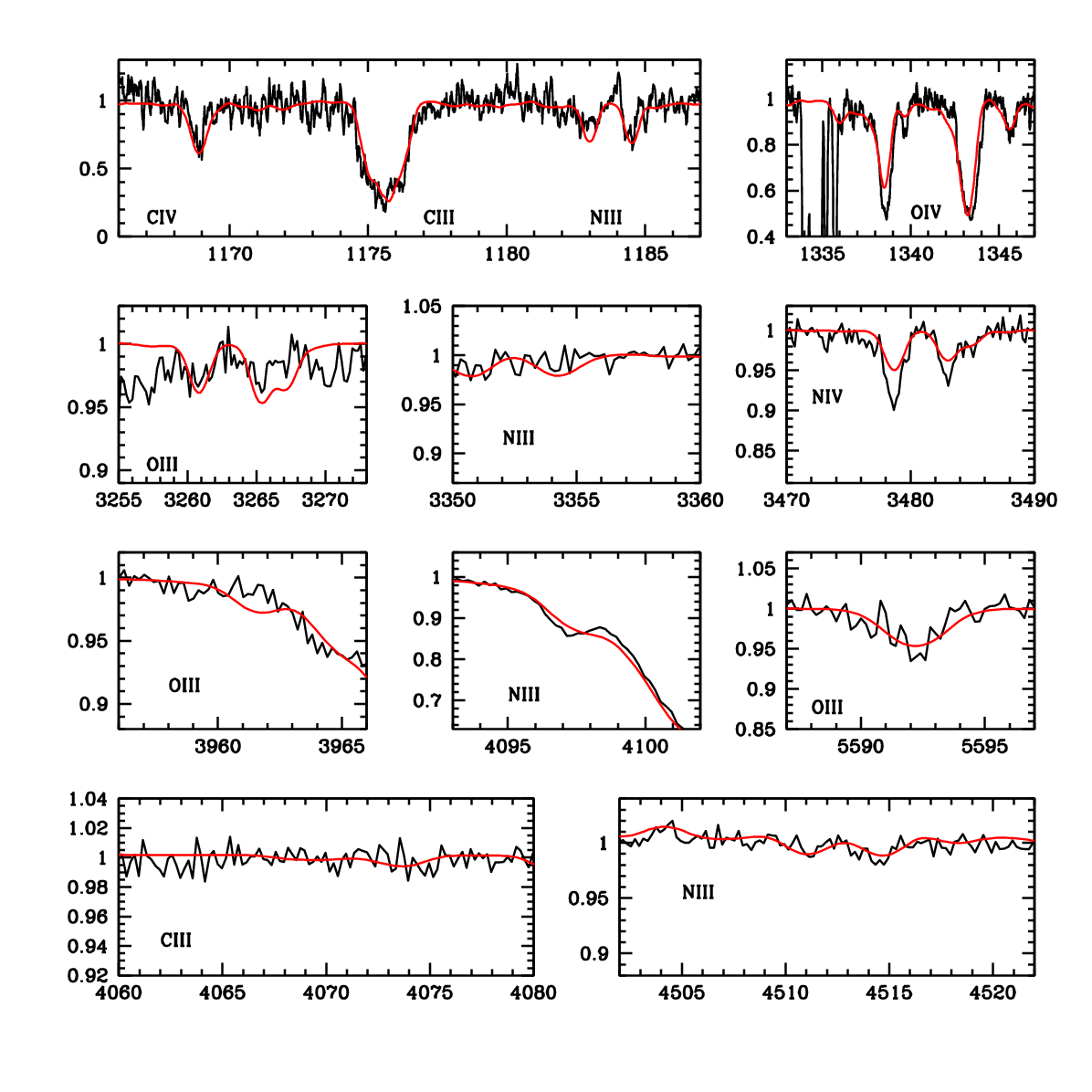}
\caption{Best fit model (red) of the spectrum (black) of AzV15. The top left (top right) panel shows the UV (optical) range. The bottom panel shows a zoom on CNO lines.} 
\label{fit_av15}
\end{figure*}

\begin{figure*}[ht]
\centering
\includegraphics[width=0.49\textwidth]{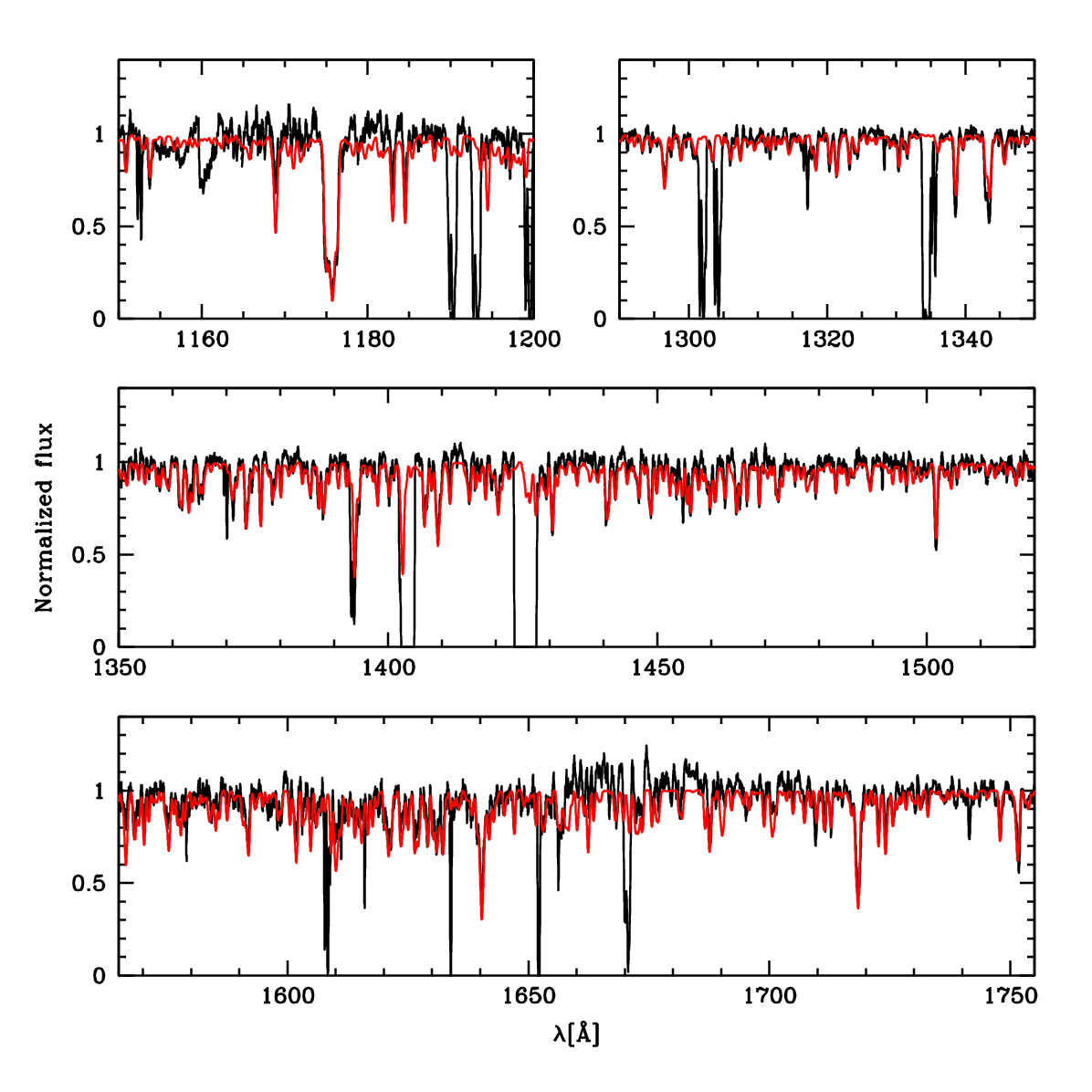}
\includegraphics[width=0.49\textwidth]{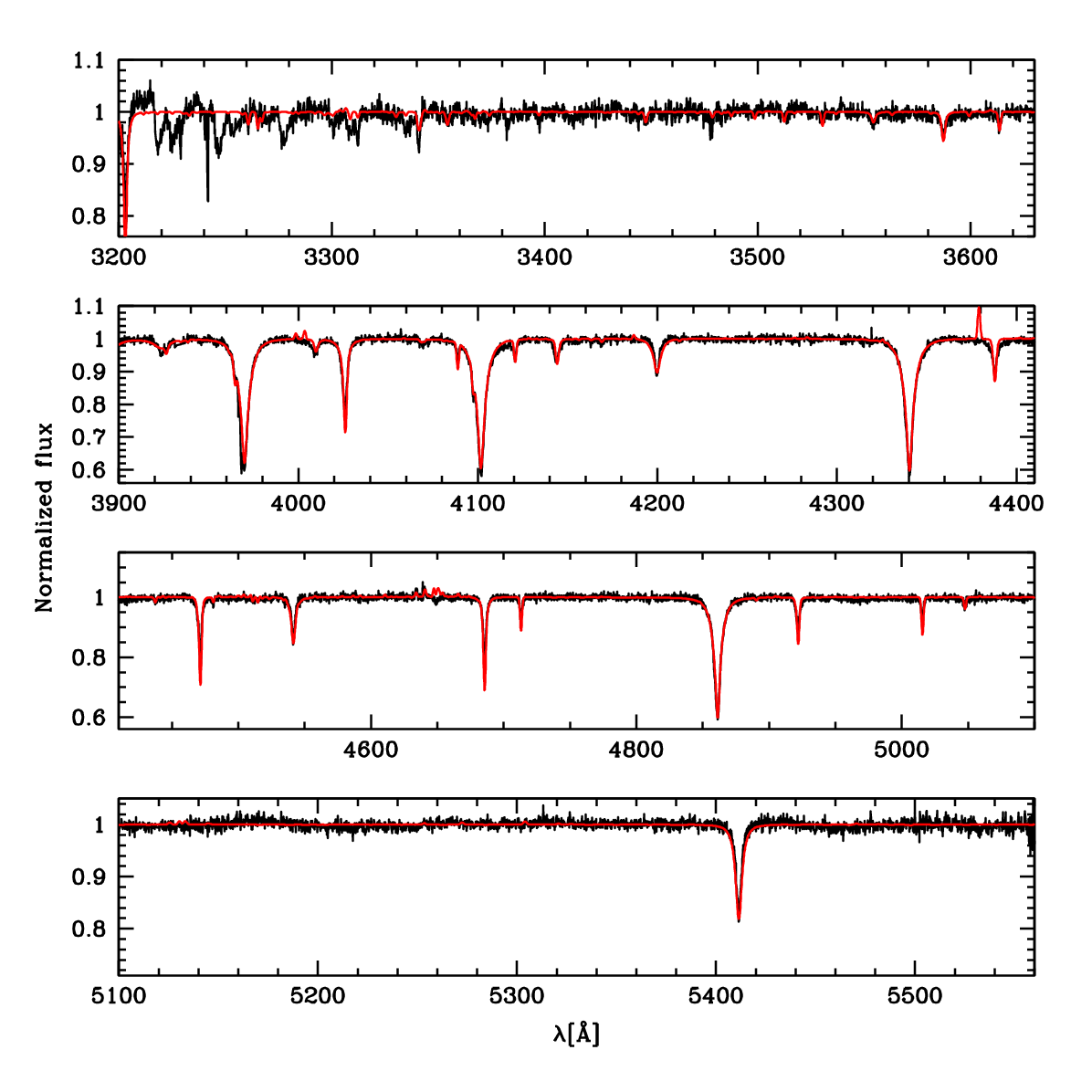}\\
\includegraphics[width=0.75\textwidth]{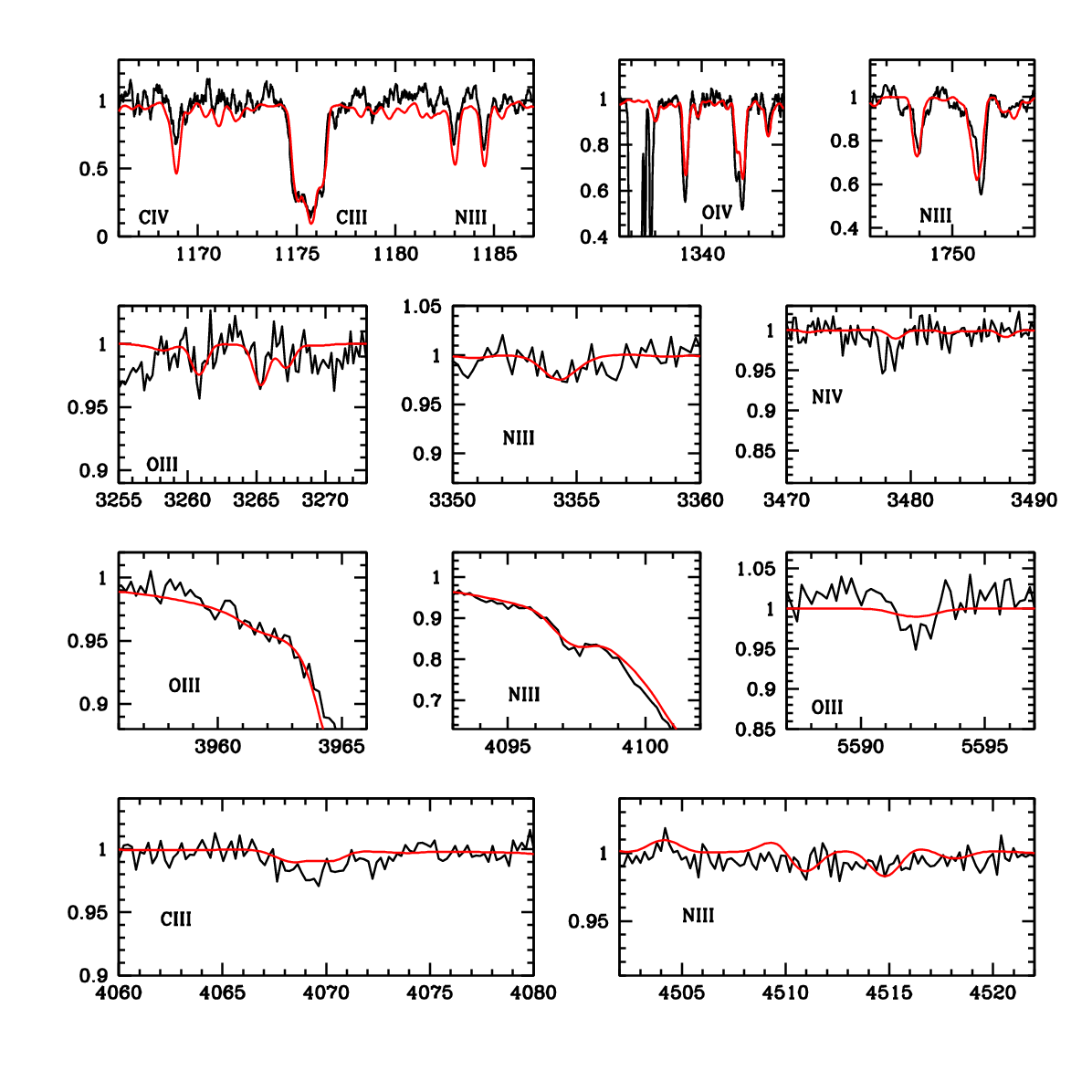}
\caption{Same as Fig.~\ref{fit_av15} but for AzV47.} 
\label{fit_av47}
\end{figure*}

\begin{figure*}[ht]
\centering
\includegraphics[width=0.49\textwidth]{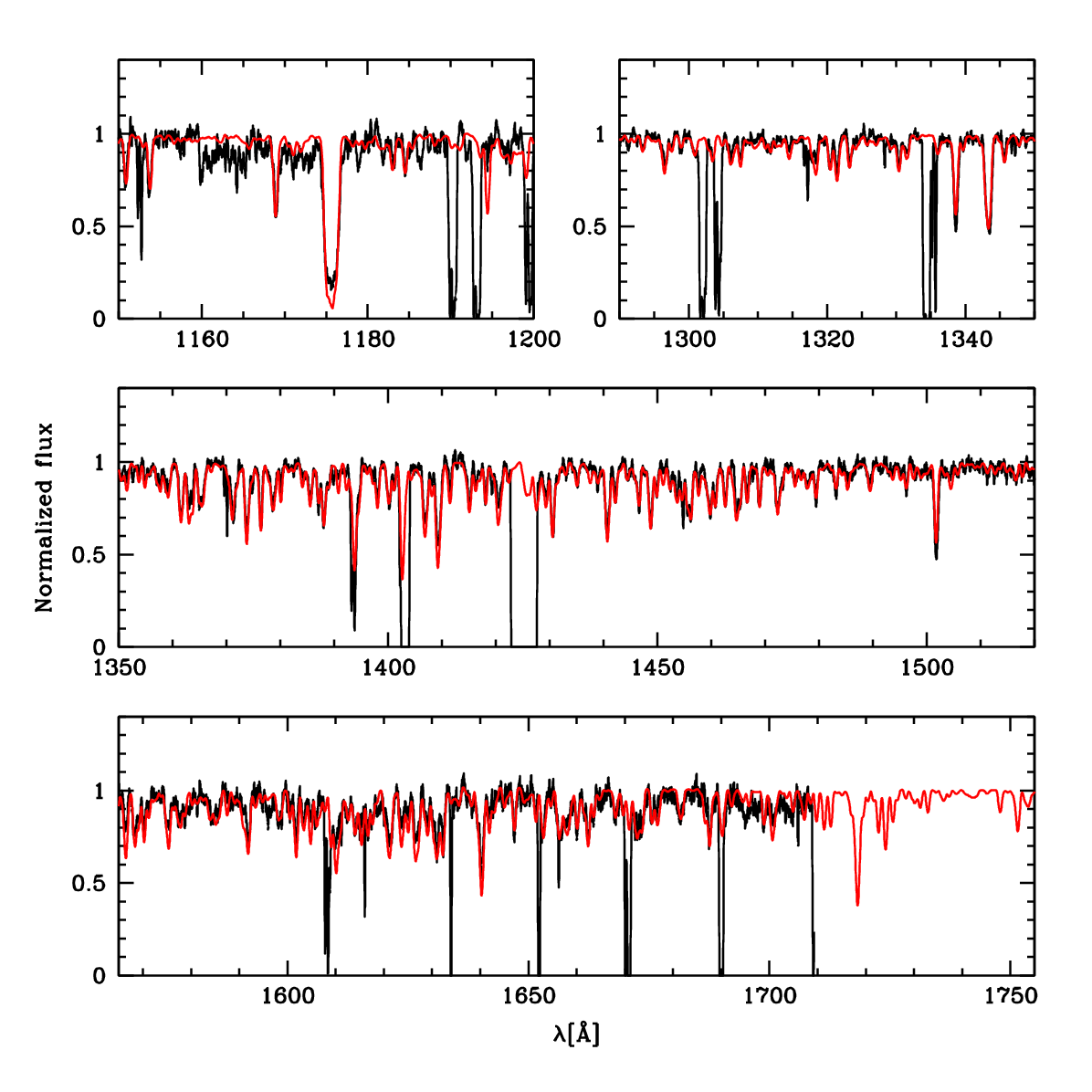}
\includegraphics[width=0.49\textwidth]{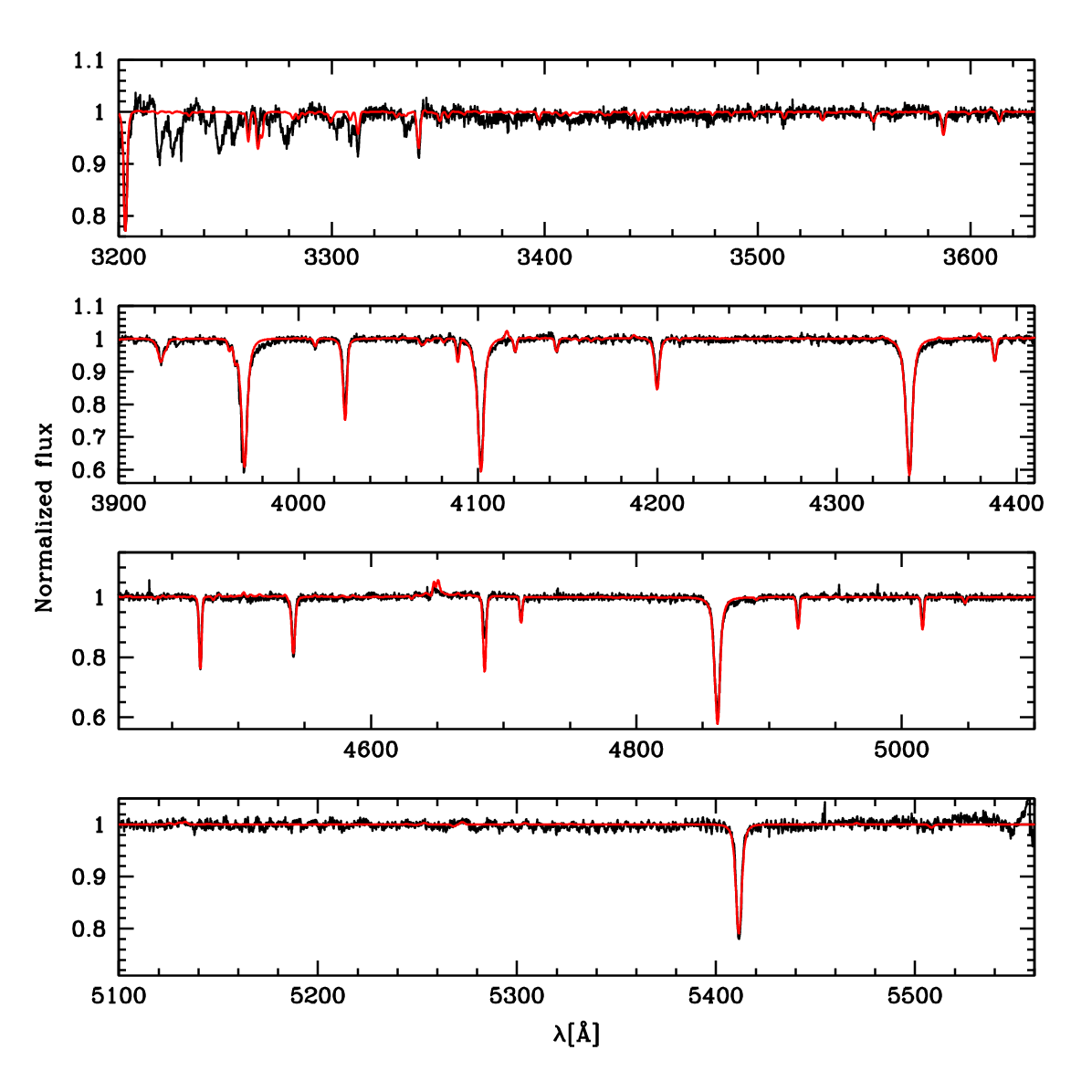}\\
\includegraphics[width=0.75\textwidth]{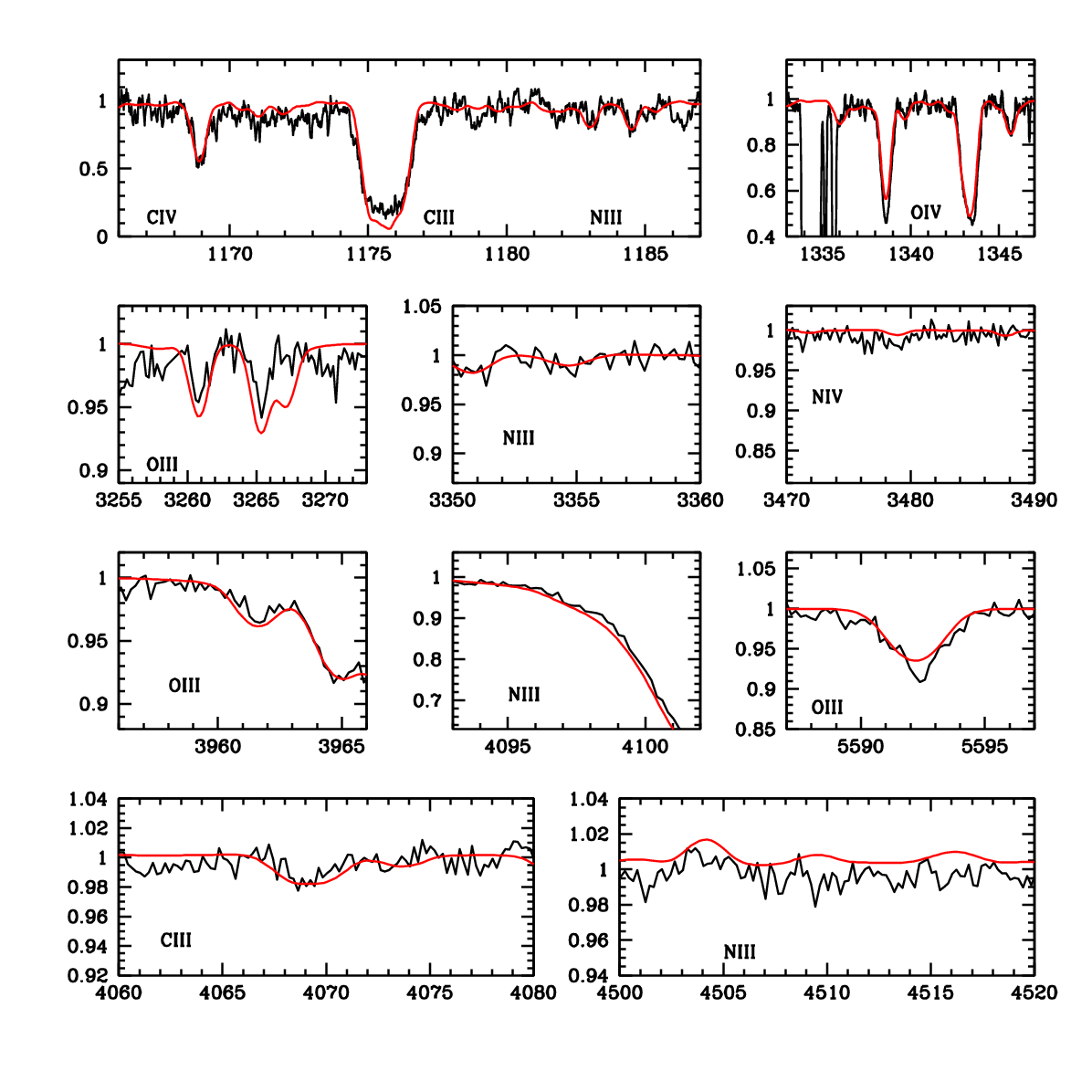}
\caption{Same as Fig.~\ref{fit_av15} but for AzV69.} 
\label{fit_av69}
\end{figure*}

\begin{figure*}[ht]
\centering
\includegraphics[width=0.49\textwidth]{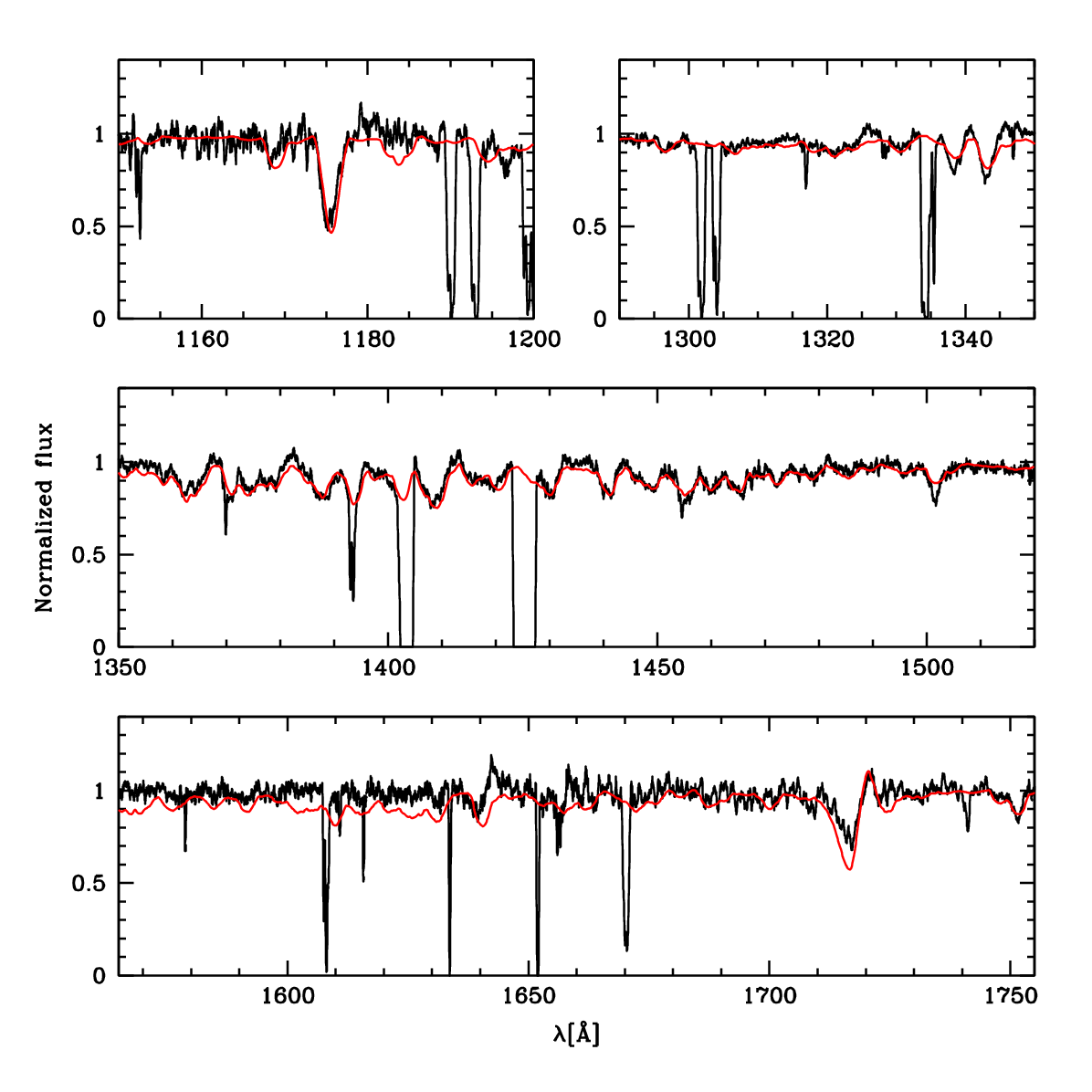}
\includegraphics[width=0.49\textwidth]{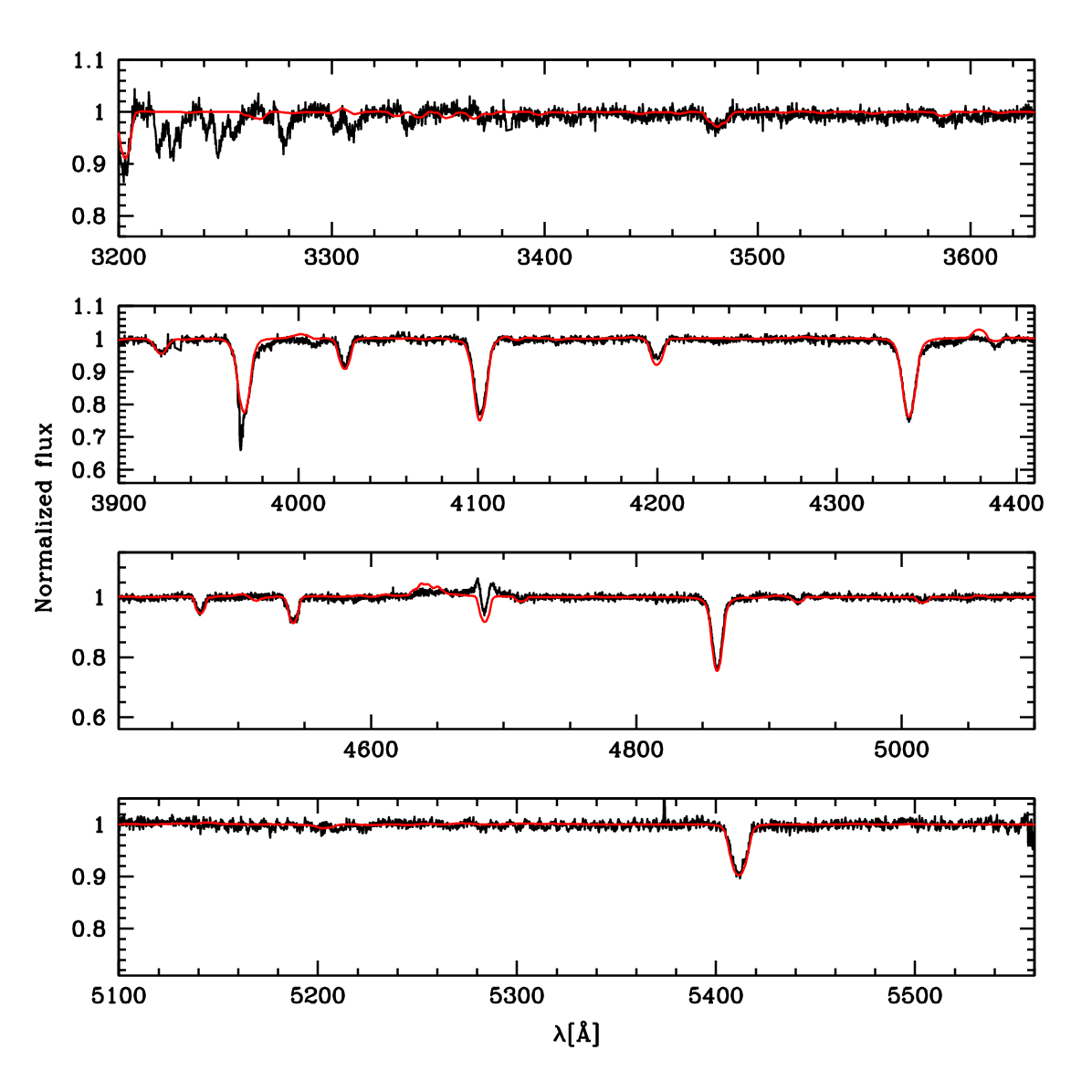}\\
\includegraphics[width=0.75\textwidth]{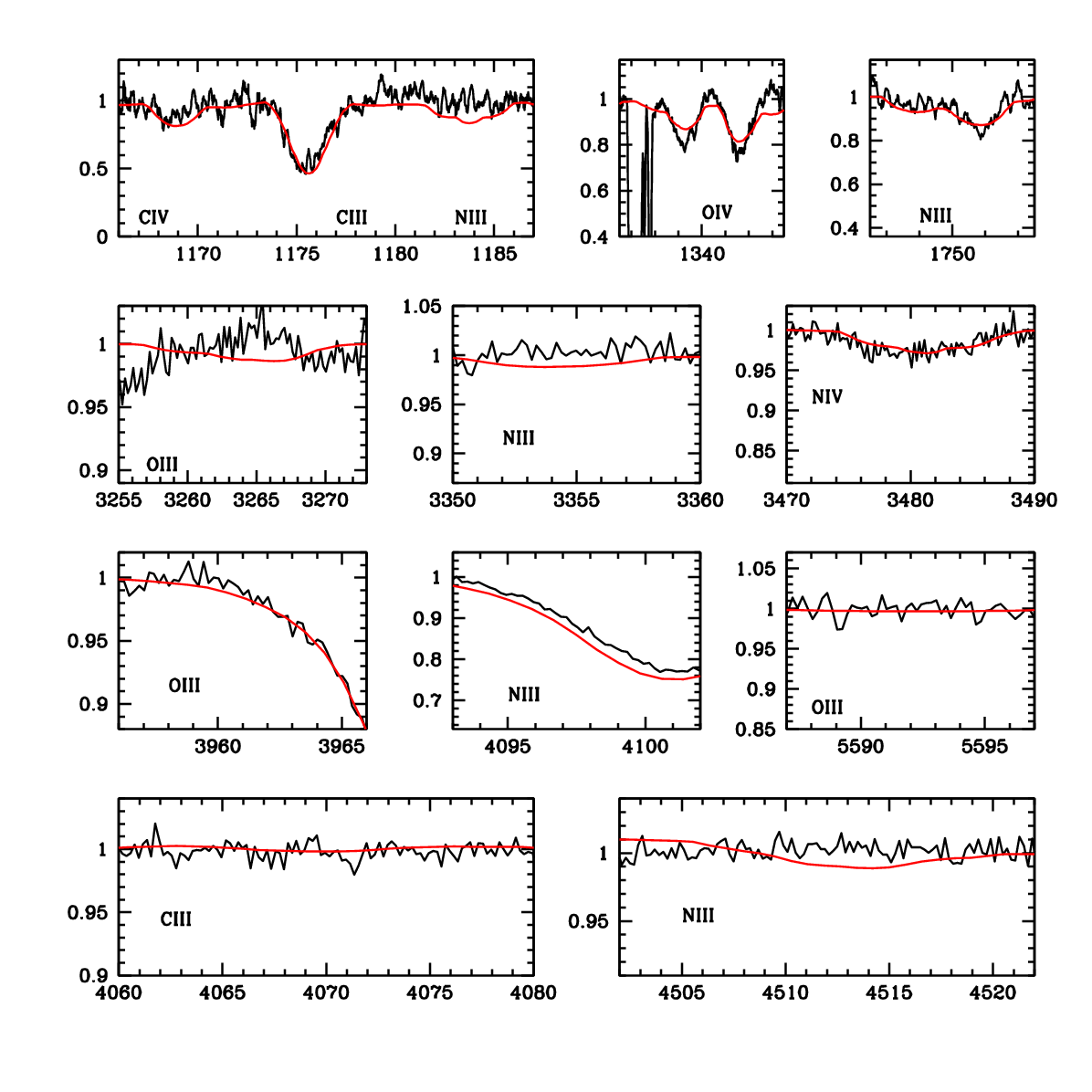}
\caption{Same as Fig.~\ref{fit_av15} but for AzV80.} 
\label{fit_av80}
\end{figure*}

\begin{figure*}[ht]
\centering
\includegraphics[width=0.49\textwidth]{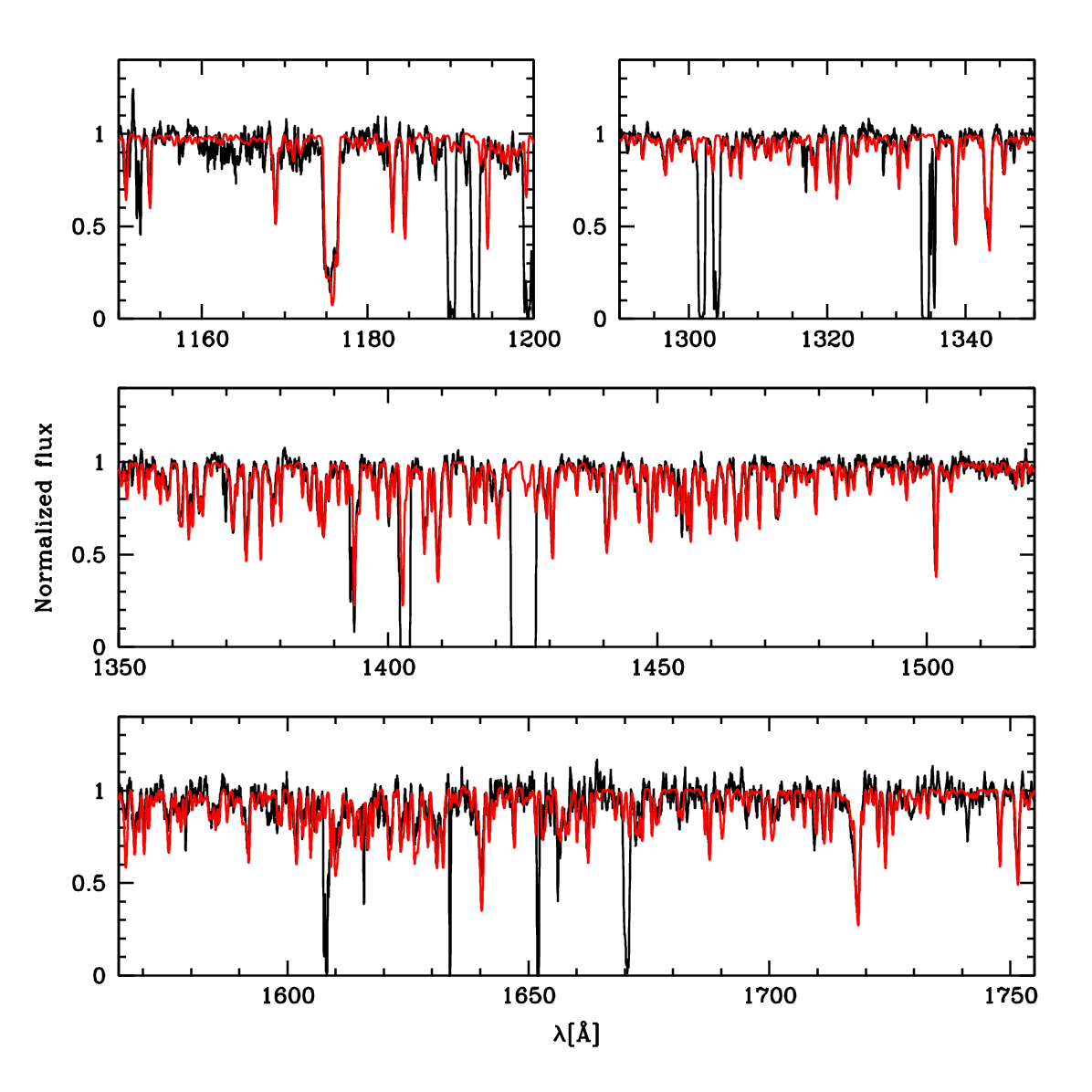}
\includegraphics[width=0.49\textwidth]{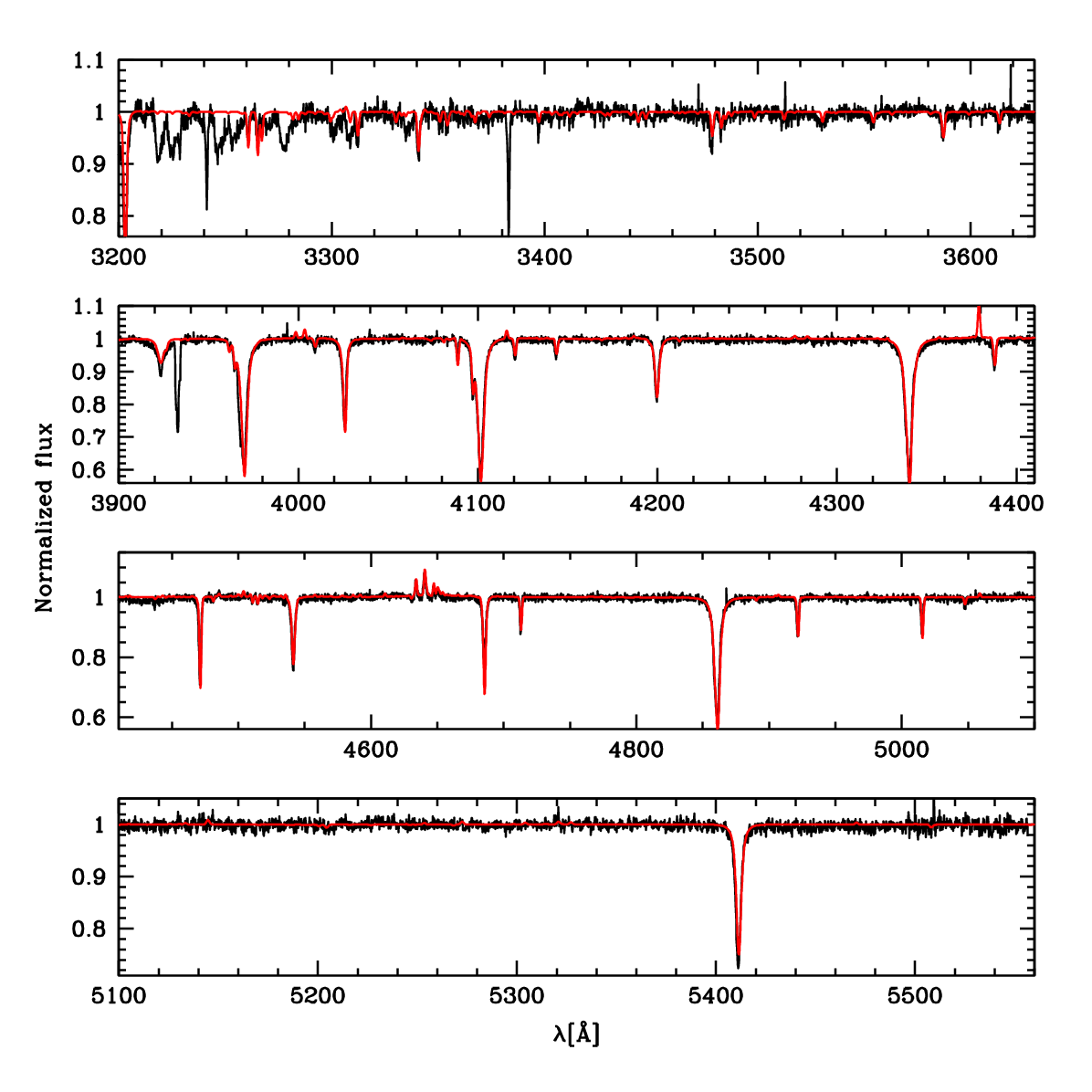}\\
\includegraphics[width=0.75\textwidth]{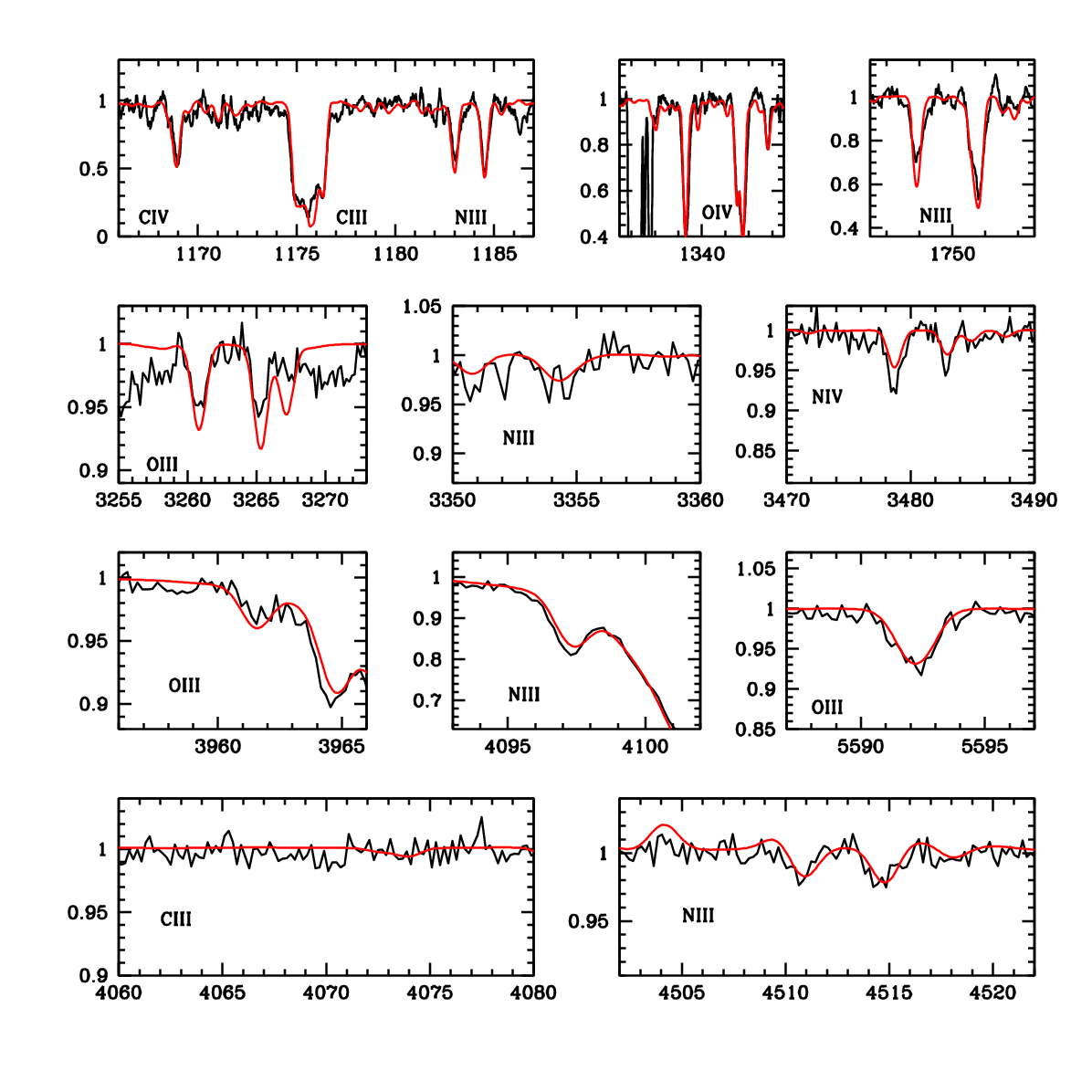}
\caption{Same as Fig.~\ref{fit_av15} but for AzV95.} 
\label{fit_av80}
\end{figure*}

\begin{figure*}[ht]
\centering
\includegraphics[width=0.49\textwidth]{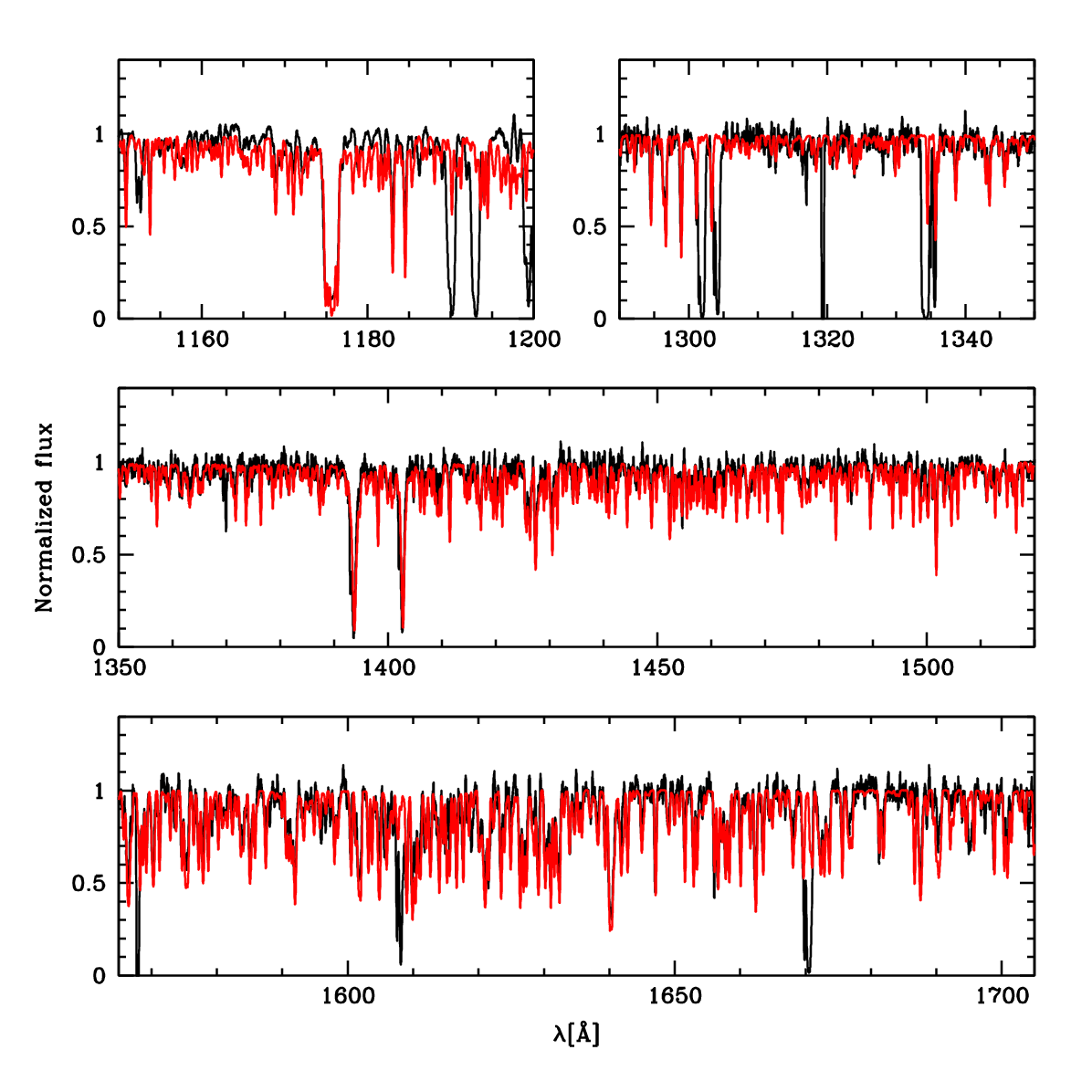}
\includegraphics[width=0.49\textwidth]{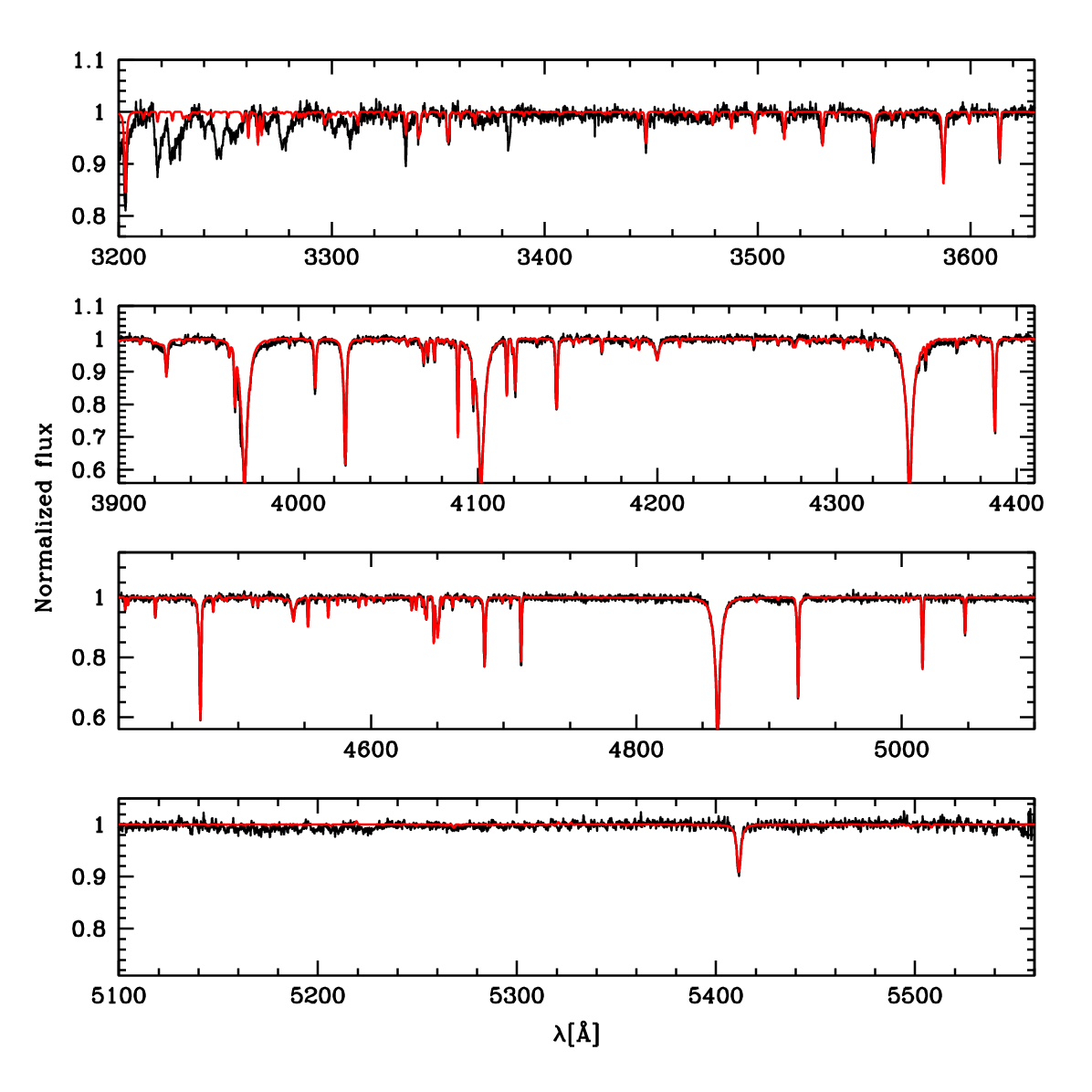}\\
\includegraphics[width=0.75\textwidth]{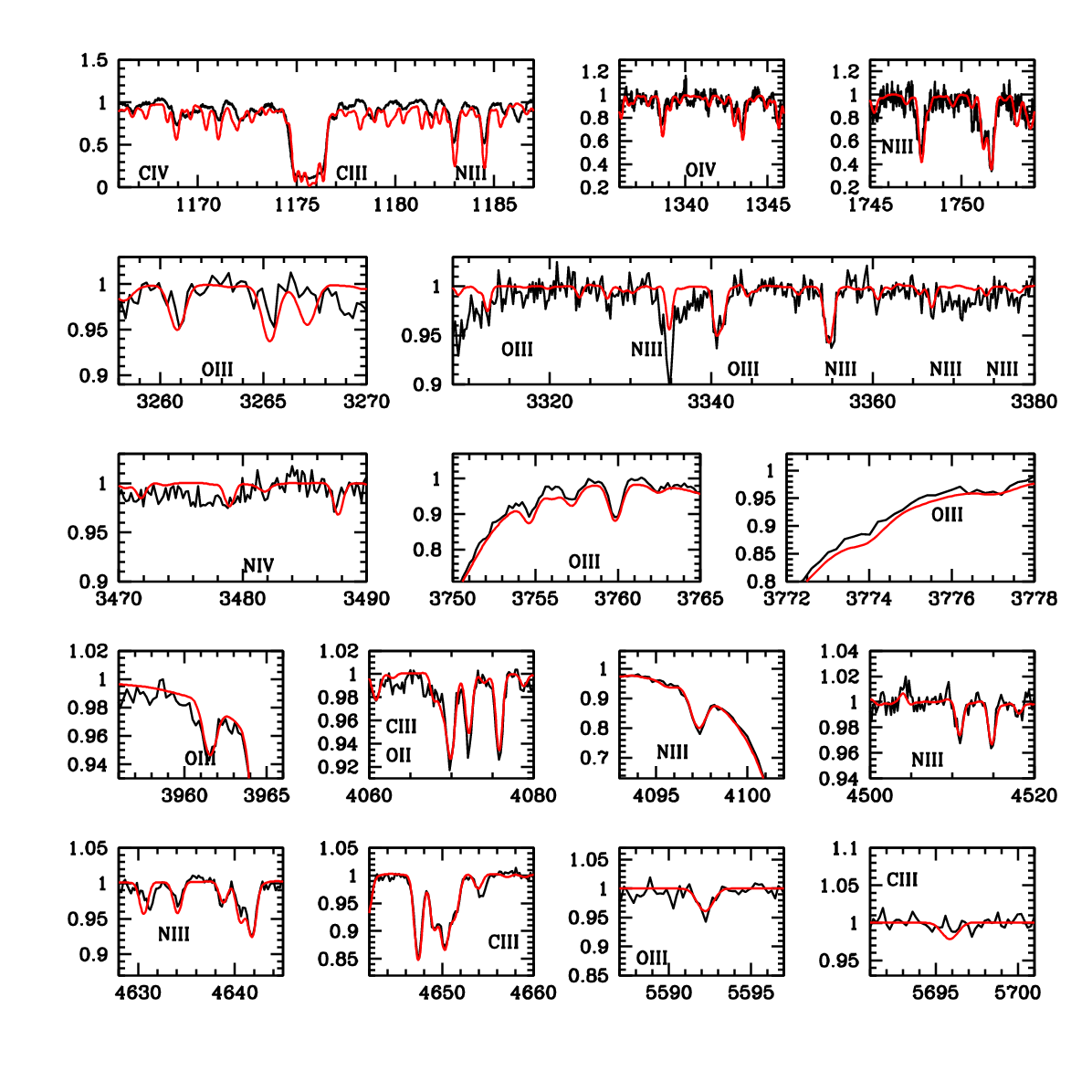}
\caption{Same as Fig.~\ref{fit_av15} but for AzV148.} 
\label{fit_av148}
\end{figure*}

\begin{figure*}[ht]
\centering
\includegraphics[width=0.49\textwidth]{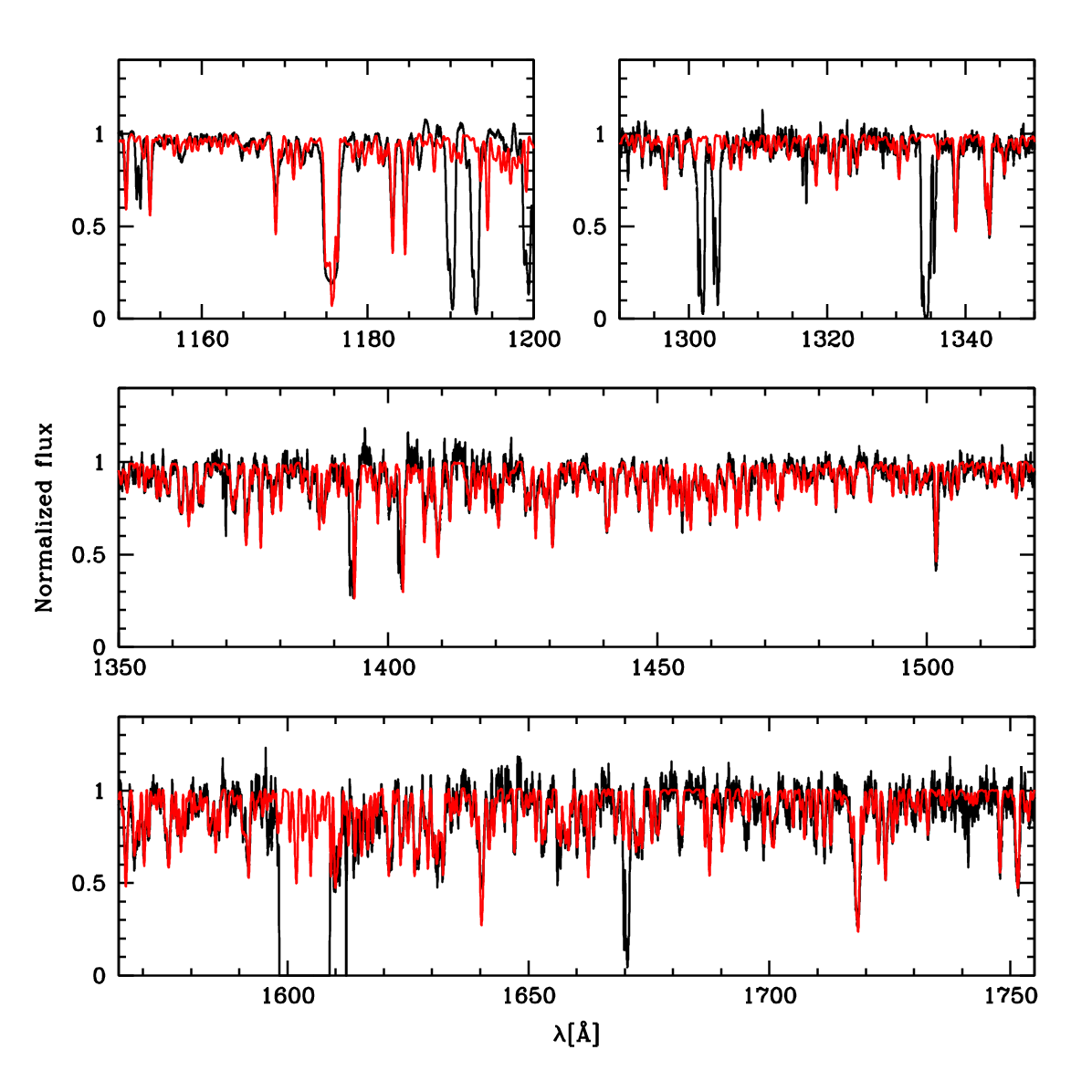}
\includegraphics[width=0.49\textwidth]{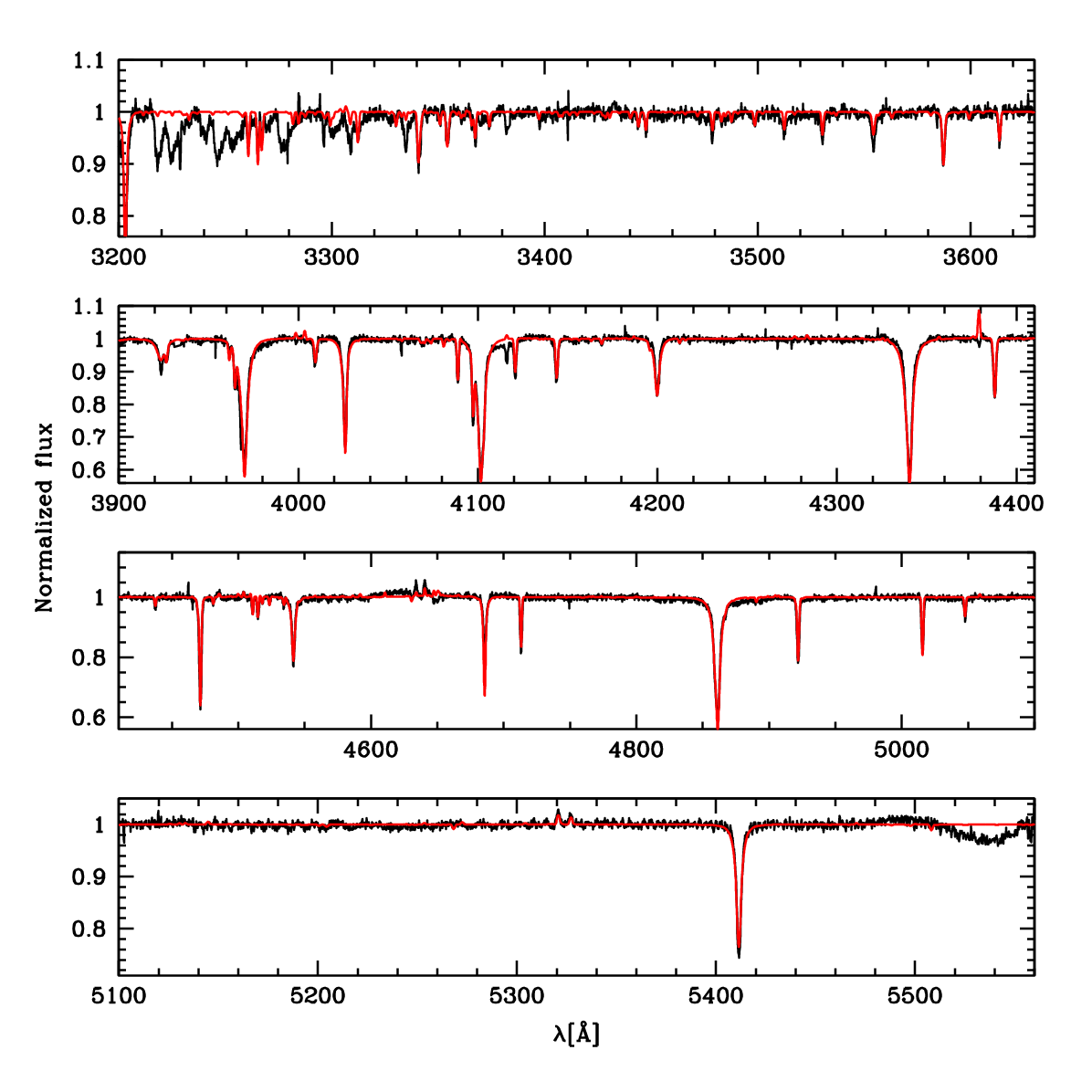}\\
\includegraphics[width=0.75\textwidth]{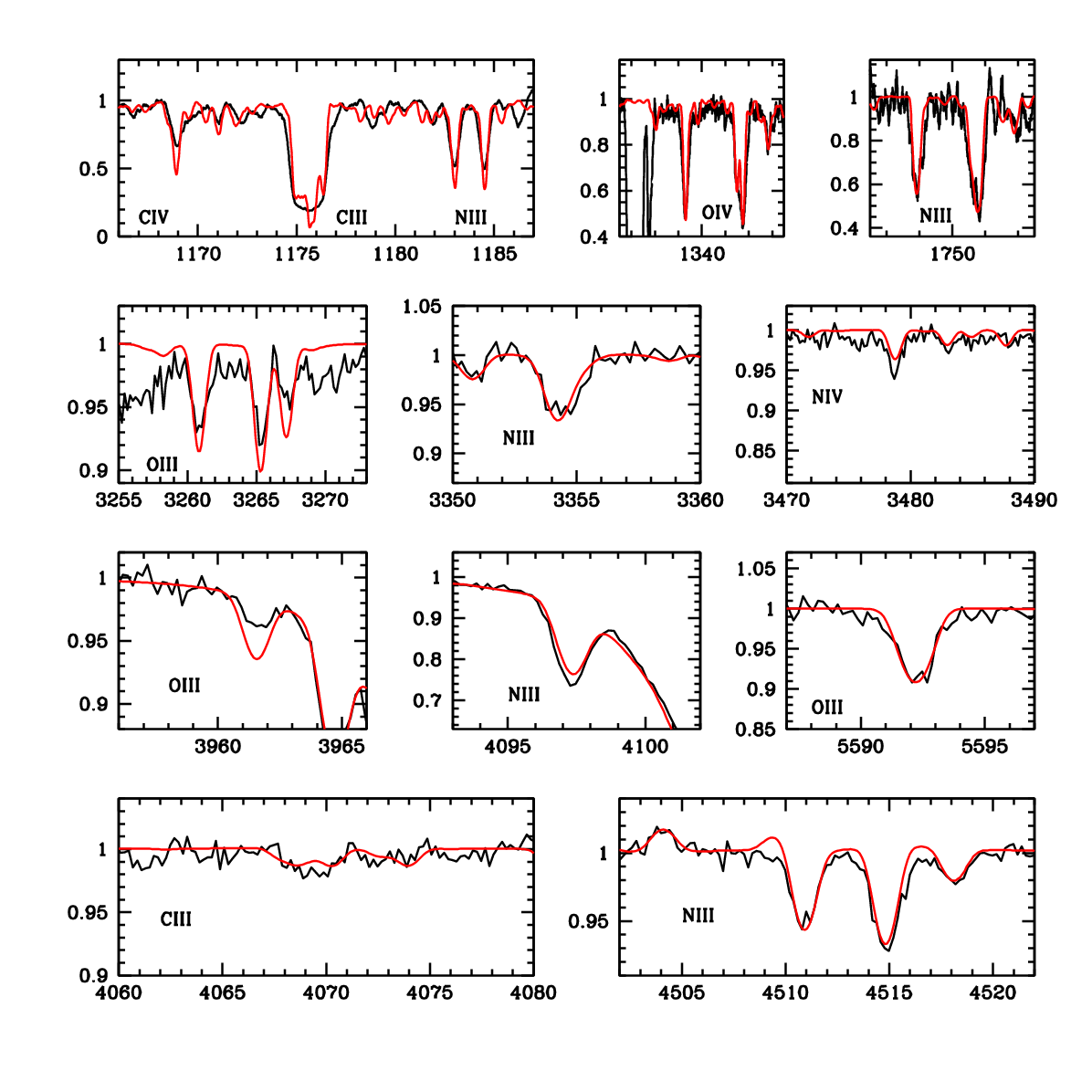}
\caption{Same as Fig.~\ref{fit_av15} but for AzV186.} 
\label{fit_av186}
\end{figure*}

\begin{figure*}[ht]
\centering
\includegraphics[width=0.49\textwidth]{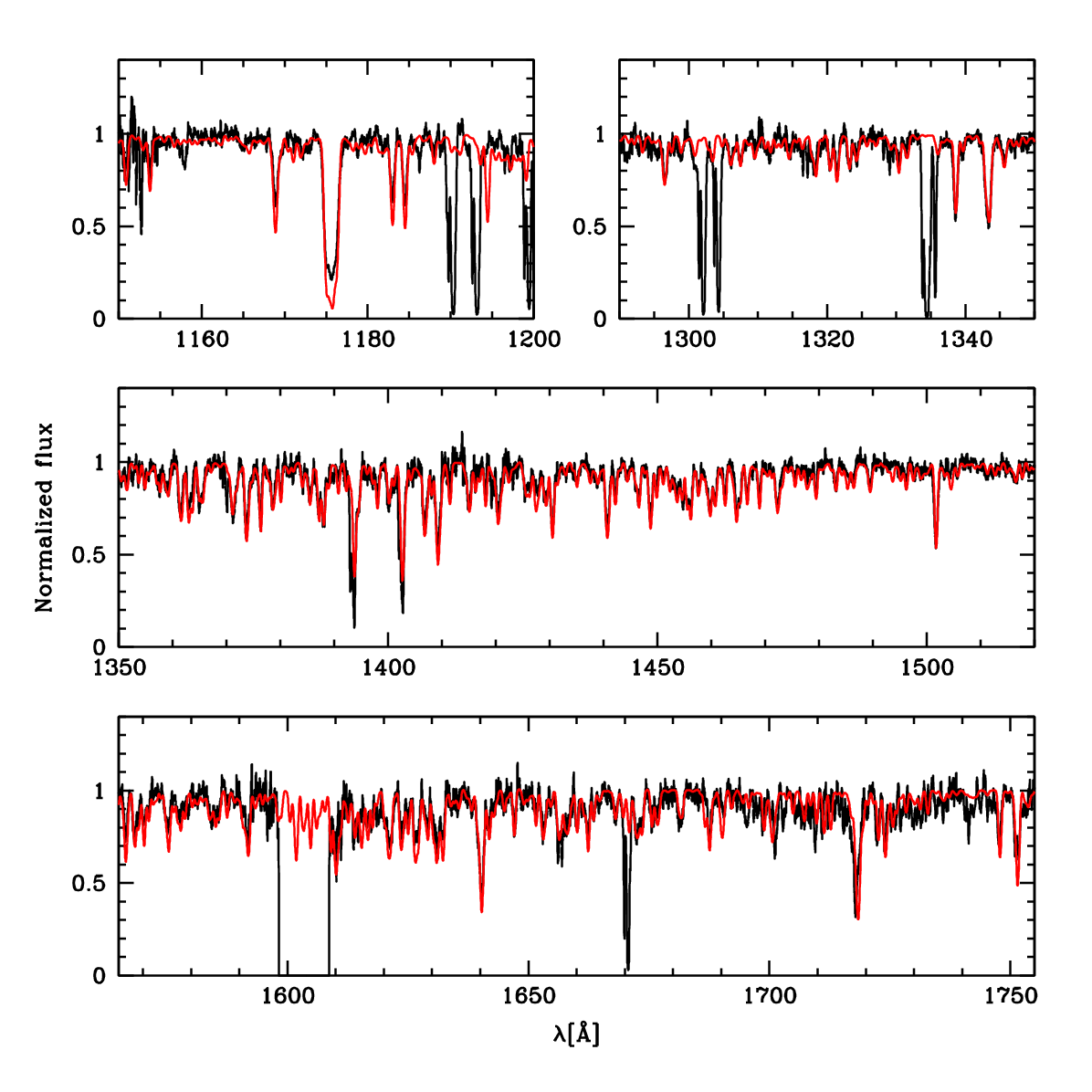}
\includegraphics[width=0.49\textwidth]{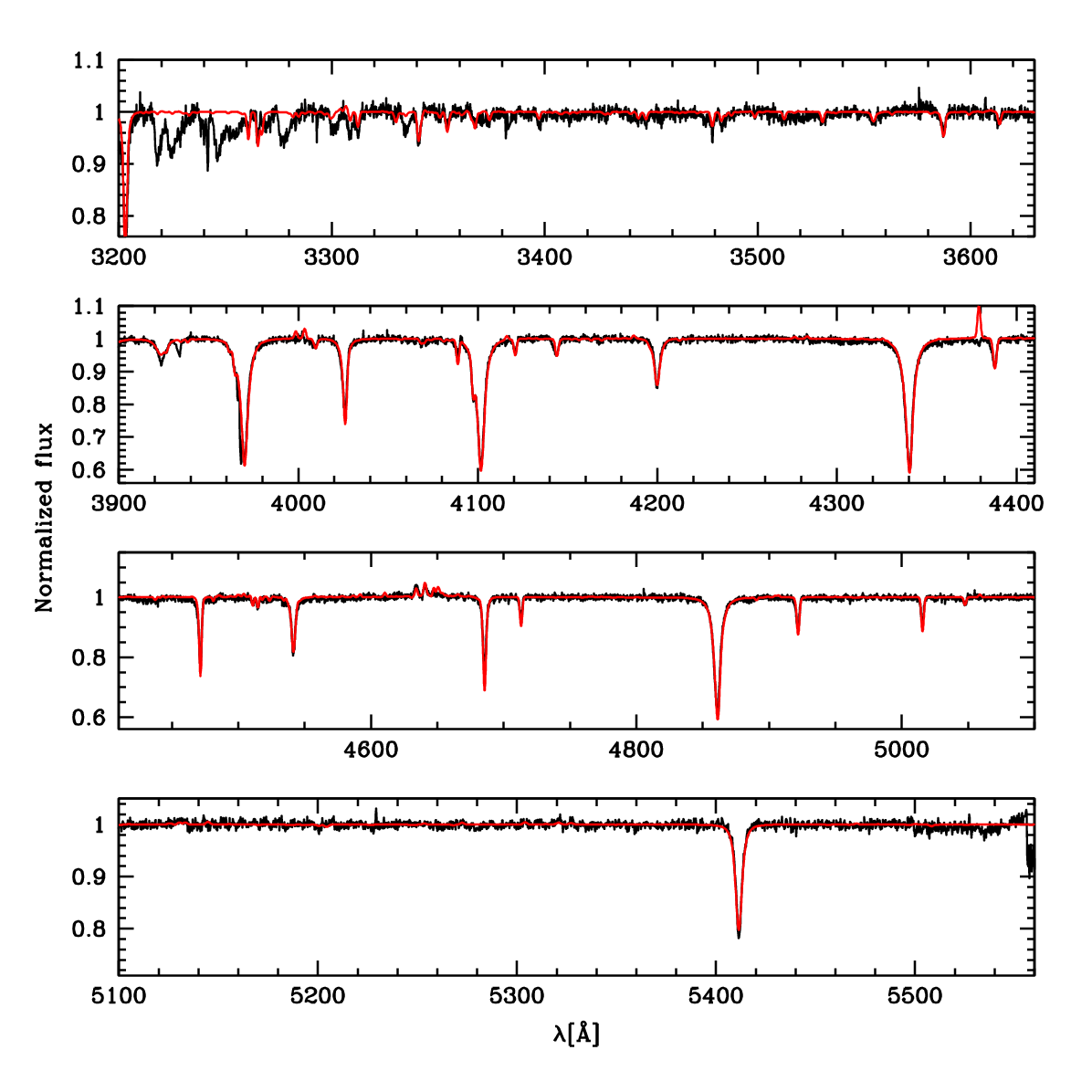}\\
\includegraphics[width=0.75\textwidth]{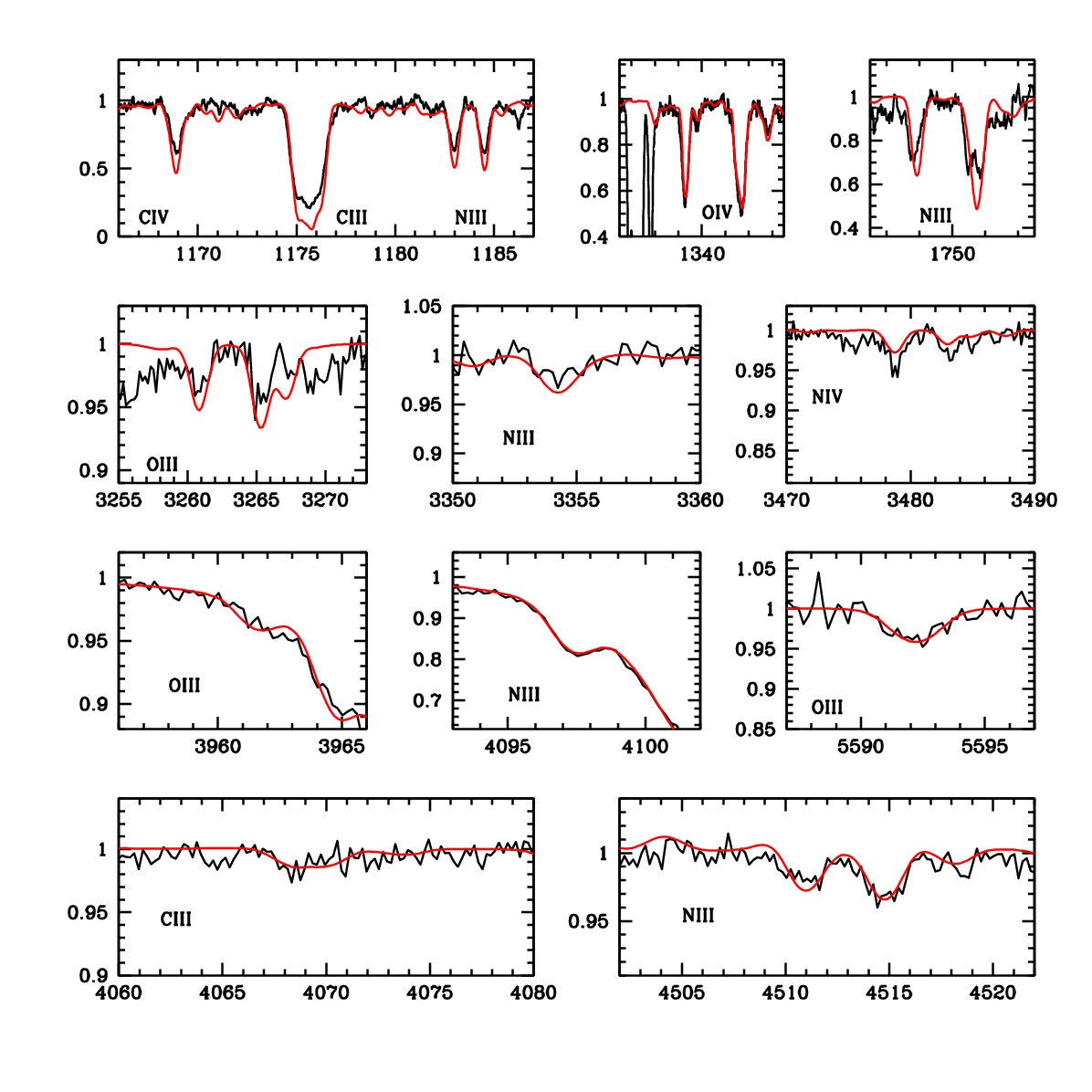}
\caption{Same as Fig.~\ref{fit_av15} but for AzV207.} 
\label{fit_av207}
\end{figure*}

\begin{figure*}[ht]
\centering
\includegraphics[width=0.49\textwidth]{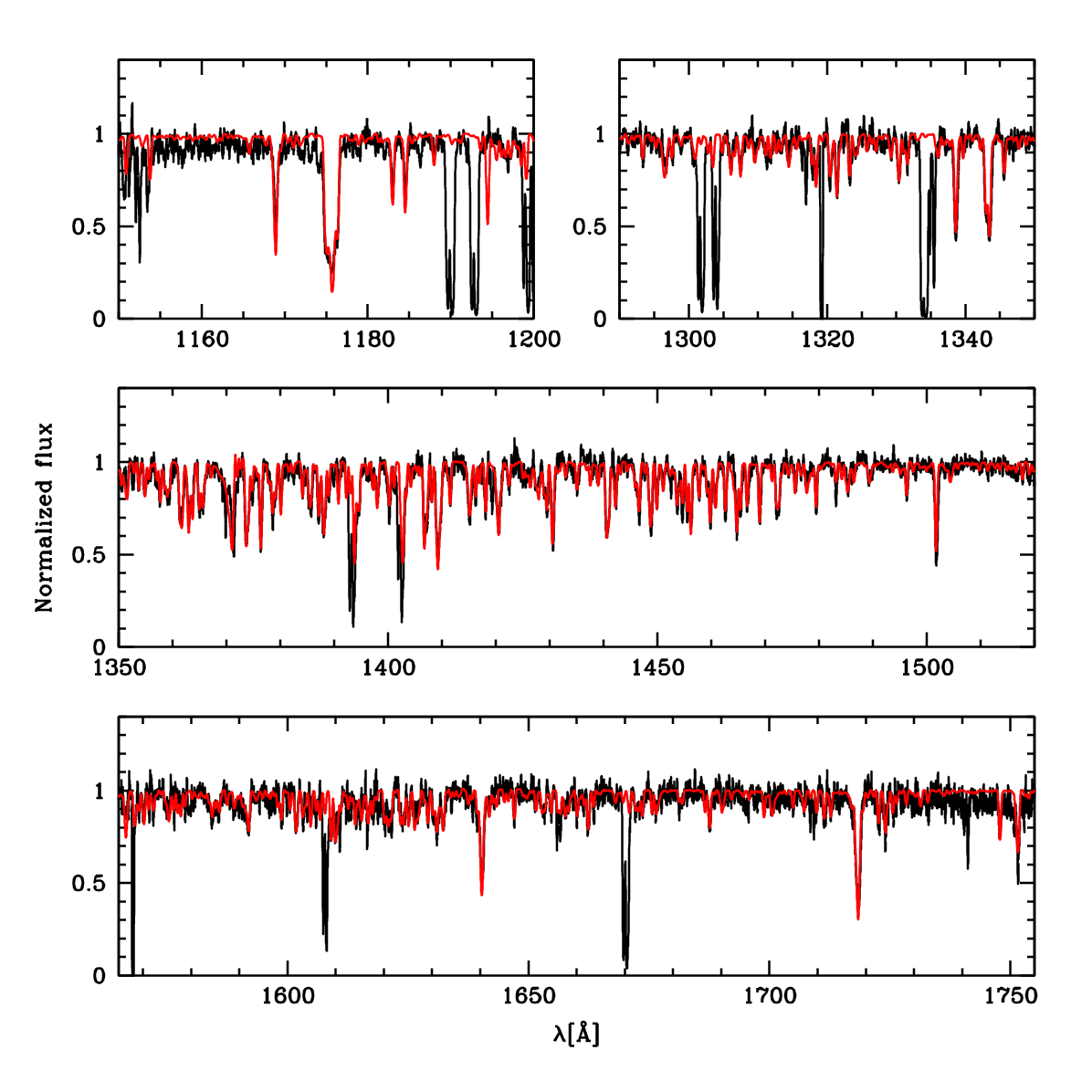}
\includegraphics[width=0.49\textwidth]{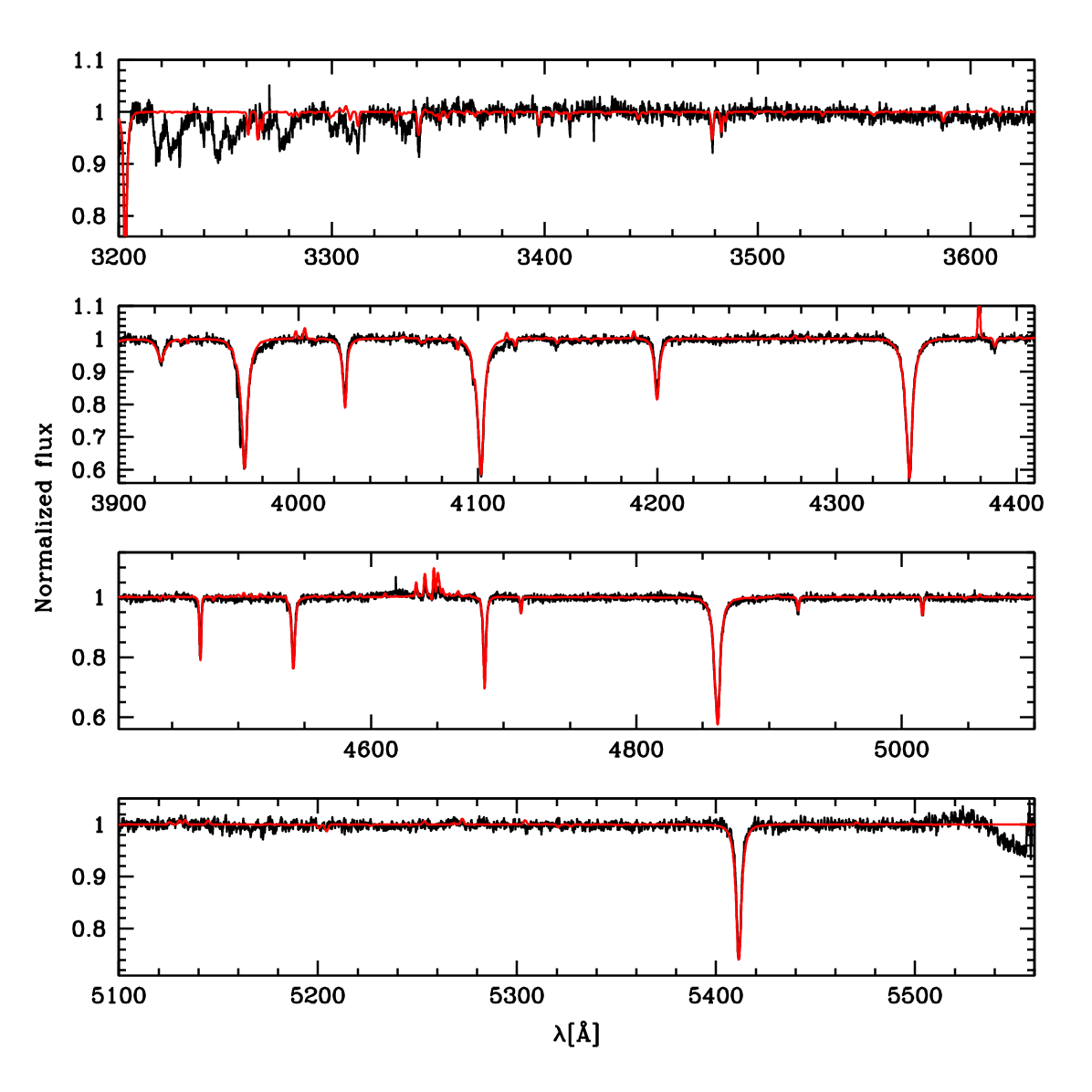}\\
\includegraphics[width=0.75\textwidth]{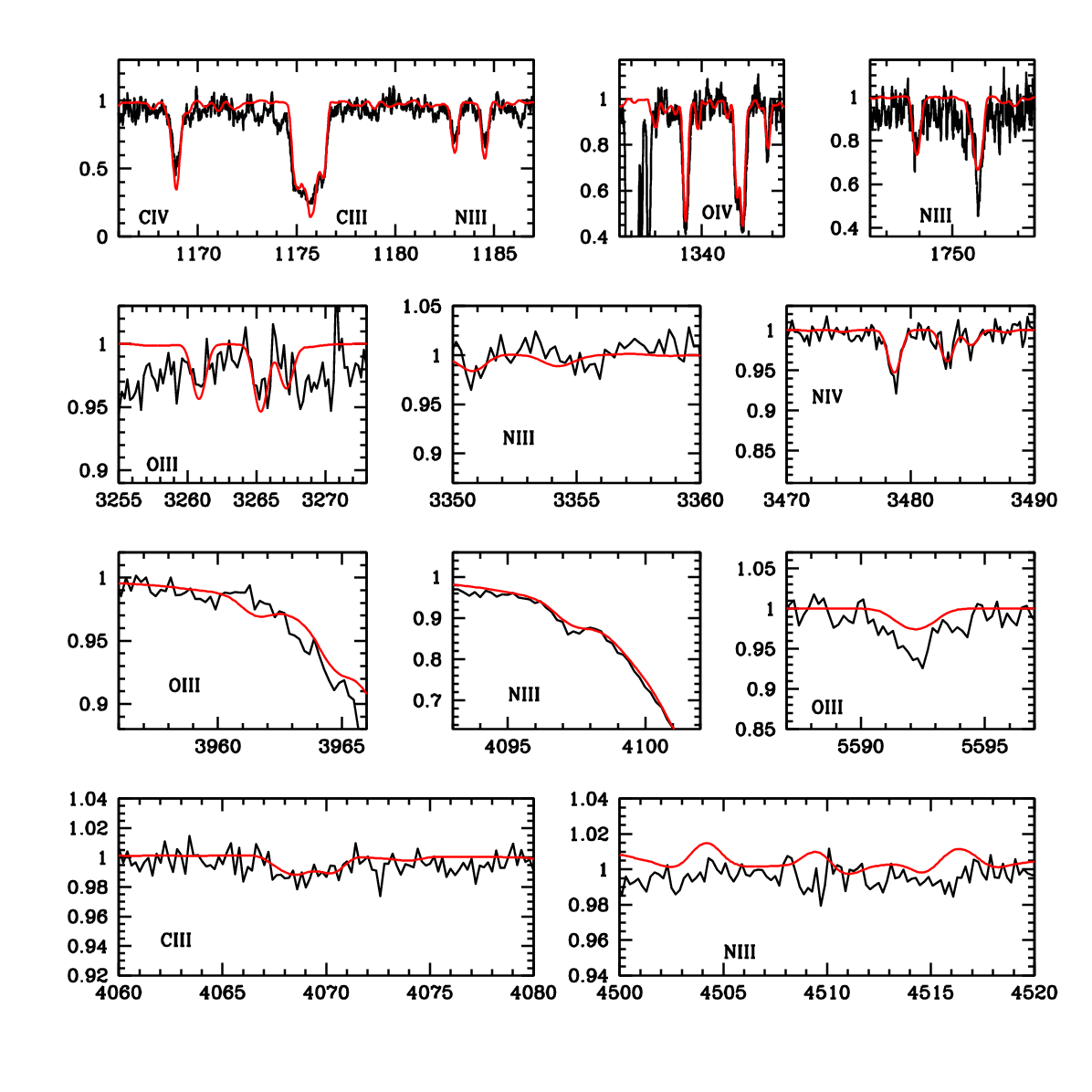}
\caption{Same as Fig.~\ref{fit_av15} but for AzV243.} 
\label{fit_av243}
\end{figure*}

\begin{figure*}[ht]
\centering
\includegraphics[width=0.49\textwidth]{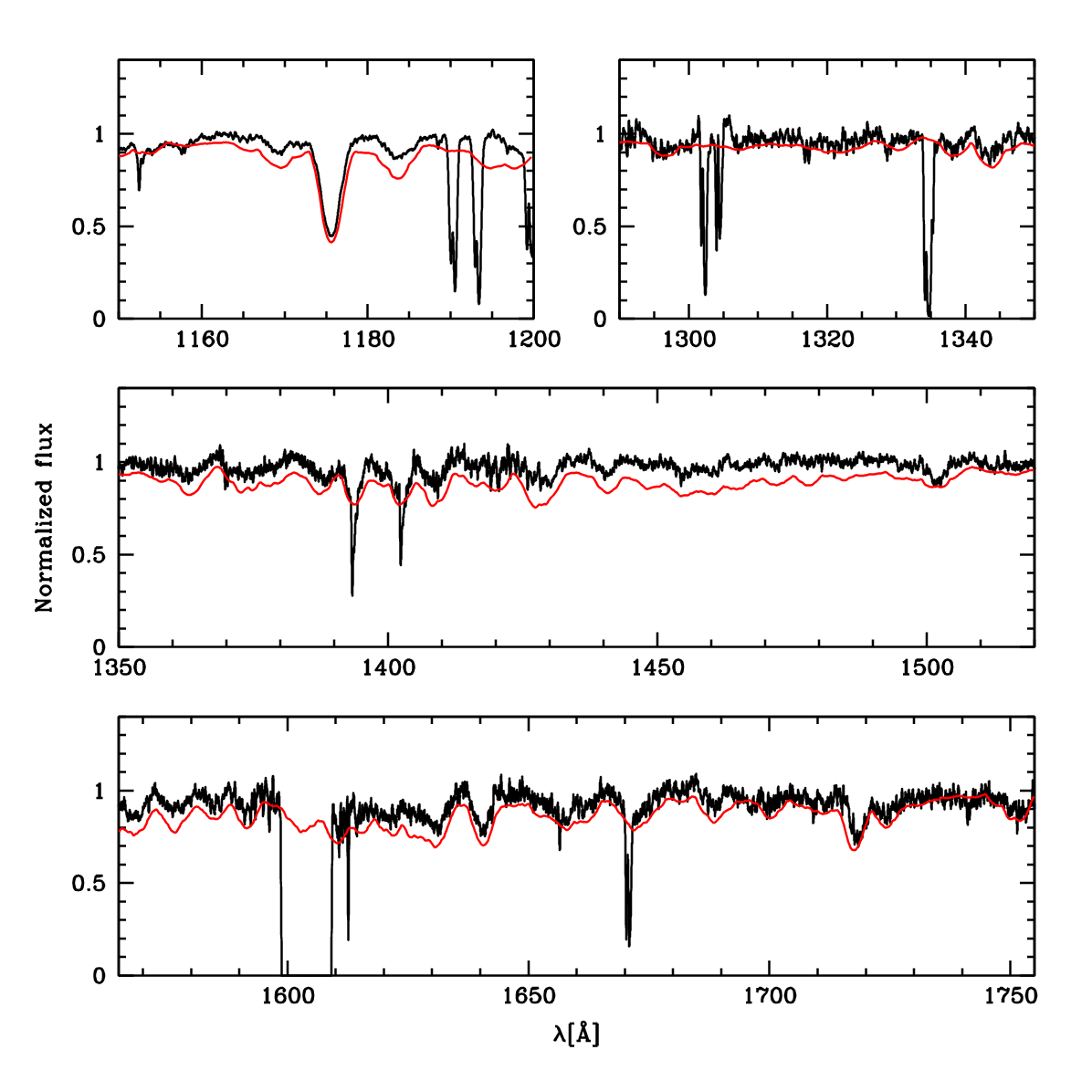}
\includegraphics[width=0.49\textwidth]{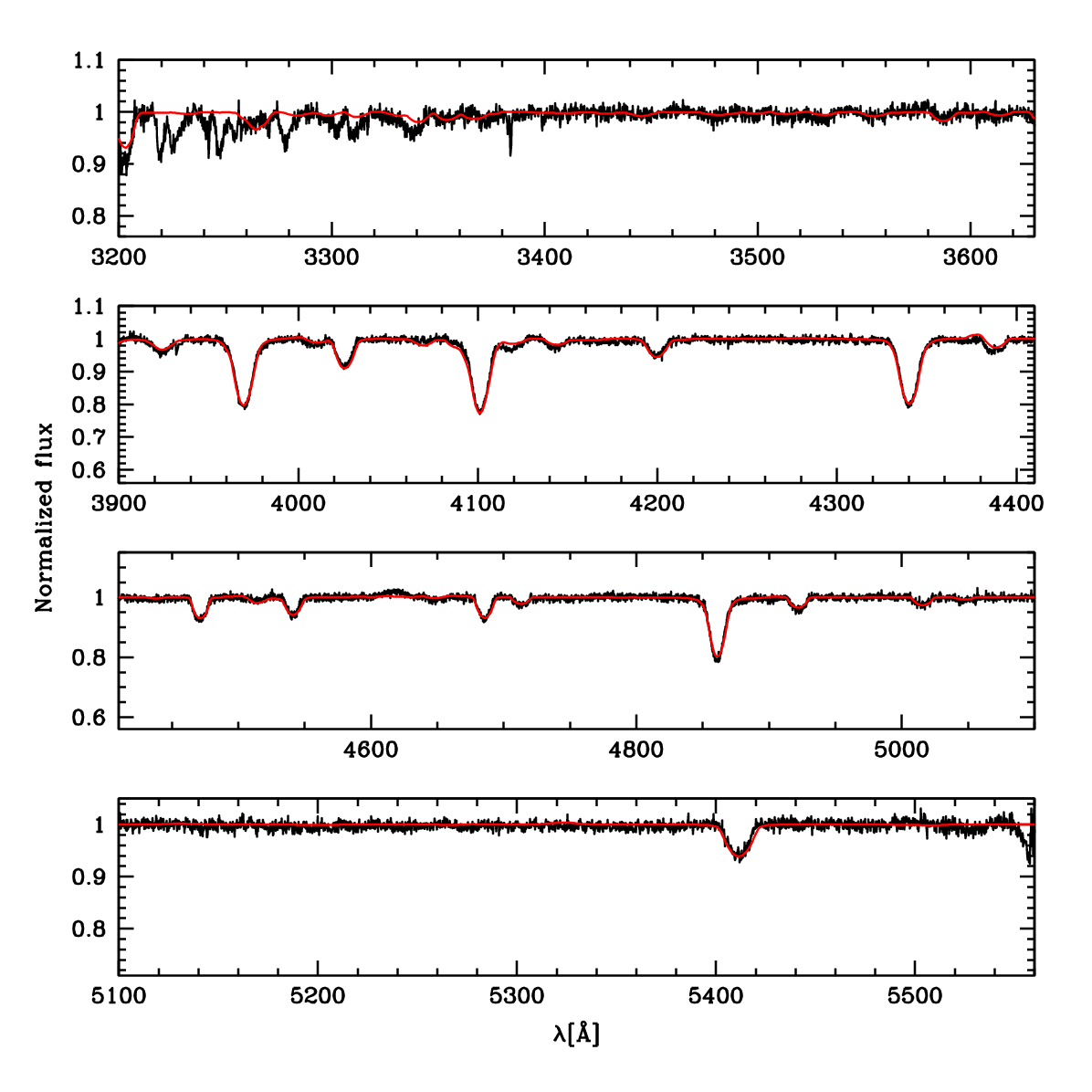}\\
\includegraphics[width=0.75\textwidth]{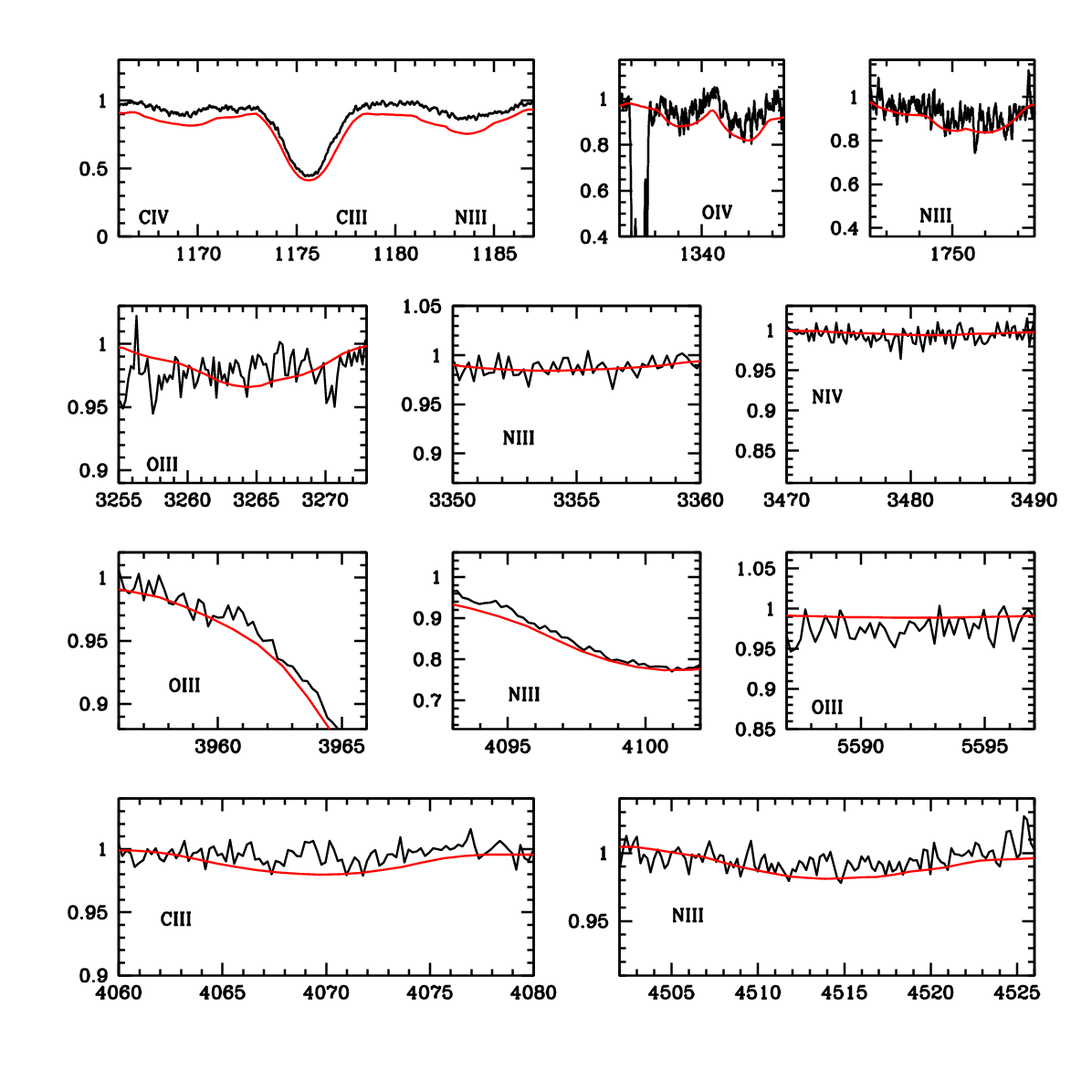}
\caption{Same as Fig.~\ref{fit_av15} but for AzV251.} 
\label{fit_av251}
\end{figure*}

\begin{figure*}[ht]
\centering
\includegraphics[width=0.49\textwidth]{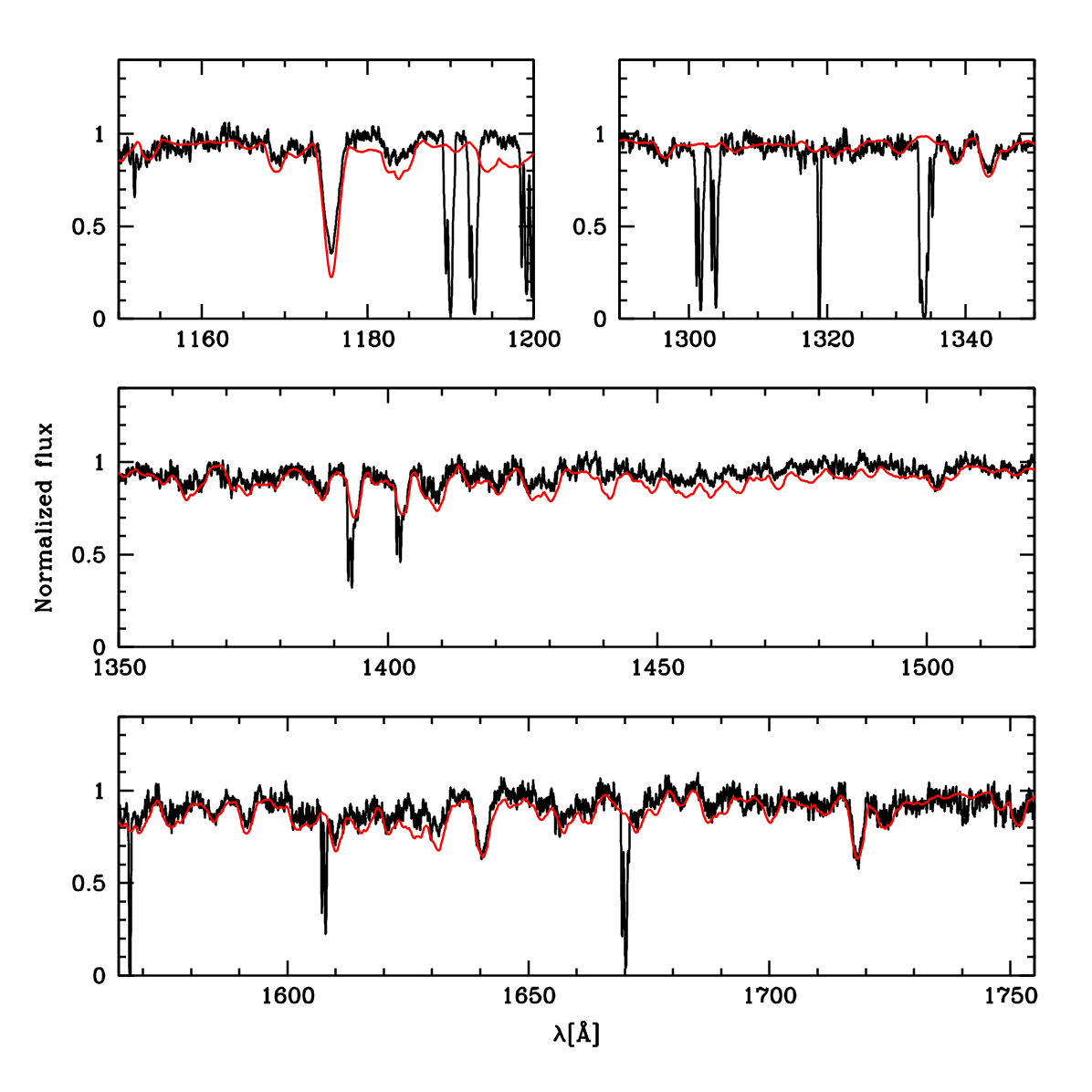}
\includegraphics[width=0.49\textwidth]{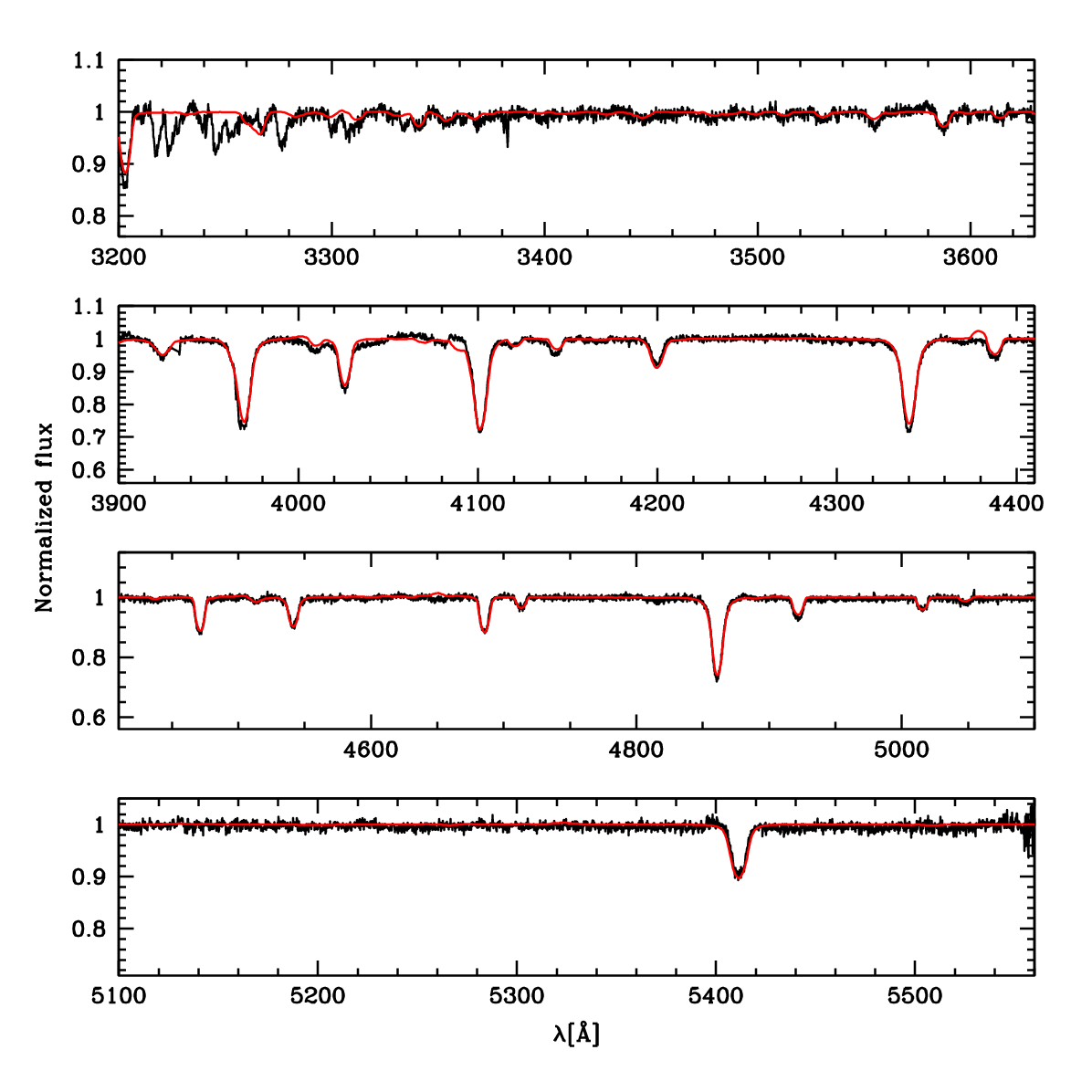}\\
\includegraphics[width=0.75\textwidth]{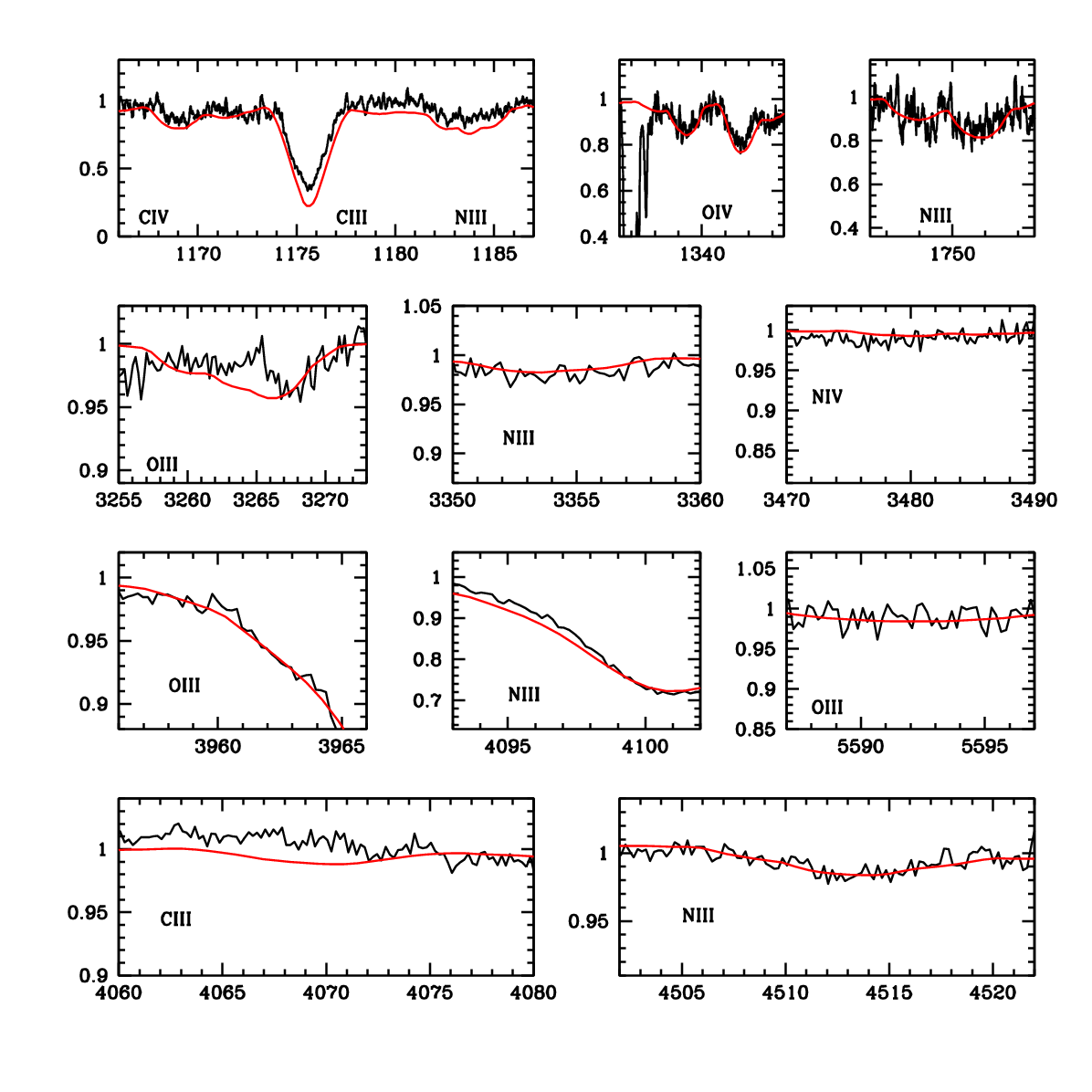}
\caption{Same as Fig.~\ref{fit_av15} but for AzV267.} 
\label{fit_av267}
\end{figure*}

\begin{figure*}[ht]
\centering
\includegraphics[width=0.49\textwidth]{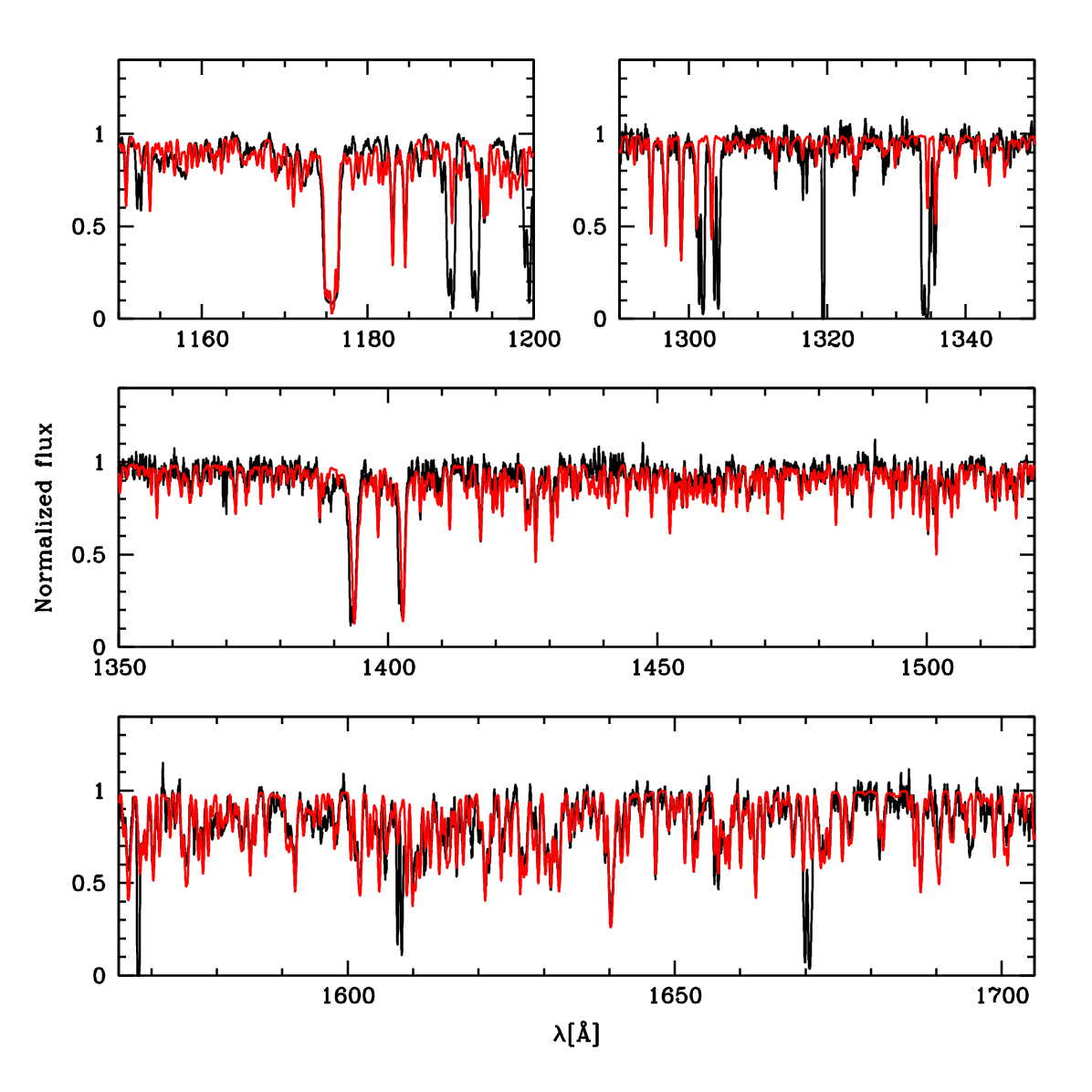}
\includegraphics[width=0.49\textwidth]{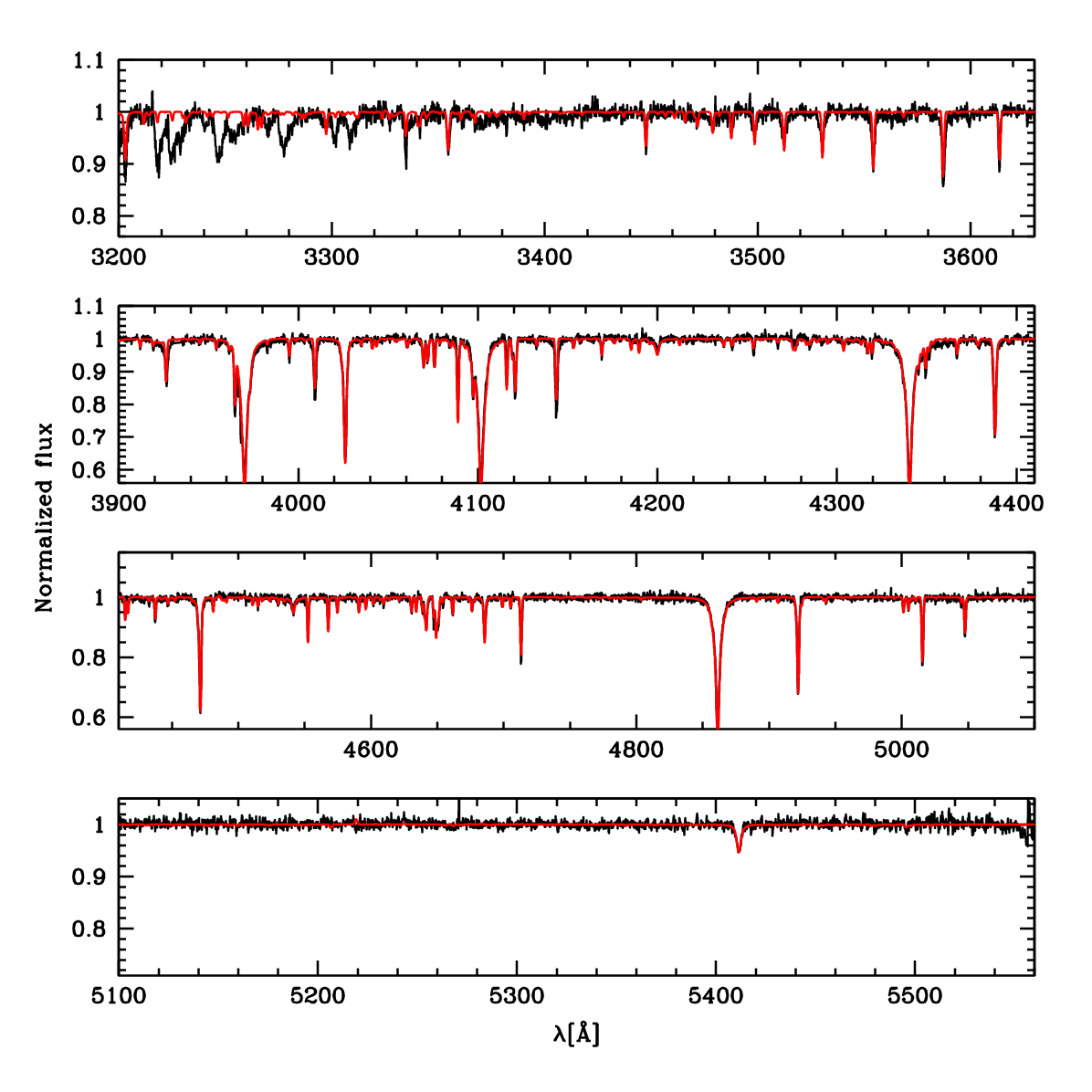}\\
\includegraphics[width=0.75\textwidth]{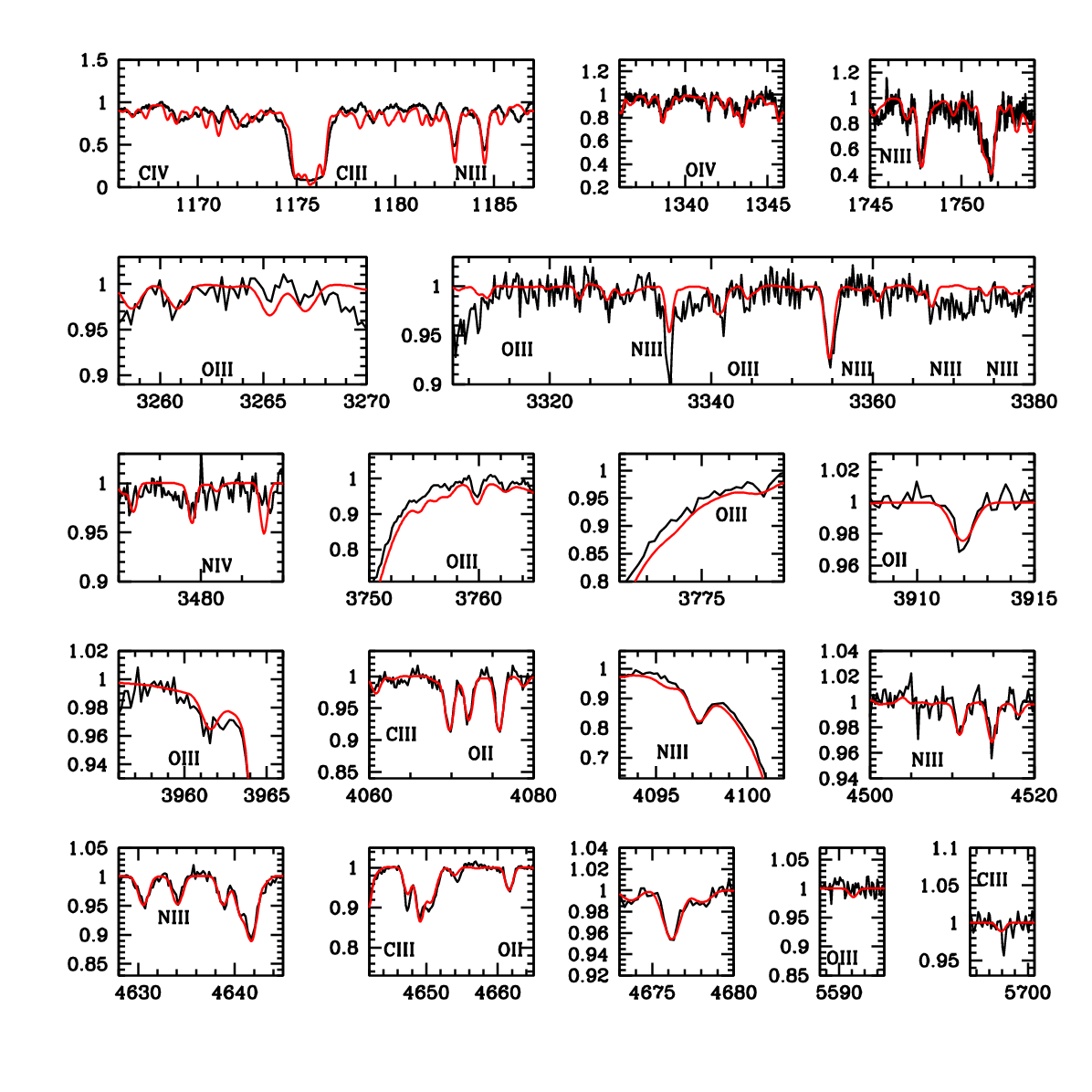}
\caption{Same as Fig.~\ref{fit_av15} but for AzV307.} 
\label{fit_av307}
\end{figure*}

\begin{figure*}[ht]
\centering
\includegraphics[width=0.49\textwidth]{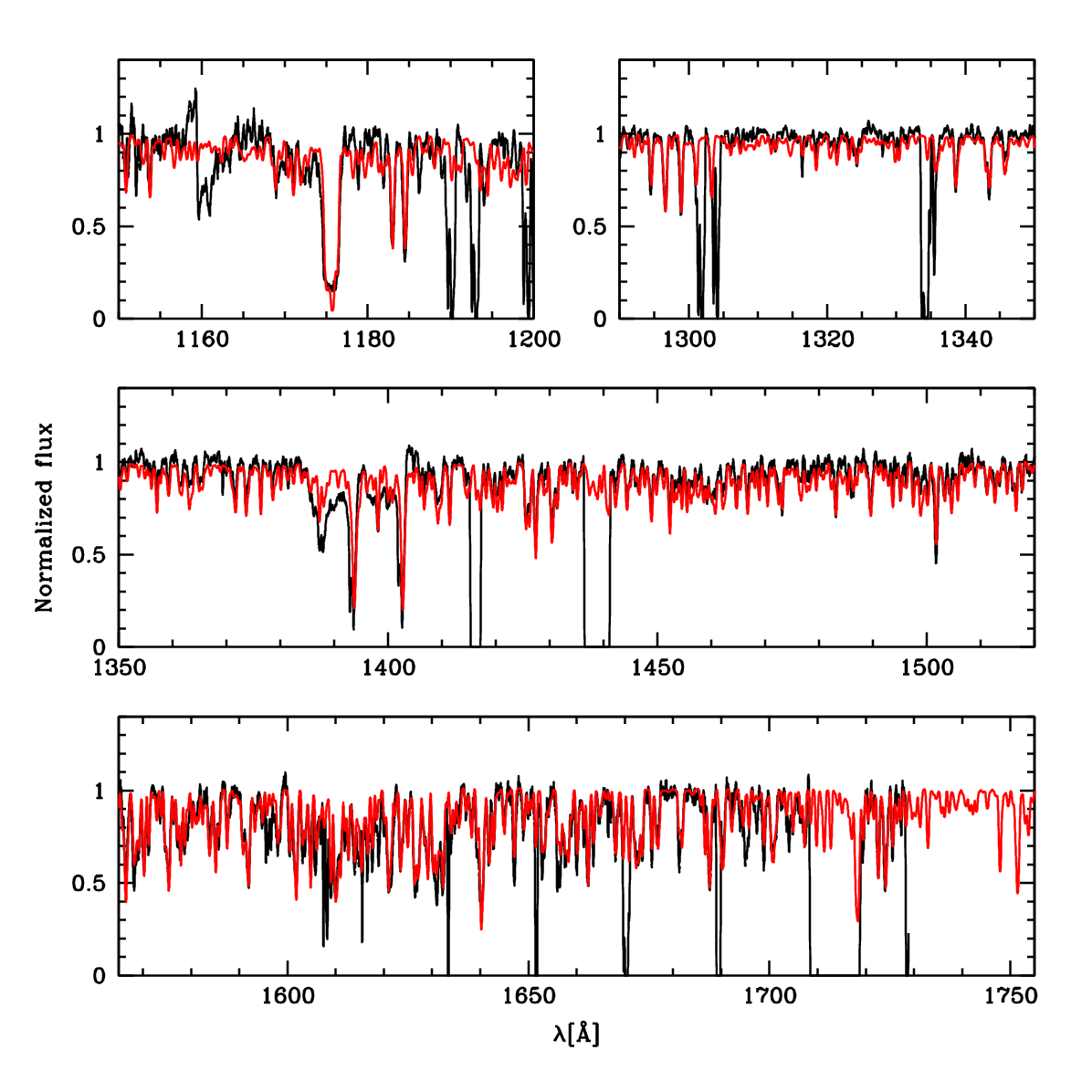}
\includegraphics[width=0.49\textwidth]{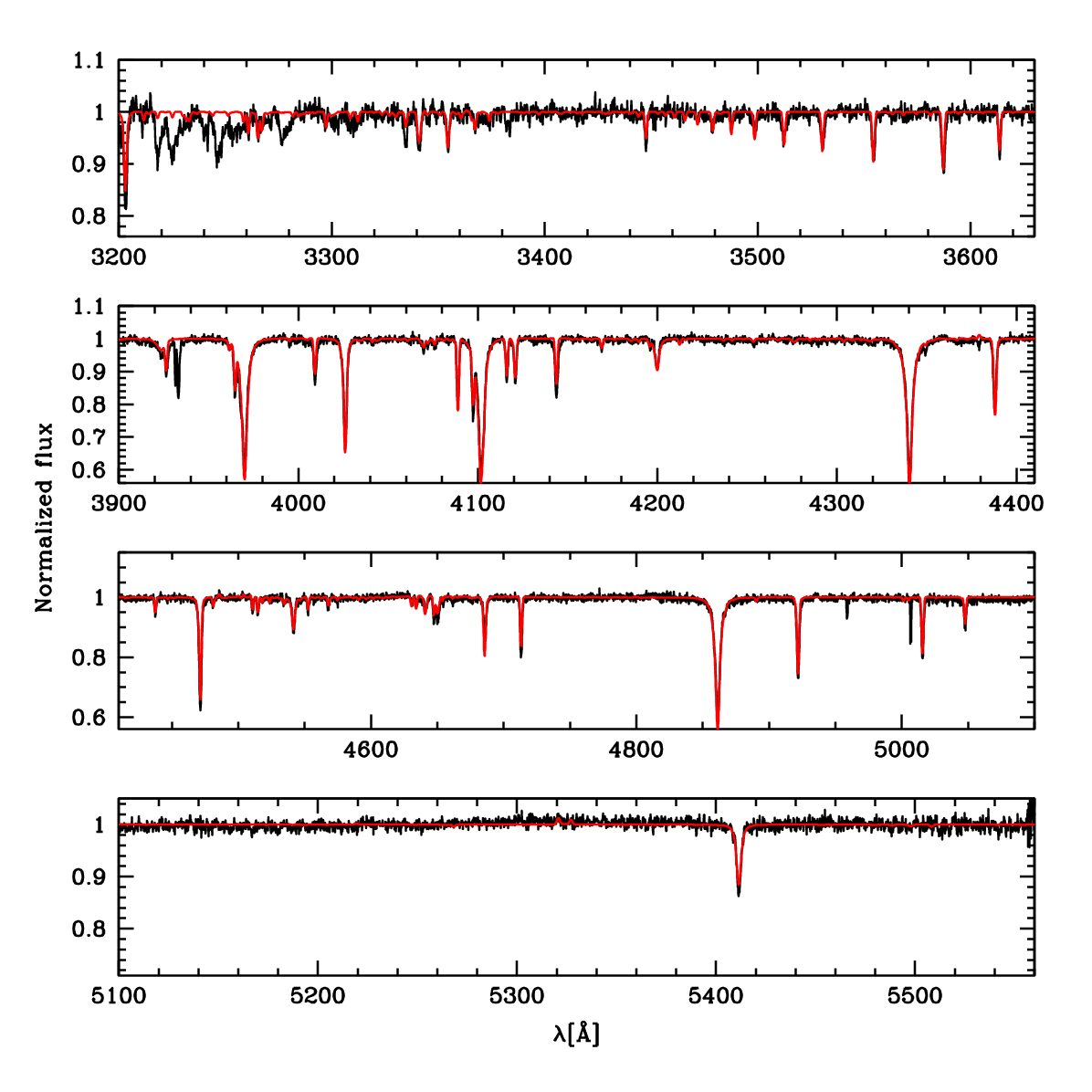}\\
\includegraphics[width=0.75\textwidth]{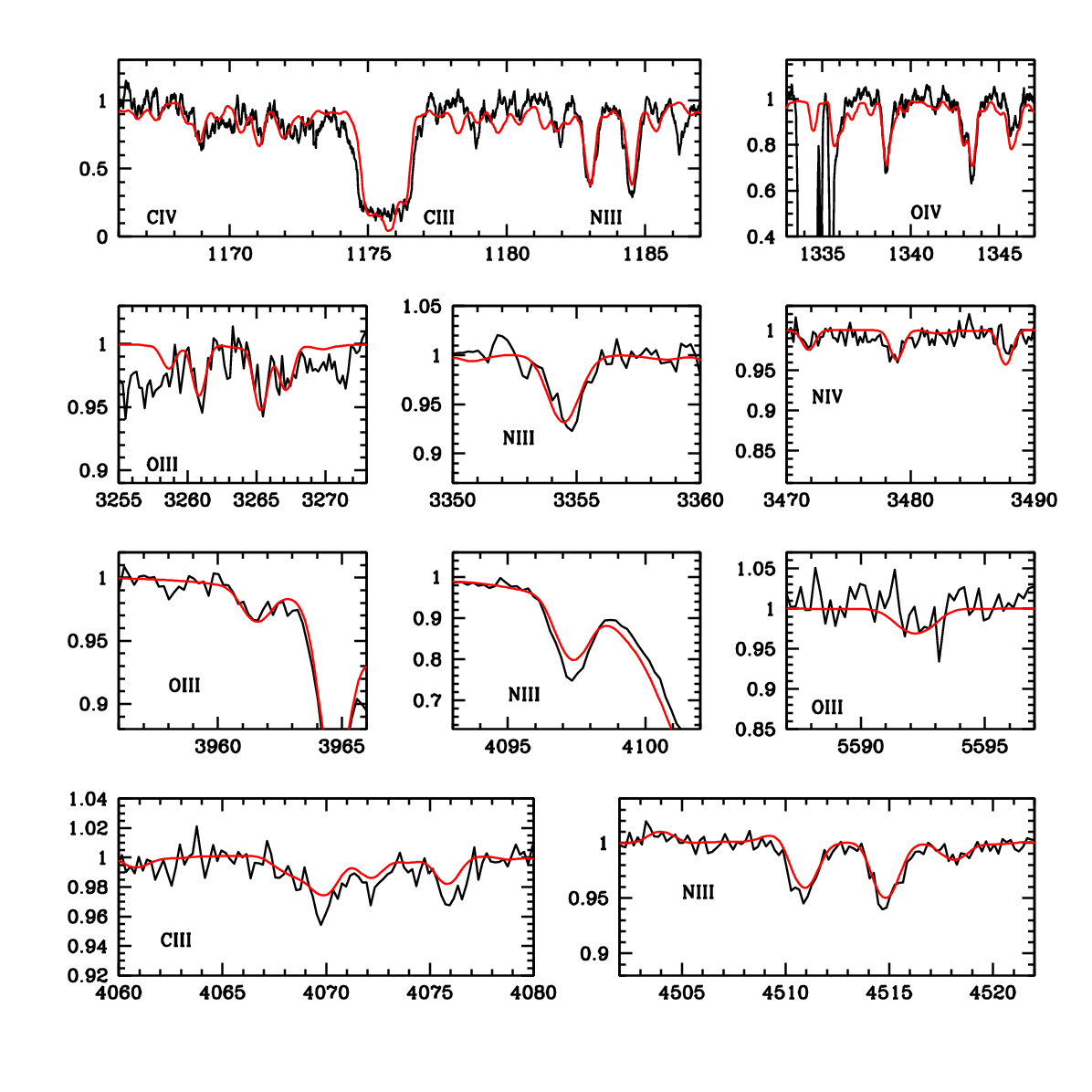}
\caption{Same as Fig.~\ref{fit_av15} but for AzV327.} 
\label{fit_av327}
\end{figure*}

\begin{figure*}[ht]
\centering
\includegraphics[width=0.49\textwidth]{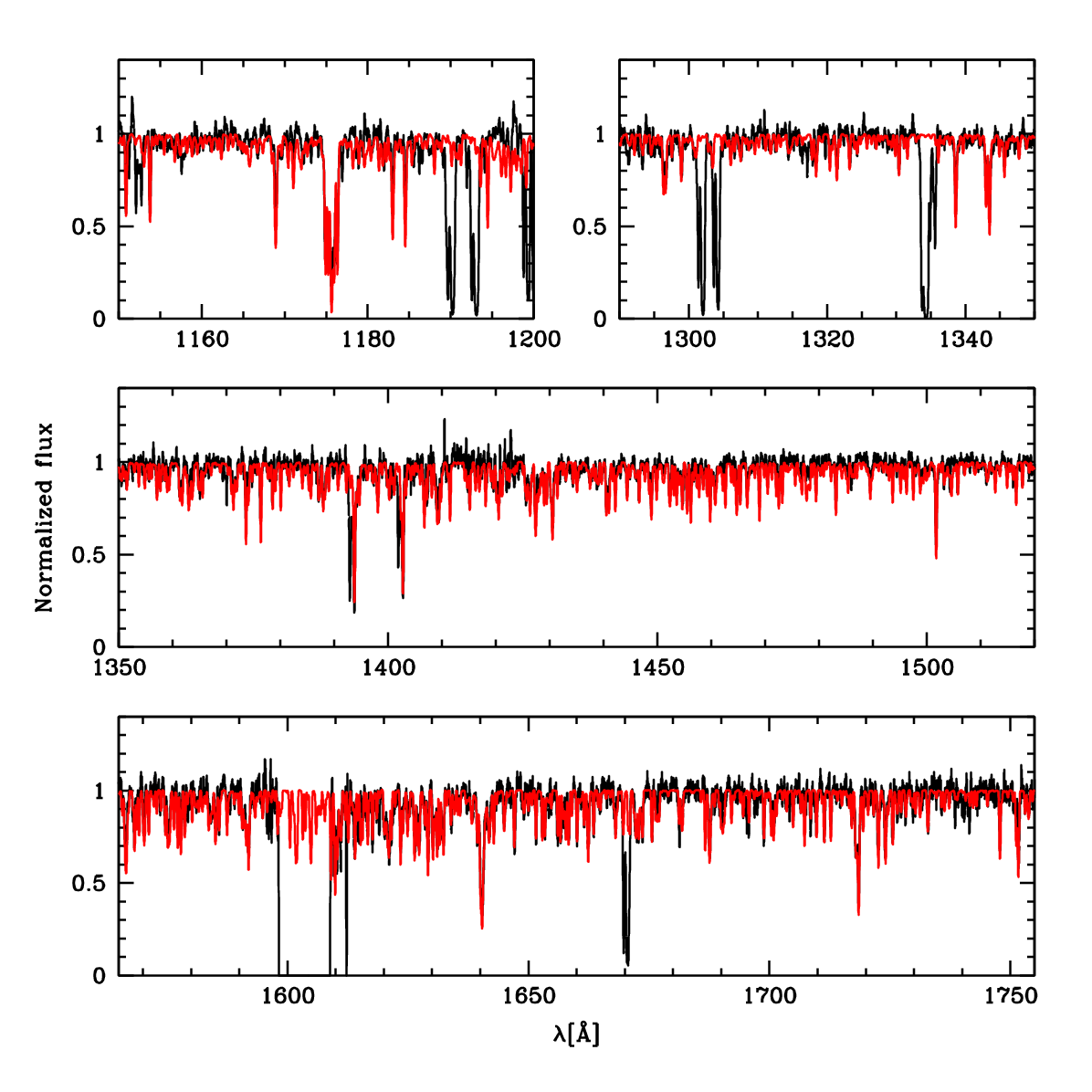}
\includegraphics[width=0.49\textwidth]{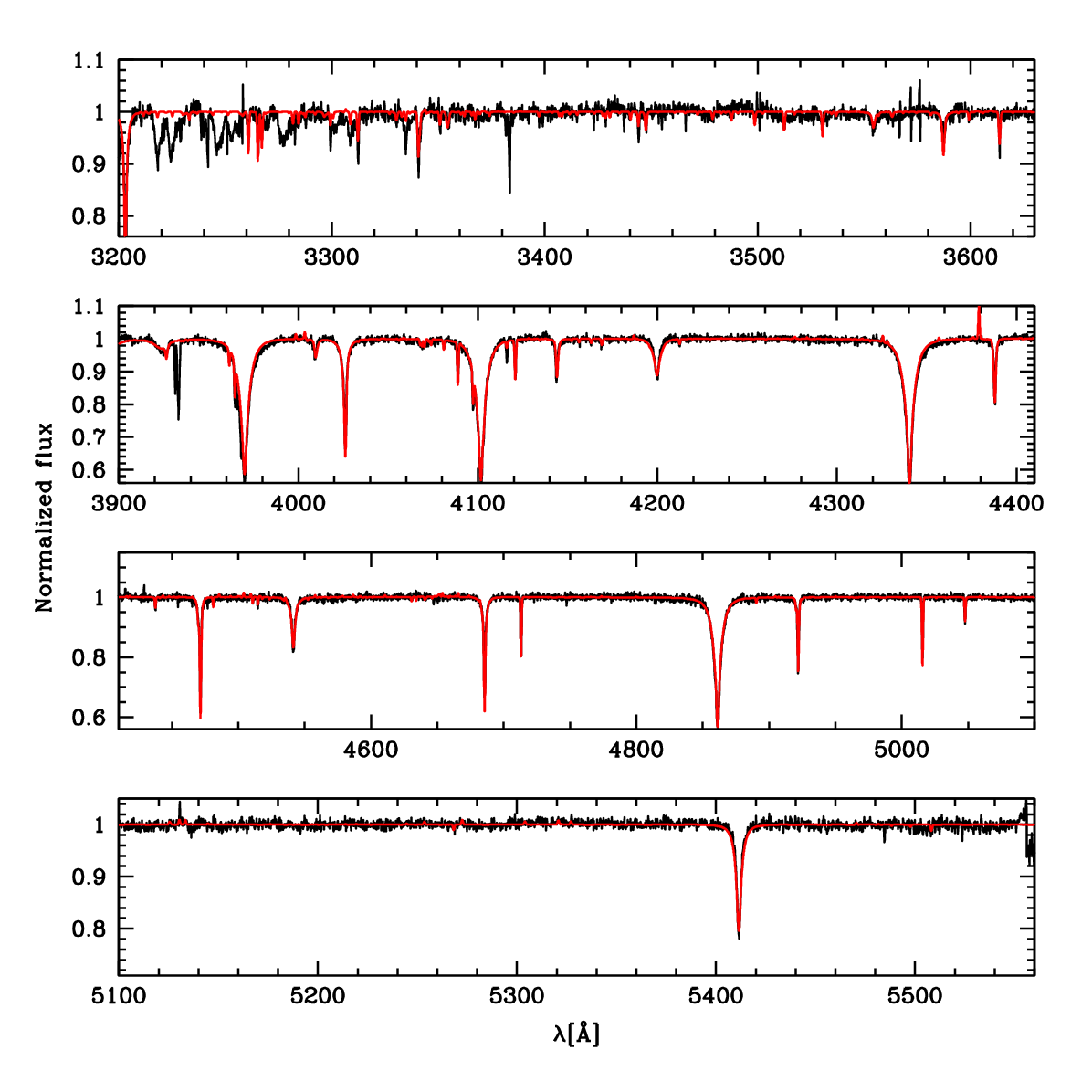}\\
\includegraphics[width=0.75\textwidth]{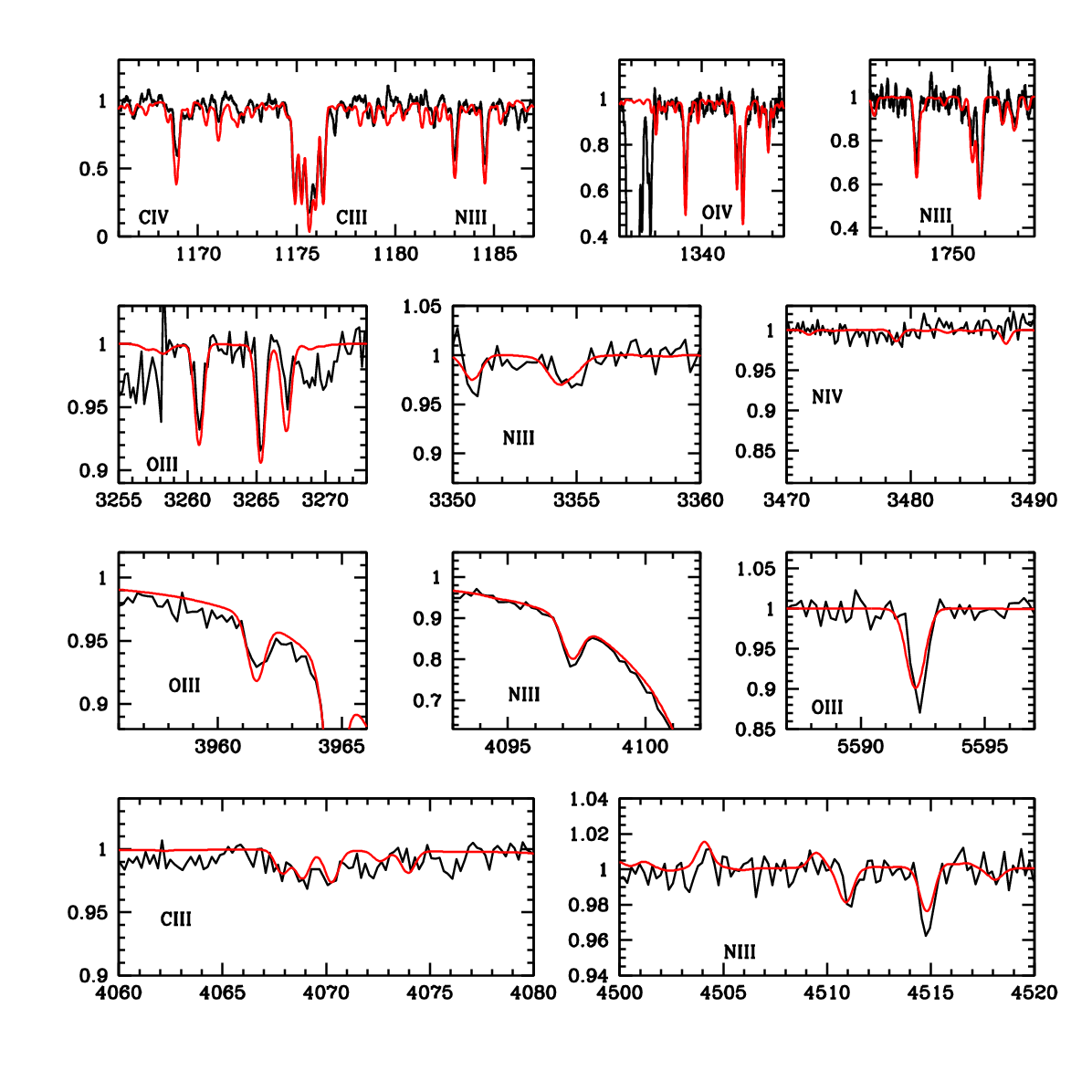}
\caption{Same as Fig.~\ref{fit_av15} but for AzV440.} 
\label{fit_av440}
\end{figure*}

\begin{figure*}[ht]
\centering
\includegraphics[width=0.49\textwidth]{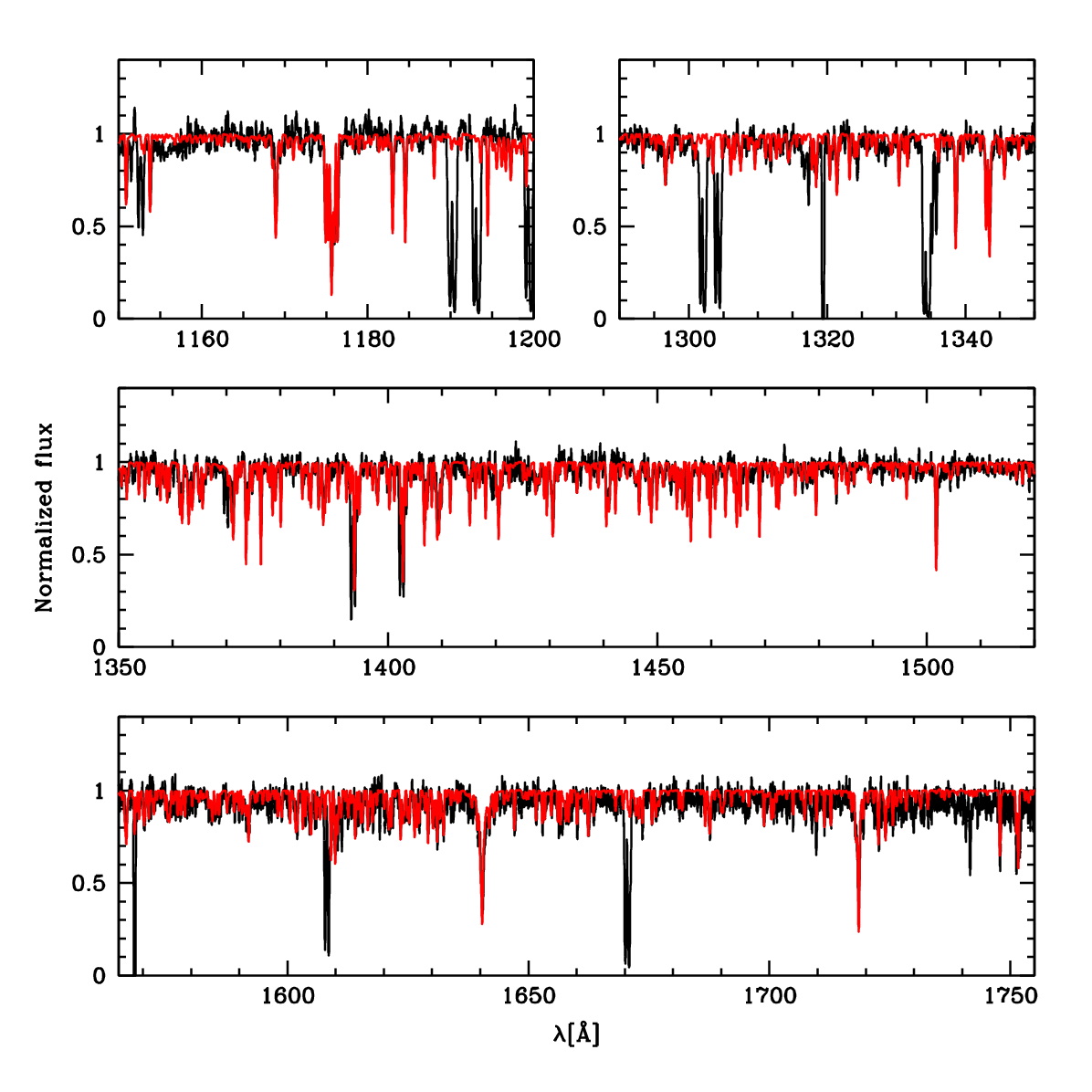}
\includegraphics[width=0.49\textwidth]{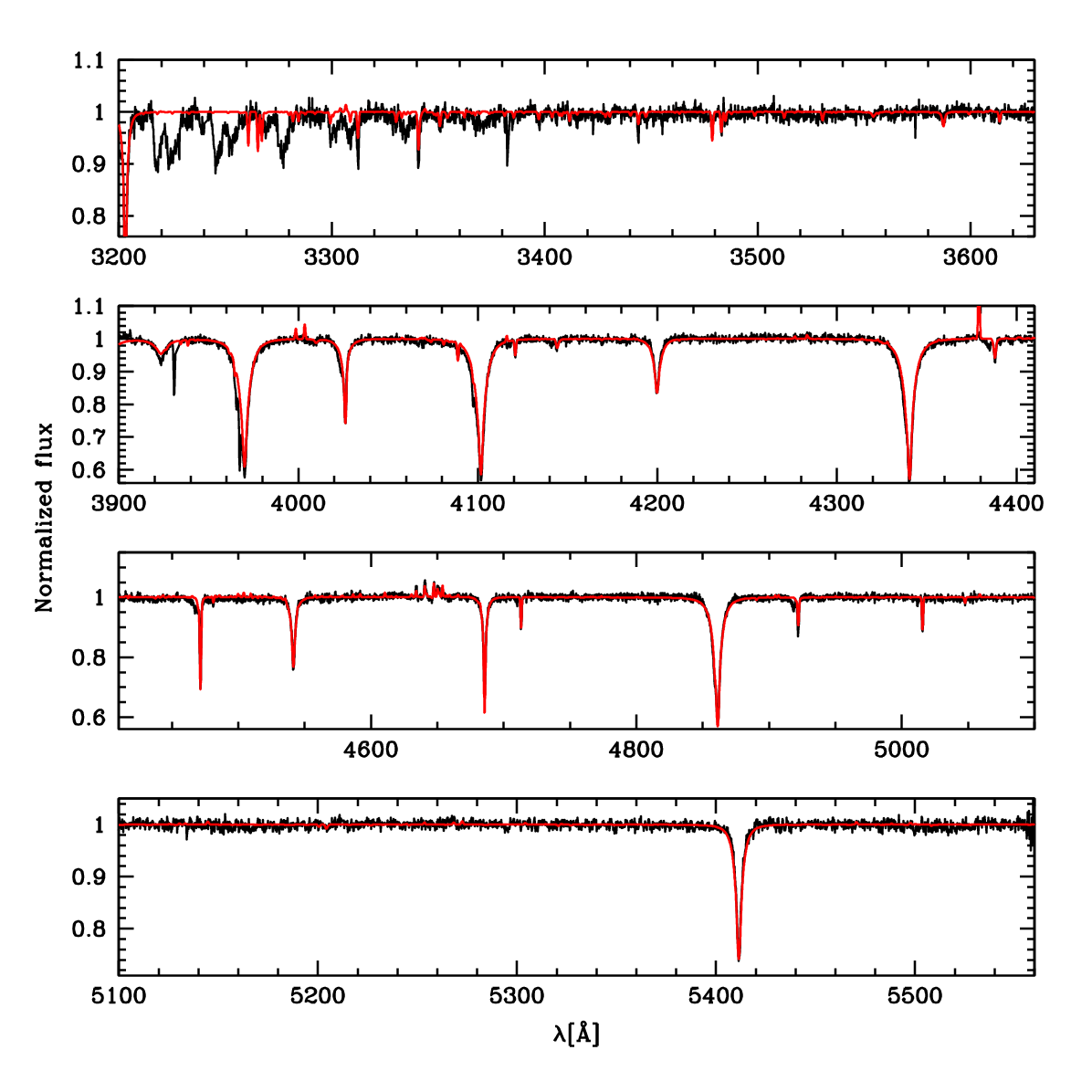}\\
\includegraphics[width=0.75\textwidth]{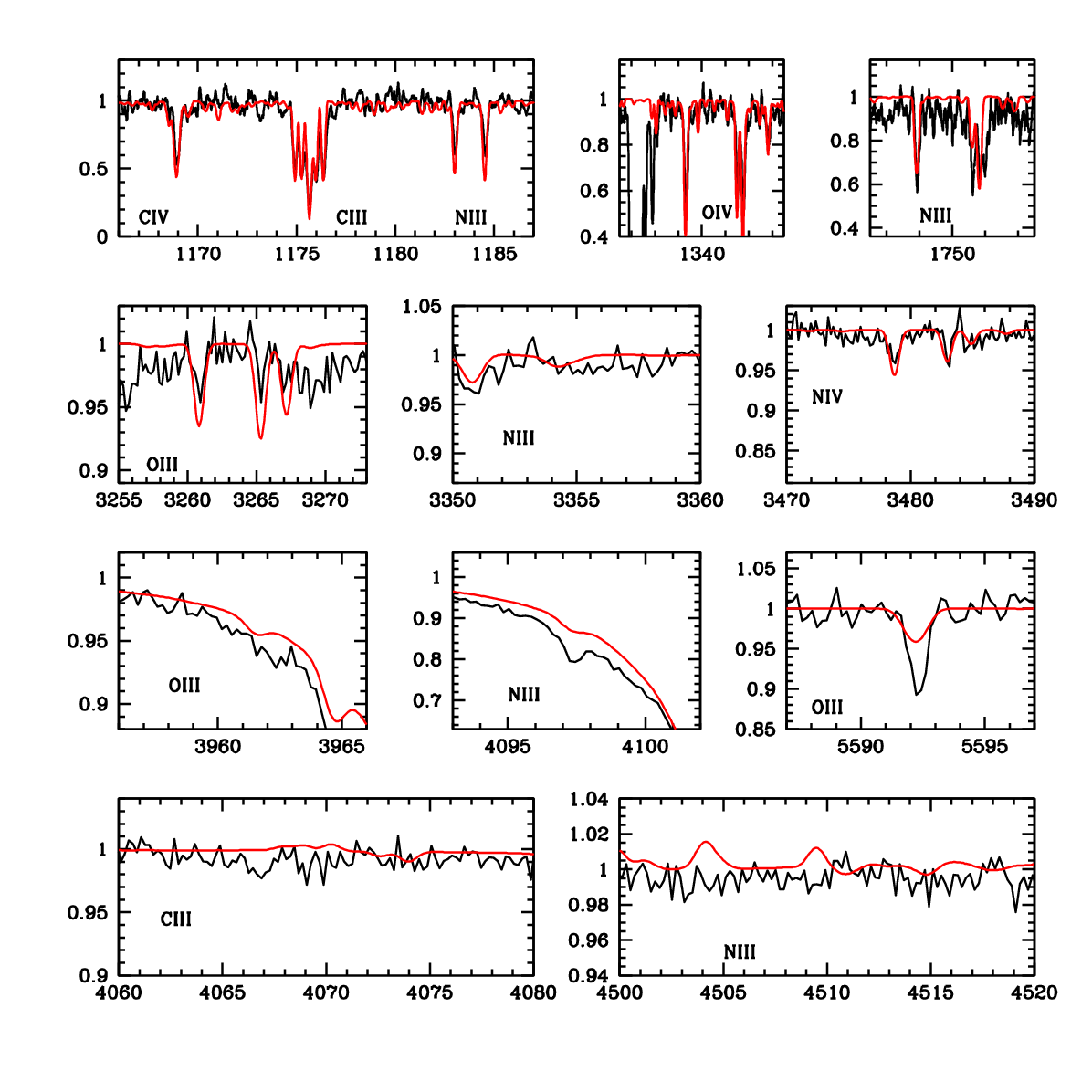}
\caption{Same as Fig.~\ref{fit_av15} but for AzV446.} 
\label{fit_av446}
\end{figure*}

\begin{figure*}[ht]
\centering
\includegraphics[width=0.49\textwidth]{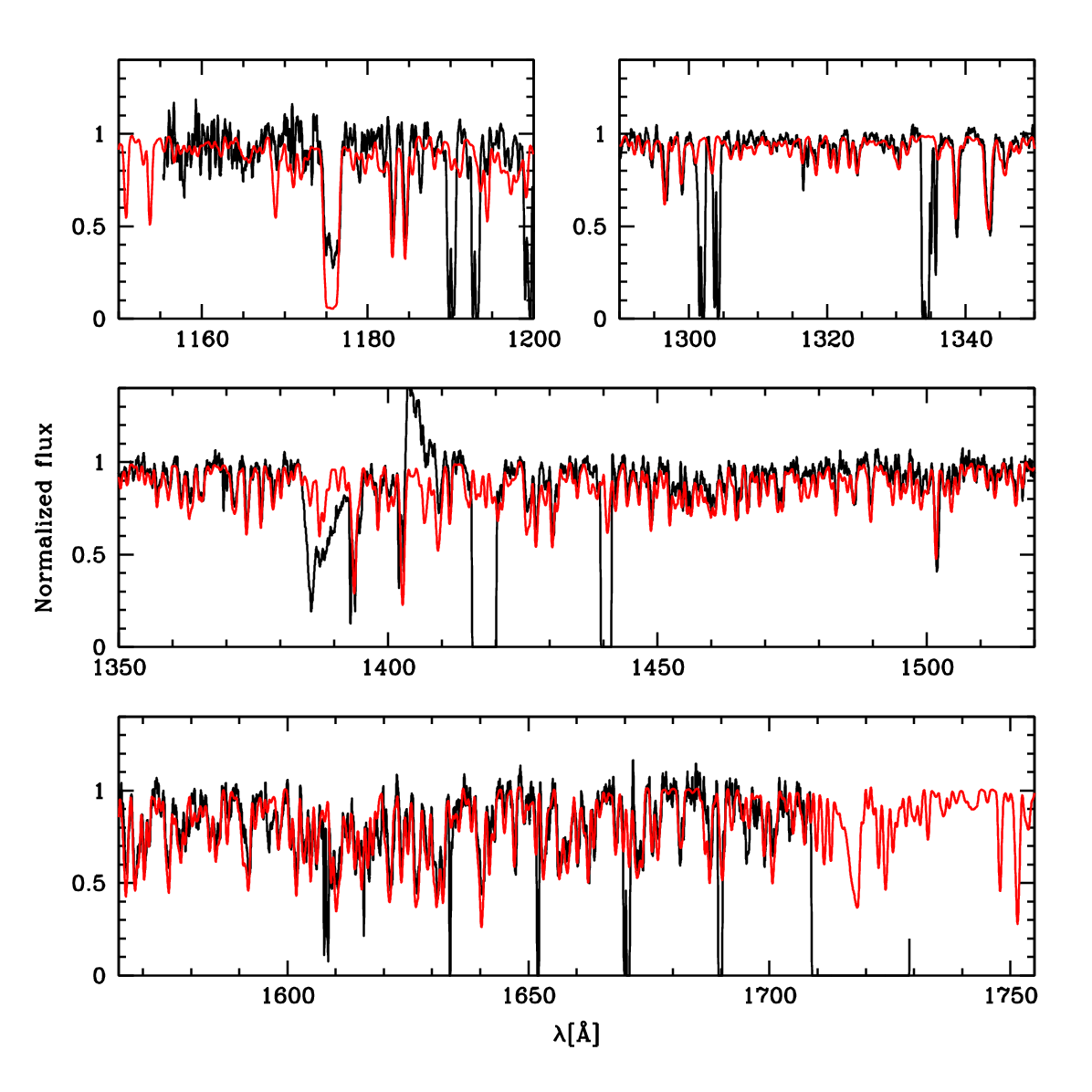}
\includegraphics[width=0.49\textwidth]{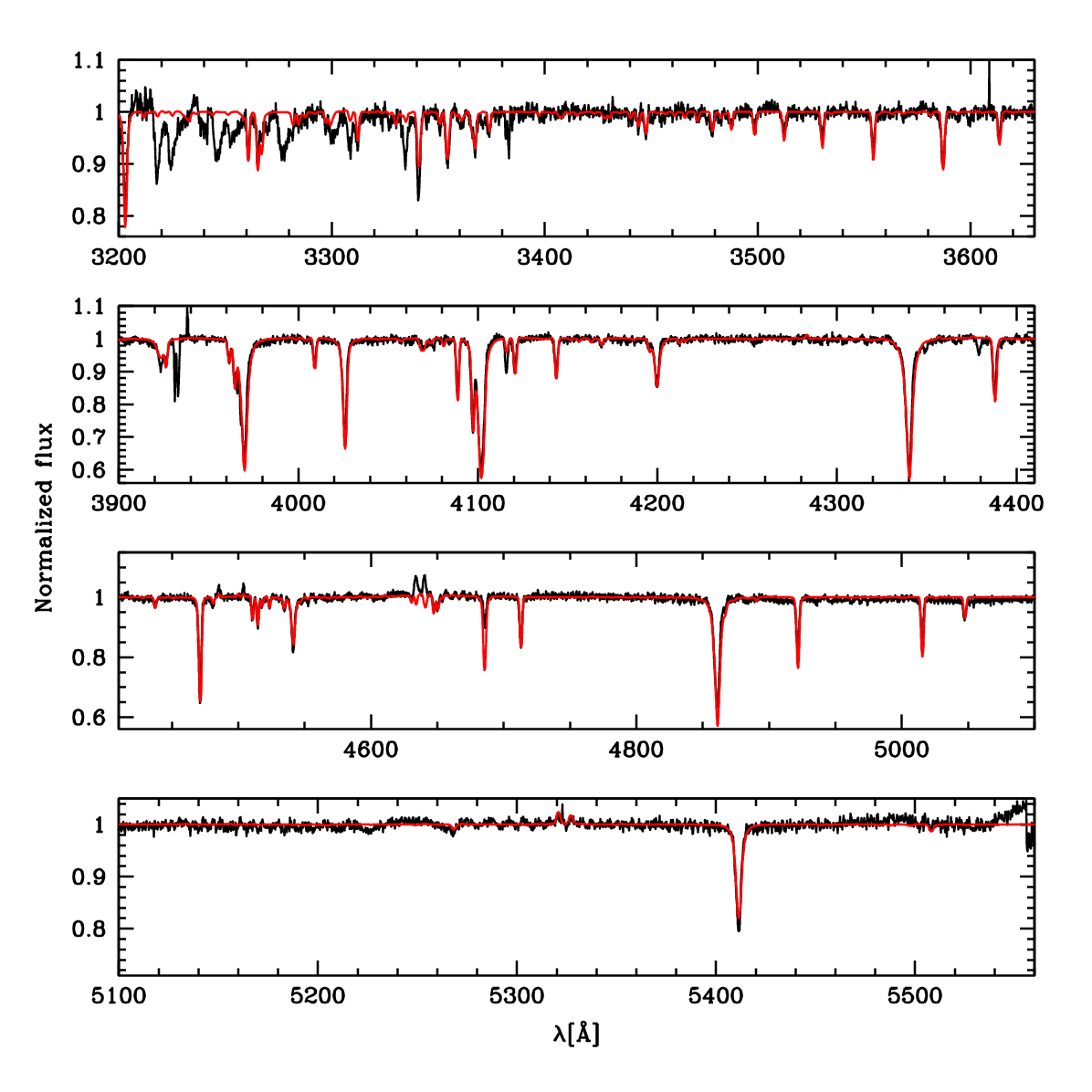}\\
\includegraphics[width=0.75\textwidth]{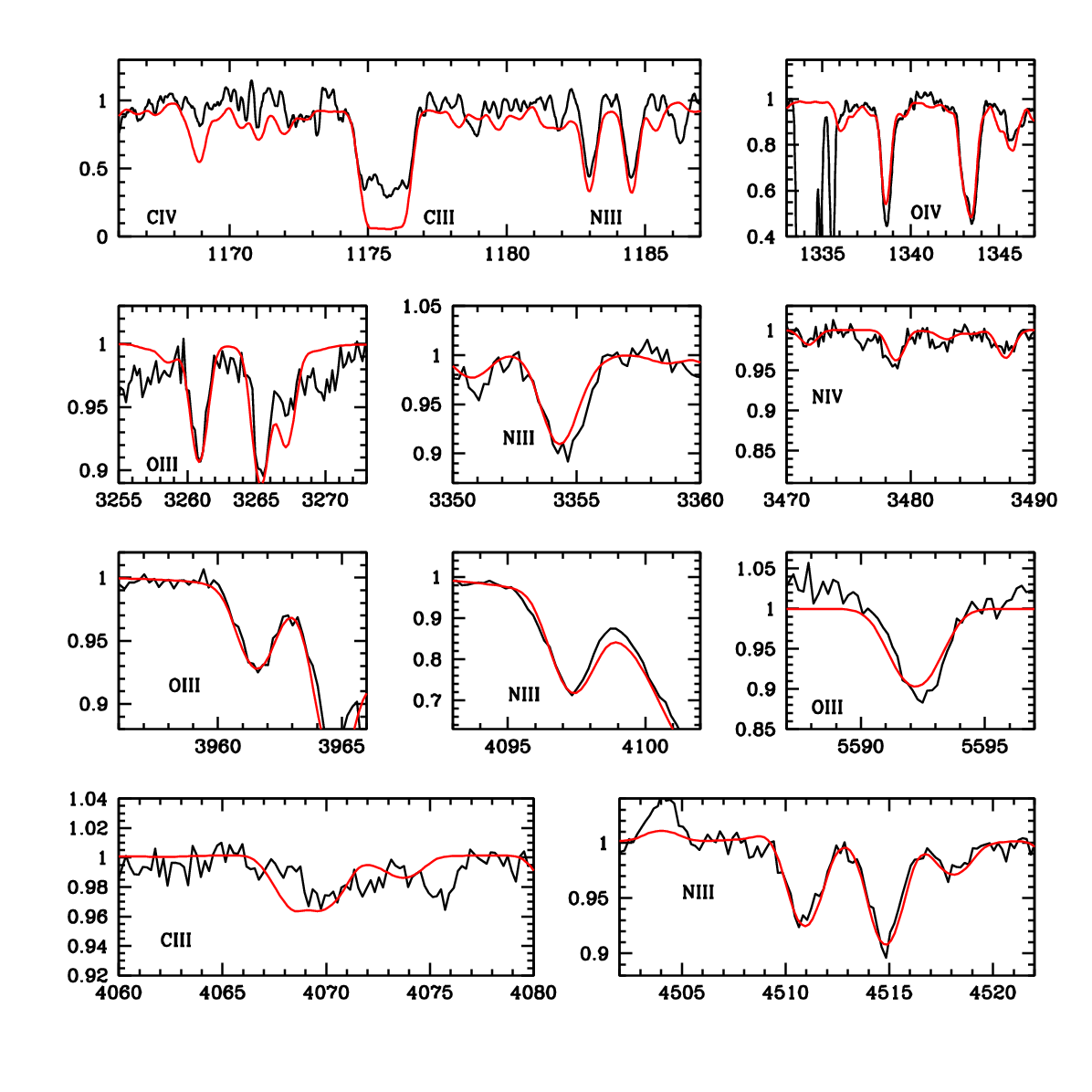}
\caption{Same as Fig.~\ref{fit_av15} but for AzV469.} 
\label{fit_av469}
\end{figure*}

\begin{figure*}[ht]
\centering
\includegraphics[width=0.49\textwidth]{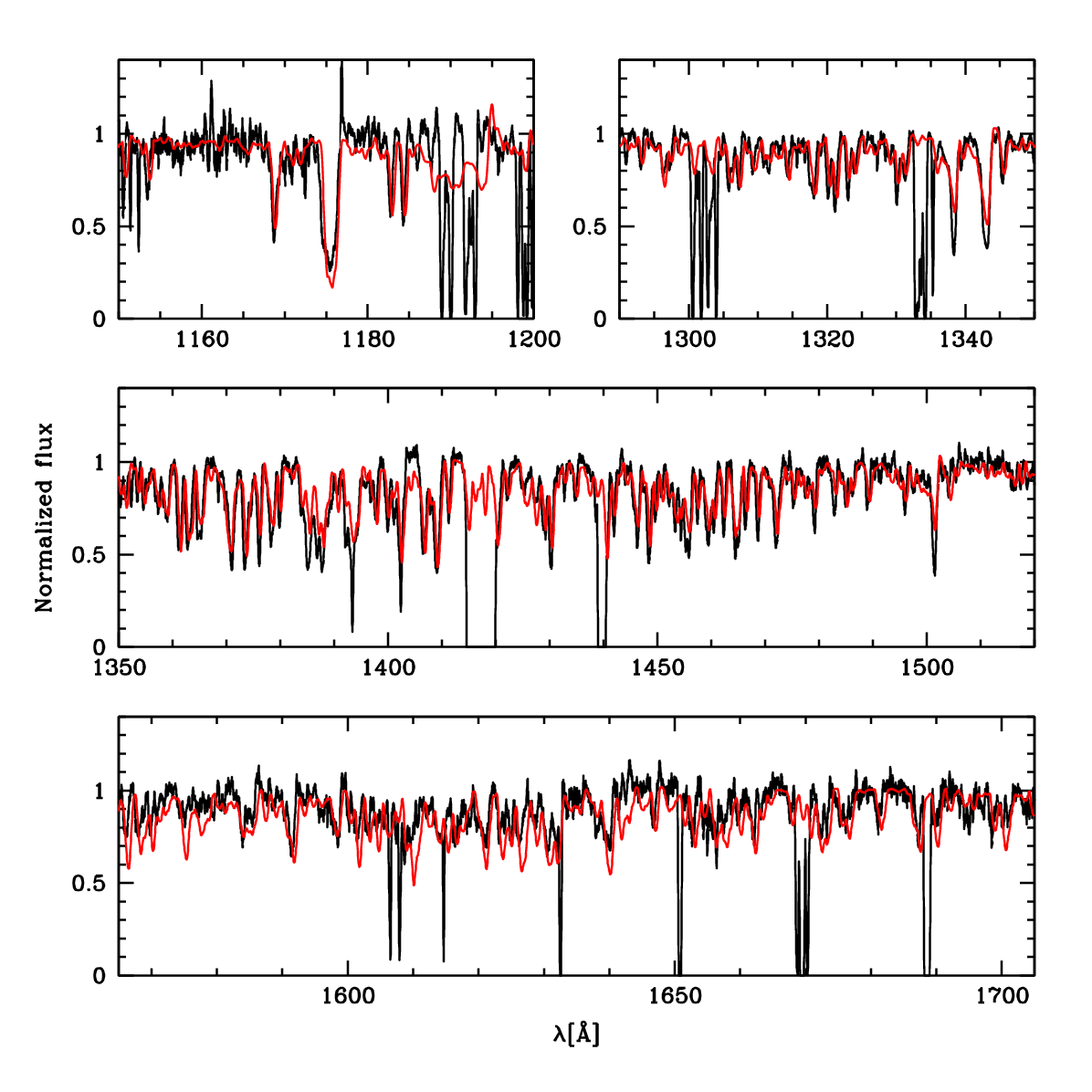}
\includegraphics[width=0.49\textwidth]{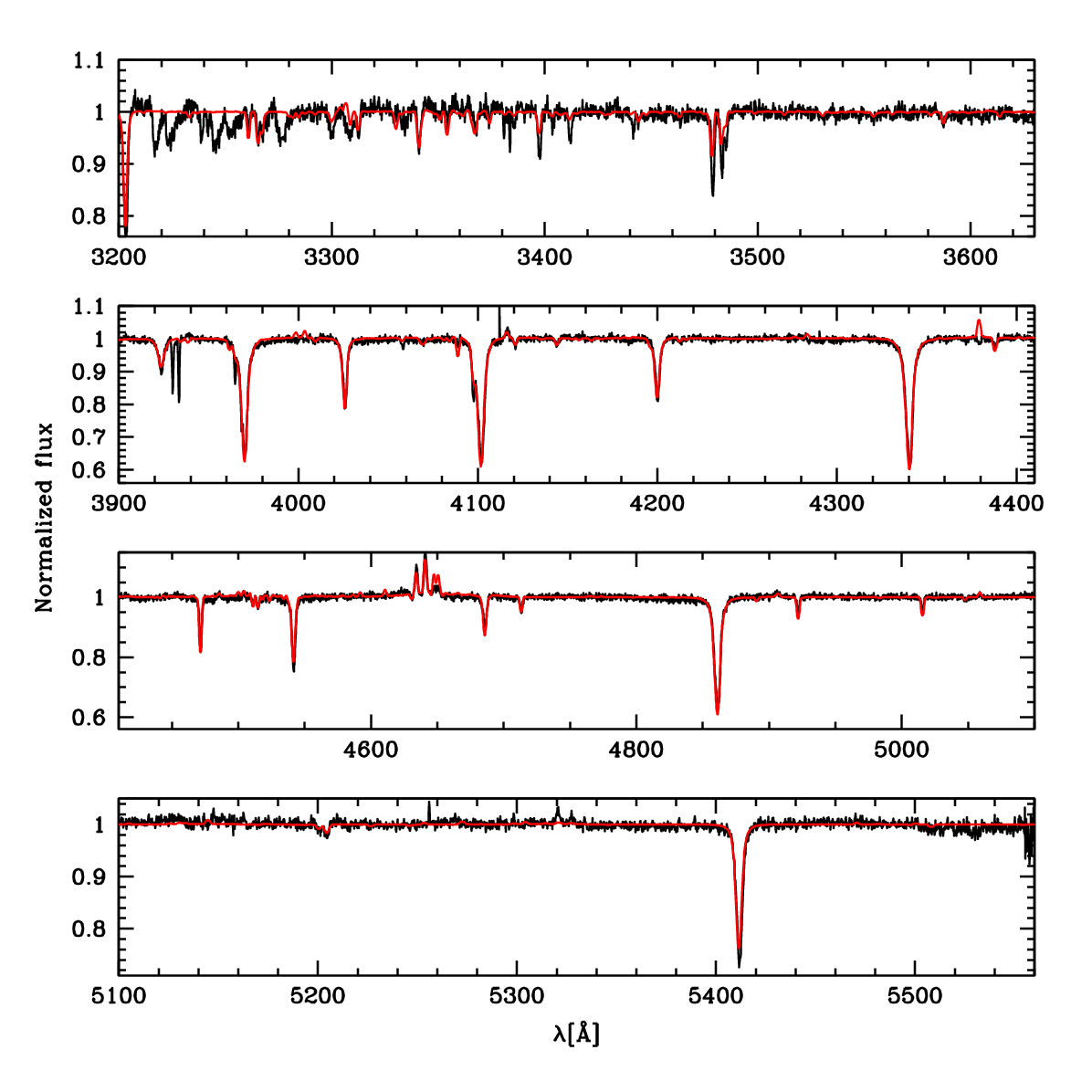}\\
\includegraphics[width=0.75\textwidth]{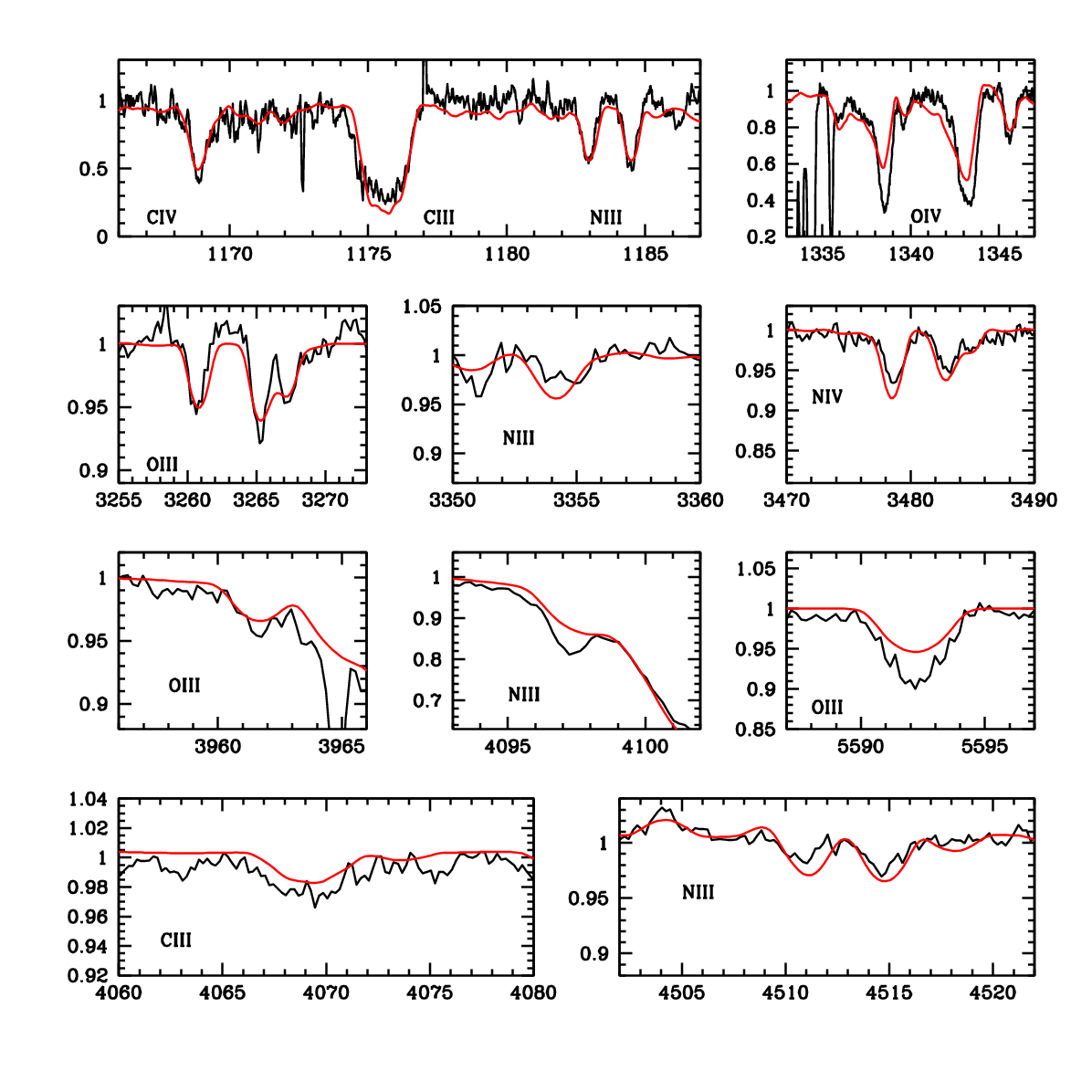}
\caption{Same as Fig.~\ref{fit_av15} but for SK~-66$^{\circ}$ 18.} 
\label{fit_skm66d18}
\end{figure*}

\begin{figure*}[ht]
\centering
\includegraphics[width=0.49\textwidth]{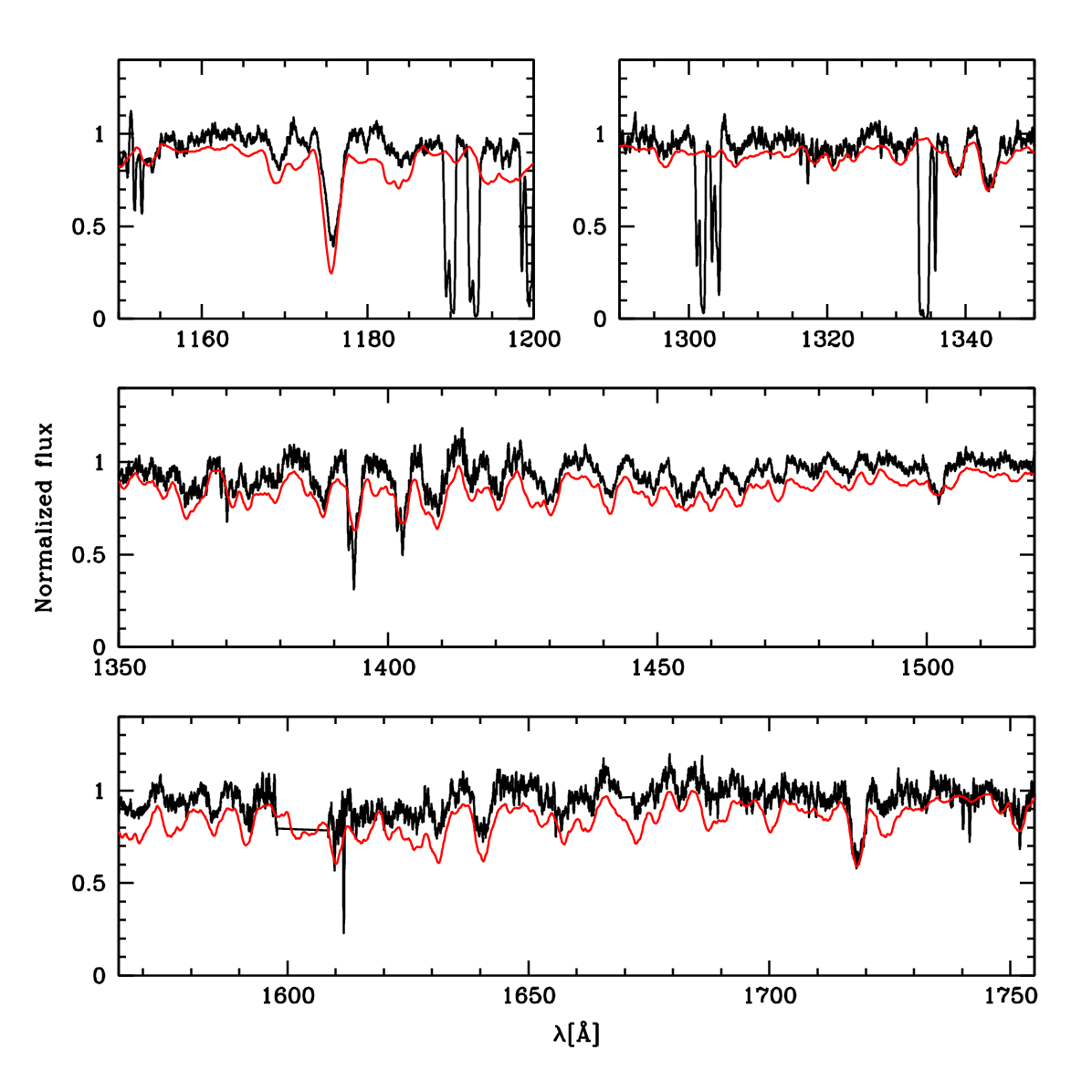}
\includegraphics[width=0.49\textwidth]{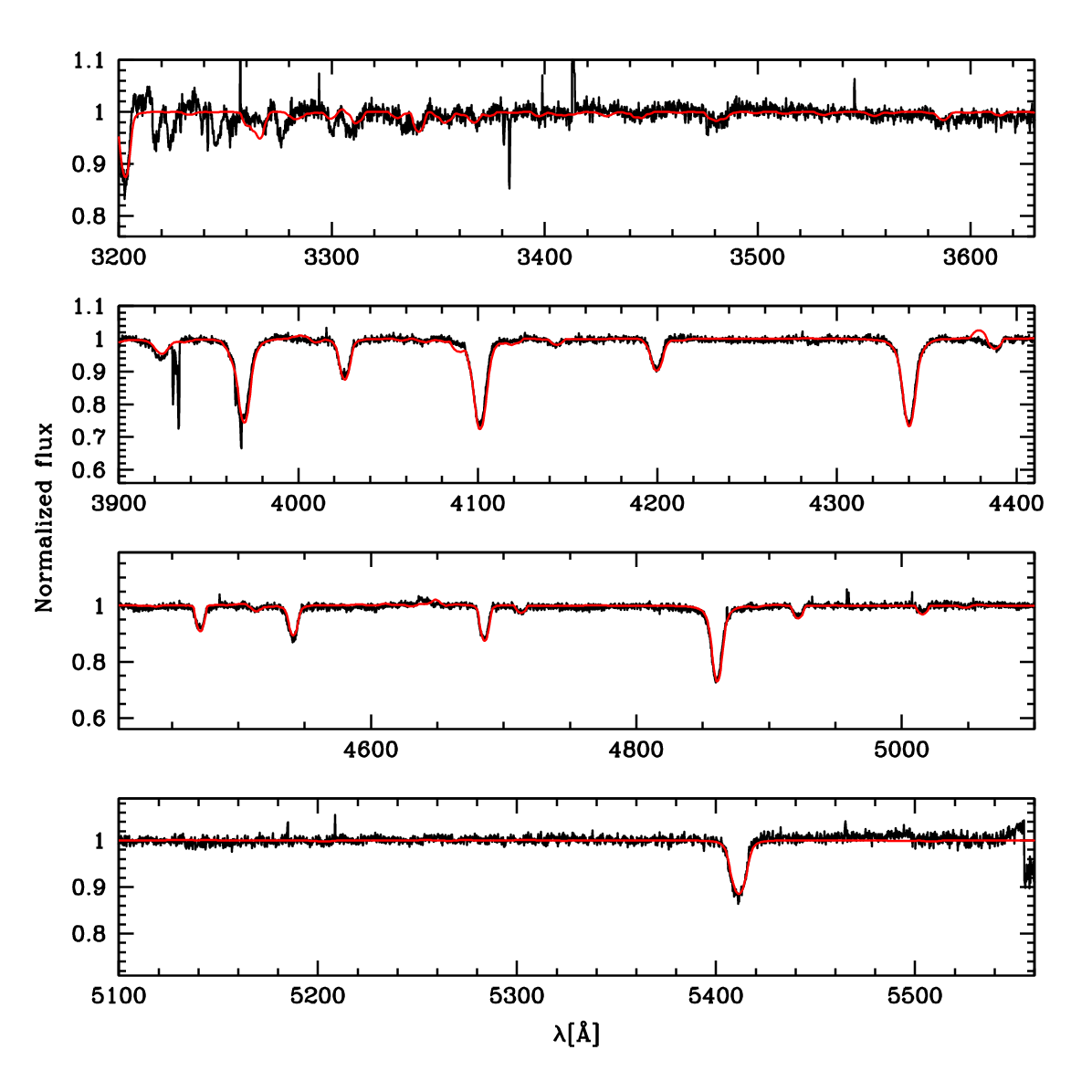}\\
\includegraphics[width=0.75\textwidth]{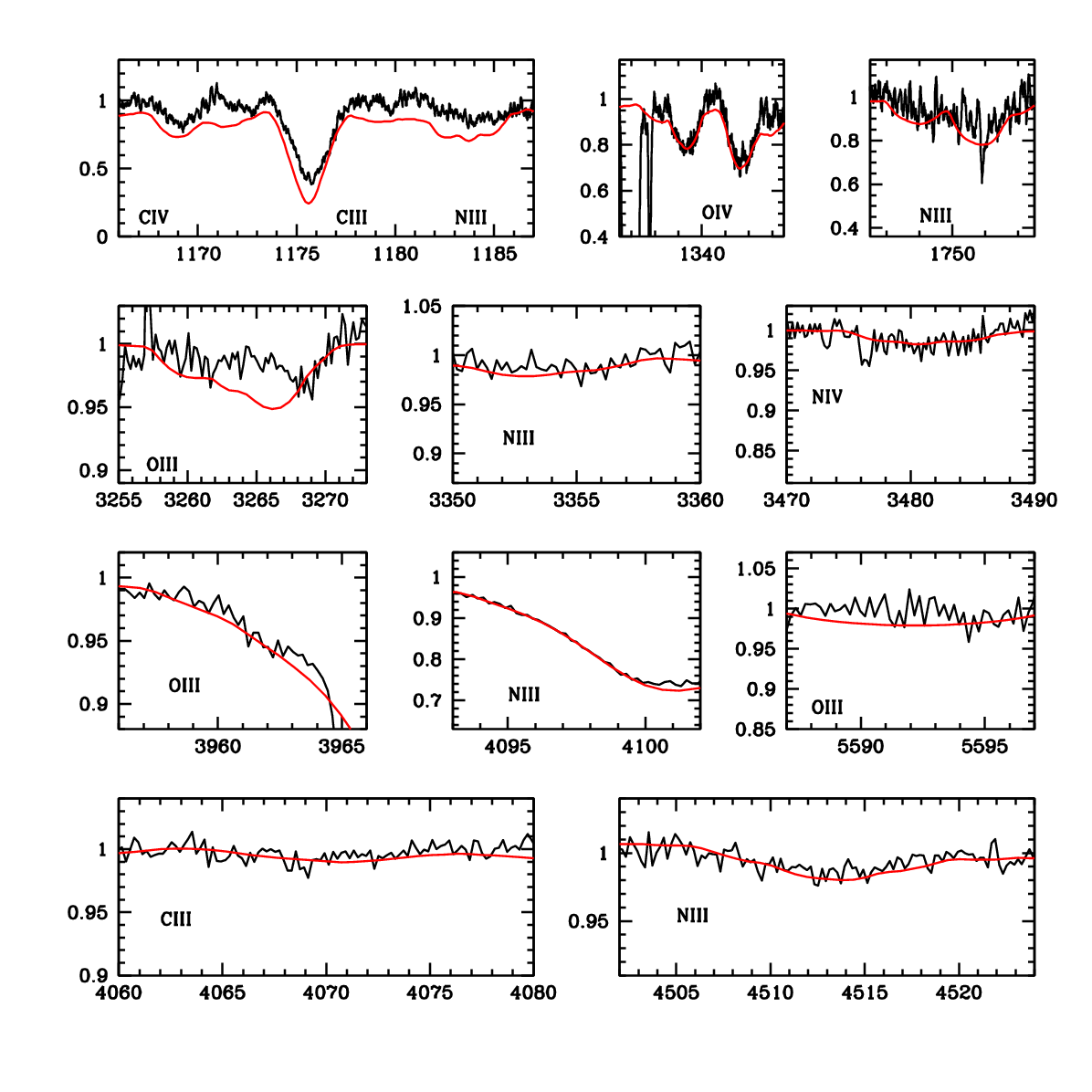}
\caption{Same as Fig.~\ref{fit_av15} but for SK~-71$^{\circ}$ 19.} 
\label{fit_skm71d19}
\end{figure*}

\begin{figure*}[ht]
\centering
\includegraphics[width=0.49\textwidth]{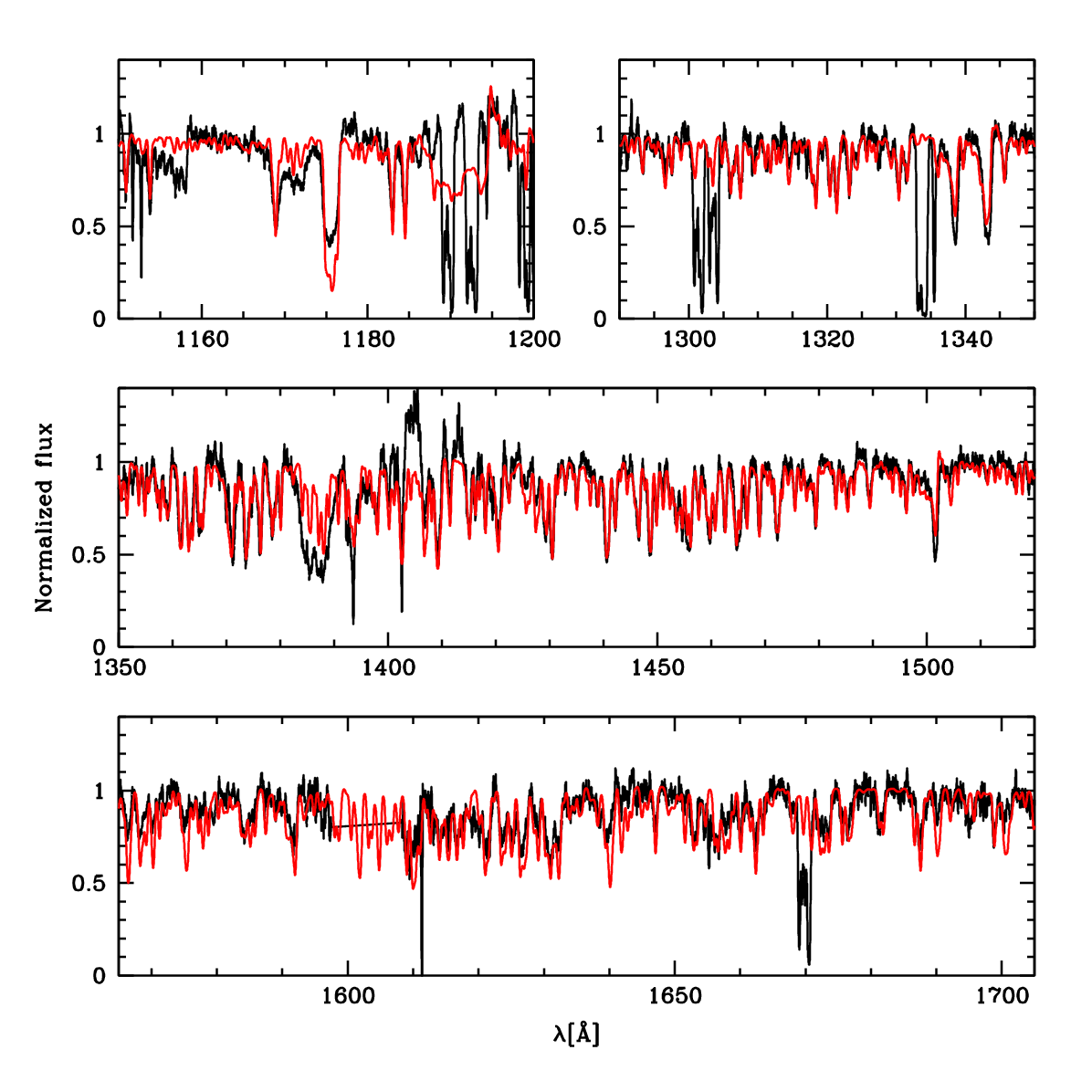}
\includegraphics[width=0.49\textwidth]{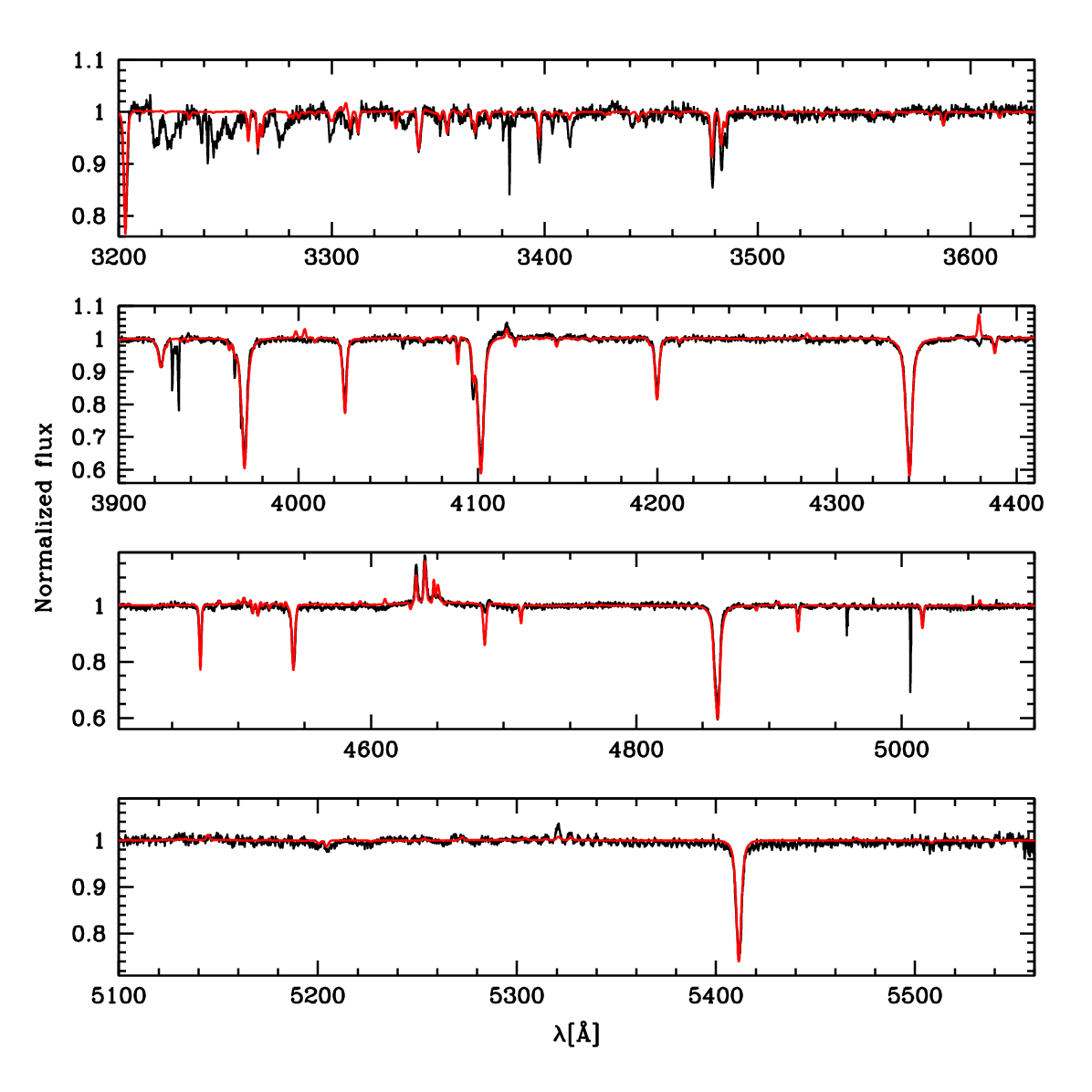}\\
\includegraphics[width=0.75\textwidth]{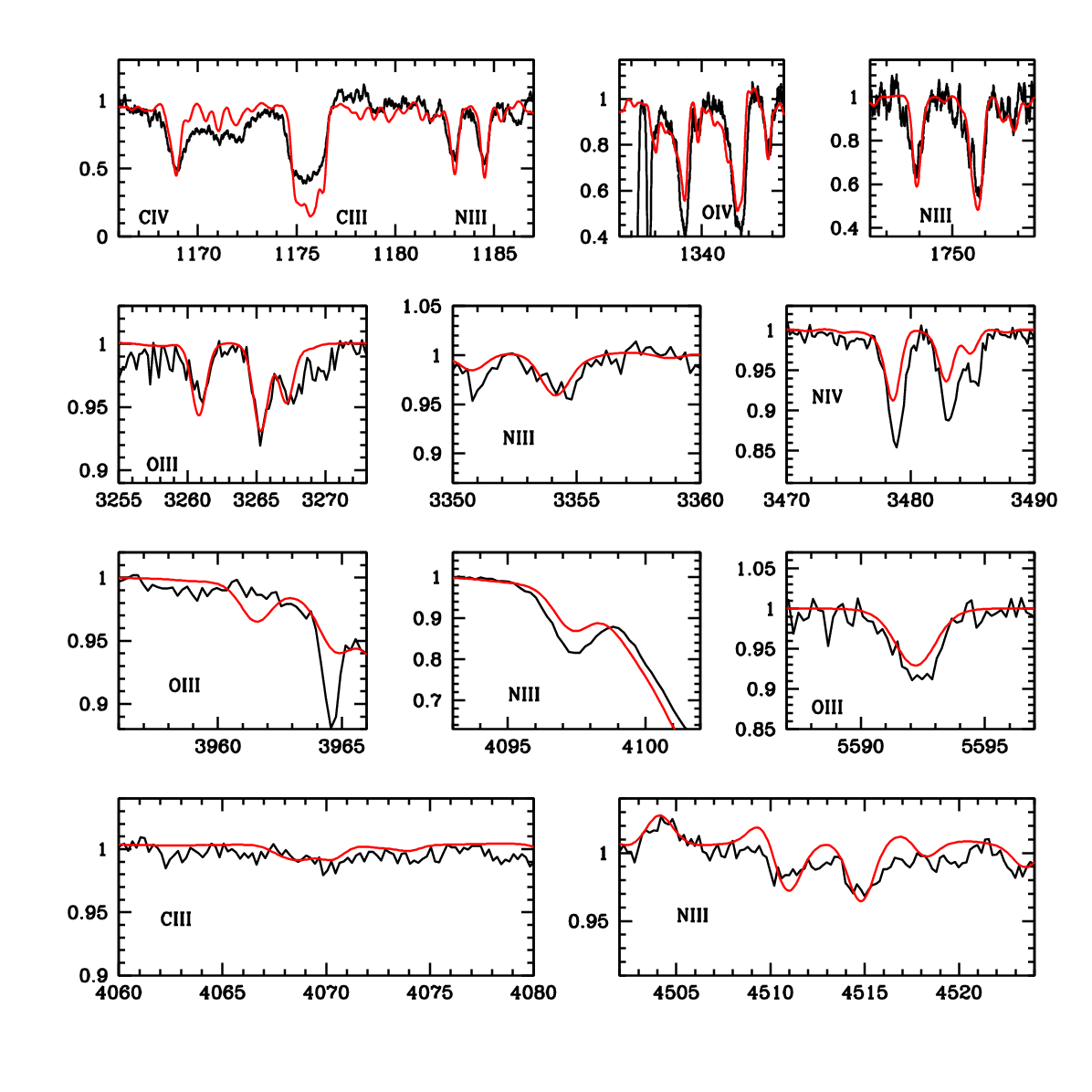}
\caption{Same as Fig.~\ref{fit_av15} but for N11 018.} 
\label{fit_N11018}
\end{figure*}

\begin{figure*}[ht]
\centering
\includegraphics[width=0.49\textwidth]{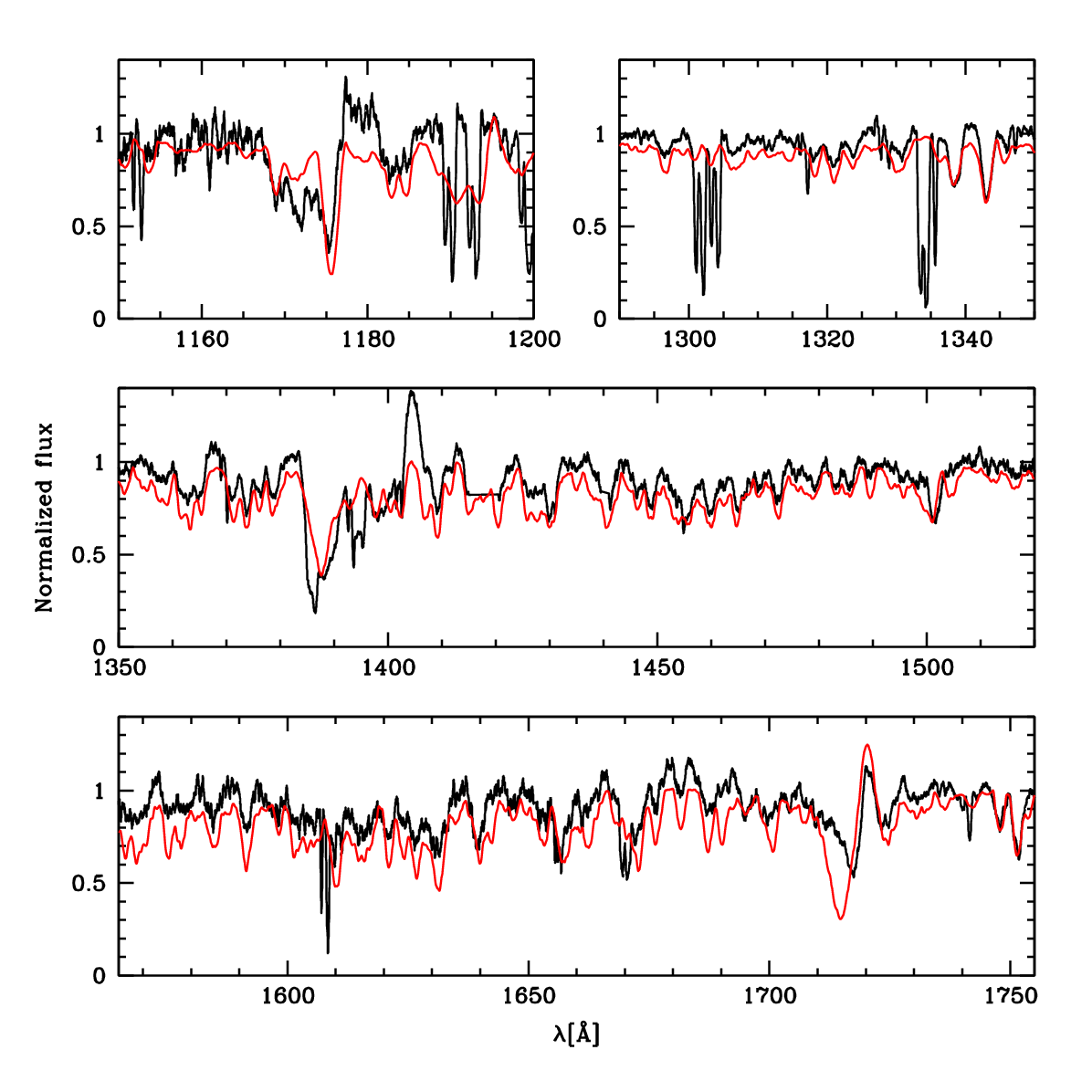}
\includegraphics[width=0.49\textwidth]{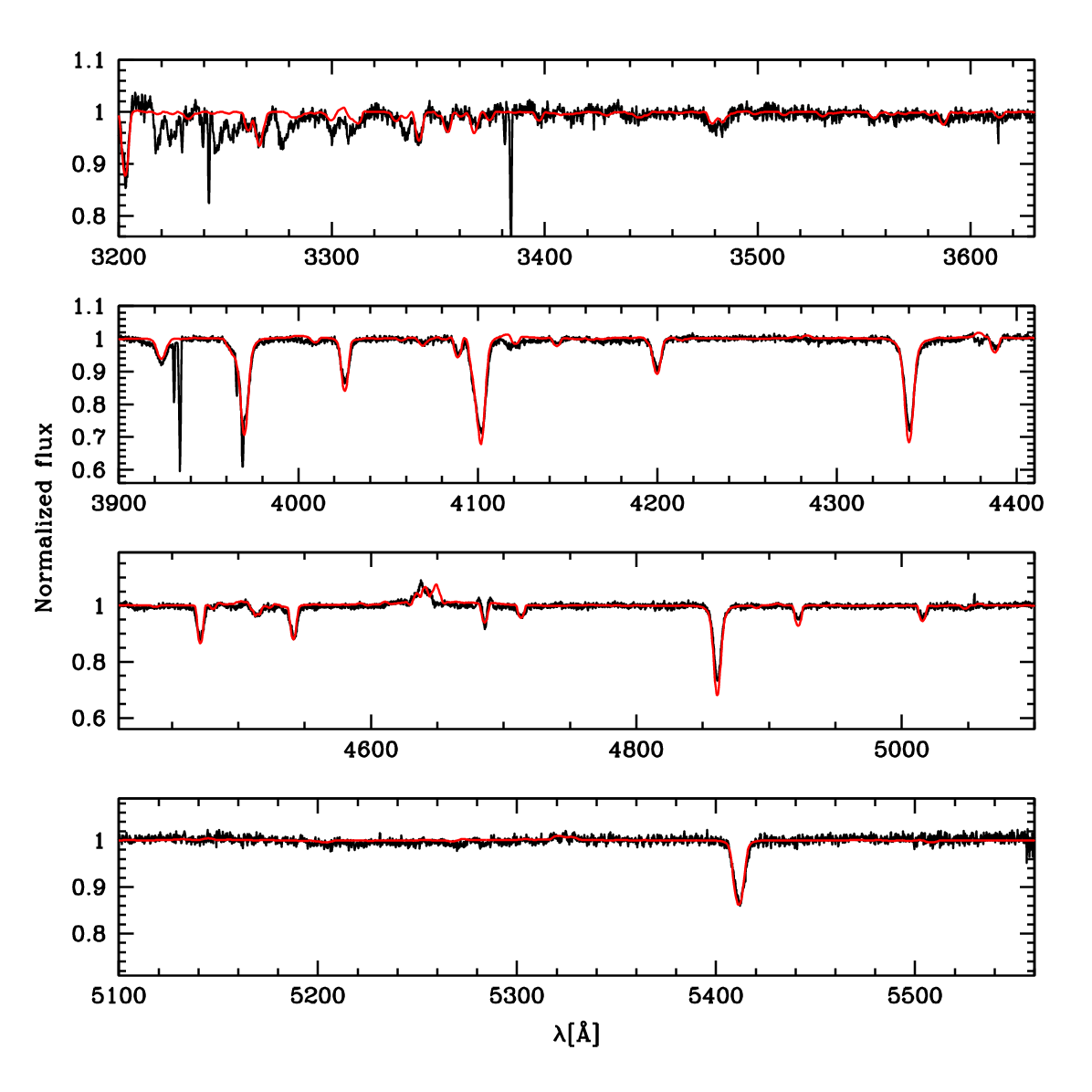}\\
\includegraphics[width=0.75\textwidth]{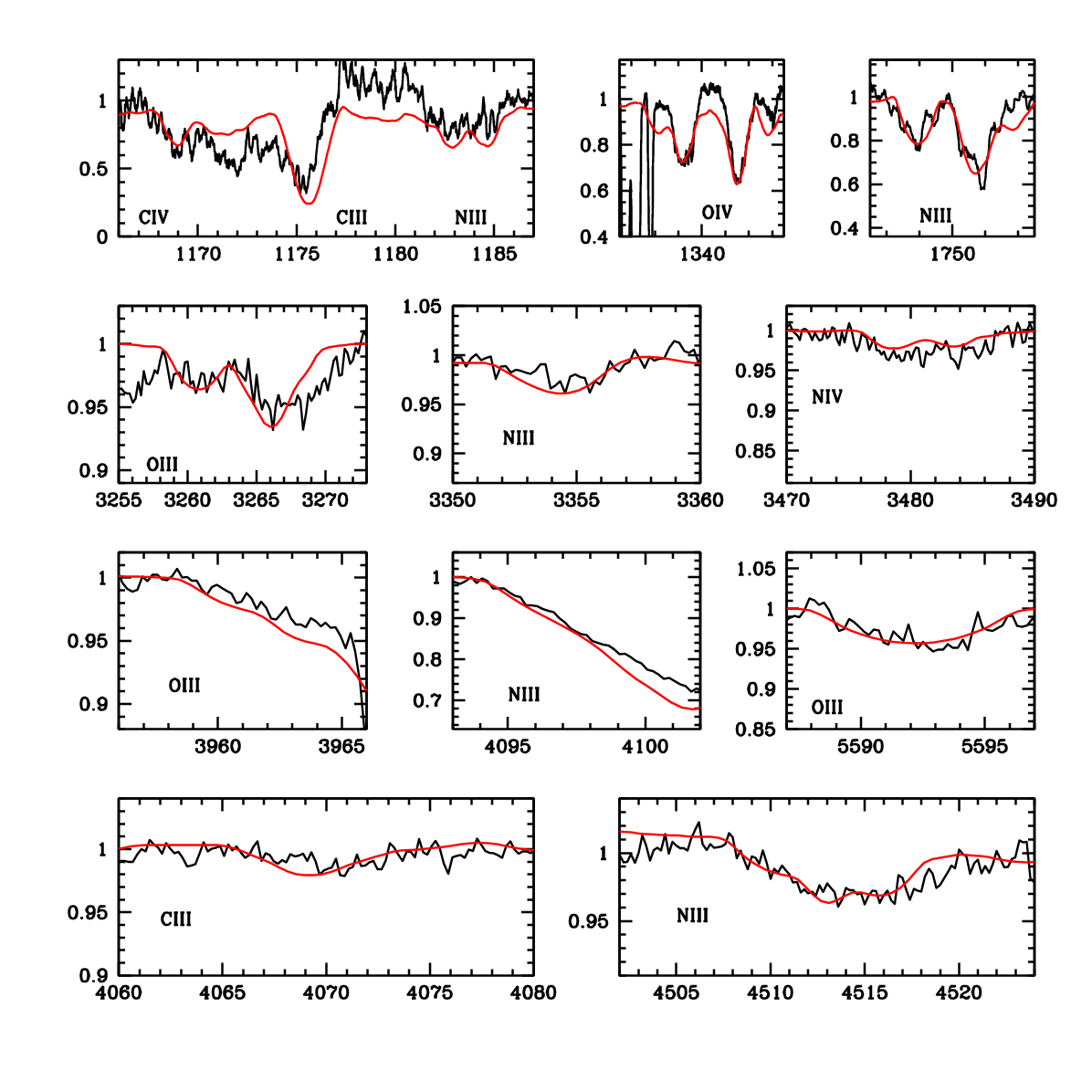}
\caption{Same as Fig.~\ref{fit_av15} but for SK~-71$^{\circ}$ 50.} 
\label{fit_skm71d50}
\end{figure*}

\begin{figure*}[ht]
\centering
\includegraphics[width=0.49\textwidth]{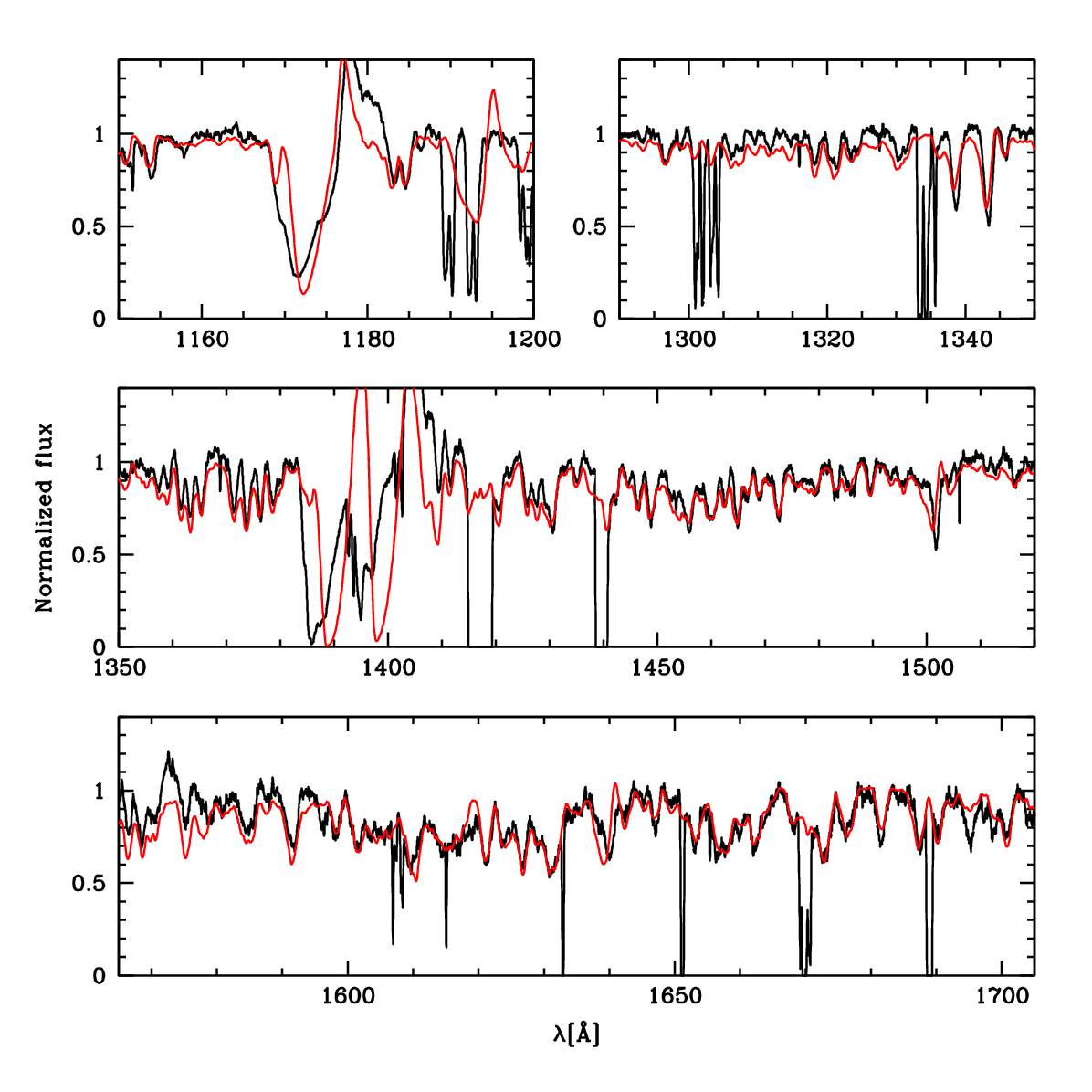}
\includegraphics[width=0.49\textwidth]{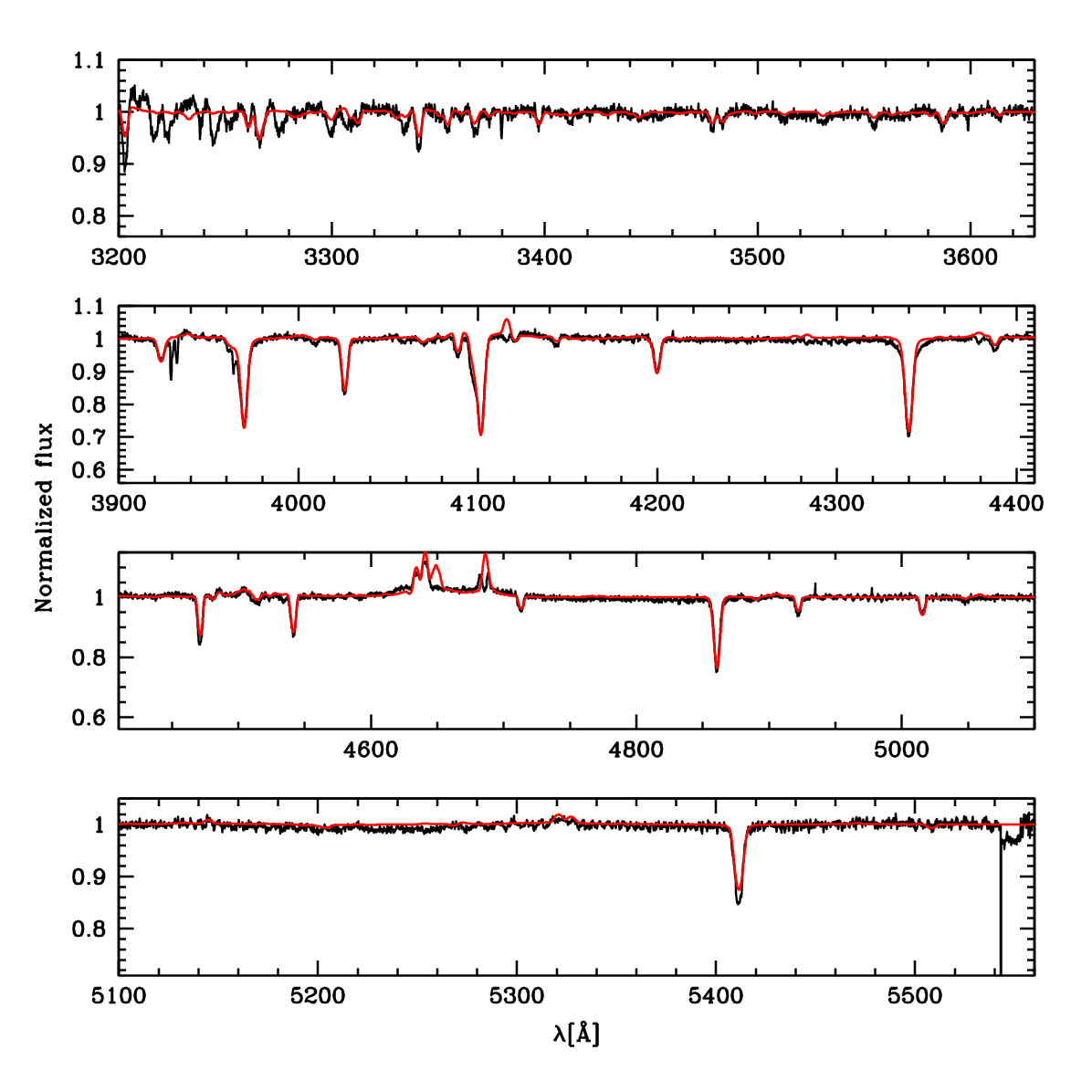}\\
\includegraphics[width=0.75\textwidth]{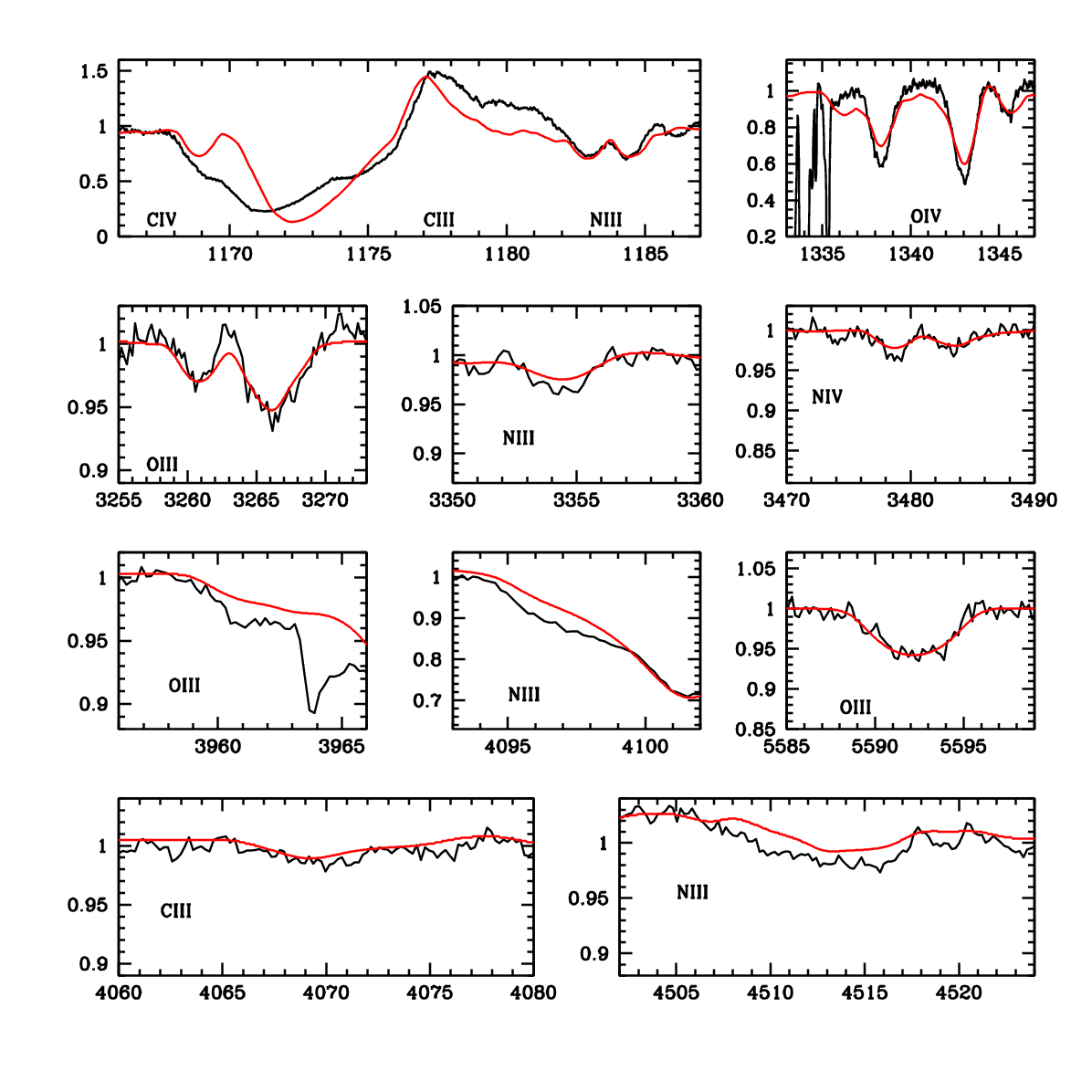}
\caption{Same as Fig.~\ref{fit_av15} but for SK~-66$^{\circ}$ 152.} 
\label{fit_skm66d152}
\end{figure*}

\begin{figure*}[ht]
\centering
\includegraphics[width=0.49\textwidth]{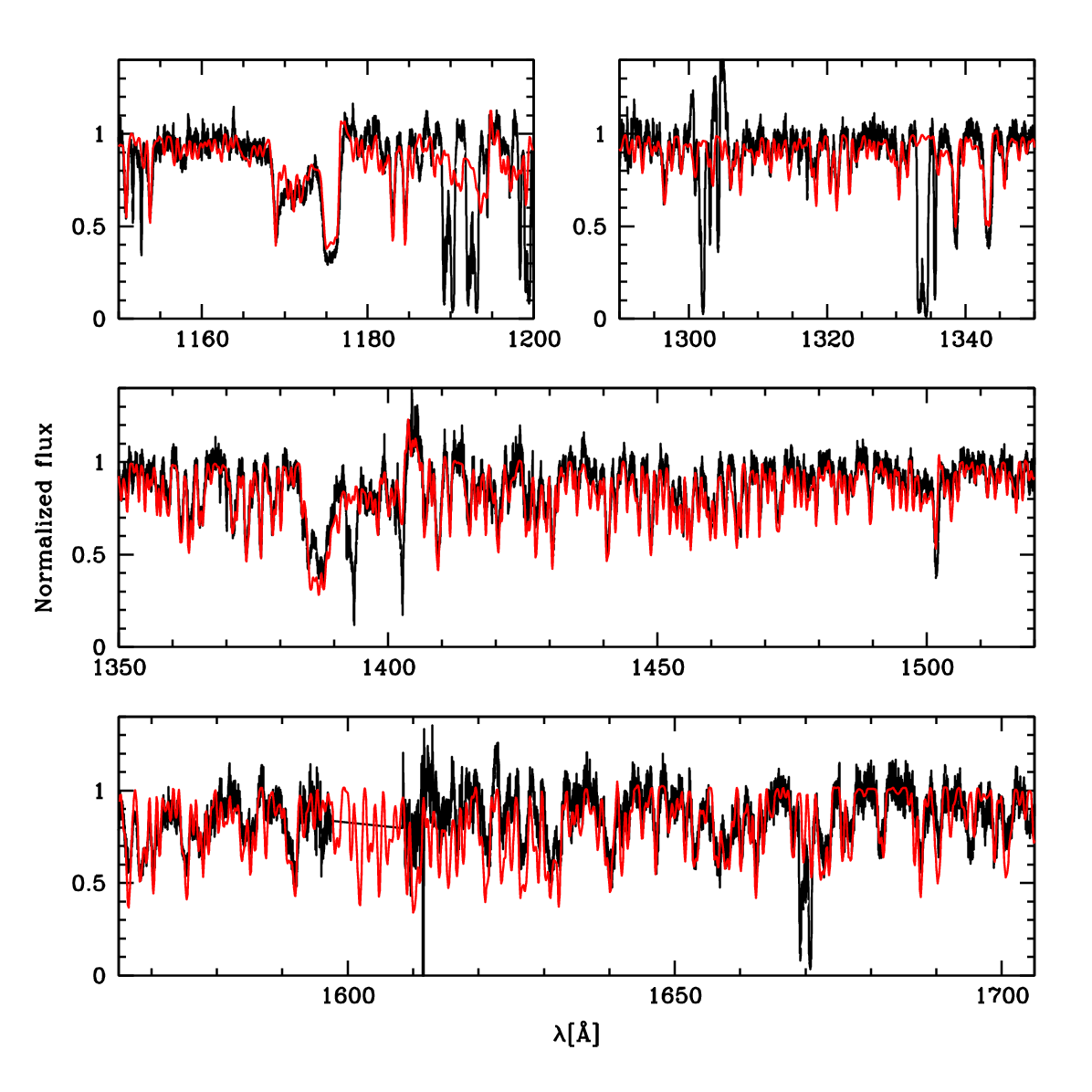}
\includegraphics[width=0.49\textwidth]{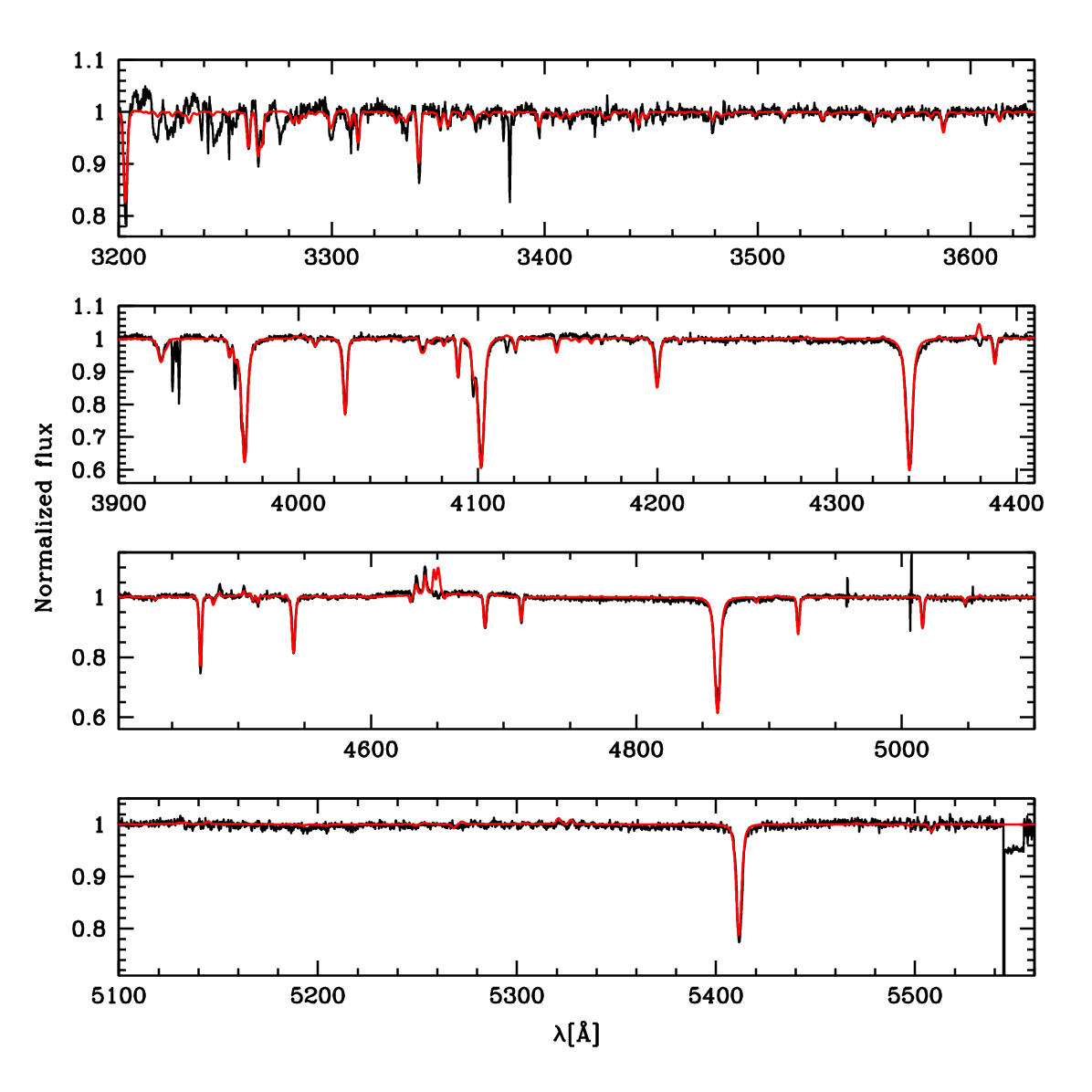}\\
\includegraphics[width=0.75\textwidth]{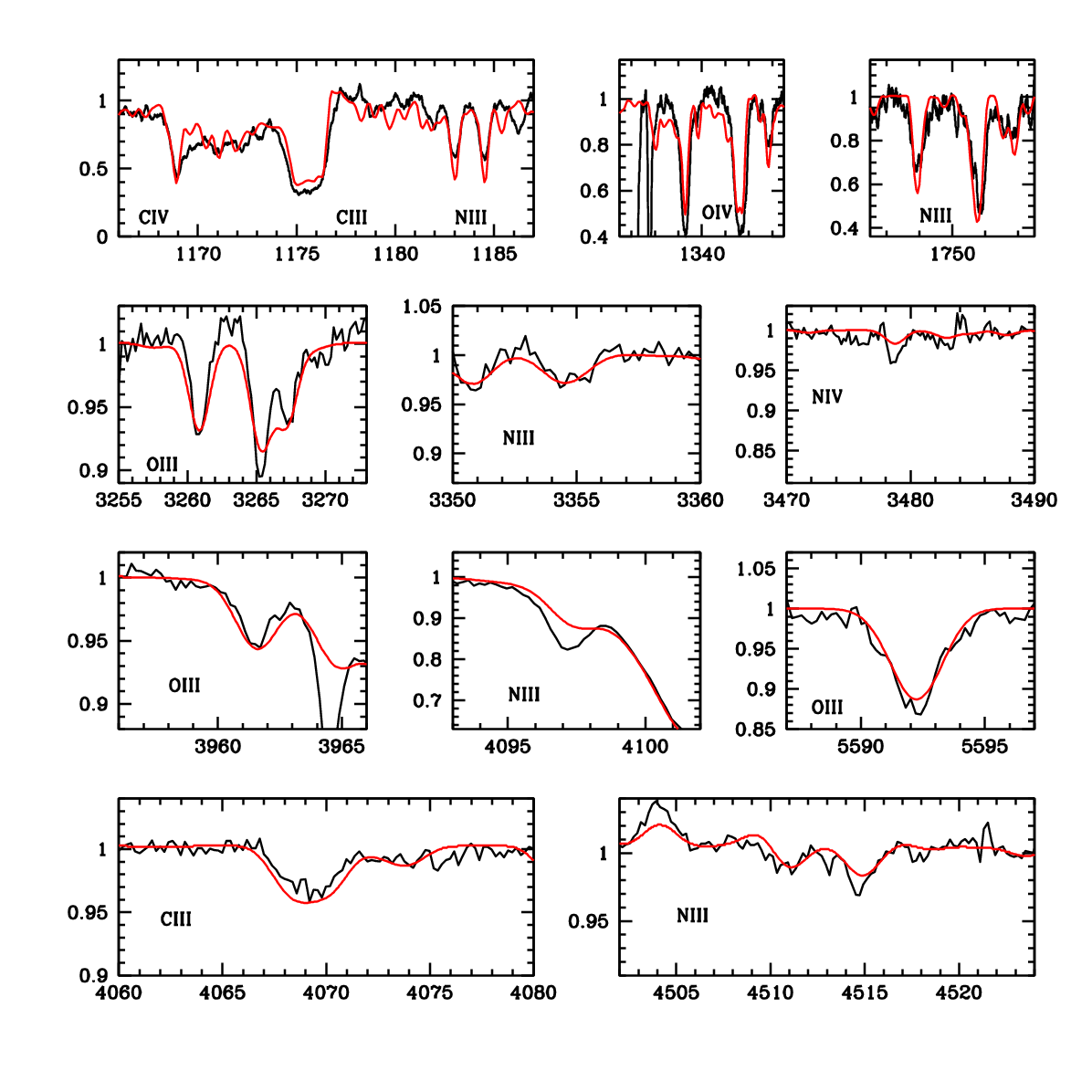}
\caption{Same as Fig.~\ref{fit_av15} but for N11~032.} 
\label{fit_n11032}
\end{figure*}

\begin{figure*}[ht]
\centering
\includegraphics[width=0.49\textwidth]{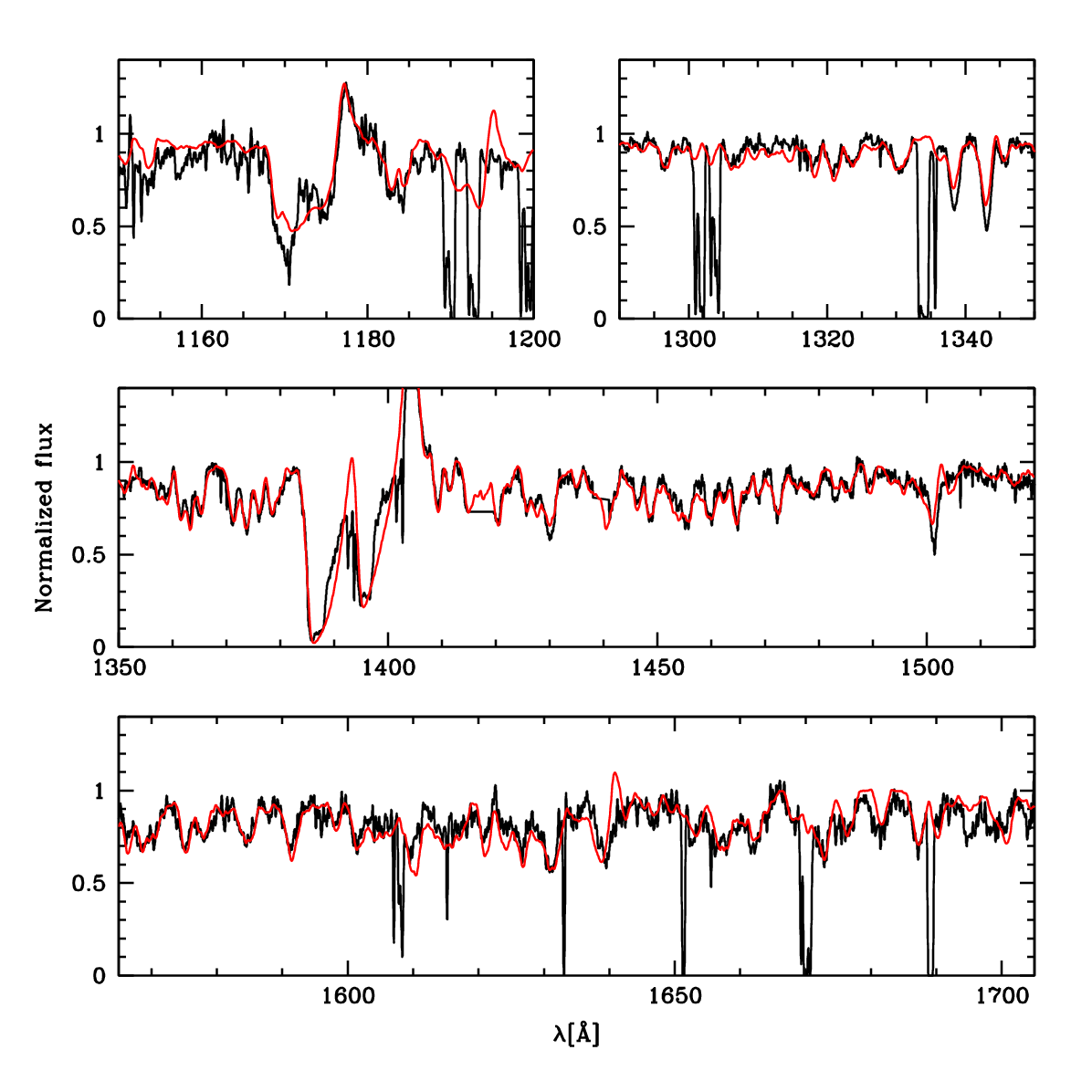}
\includegraphics[width=0.49\textwidth]{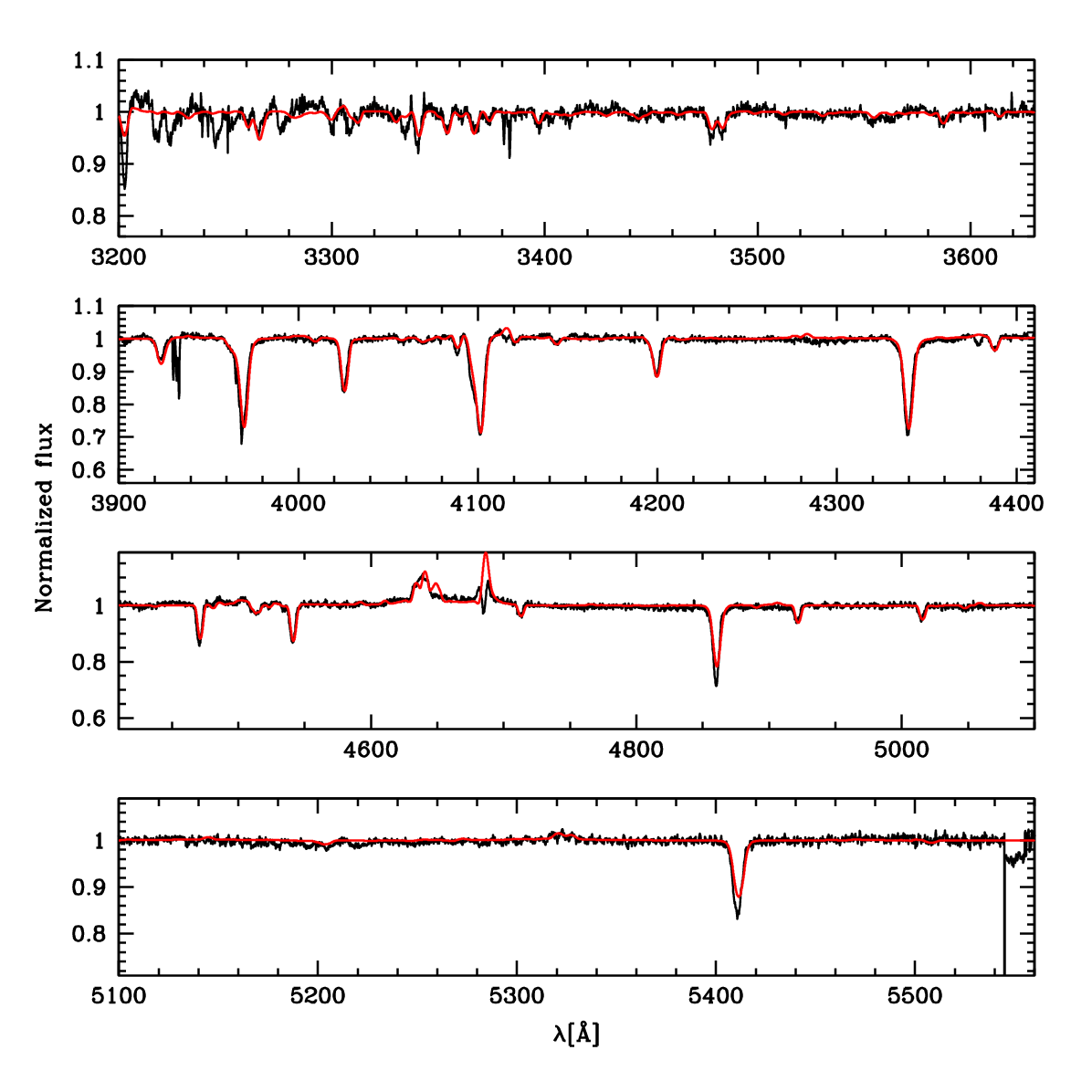}\\
\includegraphics[width=0.75\textwidth]{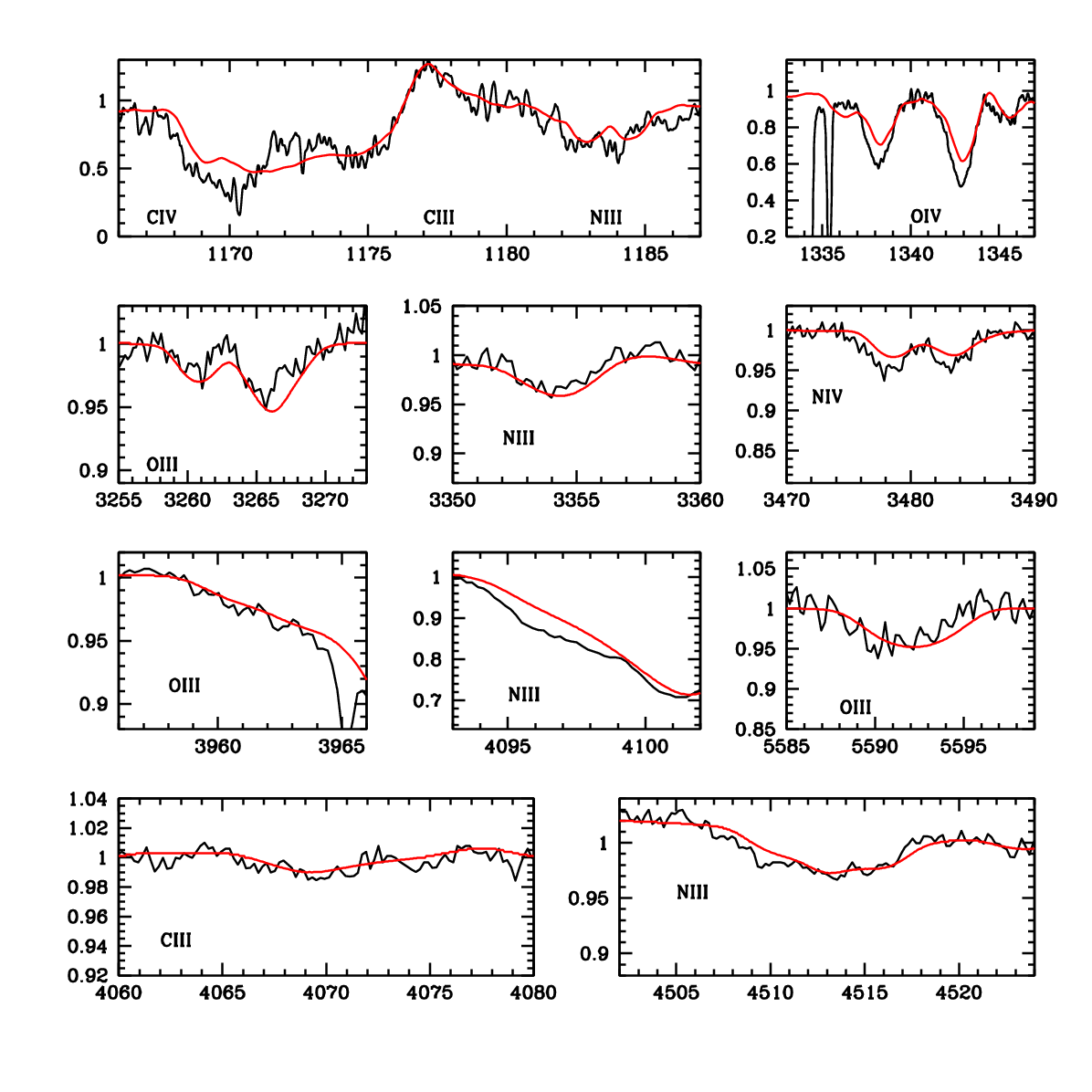}
\caption{Same as Fig.~\ref{fit_av15} but for SK~-69$^{\circ}$ 50.} 
\label{fit_skm69d50}
\end{figure*}

\begin{figure*}[ht]
\centering
\includegraphics[width=0.49\textwidth]{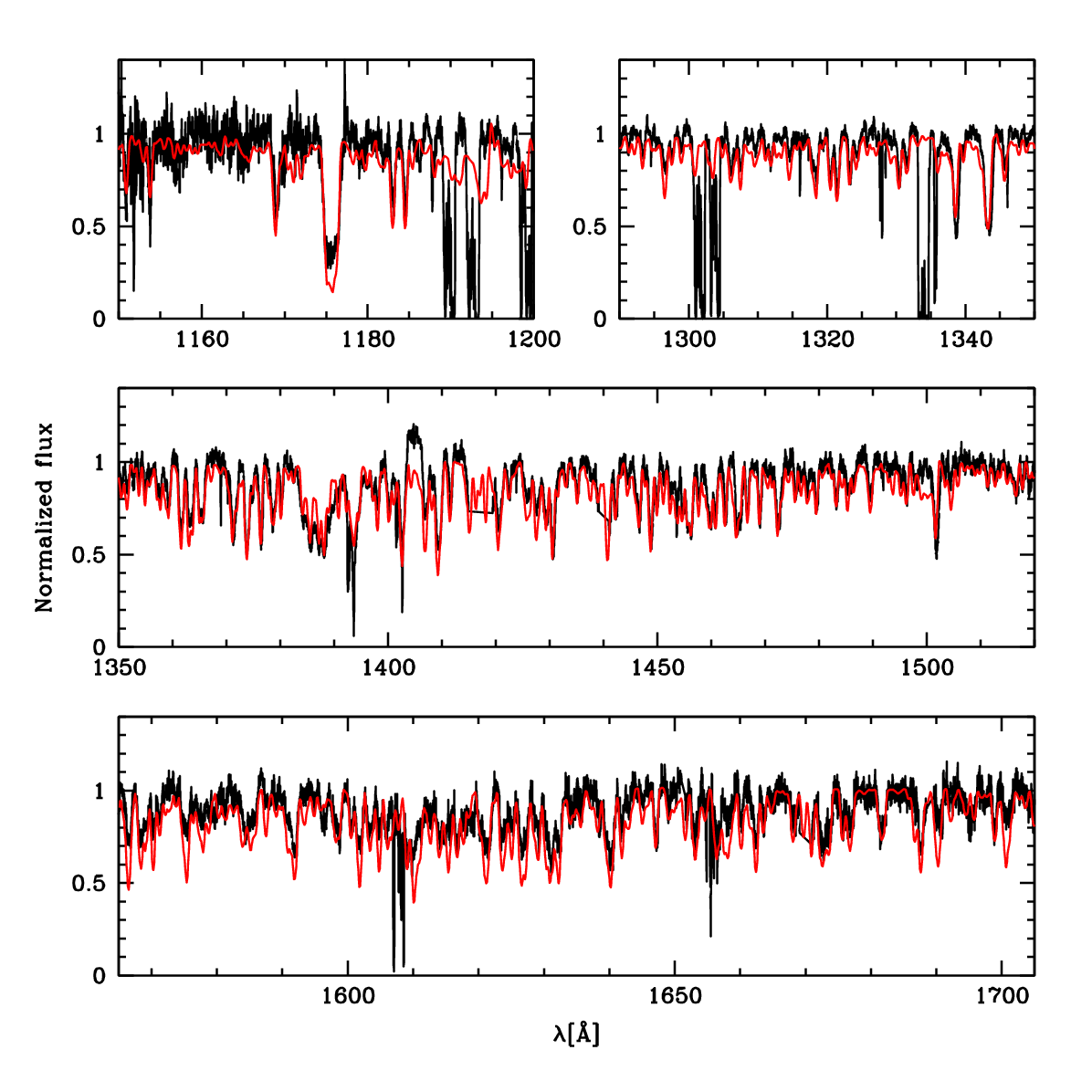}
\includegraphics[width=0.49\textwidth]{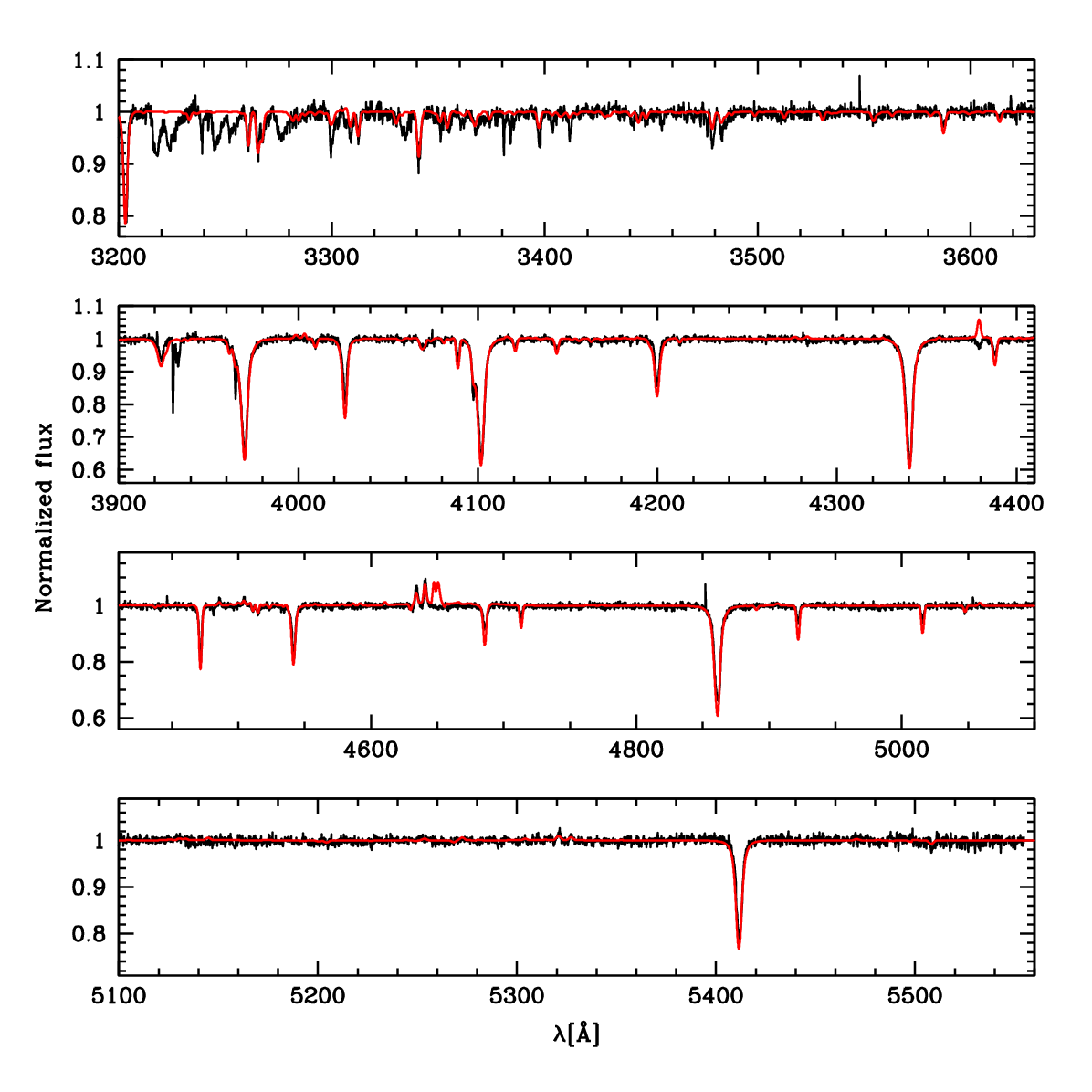}\\
\includegraphics[width=0.75\textwidth]{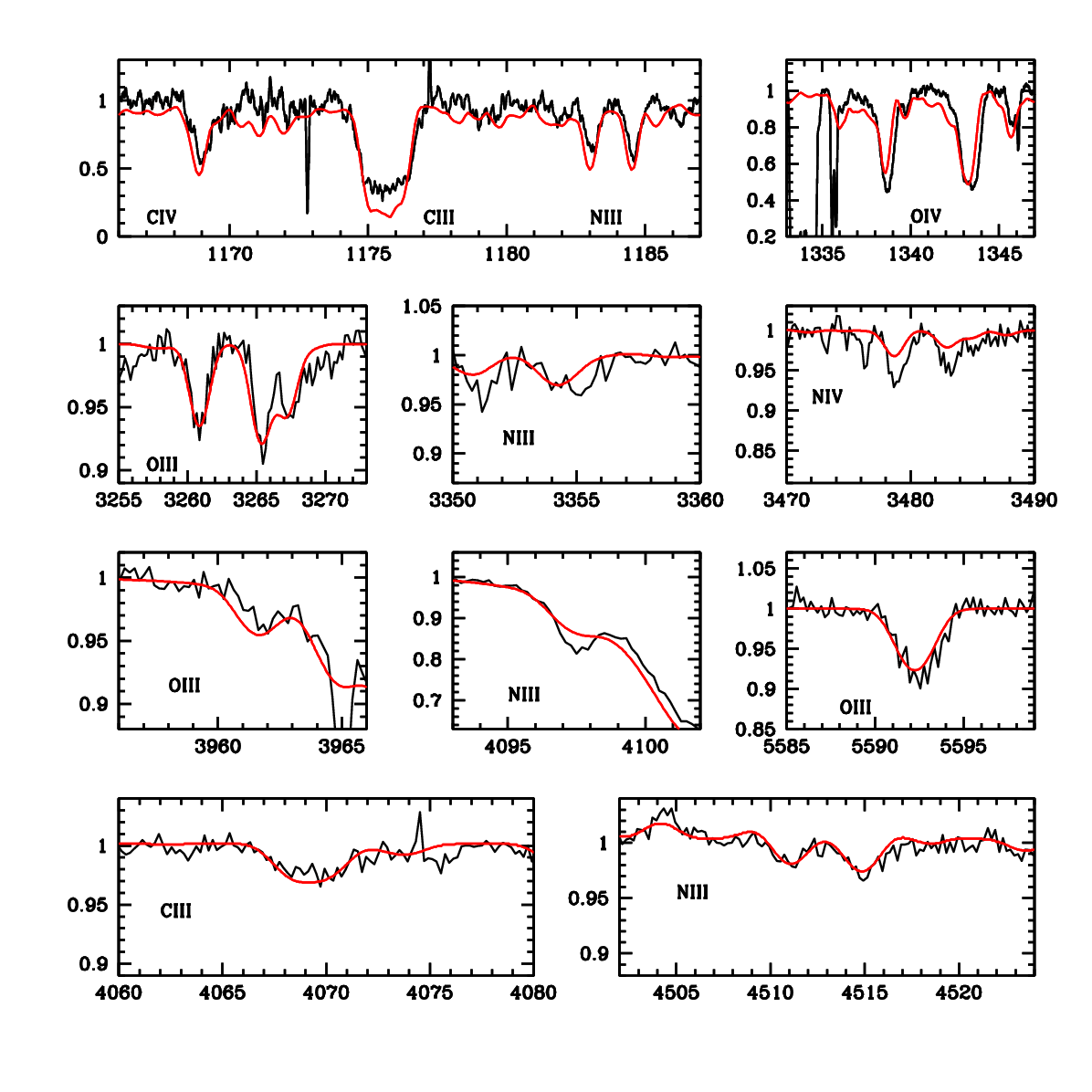}
\caption{Same as Fig.~\ref{fit_av15} but for SK~-68$^{\circ}$ 16.} 
\label{fit_skm68d16}
\end{figure*}

\begin{figure*}[ht]
\centering
\includegraphics[width=0.49\textwidth]{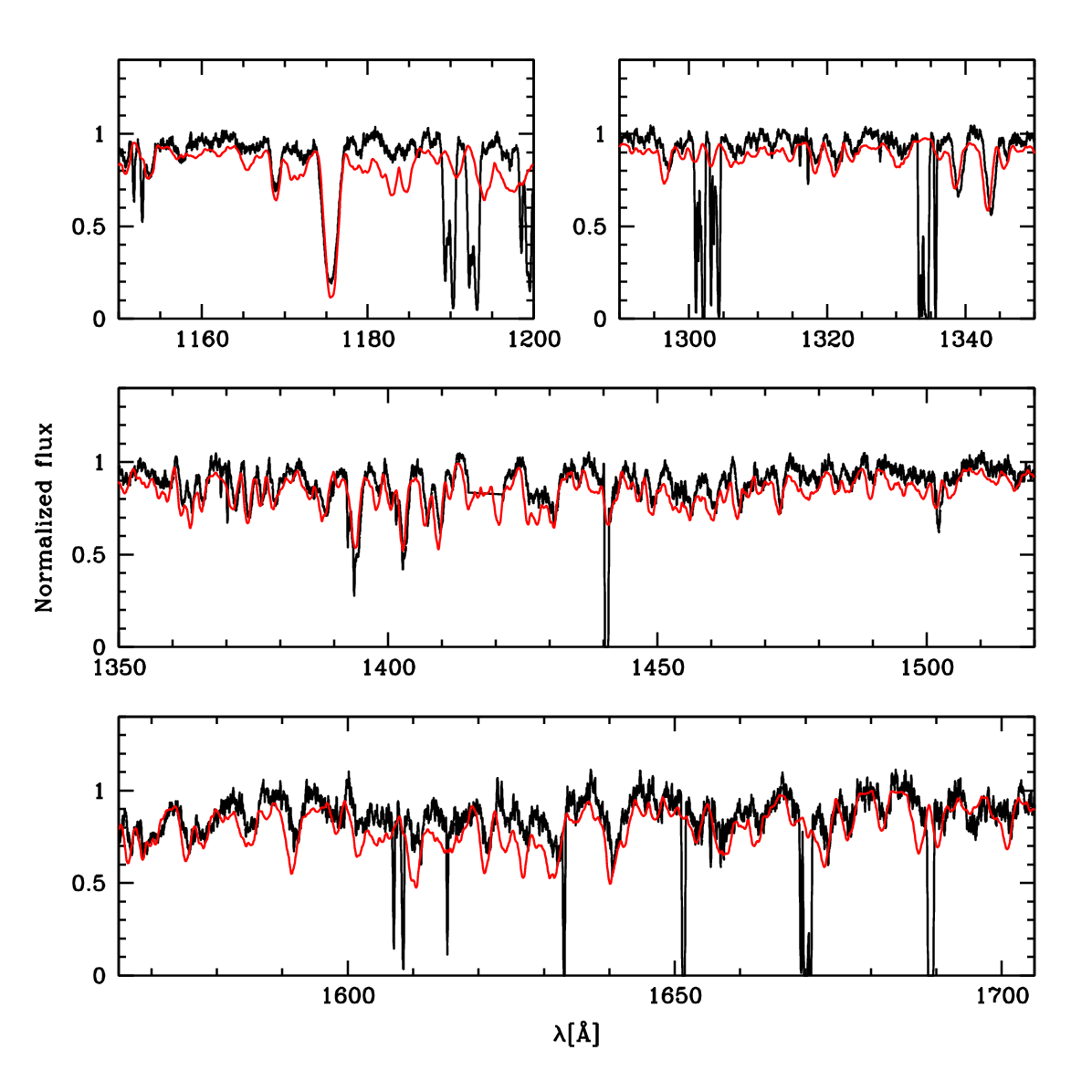}
\includegraphics[width=0.49\textwidth]{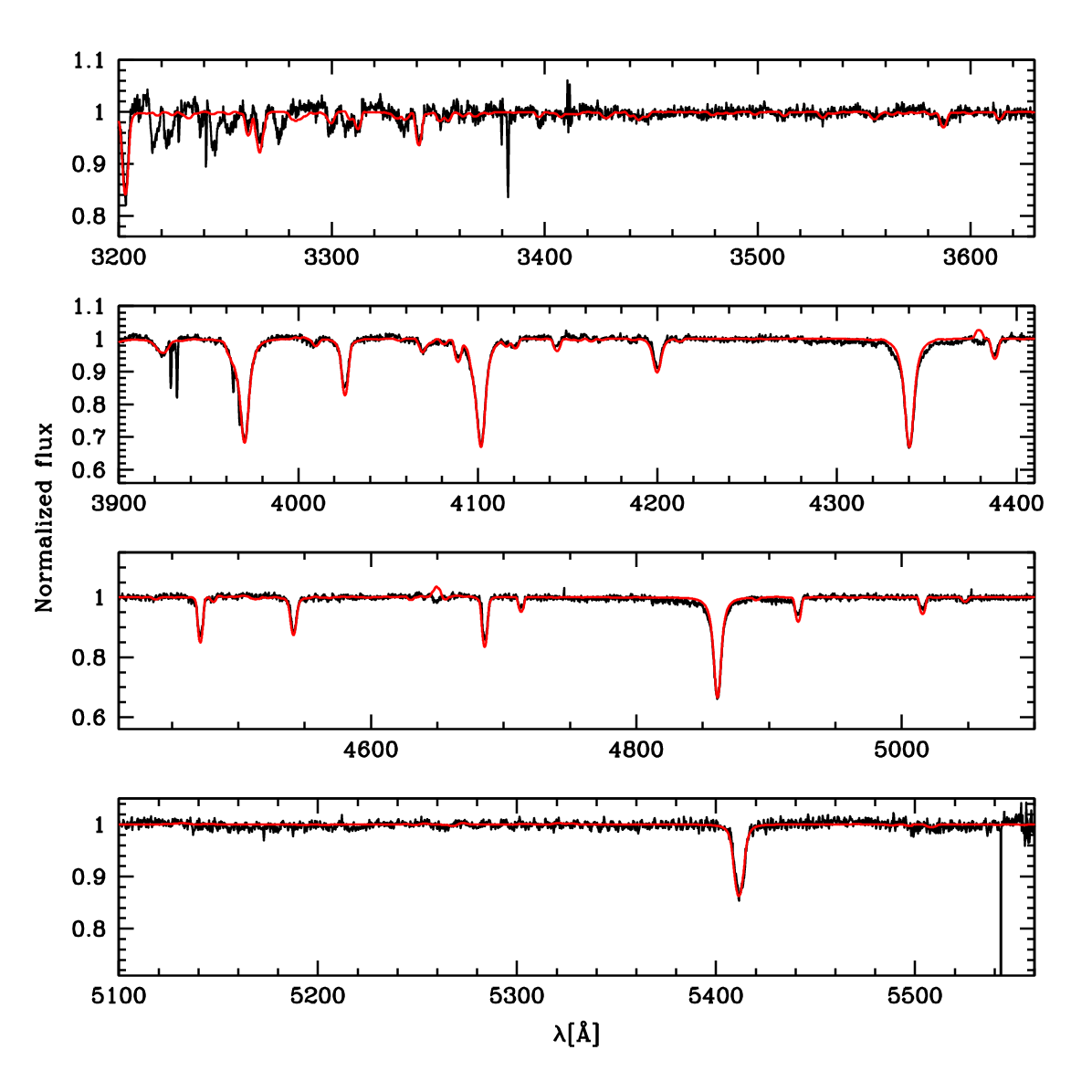}\\
\includegraphics[width=0.75\textwidth]{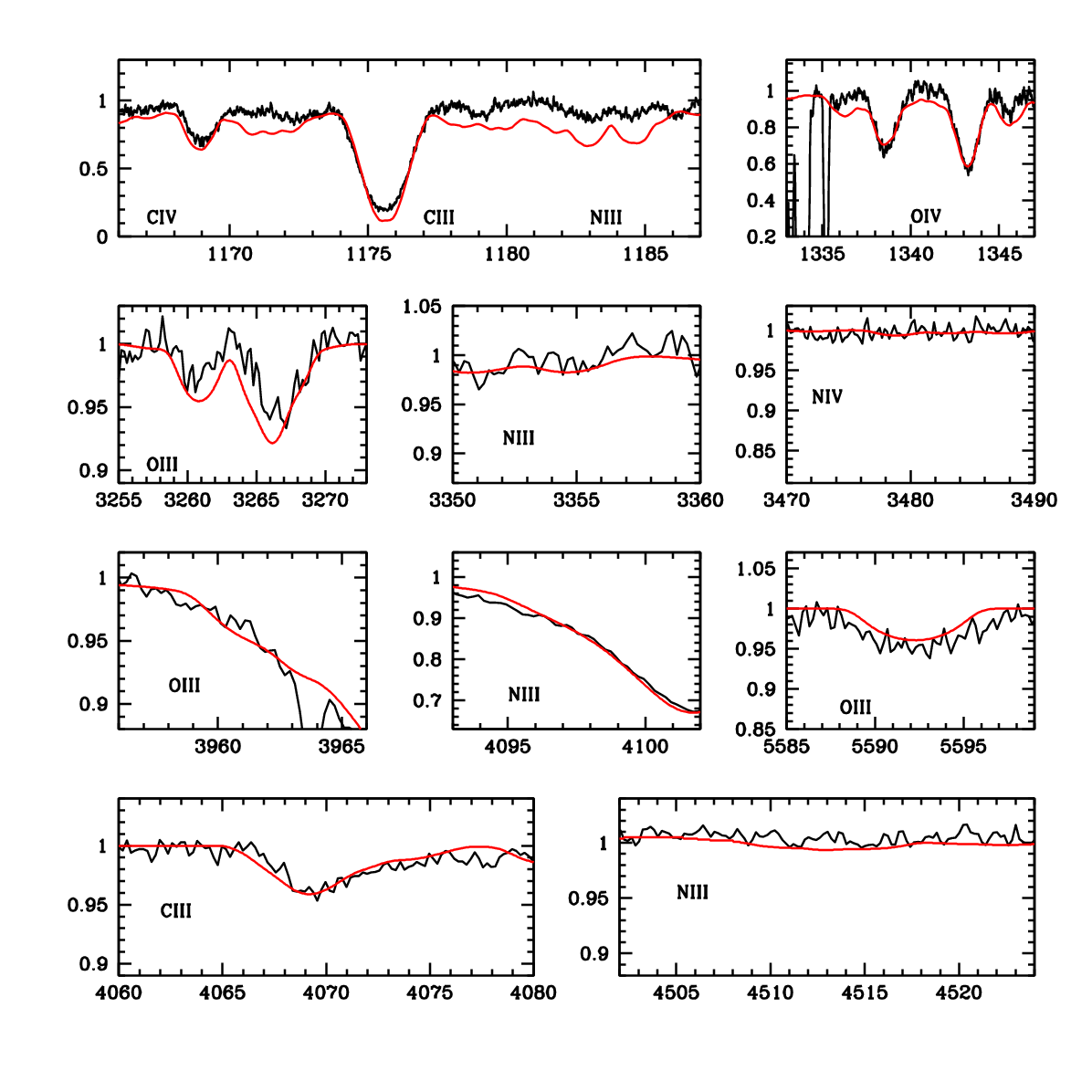}
\caption{Same as Fig.~\ref{fit_av15} but for N11~049.} 
\label{fit_n11049}
\end{figure*}

\begin{figure*}[ht]
\centering
\includegraphics[width=0.49\textwidth]{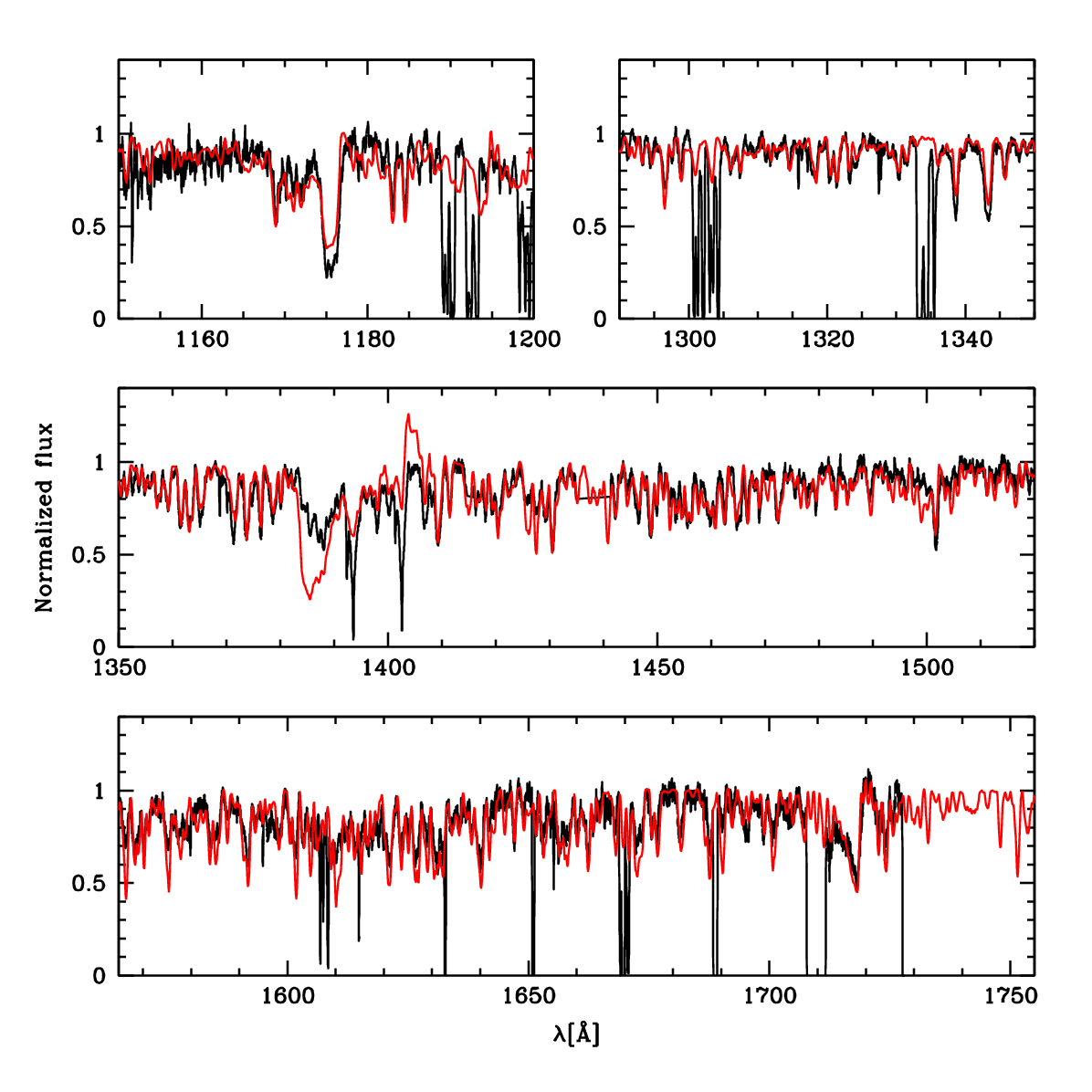}
\includegraphics[width=0.49\textwidth]{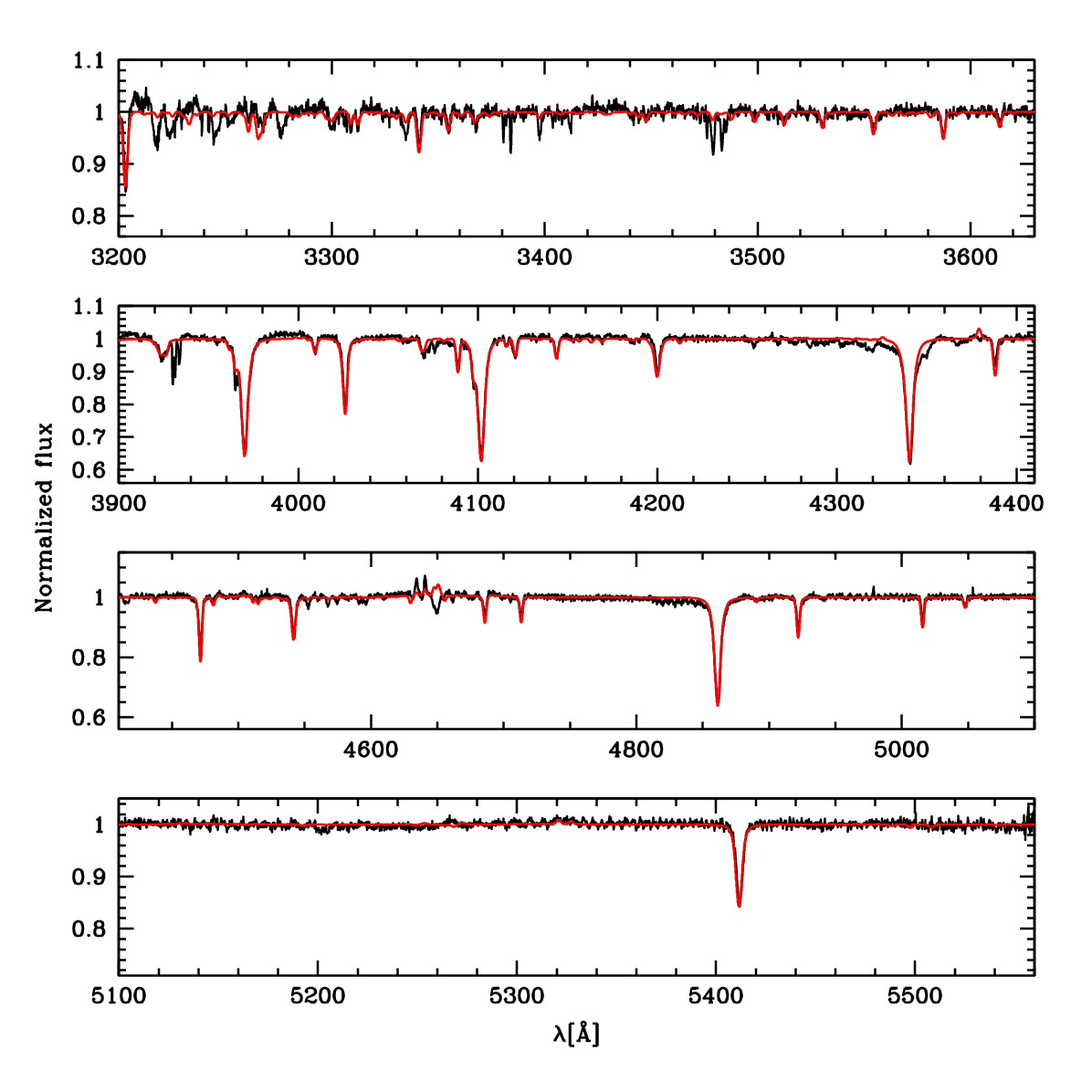}\\
\includegraphics[width=0.75\textwidth]{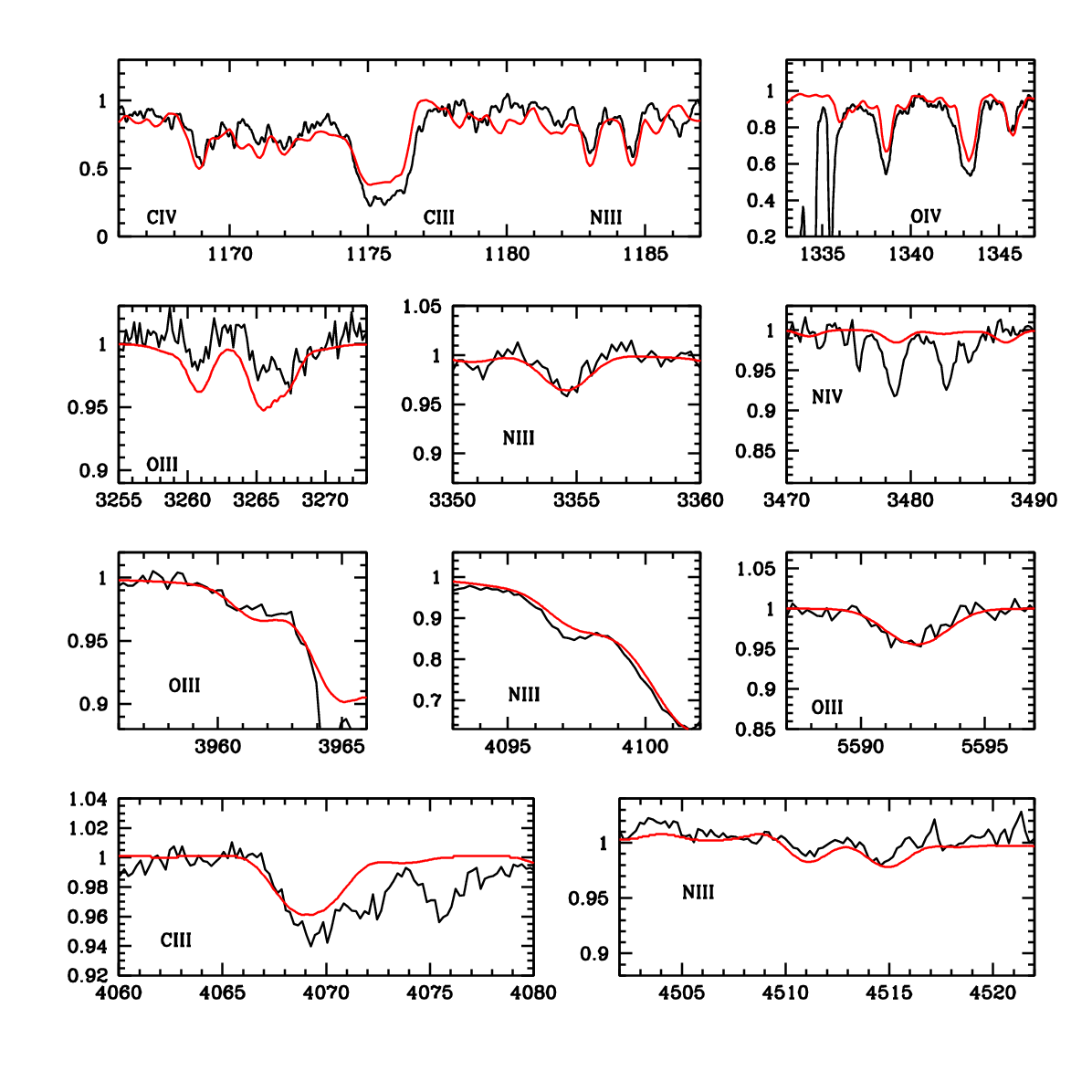}
\caption{Same as Fig.~\ref{fit_av15} but for SK~-67$^{\circ}$ 101.} 
\label{fit_skm67d101}
\end{figure*}

\begin{figure*}[ht]
\centering
\includegraphics[width=0.49\textwidth]{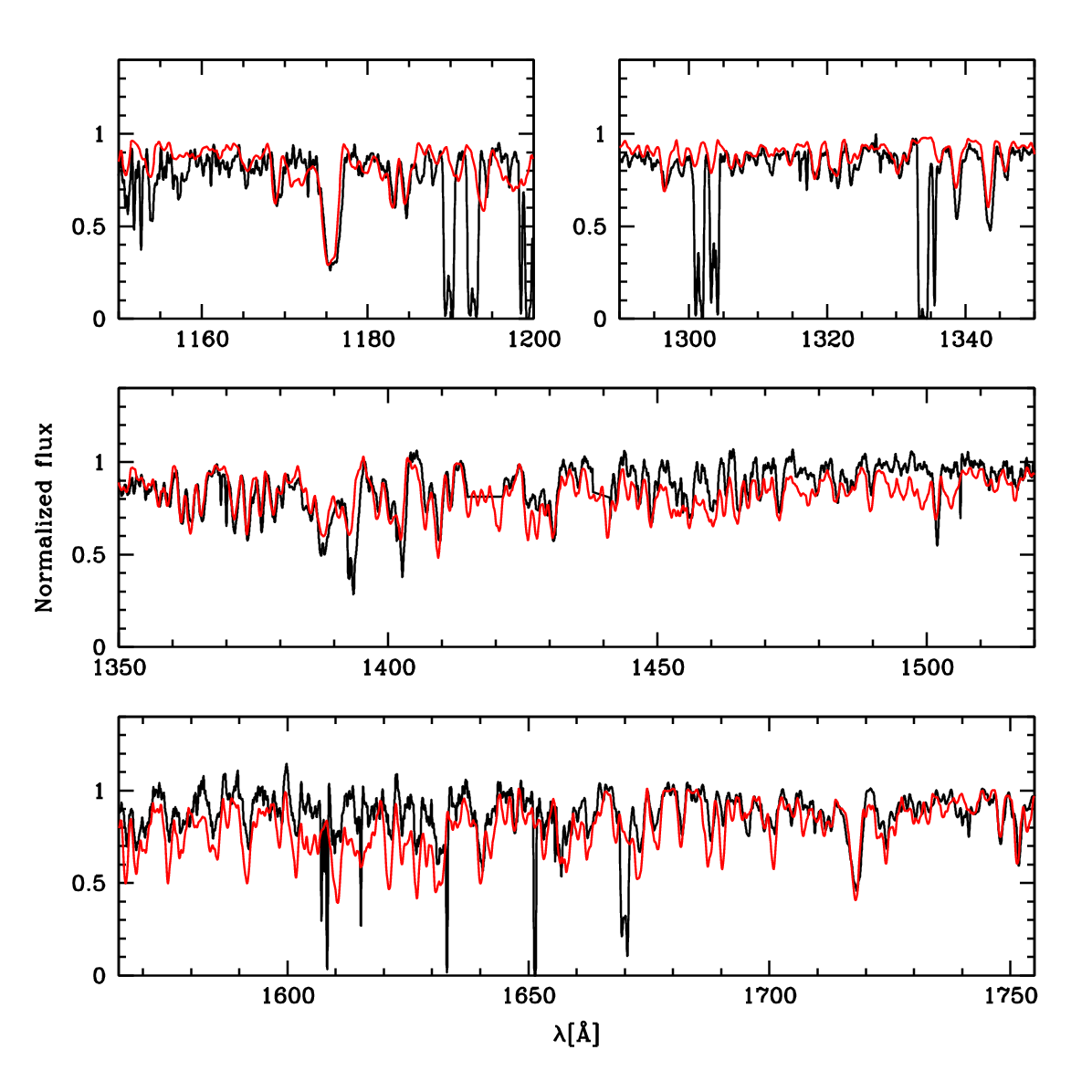}
\includegraphics[width=0.49\textwidth]{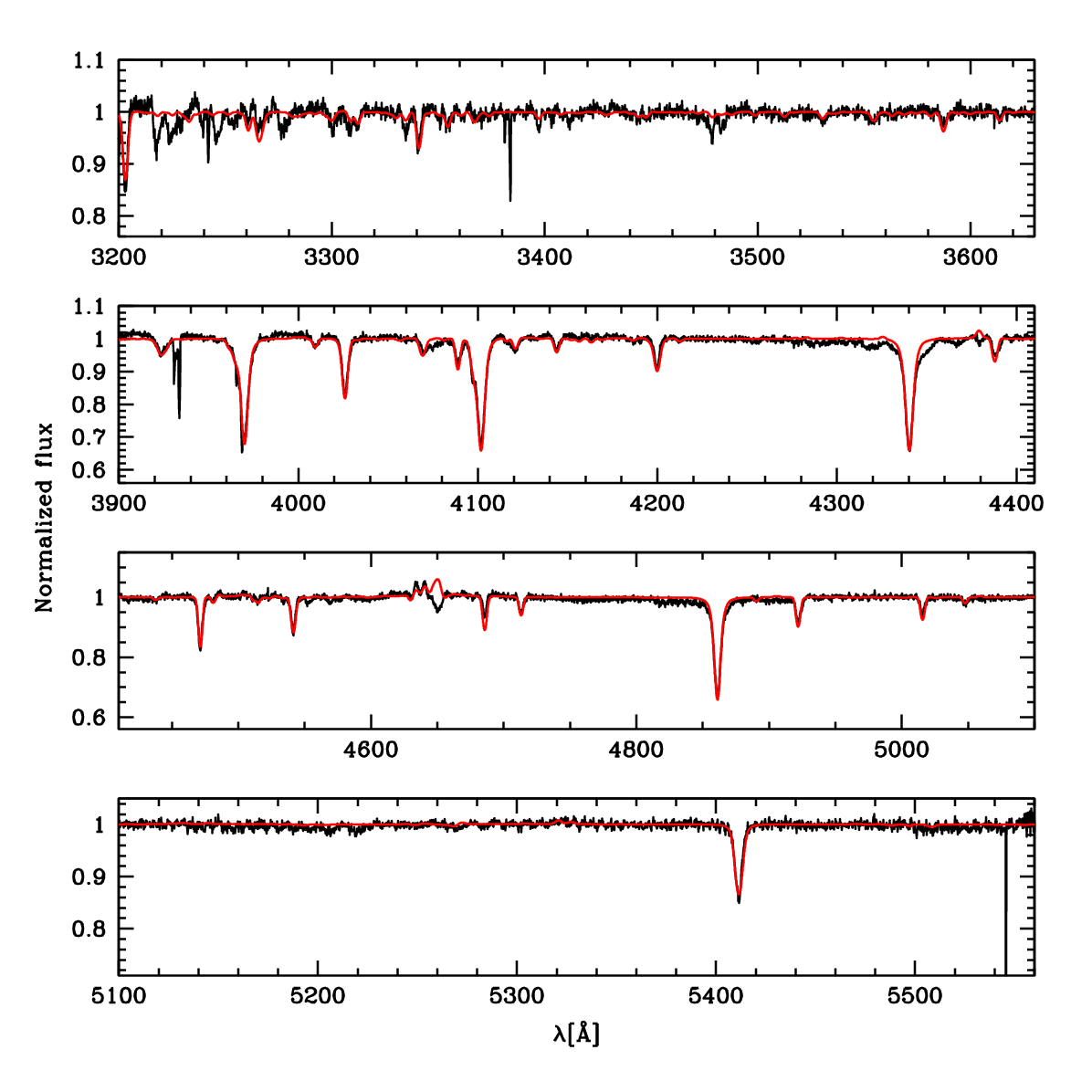}\\
\includegraphics[width=0.75\textwidth]{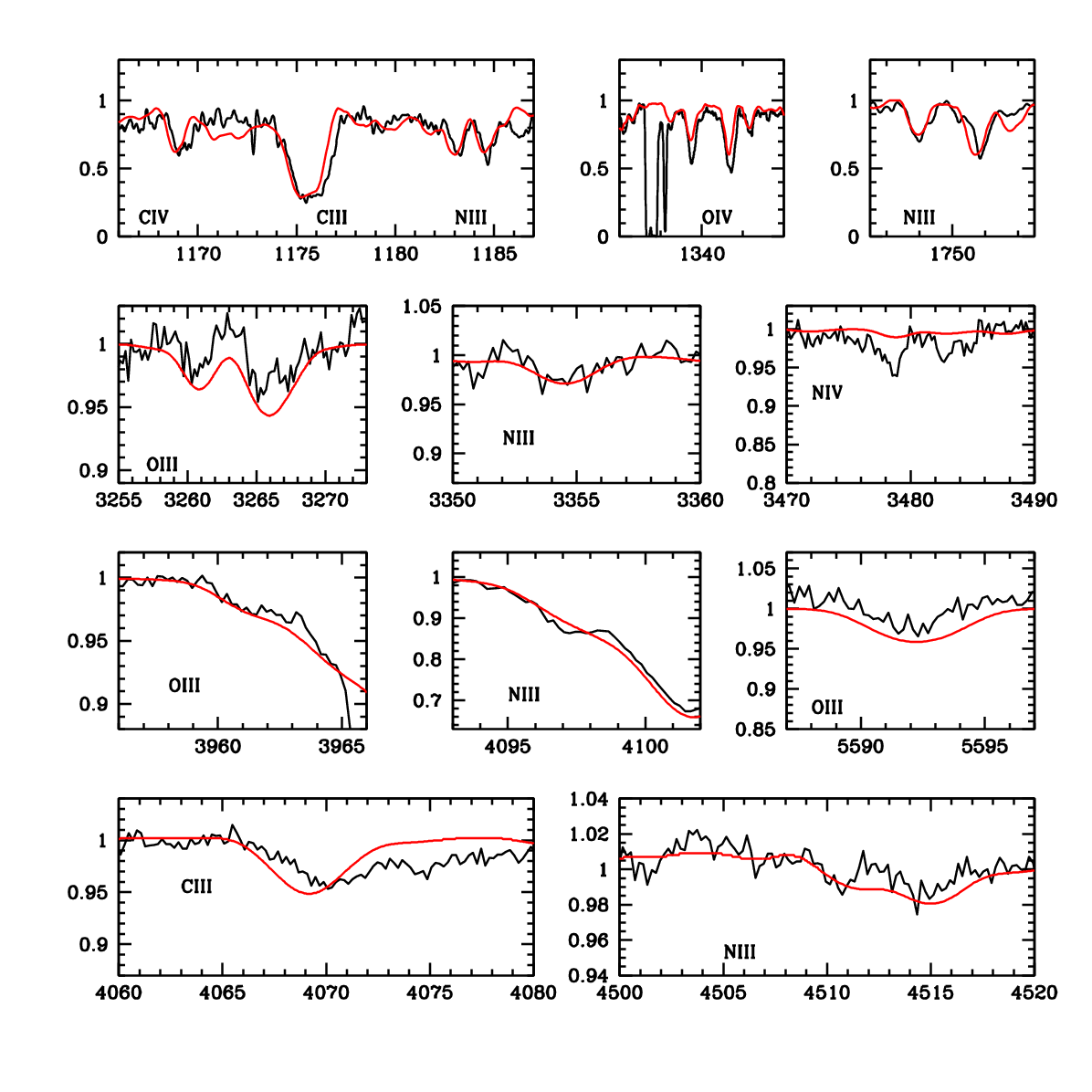}
\caption{Same as Fig.~\ref{fit_av15} but for BI~173.} 
\label{fit_bi173}
\end{figure*}

\begin{figure*}[ht]
\centering
\includegraphics[width=0.49\textwidth]{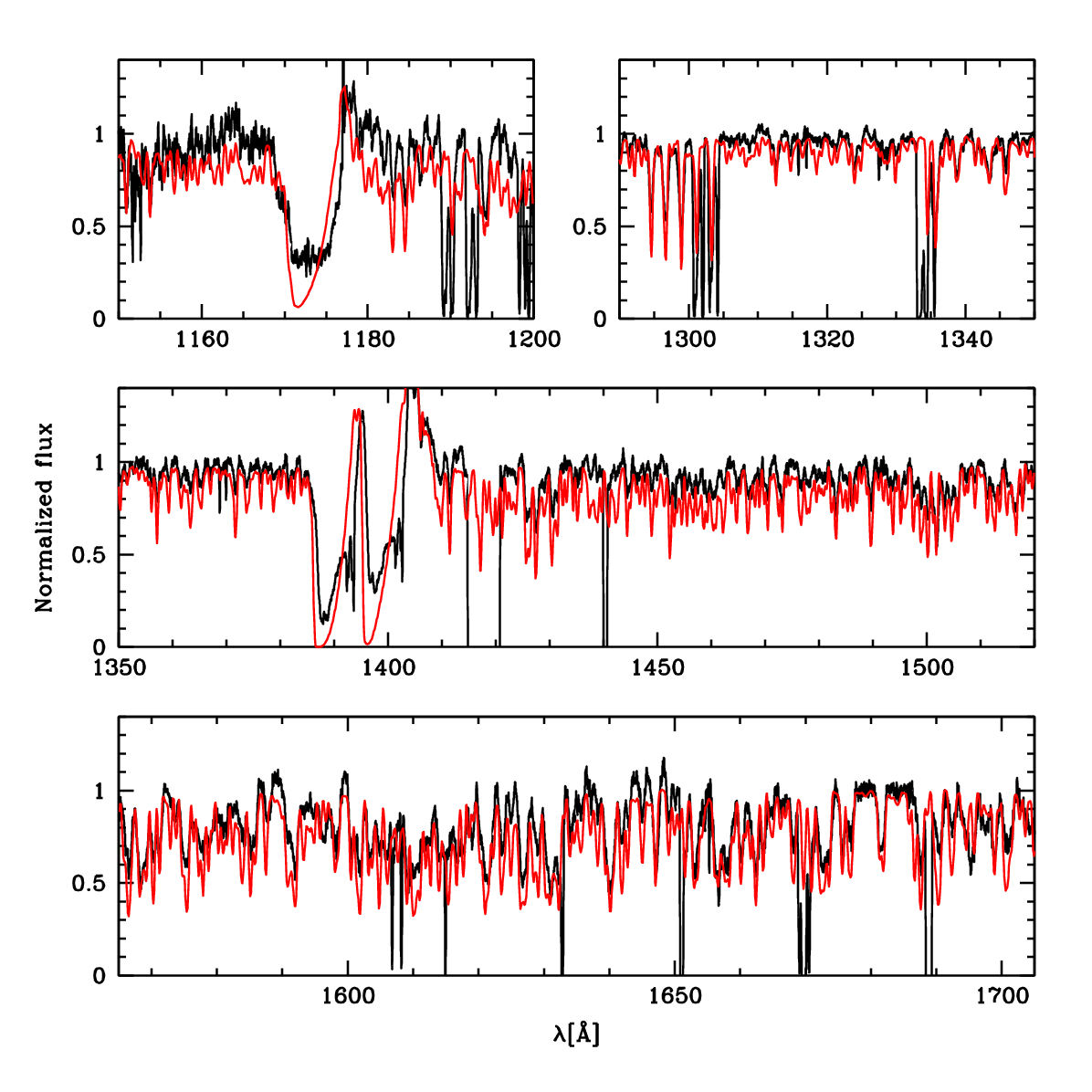}
\includegraphics[width=0.49\textwidth]{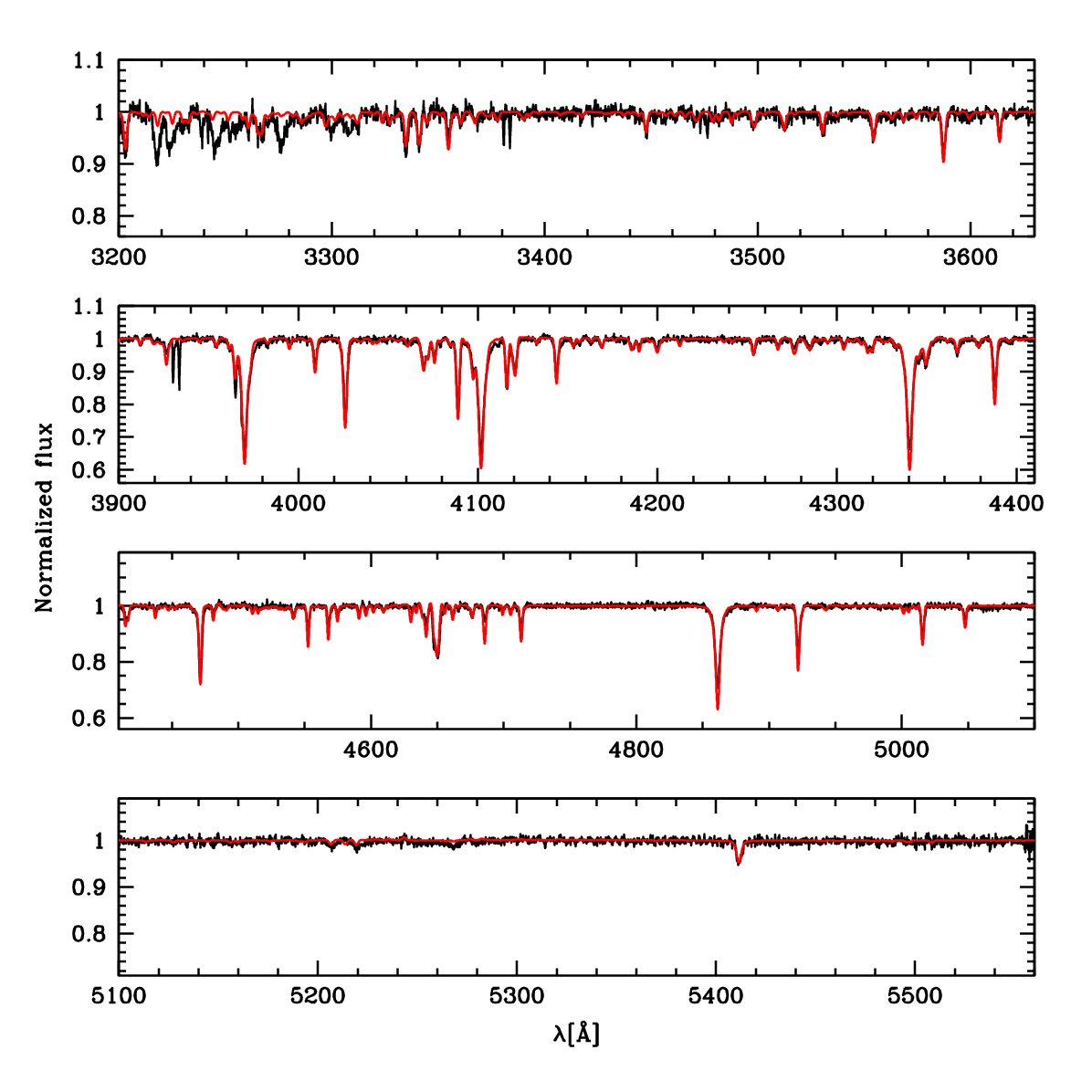}\\
\includegraphics[width=0.75\textwidth]{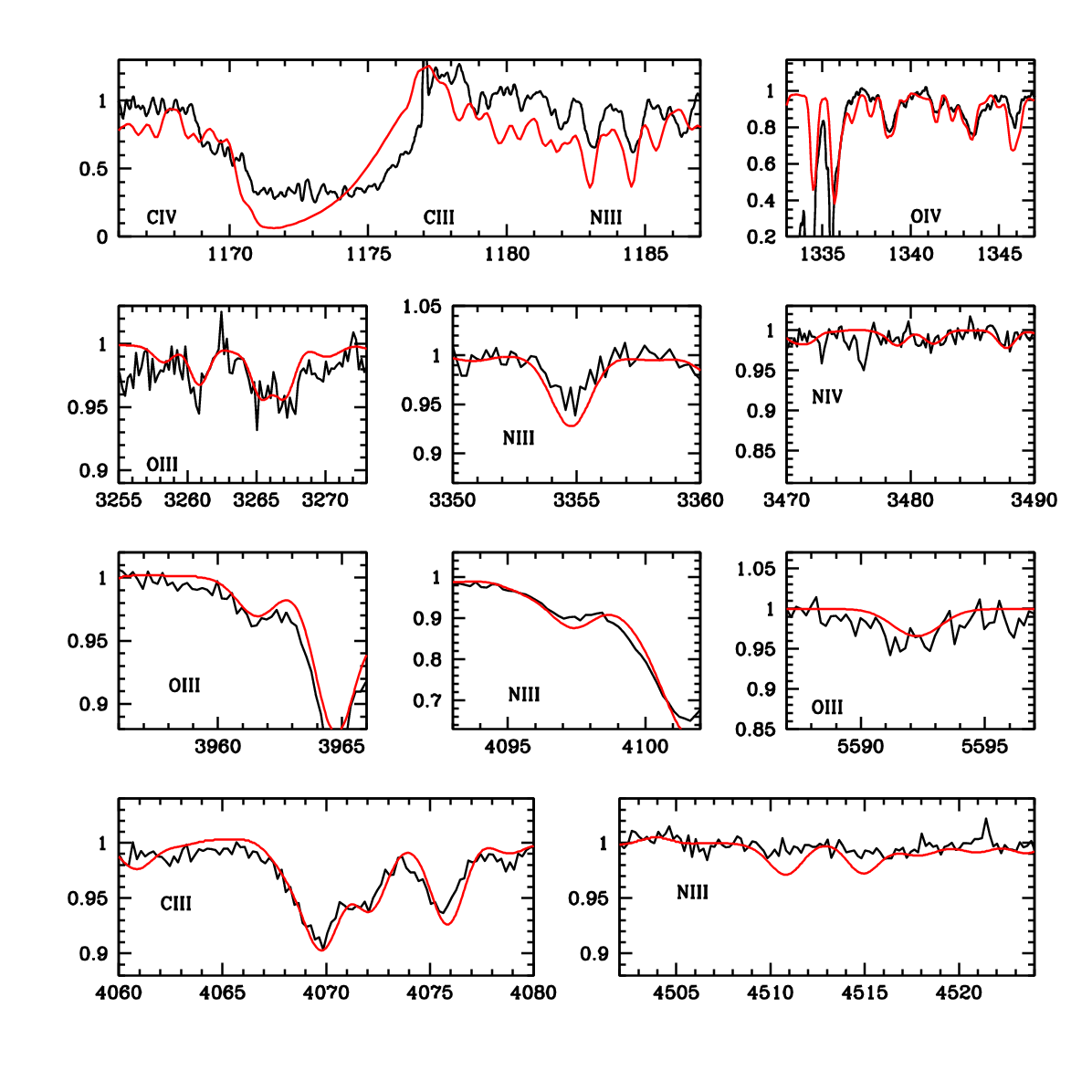}
\caption{Same as Fig.~\ref{fit_av15} but for SK~-67$^{\circ}$ 261.} 
\label{fit_skm67d261}
\end{figure*}

\begin{figure*}[ht]
\centering
\includegraphics[width=0.49\textwidth]{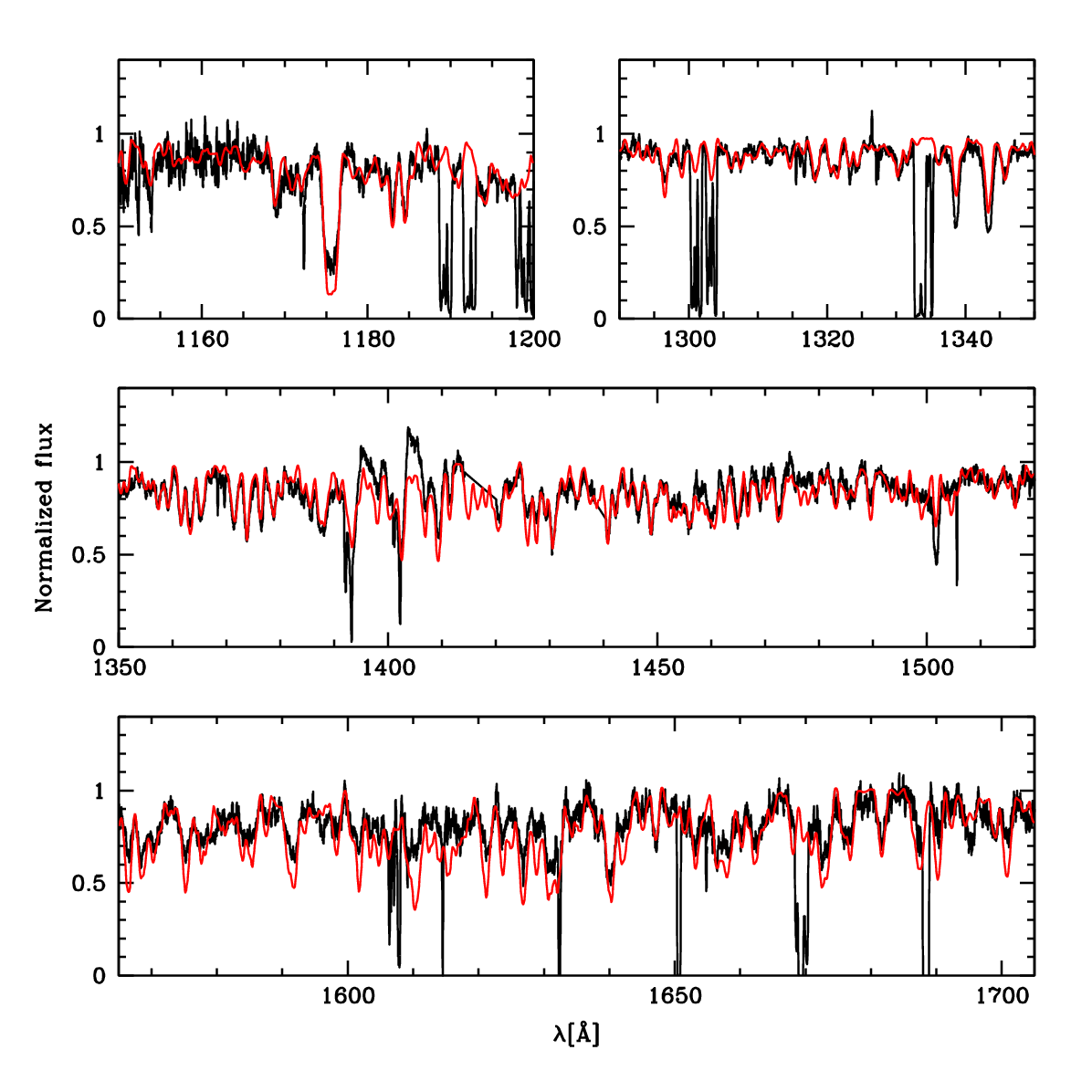}
\includegraphics[width=0.49\textwidth]{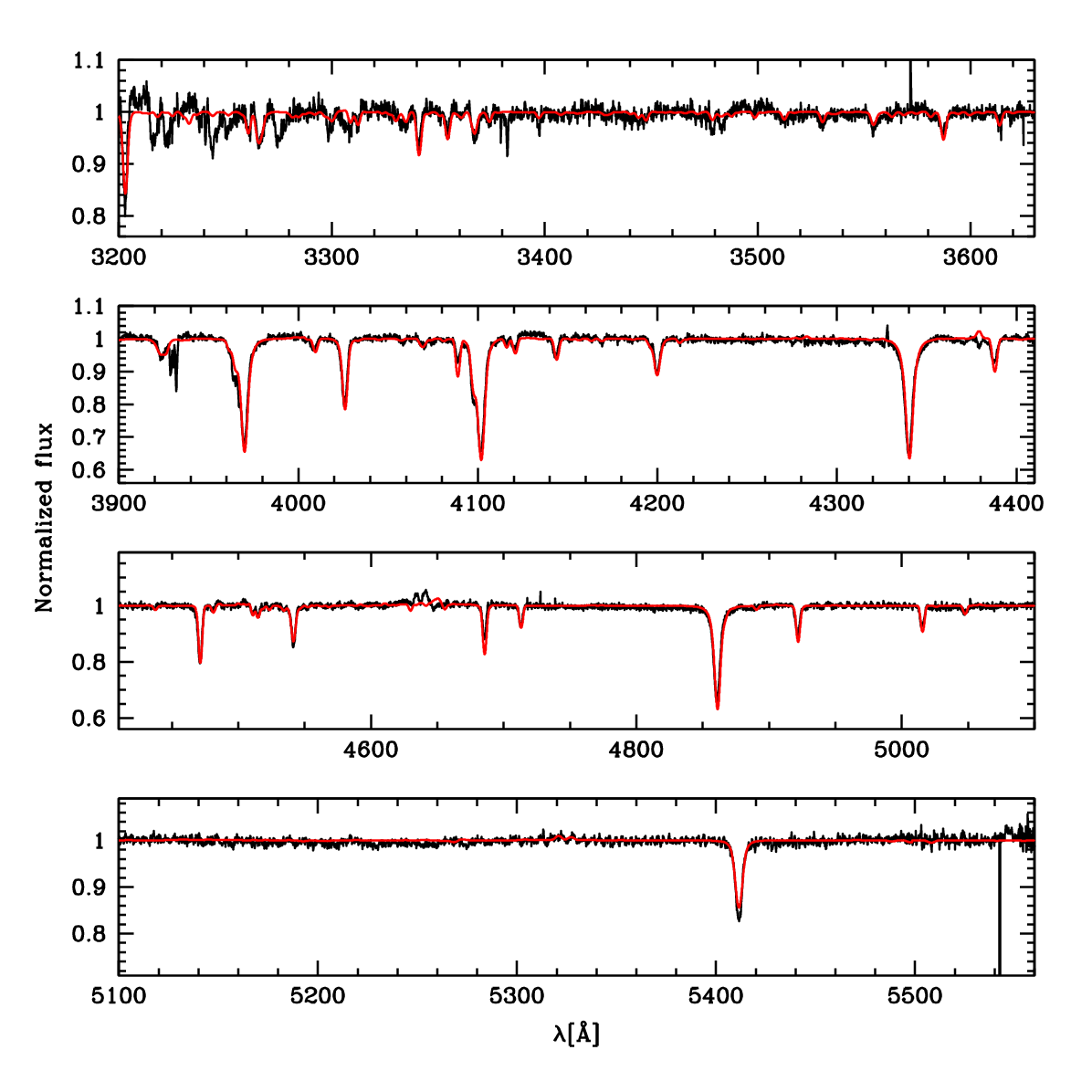}\\
\includegraphics[width=0.75\textwidth]{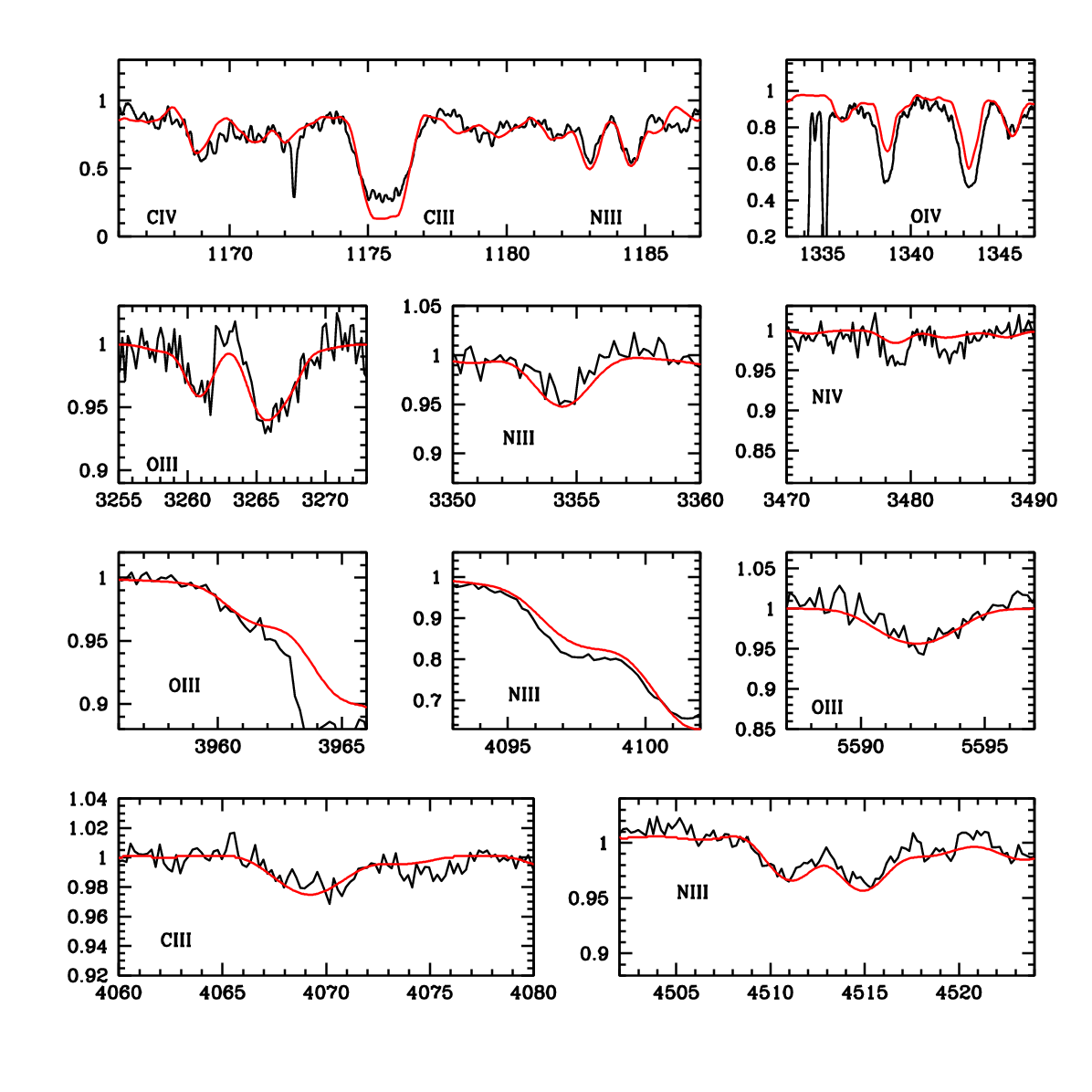}
\caption{Same as Fig.~\ref{fit_av15} but for SK~-67$^{\circ}$ 191.} 
\label{fit_skm67d191}
\end{figure*}

\begin{figure*}[ht]
\centering
\includegraphics[width=0.49\textwidth]{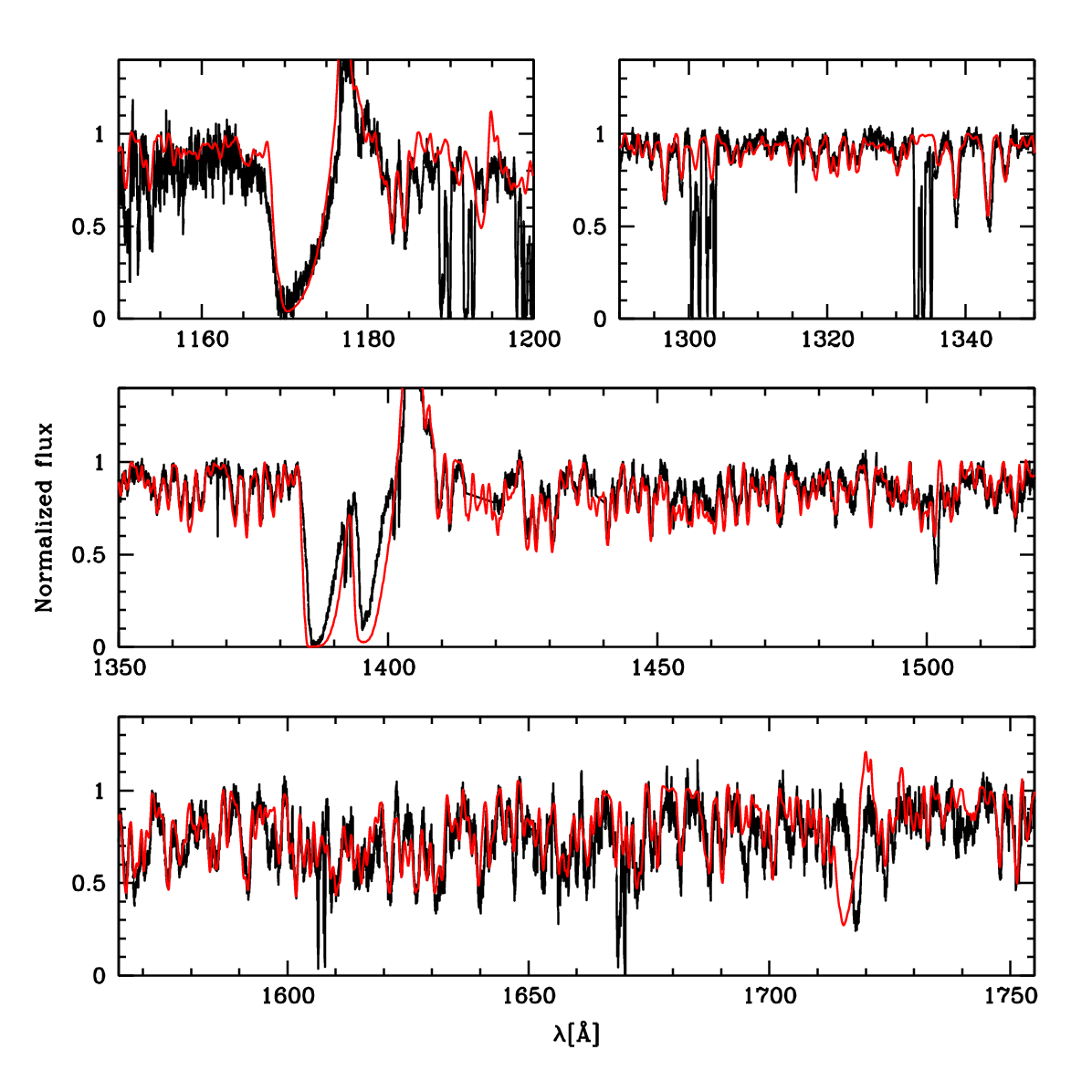}
\includegraphics[width=0.49\textwidth]{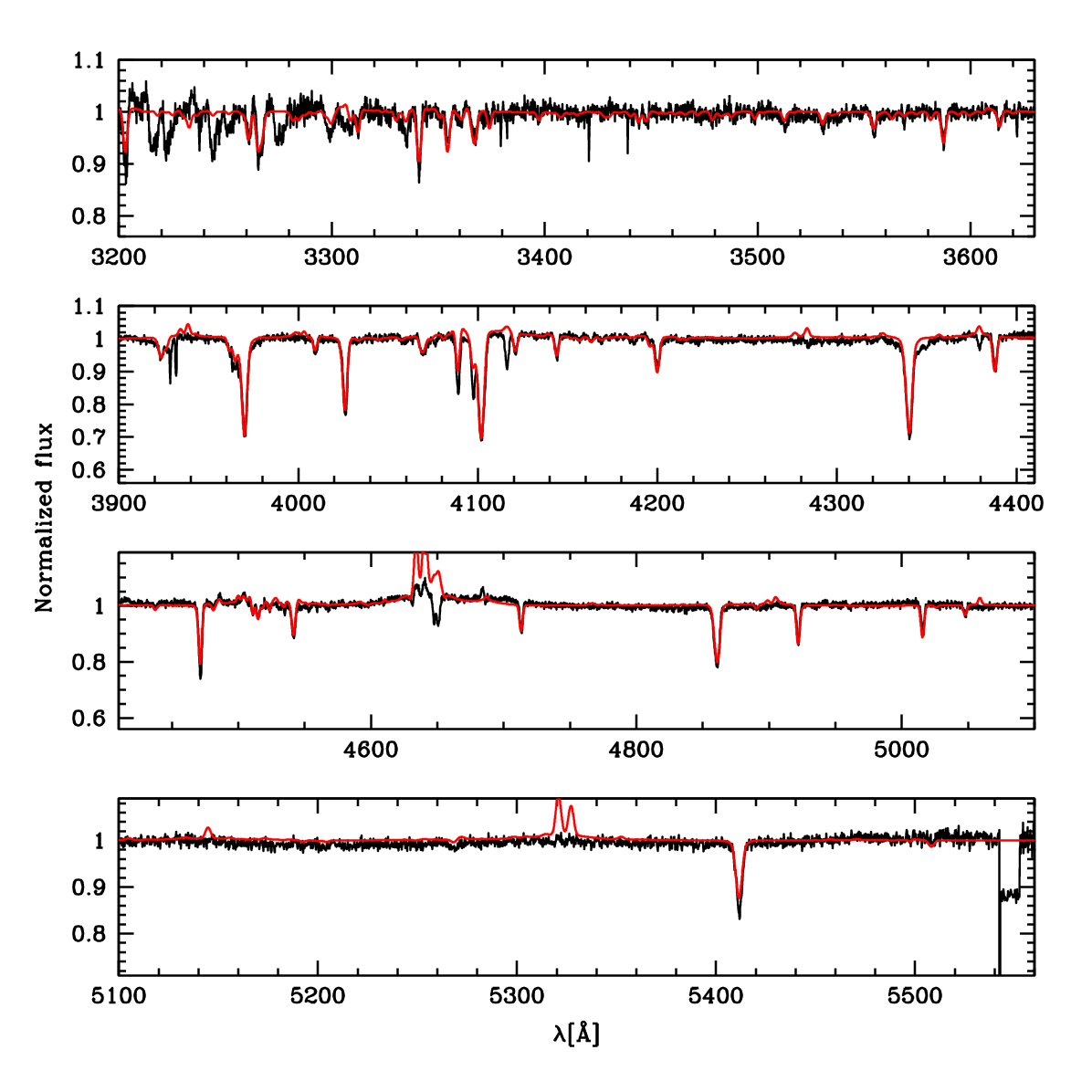}\\
\includegraphics[width=0.75\textwidth]{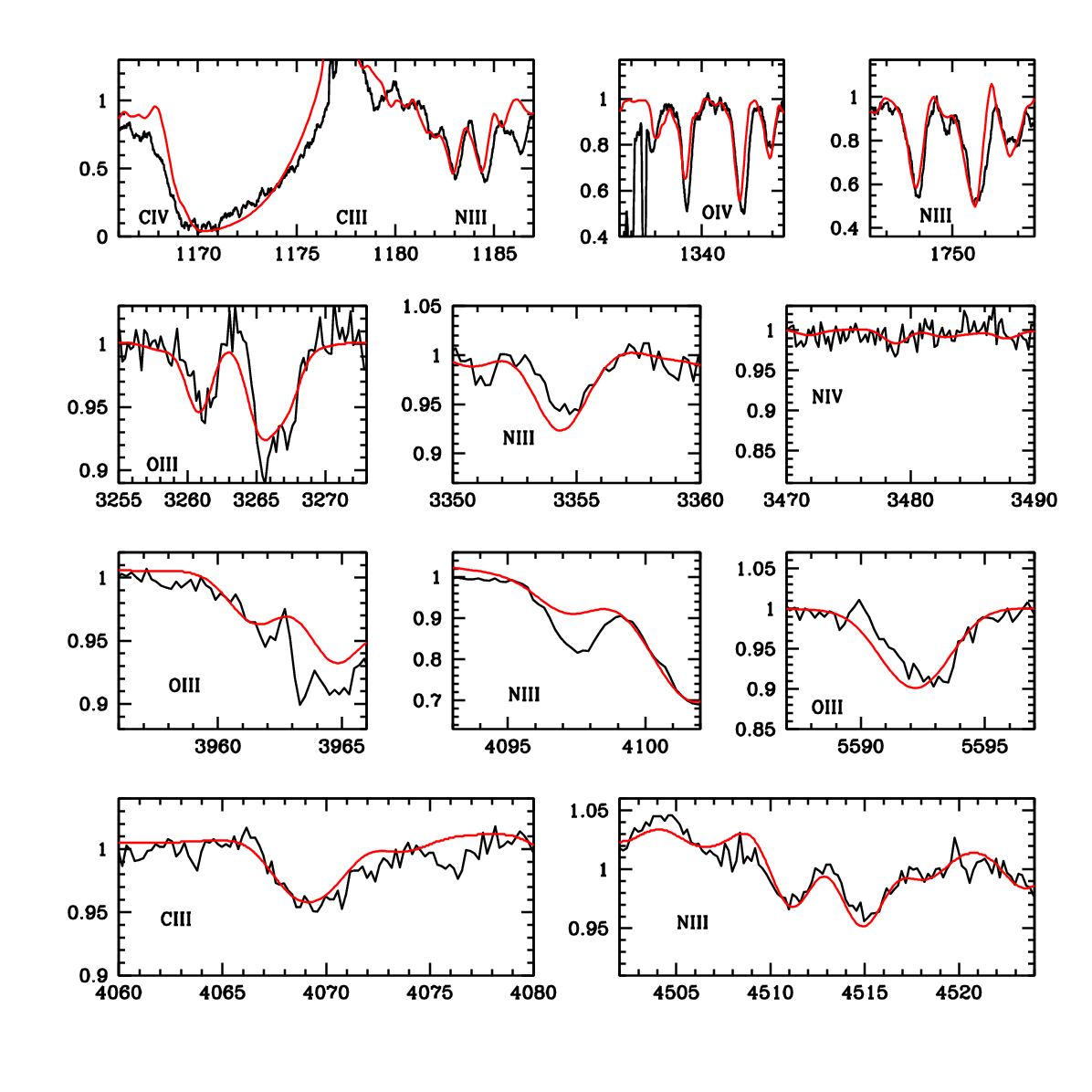}
\caption{Same as Fig.~\ref{fit_av15} but for SK~-66$^{\circ}$ 171.} 
\label{fit_skm66d171}
\end{figure*}

\begin{figure*}[ht]
\centering
\includegraphics[width=0.49\textwidth]{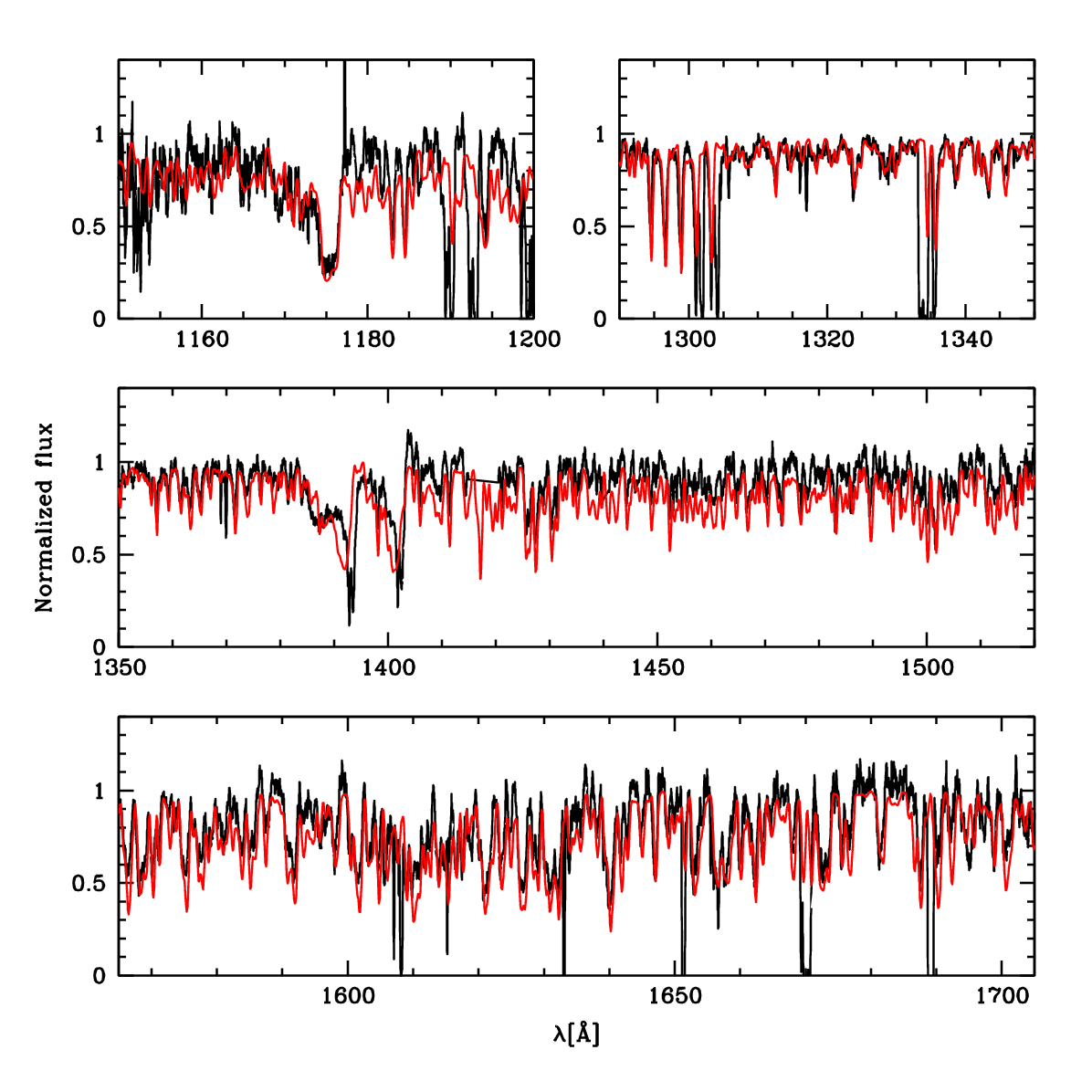}
\includegraphics[width=0.49\textwidth]{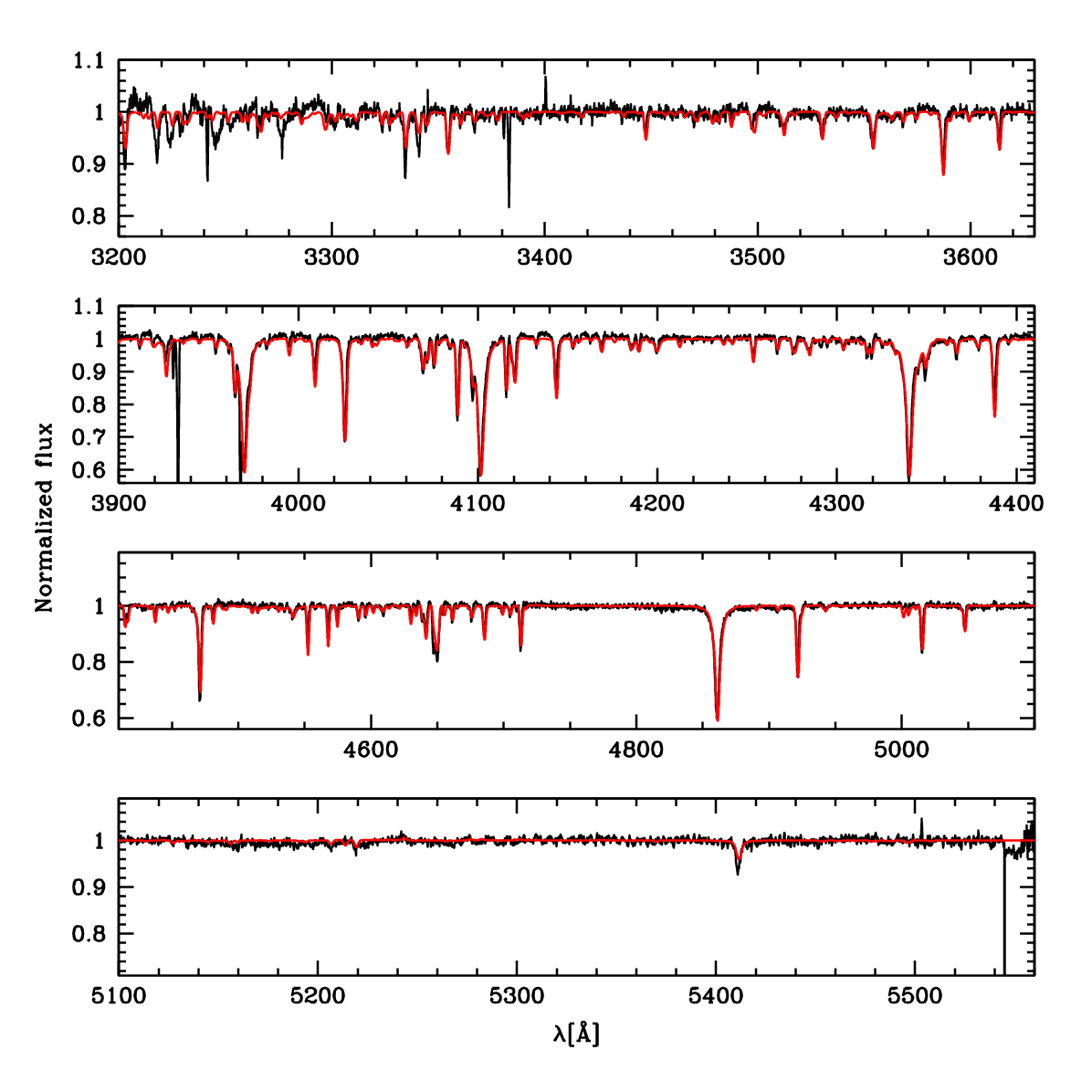}\\
\includegraphics[width=0.75\textwidth]{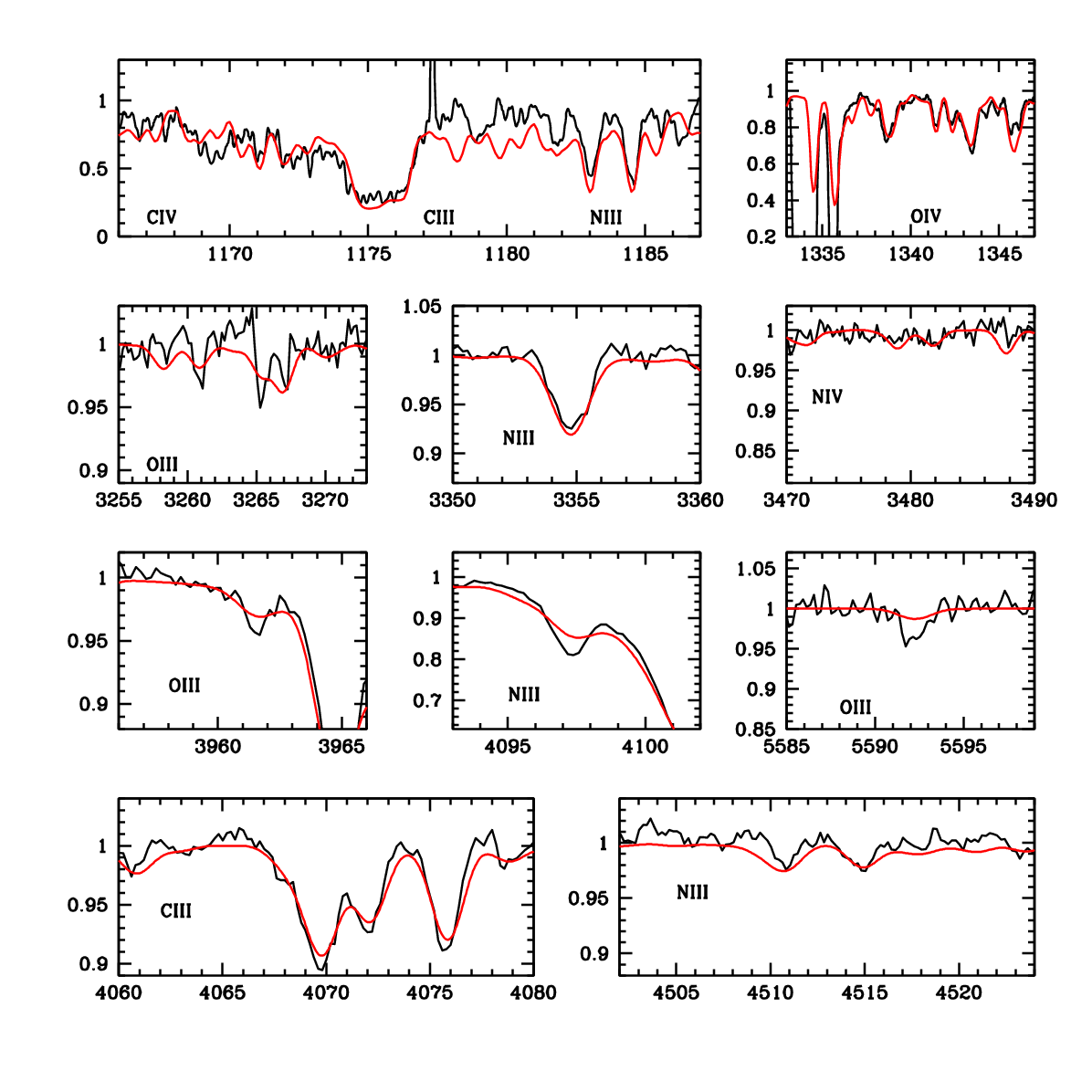}
\caption{Same as Fig.~\ref{fit_av15} but for SK~-71$^{\circ}$ 8.} 
\label{fit_skm71d8}
\end{figure*}

\begin{figure*}[ht]
\centering
\includegraphics[width=0.49\textwidth]{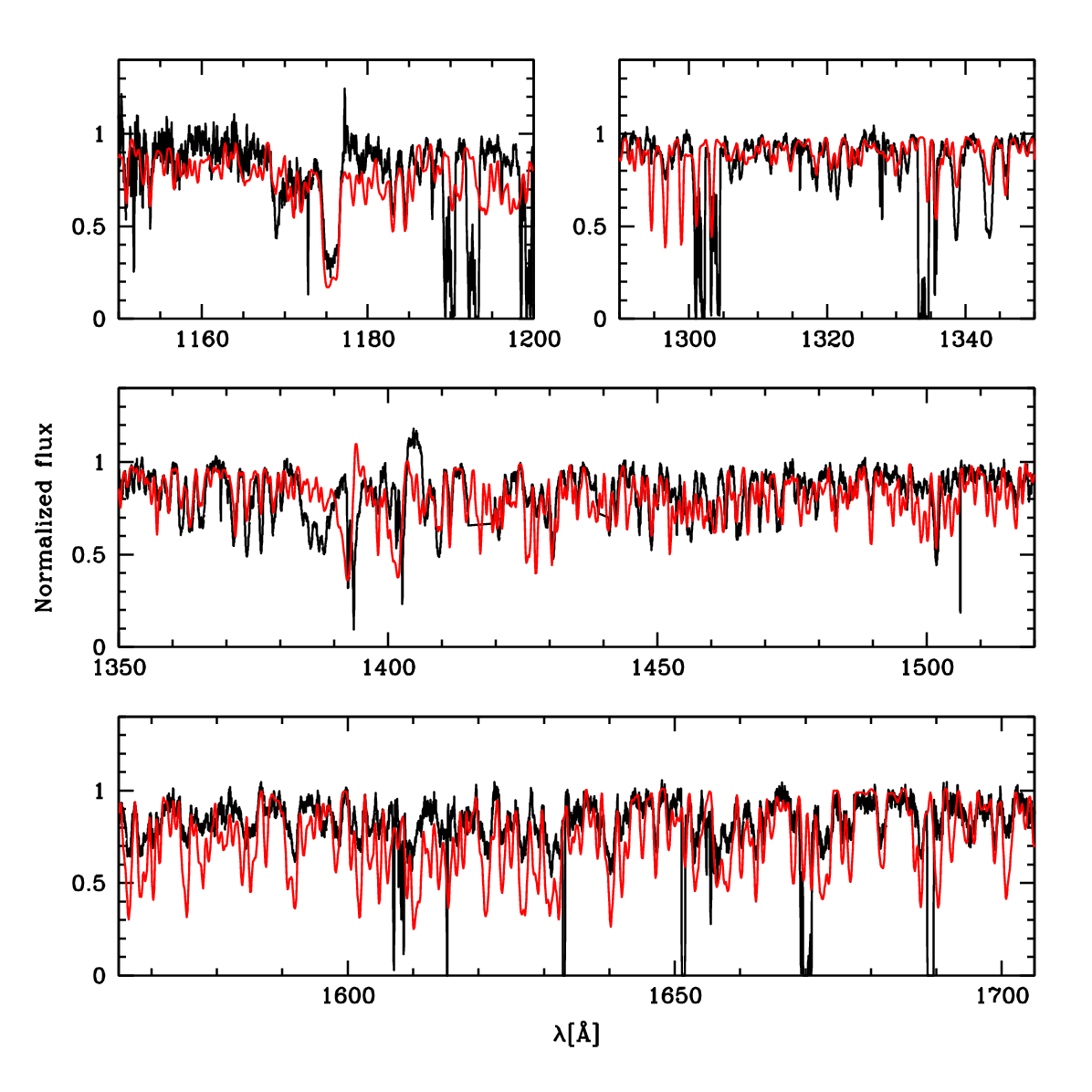}
\includegraphics[width=0.49\textwidth]{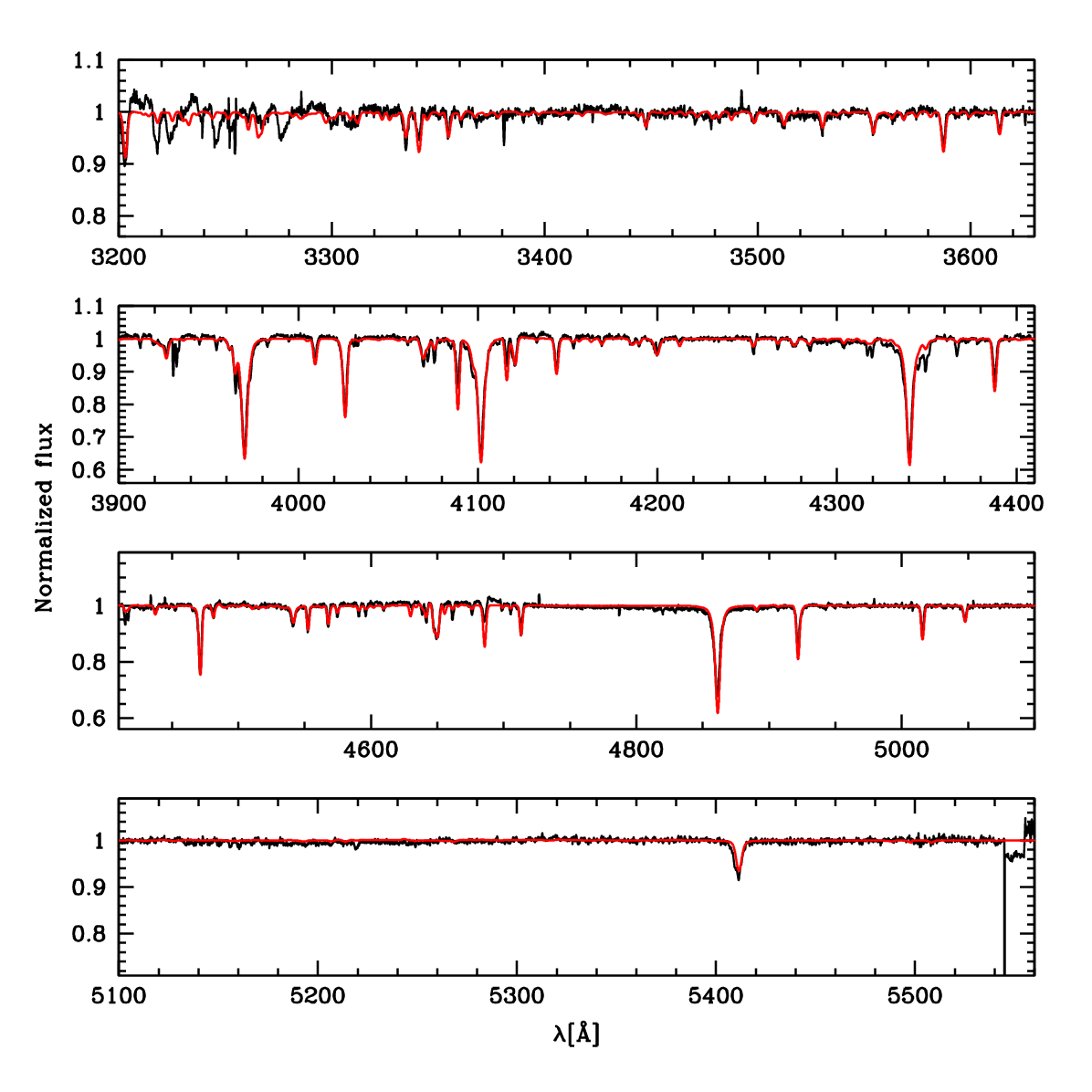}\\
\includegraphics[width=0.75\textwidth]{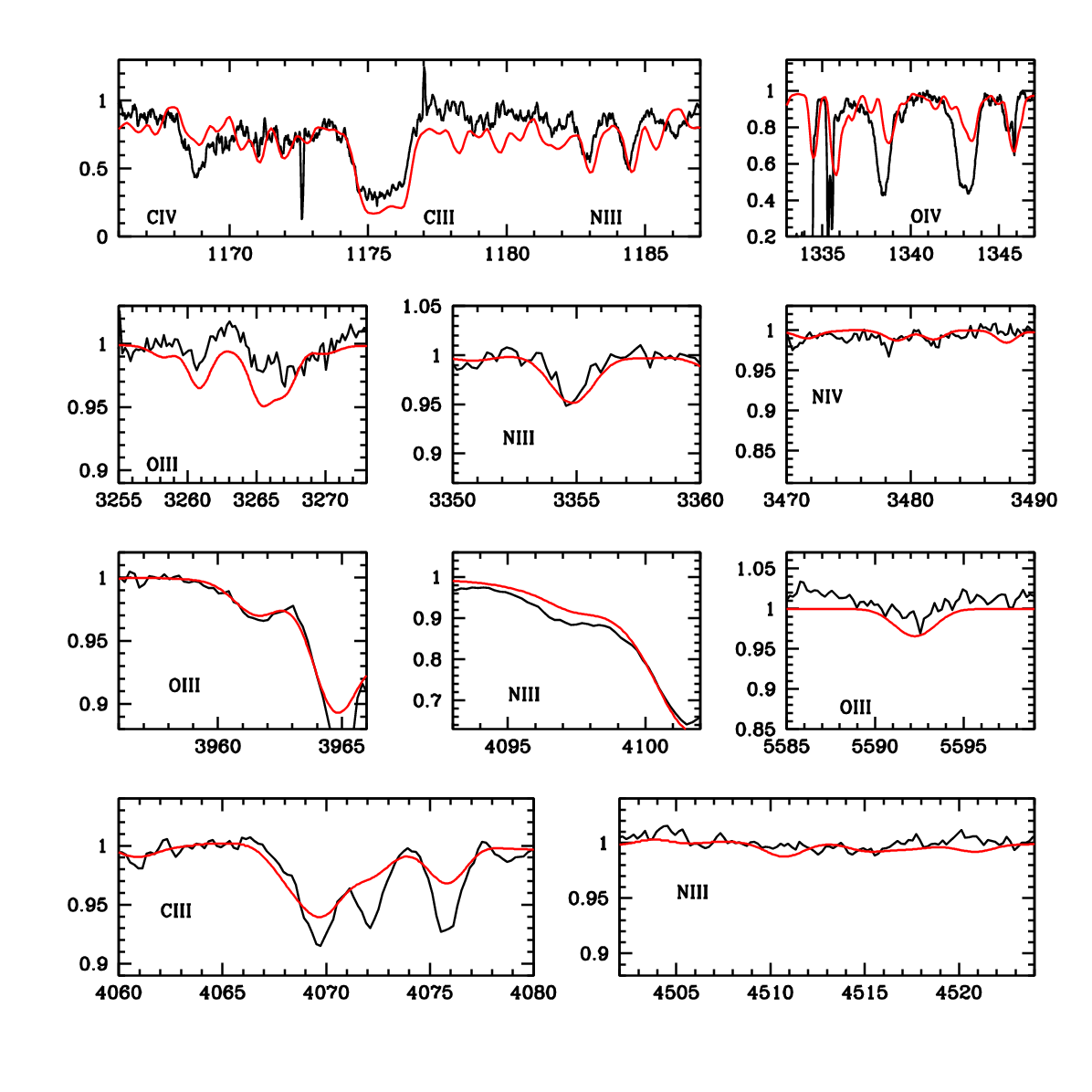}
\caption{Same as Fig.~\ref{fit_av15} but for SK~-70$^{\circ}$ 13.} 
\label{fit_skm70d13}
\end{figure*}

\begin{figure*}[ht]
\centering
\includegraphics[width=0.49\textwidth]{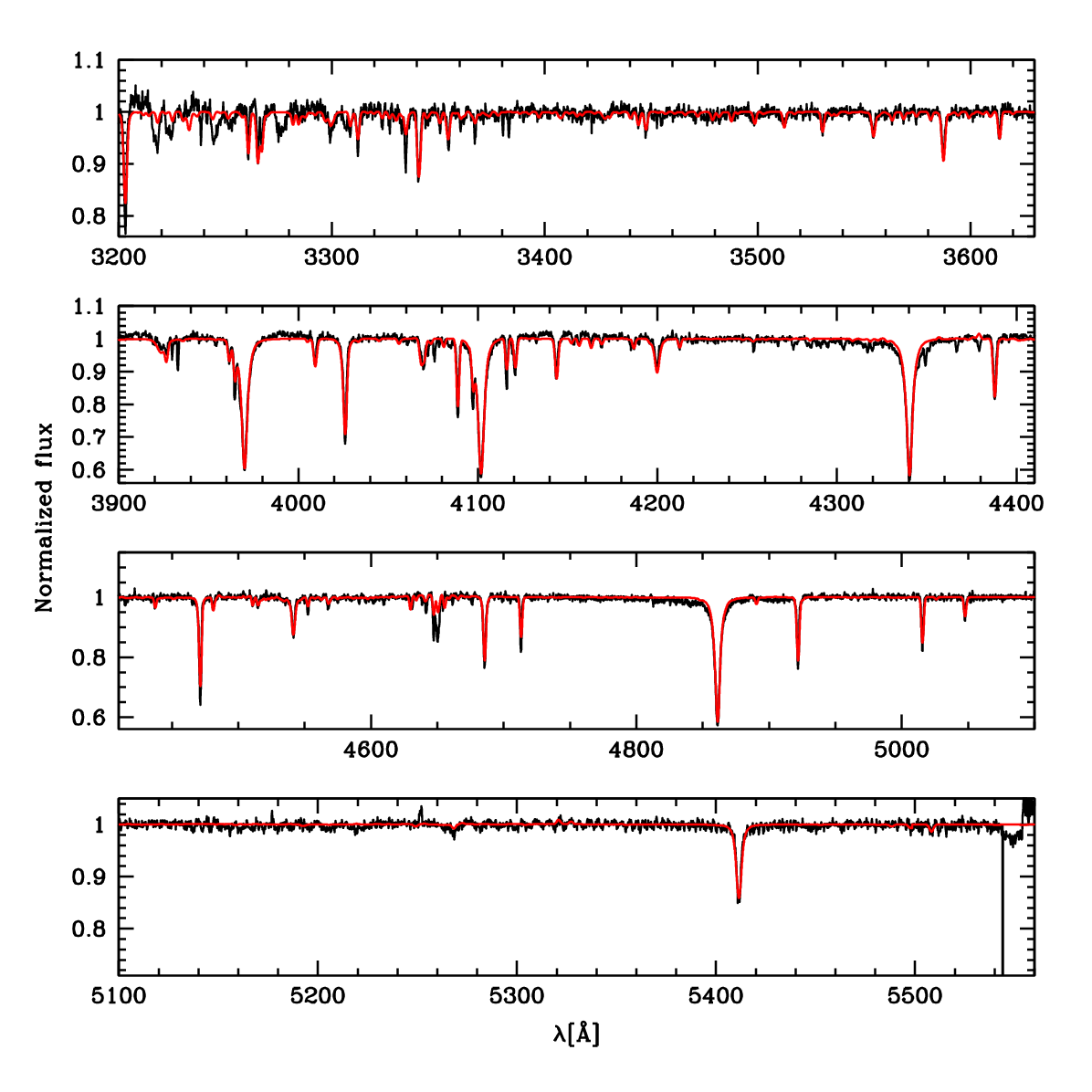}
\includegraphics[width=0.49\textwidth]{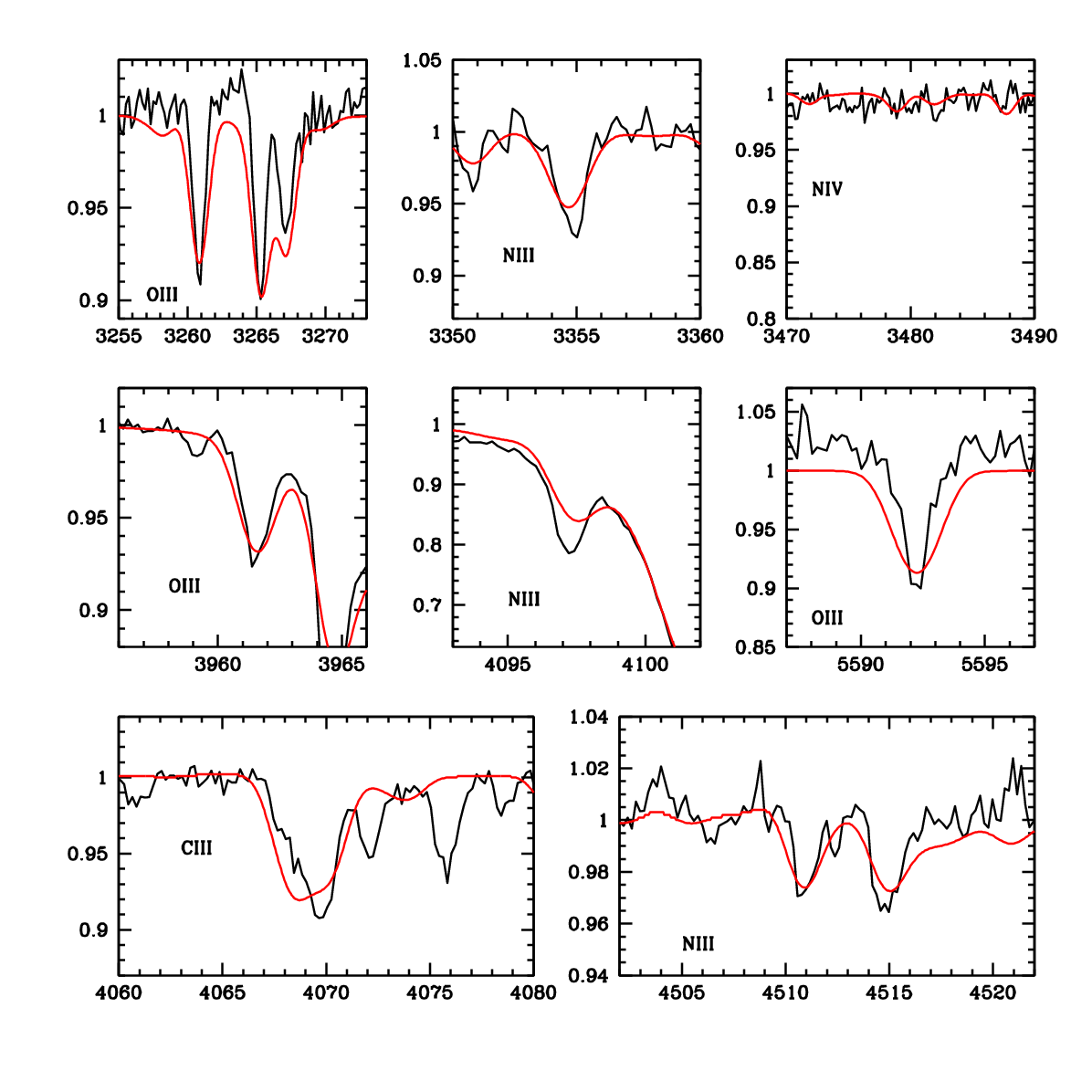}
\caption{Same as Fig.~\ref{fit_av15} but for BI~128.} 
\label{fit_bi128}
\end{figure*}

\end{appendix}

\end{document}